%% file: u2fibrations.tex
\documentclass[a4paper,11pt]{article}
\pdfoutput=1
\usepackage{a4wide}
\usepackage{amsmath,amssymb}
\usepackage{autobreak}
\usepackage{graphics,graphicx}
\usepackage{mathtools}
\usepackage{changebar}
\usepackage{color}
\usepackage{cancel}
\usepackage{blkarray}
\usepackage{multirow}
\usepackage{enumitem}
\usepackage[all,cmtip]{xy}

\usepackage{tikz,textcomp}
\usetikzlibrary{decorations.pathreplacing,calc,fadings,fit}
\usetikzlibrary{shapes,arrows,positioning,chains,matrix}
\usepackage{lscape}

\input{macrosES}

\DeclareFontFamily{U}{wncy}{}
\DeclareFontShape{U}{wncy}{m}{n}{<->wncyr10}{}
\DeclareSymbolFont{mcy}{U}{wncy}{m}{n}
\DeclareMathSymbol{\Sh}{\mathord}{mcy}{"58} 

\newcommand*{\ditto}{---\texttt{"}---}

\numberwithin{equation}{section}

\begin{document}
\thispagestyle{empty}
\begin{flushright}
UWThPh-2021-10
\end{flushright}
\begin{center}
{\LARGE\bf On genus one fibered Calabi-Yau threefolds \\\vskip 0.2cm with 5-sections}
\end{center}
\vspace{8mm}
\begin{center}
  {\large Johanna Knapp\footnote{{\tt johanna.knapp@unimelb.edu.au}}${}^{\dagger}$, Emanuel Scheidegger\footnote{{\tt esche@bicmr.pku.edu.cn}}${}^{\ddag}$
    and Thorsten Schimannek\footnote{{\tt thorsten.schimannek@univie.ac.at}}${}^{\ast}$}
\end{center}
\vspace{3mm}
\begin{center}
${}^{\dagger}$ {\em School of Mathematics and Statistics, University of Melbourne \\\vskip 0.1 cm
  Parkville VIC 3010, Australia}\\\vskip 0.1 cm 
$^\ddag${\em Beijing International Center for Mathematical Research
(BICMR) \\ \vskip 0.1 cm Peking University, 100871 Beijing, China}\\\vskip 0.1 cm 
$^\ast$ {\em Mathematical Physics Group, University of Vienna\\ \vskip 0.1 cm Boltzmanngasse 5, 1090 Vienna, Austria}
\end{center}
\vspace{15mm}
\begin{abstract}
Elliptic and genus one fibered Calabi-Yau spaces play a prominent role in string theory and mathematics.
In this article we discuss a class of genus one fibered Calabi-Yau threefolds with 5-sections from various perspectives.
In algebraic geometry, such Calabi-Yaus can be constructed as complete intersections in Grassmannian fibrations and as Pfaffian varieties.
These constructions naturally fit into the framework of homological
projective duality and lead to dual pairs of Calabi-Yaus.
From a physics perspective, these spaces can be realised as low-energy configurations ("phases") of gauged linear sigma models (GLSMs) with non-Abelian gauge groups, where the dual geometries arise as phases of the same GLSM.
Using the modular bootstrap approach of topological string theory, one can compute all-genus Gopakumar-Vafa invariants of these Calabi-Yaus.
We observe that homological projective duality acts as an element of $\Gamma_0(5)$ on the topological string partition function and the partition functions of dual geometries transform into each other.
Moreover, we study the geometries from an M-/F-theory perspective.
We compute the F-theory spectrum and show how the genus one-fibered Calabi-Yaus are connected to certain Calabi-Yaus in toric varieties via a series of Higgs transitions.
Based on the F-theory physics, we conjecture that dual geometries are elements of the same Tate-Shafarevich group.
Our analysis also leads to a classification of 5-section geometries, as well as the construction of F-theory models with charge 5 hypermultiplets. 
\noindent
\end{abstract}
\newpage
\setcounter{tocdepth}{2}
\tableofcontents
\setcounter{footnote}{0}

\newpage
\input{section_0}
\input{section_1}
\input{section_2}
\input{section_3}

\input{section_4}
\input{section_5}
\input{section_6}

\appendix
\input{appendix_0}
\input{appendix_1}
\input{appendix_2}
\input{appendix_3}

\bibliographystyle{utphys}
\bibliography{bibliography}
\end{document}

%% file: macrosES.tex
\usepackage{latexsym}
\usepackage{amsmath,amssymb,amsxtra,amscd,amsthm}
\usepackage{mathrsfs}
\newcommand{\scr}{\mathscr}
\usepackage{tikz-cd}
\usepackage{multirow}
\makeatletter%
\@ifclassloaded{amsart}%
  {}%
  {\usepackage[nottoc,notlot,notlof]{tocbibind}}%
\makeatother%
\usepackage{epsfig}
\usepackage{pdfsync}
\usepackage{thumbpdf}
\usepackage{color}
\ifx\pdfoutput\undefined
\usepackage
[dvips,
 verbose=false,
 bookmarks=true,
 colorlinks=true,
 linkcolor=webred,
 filecolor=webbrown,
 citecolor=webgreen,
 pagecolor=webblue,
 urlcolor=webblue,
 pdftitle={},
 pdfauthor={},
 pdfsubject={},
 pdfkeywords={},
 bookmarksopen=false,
 pdfpagemode=None,
 pdfview=FitH,
 pdfstartview=FitH,
 extension=pdf]{hyperref}
\else
\usepackage
[pdftex,
 verbose=false,
 bookmarks=true,
 colorlinks=true,
 linkcolor=webred,
 filecolor=webbrown,
 citecolor=webgreen,
 pagecolor=webblue,
 urlcolor=webblue,
 pdftitle={},
 pdfauthor={},
 pdfsubject={},
 pdfkeywords={},
 bookmarksopen=false,
 pdfpagemode=None,
 pdfview=FitH,
 pdfstartview=FitH,
 extension=pdf]{hyperref}
\fi
\definecolor{webred}{rgb}{.8,0,0}
\definecolor{webbrown}{rgb}{.6,0,0}
\definecolor{webgreen}{rgb}{0,0.5,0}
\definecolor{webdkgreen}{rgb}{0,0.3,0}
\definecolor{webblue}{rgb}{0,0,0.5}

\usepackage{graphics,graphicx}
\usepackage{a4wide}

\def\rE{\scr{E}}
\def\rF{\scr{F}}

\def\rI{\scr{I}}

\def\rO{\scr{O}}

\def\rV{\scr{V}}
\def\rW{\scr{W}}

\def\fH{\mathfrak{H}}

\def\fa{\mathfrak{a}}
\def\fb{\mathfrak{b}}

\def\fl{\mathfrak{l}}

\def\fs{\mathfrak{s}}

\def\mC{\mathbb{C}}

\def\mF{\mathbb{F}}

\def\mP{\mathbb{P}}
\def\mQ{\mathbb{Q}}
\def\mR{\mathbb{R}}
\def\mS{\mathbb{S}}

\def\mZ{\mathbb{Z}}

\def\tD{\mathrm{D}}

\def\tG{\mathrm{G}}
\def\tH{\mathrm{H}}

\def\tK{\mathrm{K}}
\def\tL{\mathrm{L}}

\def\tR{\mathrm{R}}
\def\tS{\mathrm{S}}

\def\tb{\mathrm{b}}

\def\th{\mathrm{h}}

\def\pG{\mathsf{G}}

\def\pP{\mathsf{P}}

\DeclareMathOperator{\chern}{ch}
\DeclareMathOperator{\ch}{c}
\DeclareMathOperator{\todd}{td}
\DeclareMathOperator{\p}{p}

\DeclareMathOperator{\rank}{rk}

\DeclareMathOperator{\Aut}{Aut}

\DeclareMathOperator{\codim}{codim}

\DeclareMathOperator{\Sym}{Sym}

\DeclareMathOperator{\Pic}{Pic}

\DeclareMathOperator{\coker}{coker}

\DeclareMathOperator{\Gr}{Gr}

\DeclareMathOperator{\Pf}{Pf}

\DeclareMathOperator{\Tot}{Tot}
\DeclareMathOperator{\Coh}{Coh}

\DeclareMathOperator{\GL}{GL}
\DeclareMathOperator{\SL}{SL}

\renewcommand{\sl}{\DeclareMathOperator{\sl}{\fs\fl}}

\def\id{\mathrm{id}} 
\def\Top{\mathrm{top}} 
 
\renewcommand{\mod}{\mathop{\mathrm{mod}}}

\def\bb1{\textup{\small{1}} \kern-3.8pt \textup{1}}
\newcommand{\diff}[2]{\textrm{d}^{#1}{#2}}

\newcommand{\n}[2]{n^{(#1)}_{#2}}

\def\PSL2Z{\mP\tS\tL(2,\mZ)}
\def\abcd{\left(\substack{ a\, b\\c\, d}\right)}

\newcommand{\legendre}[2]{\genfrac{(}{)}{}{}{#1}{#2}}

\numberwithin{equation}{section}

\providecommand{\href}[2]{#2}

\allowdisplaybreaks


%% file: section_0.tex
\section{Introduction}
Genus one fibered Calabi-Yaus play a pivotal role in string theory, most notably in F-theory, topological string theory, and various types of string dualities.
If a genus one fibration exhibits a section it is called an elliptic fibration. It is then birational to a fibration of Weierstra{\ss} curves.
On the other hand, if the fibration does not exhibit a section it will still contain a divisor that intersects the generic fiber $N$ times for some $N\in\mathbb{N}$.
Such a divisor is called an $N$-section and if $N$ is the minimal value for which such a divisor exists we refer to the geometry as a genus one fibration with $N$-sections.

The main goal of this article is a construction and in-depth analysis of genus one-fibered Calabi-Yaus with $5$-sections. While torus-fibered Calabi-Yaus with $N$-sections where $N\leq4$ have been studied in the literature before, going to $N=5$ is a conceptually non-trivial step. The reason is that, in contrast to the case $N\leq 4$~\cite{Braun:2014oya,Klevers:2014bqa,Braun:2014qka}, such Calabi-Yaus cannot be constructed by means of toric geometry.

Leaving the realm of toric geometry not only means that certain standard mathematical tools for working with Calabi-Yaus do not apply, but also that Calabi-Yaus of this type will not be part of the vast list of known examples of genus one fibered Calabi-Yau threefolds. These new Calabi-Yaus come with a multitude of new features that one can study from various angles, ranging from algebraic geometry to supersymmetric gauge theory to topological string theory and M/F-theory. The sections of this paper will take up all of these perspectives.
As a byproduct of our analysis we will also construct the first examples of elliptically fibered Calabi-Yau manifolds that realize charge $5$ matter in F-theory.

We will start off in Section~\ref{sec:geometry} by giving constructions of genus one fibered Calabi-Yaus with $5$-sections in algebraic geometry.
To construct such Calabi-Yaus, we consider fibrations of Pfaffian curves and complete intersections in Grassmannians over some base manifold.
For simplicity we will often consider threefolds and choose the base to be $\mathbb{P}^2$.
Since we cannot rely on toric geometry, it is a non-trivial task to determine the topological characteristics of these Calabi-Yaus.
The necessary methods exist and are textbook material in algebraic geometry, but they are not widely used in a string theory context.
Therefore we will spend some time outlining how to compute the Hodge numbers and the intersection rings for this more general class of examples.
While our focus is more on discussing prototypical examples, the construction is fairly general and could potentially lead to a large number of new Calabi-Yaus.
One problem that remains in the geometric construction is to check whether a given choice of bundles leads to a smooth genus one fibration.
We will circumvent this problem later in Section~\ref{sec:classification}, by using Higgs transitions in F-theory to obtain the data of 23 smooth genus one fibered Calabi-Yau threefolds with $5$-sections over $\mathbb{P}^2$.

From a physics perspective, one can obtain Calabi-Yaus as the vacuum configurations (``phases'') of certain supersymmetric gauge theories in two dimensions -- the gauged linear sigma models (GLSMs)~\cite{Witten:1993yc}. The Calabi-Yaus we are considering arise as phases of GLSMs with non-Abelian gauge groups.
Section~\ref{sec:glsm} is concerned with this point of view. We show that the examples constructed in Section~\ref{sec:geometry} indeed can be realised as phases of non-Abelian GLSMs. In particular, we find that certain examples correspond to different phases of the same GLSM, i.e.~ that they arise at different limiting points of the same stringy K\"ahler moduli space.
This leads to the expectation that the Calabi-Yaus are related by (relative) homological projective duality and that the associated D-brane categories are equivalent.

Furthermore, the GLSM provides tools to extract further information about the Calabi-Yaus. The non-Abelian duality discovered in \cite{Hori:2011pd} allows us to to analyze the strongly coupled phases of the GLSM via the weakly coupled duals. This simplifies the identification of those geometries that are determinantal varieties. Moreover, we can obtain the discriminant locus by analysing the Coulomb and mixed Coulomb/Higgs branches of GLSMs. Finally we make use of the sphere partition function of the GLSM \cite{Benini:2012ui,Doroud:2012xw,Jockers:2012dk} to extract the periods and Picard-Fuchs systems associated to the Calabi-Yaus.

Section~\ref{sec:gvandmod} is concerned with the enumerative invariants and the modular properties of the topological string partition function associated to the Calabi-Yaus.
We generalise the modular bootstrap approach pioneered in~\cite{Huang:2015sta,Cota:2019cjx} to genus one fibrations with general $N$-sections and use this to compute higher-genus Gopakumar-Vafa invariants of our $5$-section geometries. As expected, the modular properties of the topological string partition functions are governed by the modular forms for $\Gamma_1(5)$.
However, we observe a highly non-trivial relation between the topological string partition functions of Calabi-Yaus that appear as phases of the same non-Abelian GLSM.
It turns out that the partition function of one geometry can always be obtained by acting with a $\Gamma_0(5)$-transformation on the modular parameter in the partition function of the other.
By performing a numerical analytic continuation of the periods of the generic fiber, we show that such a transformation appears as part of the transfer matrix that transports the brane charges between the two associated large volume limits in the moduli space.

In Section~\ref{sec:ftheory} we discuss our Calabi-Yaus from the viewpoint of M- and F-theory. By studying extremal transitions, using the physics of F-theory, we show how our genus one fibrations fit into the large web of Calabi-Yaus that is connected via extremal transitions.
To this end we first consider a toric degeneration of the Grassmannian complete intersection curve~\cite{sturmfels1996grobner,Batyrev:1998kx}.
Resolving the singularities leads to a family of elliptic curves that is realized as a codimension five complete intersection in a toric ambient space.
These can be used as fibers to construct a large class of elliptically fibered Calabi-Yau $d$-folds that via F-theory lead to effective supergravities with gauge group $U(1)\times U(1)$.
We apply a recently developed technique based on the so-called fiber Gopakumar-Vafa invariants~\cite{Paul-KonstantinOehlmann:2019jgr} to determine the spectrum of matter representations and show
that the resulting theories allow a Higgs transition
\begin{align}
	U(1)\times U(1)\rightarrow U(1)\rightarrow \mathbb{Z}_5\,.
	\label{eqn:2stephiggs}
\end{align}
The $\mathbb{Z}_5$-theories are realized by F-theory on genus one fibered Calabi-Yau manifolds with $5$-sections and geometrically the Higgs transition corresponds to an extremal transition that connects the associated elliptic and genus one fibrations.
Using this transition we also obtain base independent expressions for the multiplicty of matter representations in F-theory compactifications on generic genus one fibered Calabi-Yau manifolds with $5$-sections.

After performing the first step of the Higgs transition~\eqref{eqn:2stephiggs} one obtains a theory with gauge group $U(1)$ and matter with charges $q=1,\dots,5$. 
This transition is already non-toric and does not provide information on the intermediate geometry.
To remedy this situation, we engineer a family of codimension 3 complete intersections in a toric ambient space that realize F-theory vacua with gauge group $U(1)$ and charge $5$ matter.
While charge $5$ matter in Type IIB compactifications has been discussed in~\cite{Cianci:2018vwv}, this is, to our knowledge, the first construction of such theories in F-theory.
Using the base independent expressions for the matter representations associated to the elliptic and genus one fibrations we show that, in the case of Calabi-Yau threefolds, all of the anomalies of the associated 6d supergravities cancel.

Based on our knowledge about the derived equivalence and the massless matter spectra we then discuss the Tate-Shafarevich group associated to the genus one fibrations.
We argue that the two fibrations that arise as different geometric limits in the same moduli space lead to equivalent F-theory vacua.
Moreover, we find strong evidence that when equipped with actions of the Jacobian fibrations, they realize all four of the non-trivial elements in the $\mathbb{Z}_5$ Tate-Shafarevich group.

Both the codimension 5 and codimension 3 elliptic fibers are then used in Section~\ref{sec:classification} to obtain a classification of genus one fibrations over $\mathbb{P}^2$.
The idea is to determine all fibrations of the complete intersection fibers over $\mathbb{P}^2$ that do not exhibit unresolved singularities and that at the same time produce matter spectra
that can be used to construct a D-flat direction of the superpotential and break the gauge group to $\mathbb{Z}_5$.
In this way we obtain a list of topological invariants, enumerative invariants and fundamental periods for 23 genus one fibered Calabi-Yau threefolds over $\mathbb{P}^2$.
For every example we can explicitly construct the geometry using the techniques from Section~\ref{sec:geometry} as well as a GLSM that reproduces the corresponding data upon localization.
In Appendix~\ref{sec:estrings} we perform a similar systematic construction of fibrations over the Hirzebruch surface $\mathbb{F}_1$ and calculate the corresponding elliptic genera of E-strings, that arise in the F-theory compactification.
It turns out that they can be expressed in terms of the elliptic genus of the ordinary E-string with special values for the mass parameters.
An analogous observation in the context of genus one fibrations with $N$-sections for $N=1,\ldots,4$ has been made in~\cite{Cota:2019cjx}.\\\\
{\bf Acknowledgments:} We would like to thank David Erkinger, Cesar Fierro Cota, Albrecht Klemm, Tianle Liu, Paul Oehlmann and Eric Sharpe for discussions and collaborations on related projects. We also thank Daniel Kl\"awer for informing us about his upcoming work~\cite{Klaewer:2021ab}. E.S. wishes to thank the University of Melbourne for having him as a guest for the entire year of 2020 during which most of the work for this project has been done.
The work of T.S. is supported by the Austrian Science Fund (FWF):P30904-N27.


%% file: section_1.tex
\section{Geometry}
\label{sec:geometry}

\subsection{Normal forms of genus one curves}
\label{sec:normalcurves}
Genus one fibrations with $N$--sections can be constructed using the normal forms of degree $N$ curves.
It is well known that every elliptic curve can be mapped into the Weierstrass form~\footnote{See e.g. lecture 26 of~\cite{arithmeticMIT} for a nice review.}
\begin{align}
y^2=x^3+fxz^4+gz^6\,,
\end{align}
with $[x:y:z]$ being homogeneous coordinates on the weighted projective space $\mathbb{P}_{231}$.
This can also be applied to the fibers of elliptic fibrations, which are birational to the associated fibrations of Weierstrass models.
More generally, a genus one fibration with 2-sections can be mapped into a fibration of degree $4$ hypersurfaces in $\mathbb{P}_{112}$,
a fibration with 3-sections can be mapped into a fibration of cubic hypersurfaces in $\mathbb{P}^2$ and if the fibration has 4-sections then one can realize the generic fibers as complete intersections
of two quadrics in $\mathbb{P}^3$~\cite{Braun:2014oya}.
Note that in all of these cases the normal form of the fiber is a complete intersection in a toric ambient space.

Let us briefly outline the algorithm that can be used to obtain the corresponding normal form at the example of a genus fibration with three sections.
The $3$-section $S$ induces a divisor of degree $3$ on a generic fiber $C$ and the associated line bundle on the fiber admits three sections which we denote by
\begin{align}
	\{x,y,z\}\in\tH^0(C,\mathcal{O}_C(S))\,.
\end{align}
These sections embed the fiber into $\mathbb{P}(\tH^0(C,\mathcal{O}_C(S))=\mathbb{P}^2$.
To realize the image of this embedding as a hypersurface, note that by
Riemann--Roch, the number of sections of a line bundle on a genus one curve is equal to the degree of the bundle.
The line bundle $\mathcal{O}_C(3S)$ thus admits nine independent global sections but there are $10$ monomials of degree three in $x,y,z$.
The corresponding relation is a cubic equation that realizes the fiber as a hypersurface in $\mathbb{P}^2$.
Performing this procedure for every fiber leads to a fibration of cubic curves that is birationally equivalent to the original genus one fibration.

A similar procedure can also be applied to genus one curves of degree $5$ and thus to the fibers of a genus one fibrations with $5$-sections.
The sections of the corresponding degree $5$ line bundle embed the curve into $\mathbb{P}^4$ but the image of this embedding is neither a hypersurface nor a complete intersection.
To see this, note that the second power of the bundle has $10$ sections but there are $15$ quadratic monomials in the five homogeneous coordinates.
The curve is therefore the vanishing locus of five quadratic polynomials in $\mathbb{P}^4$.
One can show that these polynomials are the $4\times 4$ Pfaffians of a $5\times 5$ skew-symmetric matrix with entries that are linear in the homogeneous coordinates~\cite{fisher1}.
The normal form for genus one curves of degree $5$ is therefore a Pfaffian curve in $\mathbb{P}^4$.

However, there is a second way to construct this curve.
The Grassmannian $\text{Gr}(2,5)$ of 2-planes in a five-dimensional vector space $V$ admits a Pl\"ucker embedding into $\mathbb{P}(\wedge^2V)=\mathbb{P}^{9}$.
The image is also defined by quadrics and those are again the five $4\times 4$ Pfaffians of a $5\times 5$ skew-symmetric matrix.
Moreover, the variety that is defined by the Pfaffians of any generic $5\times 5$ skew-symmetric matrix with entries linear in the homogeneous coordinates of $\mathbb{P}^9$ is isomorphic to $\text{Gr}(2,5)$.
One can now consider $\mathbb{P}^4$ as a complete intersection in
$\mathbb{P}^9$ and the intersection of this $\mP^4$ with the image of $\text{Gr}(2,5)$ realizes the curve as a codimension five complete intersection in the Grassmannian.
As we will discuss in Section~\ref{sec:homol-proj-dual}, the relation between the Pfaffian curve and the complete intersection in a Grassmannian is a simple example of homological projective duality.
However, while the duality reduces to an isomorphism at the level of
the curves, it turns out that the corresponding relative homological
projective duality between the fibrations leads to a non--trivial
derived equivalence betweenn non--isomorphic genus one fibered Calabi-Yau manifolds.

To construct genus one fibration with $5$-sections, we can therefore
start by considering either $\mathbb{P}^4$ or $\text{Gr}(2,5)$ bundles
over a given base variety~\footnote{A similar construction has been
  suggested in~\cite{Kimura:2019bzv}.}.
In the case of projective bundles we can then consider skew-symmetric
morphisms between rank $5$ vector bundles and obtain fibrations of
genus one curves by considering the associated Pfaffian ideals, see Section~\ref{sec:pfaffian-calabi-yau}.
On the other hand, approriate rank $5$ bundles can be used to define
codimension five complete intersections in Grassmann bundles that also
exhibit a genus one fibration with a $5$-section, see Section~\ref{sec:compl-inters-calabi}.
In the following we will make both of these constructions precise and
also introduce the necessary machinery to impose the Calabi-Yau
condition and calculate the Hodge numbers as well as the intersection
numbers. 

\subsection{Homogeneous vector bundles on Grassmann bundles}
\label{sec:homog-vect-bundl}
In this section we introduce the main mathematical concepts that we
need to construct the varieties that will serve as ambient spaces for
the Calabi--Yau varieties that are realized either as complete intersections or Pfaffian
subvarieties. We present these concepts in a large generality to allow
for applications in a wider context than genus one fibrations.

\subsubsection{Grassmann bundles}
\label{sec:some-title}

Let $P$ be a smooth complex projective variety of dimension $d$, and
$F$ a vector bundle of rank $n$ over $P$. For $0 \leq k \leq n$ we
consider the Grassmann bundle (also called relative Grassmannian) $\Gr_k(F)$ over $P$ whose fiber over $p
\in P$ is the Grassmannian $\Gr_k(F_p)$ of $k$--planes in the fiber
$F_p$ of $F$. We denote the total space of $\Gr_k(F)$ by $G$ and the
projection to $P$ by $\pi$. The case $k=1$ corresponds to the
projective bundle of $F$, also denoted by $\mP(F)$. The dimension of
$G$ is $k(n-k)+d$.

The Grassmann bundle $G=\Gr_k(F)$ comes with a short exact sequence of
vector bundles on $G$:
\begin{equation}
  \label{eq:tautological}
  0 \to S \to \pi^*F \to Q \to 0,
\end{equation}
where $S$ is the tautological rank $k$ subbundle and $Q$ is the
tautological quotient bundle of rank $n-k$. 
We write $\rO_{G/P}(-1)$ for the Grothendieck line bundle $\det S = \wedge^kS$ and $\sigma_1 = -\ch_1(\det S)
= -\ch_1(S)$ for its first Chern class.  These bundles, i.e. $S$, $Q$
and $\rO_{G/P}(1)$, form the fundamental constituents of the (relative)
homogeneous vector bundles on $P$ as will be reviewed in
Section~\ref{sec:homog-vect-bundl-1}. We have $\pi_*\rO_{G/P}(1) \cong
F\spcheck$ and hence the space of global sections of $\rO_{G/P}(1)$ is $\tH^0(G,\rO_{G/P}(1))
= \tH^0(P,F\spcheck)$. The Pl\"ucker embedding $\Gr_k(F) \hookrightarrow \mP(\wedge^2F)$ is given
by these global sections.

Dualizing this construction, i.e.~starting
from $F\spcheck$ and considering the Grassmannian
$\Gr_{n-k}(F_p\spcheck)$ of $(n-k)$--planes in the fiber $F_p\spcheck$, we
obtain an isomorphism $\delta:\Gr_{n-k}(F\spcheck) \cong \Gr_k(F)$
sending a subbundle $W \subset F$ to its annihilator $\mathrm{Ann}\; W = \{ \ell \in
F\spcheck \mid \ell(w) = 0 \;\forall w \in W\} \subset
F\spcheck$. This entails $\delta^*S \cong Q\spcheck$ and $\delta^*Q \cong S\spcheck$.  This will
allow us to trade the bundle $Q$ for the bundle $S\spcheck$ on the
dual Grassmann bundle, and vice versa.

The relative tangent bundle $T_{G/P}$ is defined by
\begin{equation}
  \label{eq:66}
  0 \to T_{G/P} \to T_G \to \pi^*T_P \to 0,
\end{equation}
and there is an isomorphism
\begin{equation}
  \label{eq:65}
  T_{G/P}  \cong S\spcheck \otimes Q.
\end{equation}
One can derive a formula for the canonical bundle of $G$
from~\eqref{eq:65} and the two short exact
sequences~\eqref{eq:tautological} and~\eqref{eq:66} as follows.
The sheaf of holomorphic $1$--forms is $\Omega^1_{G} =
T_{G}\spcheck$ and the canonical bundle is defined as $\omega_{G} =
\det \Omega^1_{G}   = \wedge^{\dim G} \Omega^1_{G}$. From the dual
of~\eqref{eq:66} we get $\det \Omega^1_{G} = \det
\Omega^1_{G/P} \otimes \pi^*\det \Omega^1_{P}$. The second factor
is the canonical bundle $\pi^*\omega_{P}$ of the base $P$. For the
first factor it follows from the dual of~\eqref{eq:65} that $\det \Omega^1_{G/P} = \det S^{\otimes (n-k)} \otimes \det
Q\spcheck{}^{\otimes k}$. Finally, from~\eqref{eq:tautological} we
have $\det Q\spcheck = \pi^*\det F\spcheck \otimes \det S$. Hence, we conclude that 
\begin{equation}
  \omega_G = \pi^*\left(\omega_P \otimes \det(F\spcheck)^{\otimes k} \right)\otimes \rO_{G/P}(-n).
\end{equation}
From the tautological short exact sequence~\eqref{eq:tautological} we can also determine the
intersection ring of the Grassmann bundle $G$ as follows~\cite{Fulton:1998ab}:
\begin{equation}
  \label{eq:88}
  \tH^*(G,\mQ) \cong \tH^*(P,\mQ)[s_1,\dots,s_k,q_1,\dots,q_{n-k}] /
  \fa \,,
\end{equation}
where the ideal $\fa$ is defined by the homogeneous components of the relation
\begin{equation}
  \label{eq:70}
  \left(1+\sum_{i=1}^ks_i\right)\left(1+\sum_{j=1}^{n-k} q_{j}\right) - \pi^*\left(1+\sum_{j=1}^n\ch_j(F)\right) \,.
\end{equation}
The isomorphism~\eqref{eq:88} sends $s_i$ to $\ch_i(S)$ and $q_i$ to $\ch_i(Q)$. The
variables $q_i$, $i=1,\dots,n-k$, can be eliminated to yield
\begin{equation}
  \label{eq:12}
  \tH^*(G,\mQ) = \tH^*(P,\mQ)[s_1,\dots,s_k] / \fb
\end{equation}
for some ideal $\fb$ generated by $k$ relations in degrees
$n-k+1,\dots,n$. The cohomology classes of $G$ can be conveniently
expressed in terms of (Poincar\'e duals of) relative Schubert varieties~\cite{Trautmann:2007ab}
$\sigma_\lambda \in \tH^*(G,\mQ)$, where $\lambda$ is a partition,
represented by a Young diagram that fits into a rectangular grid of
size $k\times(n-k)$. In particular, we identify
\begin{equation}
  \label{eq:44}
  \begin{aligned}
    \ch_r(S\spcheck) &= \sigma_{(1^r)}, & r = 1,\dots,k\\
    \ch_r(Q - \pi^*F) &= \sigma_{(r)}, & r=1,\dots,n-k.\\
  \end{aligned}
\end{equation}
An important notion in the birational geometry of a Grassmann bundle
is its ample or, dually, its Mori cone. We are not aware of a description of these
cones in general. If $G$ is a projective bundle, i.e. $k=1$, over a toric variety
$P$, then $G$ is also a toric variety, and the Mori cone can be
determined by combinatorial methods. More generally, the projection
$\pi: G \to P$ corresponds to an extremal ray of the Mori cone. As we
are lacking a description of these cones, we cannot decide whether
$\omega_G^{-1}$ is nef or not and which of these Grassmann bundles are Fano varieties.

\subsubsection{Homogeneous vector bundles}
\label{sec:homog-vect-bundl-1}

In this section, we review properties of homogeneous vector bundles on
Grassmann bundles by viewing the latter as (relative) homogeneous
spaces $\pG/\pP$. We will work with (relative) homogeneous vector bundles on (relative) Grassmannians for two reasons:
One is that the Calabi--Yau threefolds of our interest will be defined as complete
intersections or as Pfaffian subvarieties given by sections or maps,
respectively, of such bundles. For those most of their geometric properties can be determined in
terms of the representation theory of $\pG$ and $\pP$. In particular, we
have the powerful theorem of Borel--Weil--Bott at our disposal, see
Section~\ref{sec:borel-weil-bott}, that reduces the computation of cohomology groups to representation theoretic
calculations. The second reason is that such Calabi--Yau threefolds appear as geometric phases of GLSMs whose defining data is (the
weights of) a representation of a compact subgroup of $G$, to be
discussed in Section~\ref{sec:glsmreview}.

We first review the case for the Grassmannian and then discuss the
modifications for the Grassmann bundles.
Let $\pG$ be a semisimple complex Lie group, and $\pP$ a maximal
parabolic subgroup. We can view $\pG$ as a principal $\pP$--bundle
over $\pG/\pP$, i.e. we have a right action $g \mapsto gp^{-1}$, $g\in
\pG, p \in \pP$. Moreover, there is a left action $\pG \to
\Aut(\pG/\pP)$ which maps $x\pP \mapsto gx\pP$, $g \in
\pG$. Let $V$ be the standard representation of $\pG$ of dimension $n$, and $W$ any representation of $\pP$. Then we can form the associated vector bundle over $\pG/\pP$
\[
  \begin{aligned}
  \rW &= \pG \times W /_\sim \to \pG/\pP
  \end{aligned}
\]
where the quotient is taken with respect to the equivalence relation $(g,w) \simeq (gp^{-1},pw)$, $p \in \pP$. Such a vector bundle over $\pG/\pP$ is called homogeneous with respect to $\pG$.
Similarly, restricting the representation $V$ to $\pP$, $\rV$ will be the associated vector bundle to $V$. Therefore, in this context, one can define irreducible and
indecomposable homogeneous bundles, in analogy with the definitions in
representation theory. In the case of the Grassmannian, after choosing a
basis $V\cong \mC^n$, we have $\Gr_k(V) = \pG/\pP$ with
$\pG=\SL(V)\cong\SL(n,\mC)$ and $\pP = \{g\in \SL(n,\mC)\mid g=\left(
  \substack{A\;*\\0\;B} \right), A \in  \GL(k,\mC), B \in \GL(n-k,\mC)
\}$. If we take $W\subset V$ to be a subspace of dimension $k$, then $\rV$ and $\rW$ correspond
to the trivial and to the tautological subbundle of $\Gr_k(V)$,
respectively. 

For more general bundles, we need to review some facts about the relation
between the representation theory of $\pP$ and the one of
$\SL(n,\mC)$, see e.g.~\cite{Fulton:1991ab,Kuechle:1995ab}. 
An irreducible homogeneous vector bundle $\rW$ over $\Gr_k(V)$,
homogeneous with respect to $\SL(V)$ corresponds to a weight of
$\SL(n,\mC)$ that is dominant for the subgroup $\SL(k,\mC) \times
\SL(n-k,\mC)$. This weight will be the highest weight of the
irreducible representation $W$ of $\pP$ defining $\rW$.

We will denote such a weight by
$\beta=(\beta_1,\dots,\beta_k;\beta_{k+1},\dots,\beta_n) =
(\beta_1,\dots,\beta_n)$ with respect to the basis $e_1,\dots,e_n$ of
$\mR^n$ underlying the root system of $A_{n-1}$ with positive roots
$e_i-e_j$, $1 \leq i < j \leq n$. The semicolon in the first equality
refers to the concatenation of the weights $(\beta_1,\dots,\beta_k)$
and $(\beta_{k+1},\dots,\beta_n)$ for the subgroups $\SL(k,\mC)$ and 
$\SL(n-k,\mC)$, respectively. 
Since the Weyl group of $\pP$ is $S_{k} \times S_{n-k}$, the product of
the Weyl groups of $\SL(k,\mC)$ and $\SL(n-k,\mC)$, it acts by
permutations among the first $k$ and the last $n-k$ entries of
$\beta$. The vector bundle $\rW$ is then globally generated if and only if the highest
weight $\beta$ satisfies $\beta_i \geq \beta_j$ for $1 \leq i < j \leq n$.

The weight lattice of $\GL(V)$ is isomorphic to $\mZ^n$ via the map of 
taking the $d$-th fundamental weight,  i.e.  the  highest  weight  of
$\wedge^d V$,  to  the  sum  of  the  first $d$ basis  vectors  of
$\mZ^n$.  Under this isomorphism, the dominant integral weights of
$\GL(V)$ correspond to nonincreasing sequences of integers $\beta=
(\beta_1,\dots,\beta_n) \in \mZ^n$. For  such a sequence $\beta$, we denote by
$\mS_\beta V$ the corresponding irreducible representation of
$\GL(V)$ of highest weight $\beta$.  The assignment $V \mapsto
\mS_\beta V$ for a dominant integral weight $\beta$ is known as a
Schur functor. This assignment globalizes to (arbitrary) homogeneous vector bundles $\rW$ over a homogeneous variety $\pG/\pP$, i.e. we form the associated vector bundle $\mS_\beta\rW = \pG \times \mS_\beta W/_\sim$. The two simplest examples are the symmetric powers $\mS_{(m)}\rW = \Sym^m\rW$ and the antisymmetric powers  $\mS_{(1^m)}\rW = \wedge^m \rW$. 

It remains to specify the weights for the most important representations. 
The only change for $\pG=\SL(V)$ is that the weight lattice is isomorphic to the quotient $\mZ^n/\mZ(1,\dots,1)$. 
The trivial representation of $\SL(V)$ of dimension 1 has highest weight
$\beta=(j,\dots,j)$ for any $j\in\mZ$. It can be used to renormalize the weights to
$\beta_n=0$, i.e. if $\beta=(\beta_1,\dots,\beta_n)$ and $\alpha = \beta + (j^n)$ then
\begin{equation}
  \label{eq:23}
  \mS_\alpha \rW \cong \mS_\beta \rW \otimes \det \rW^{\otimes j}.  
\end{equation}
In this way, a highest weight can be identified with a
partition of a positive integer. In this case, we will write $\lambda$
for the weight $\beta$. The tautological quotient bundle $Q$
and the tautological subbundle $S$ come from the representations with
highest weights $\beta=(1,0,\dots,0)$ and
$\beta=(0,\dots,0;1,0,\dots,0)$, respectively. The tangent bundle
$T_G$ is irreducible with highest weight $\beta=(1,0,\dots,0,-1)$.
Finally, if the bundle $\rW$ comes from a representation of highest
weight $\beta=(\beta_1,\dots,\beta_n)$, then the dual bundle
$\rW\spcheck$ comes from a dual representation of highest weight
$\beta\spcheck=(-\beta_k,\dots,-\beta_1;-\beta_n,\dots,-\beta_{k+1})$,
i.e.
\begin{equation}
  \label{eq:35}
  \mS_\beta \rW\spcheck \cong \mS_{\beta\spcheck}\rW.
\end{equation}
It follows that $\wedge^mS\spcheck$ comes from a representation of
highest weight $(1^m,0^{k-m};0,\dots,0)$. In particular
$\rO_{\Gr_k(V)}(1)$ comes from $(1^k;0,\dots,0)$. Moreover, we have
\begin{equation}
  \label{eq:36}
  \wedge^m\rW \cong \wedge^{n-m}\rW\spcheck\otimes \det \rW^{\otimes n}.
\end{equation}

So far, we have given a description of homogeneous vector bundles over
Grassmannians. For our purposes we need the generalization to the
relative version for Grassmann bundles over a projective variety
$P$. We start with an alternative description of the Grassmann
bundle. Given a vector bundle $F$ over $P$ of rank $n$, we let $\rE$
be the principal $\pG$--bundle over $P$ induced by the frame bundle of
$F$, here $\pG=\SL(V)$ as above. Then we can form the associated
bundle $\pi: G = \rE \times \pG/_\sim \to P$  where the quotient is
with respect to the equivalence relation $(e,g) \simeq (eh^{-1},hg)$,
$h\in \pG$.  The fiber of $G$ is $\pG/\pP$. With our specific choice for $F$,
$\pG$ and $\pP$, $G$ is isomorphic to the Grassmann bundle $\Gr_k(F)$
over $P$. Here, we have used the right action of $\pG$ on $\pG$ viewed
as $\pP$ principal bundle. Combining this with the left action by
$\pP$ we can form the associated vector bundle
\begin{equation}
  E  = \rE \times \pG\times W/_\sim \to P,
  \label{eq:89}
\end{equation}
where the quotient is with respect to the
equivalence relation $(e,g,w) \simeq (eh^{-1},hgp^{-1},pw)$ for $h\in
\pG$ and $p\in\pP$. The projection to the first two factors induces a
map to the Grassmann bundle $G$ whose fiber is the vector space
$W$. Moreover, $\pi^{-1}(p) \cong \Gr_k(V)$ and $E|_{\pi^{-1}(p)}
\cong \rW$ for all points $p\in P$. Replacing $W$ by $\mS_\beta W$ as
above yields the relative version of homogeneous vector bundles
$\mS_\beta E$ over the Grassmann bundle $G$. 

In summary, an irreducible globally generated homogeneous vector bundle $E$ on a Grassmann bundle $G$ will
be of the form
\begin{equation}
  \label{eq:1}
  E = \mS_\lambda S\spcheck \otimes \mS_\mu Q \otimes \rO_{G/P}(p) 
\end{equation}
for some partitions $\lambda$, $\mu$ such that $\lambda_k=0$,
$\mu_{n-k}=0$, and $p \geq 0$. Finally, all our vector bundles on $G$ will be of the form
\begin{equation}
  \label{eq:87}
  E = \bigoplus_{i} E_i \otimes \pi^*E'_i
\end{equation}
where $E_i$ is of the form~\eqref{eq:1} and $E'$ is a globally generated vector bundle on $P$.

\subsubsection{The Borel--Weil--Bott theorem}
\label{sec:borel-weil-bott}

The Borel--Weil--Bott theorem gives the cohomology of all irreducible
homogeneous vector bundles over $\pG/\pP$ in terms of $\pG$
representations, e.g.~it gives the cohomology of all irreducible
homogeneous vector bundles over $\Gr_k(V)$ in terms of $\SL(V)$
representations. More generally, it allows us to compute the
higher direct image sheaves $\tR^s\pi_*E$ of homogeneous vector
bundles $E$ on a Grassmann bundle $\pi: \Gr_k(F) \to P$. For details,
see e.g.~\cite{Weyman:2003ab}. Roughly speaking,
the higher direct image sheaf $\tR^s\pi_*E$ is a coherent sheaf over $P$ whose stalk
$(\tR^s\pi_*E)_p$ over the point $p\in P$ is the cohomology group
$\tH^s(\pi^{-1}(p), E|_{\pi^{-1}(p)}) = \tH^s(\Gr_k(V),\rW)$. Here,
$E$ and $\rW$ are related as in~\eqref{eq:89}.

The symmetric group $S_n$ acts on the weight lattice
$\mZ^n$ by permuting the factors. Denote by $\ell: S_n\mapsto \mZ$
the standard length function, i.e.~$\ell(\sigma)$ is the minimal
number of simple transpositions $i \leftrightarrow i+1$ that
generate the permutation $\sigma$. We say $\beta \in \mZ^n$ is regular if
all of its components are distinct; in this case, there is a unique
$\sigma \in S_n$ such that $\sigma(\beta)$ is a strictly decreasing
sequence. Finally, let $\rho= (n,n-1,...,2,1)\in \mZ^n$ be the sum of the fundamental weights.

Let $\lambda=(\lambda_1,\dots,\lambda_k)$ and
$\mu=(\mu_1,\dots,\mu_{n-k})$ be two sequences of nonincreasing
integers and consider their concatenation $\beta=(\lambda;\mu) \in \mZ^n$.
We consider the bundle 
\begin{equation}
  E(\beta) = \mS_\lambda S\spcheck \otimes \mS_\mu Q 
\end{equation}
over $G=\Gr_k(F)$. Note that we allow $\lambda_k=p\in\mZ$ so that the factor
$\rO_{G/P}(p)$ is included in $\mS_\lambda S\spcheck$. Moreover, if
$\mu$ is a partition then we use $\mS_{\mu}Q \cong \mS_{\mu\spcheck}Q\spcheck$.
Then the theorem asserts the following alternative: Either
$\sigma(\beta+\rho)-\rho=\beta$ for some permutation $\sigma \in S_n$
and then
\begin{equation}
  \tR^s\pi_*E(\beta) = 0,
\end{equation}
or there exists a unique $\sigma\in S_n$ such that
$\sigma(\beta+\rho)-\rho=\nu$ is a partition. In this case
\begin{equation}
  \tR^s\pi_*E(\beta) =
  \begin{cases}
    \mS_\nu F\spcheck & s = \ell(\sigma)\\
    0 & \text{otherwise}.
  \end{cases}
\end{equation}
In the special case of $\pG/\pP=\SL(n+1,\mC)/\SL(n,\mC) = \mP^n$, there is
closed formula known as the Bott formula for the dimensions of the cohomology of the sheaf
$\Omega^p_{\mP^n}$ of holomorphic $p$--forms and its twists~\cite{Okonek:2011ab}:
\begin{equation}
  \th^q(\mathbb{P}^n,\Omega^p_{\mathbb{P}^n}(k))=\left\{\begin{array}{ccc}
  \left(\begin{array}{c}k+n-p\\k\end{array}\right)\left(\begin{array}{c}k-1\\p\end{array}\right)&\quad& q=0,0\leq q\leq n,k>p\\
      1&\quad& k=0,0\leq p=q\leq n\\
      \left(\begin{array}{c}-k+p\\-k\end{array}\right)\left(\begin{array}{c}-k-1\\n-p\end{array}\right)&\quad&q=n,0\leq p\leq n,k<p-n\\
          0&\quad&\text{else.}
     \end{array}  \right.
     \label{eq:Bottformula}
\end{equation}

\subsection{Complete intersection Calabi--Yau varieties in Grassmann bundles}
\label{sec:compl-inters-calabi}

We now have all the ingredients to give a construction of Calabi-Yau
varieties as complete intersections in Grassmann bundles. After discussing the general construction, we elaborate on several prototypical examples that will be analyzed in detail in the remainder of the article.

\subsubsection{General construction}
\label{sec:general-constr-Gr}

Let $G=\Gr_k(F)$ be a Grassmann bundle as in~\ref{sec:some-title}. Let $E$ be a homogeneous vector bundle of rank $r$ on $G$, and $s=(s_1,\dots,s_r)
\in \tH^0(G,E)$ be a general section of $E$. Then its zero scheme
\begin{equation}
  X = Z(s)
\end{equation}
is a complete intersection of codimension $r$ in $G$. If $X$ is
nonempty, then it is of dimension $d+k(n-k)-r$. To determine its
canonical sheaf, one considers the standard minimal resolution of the ideal sheaf
$\rI_X$ of $X$ in terms of the Koszul complex of $E$:
\begin{equation}
  \label{eq:resIX}
  0 \to \wedge^r E\spcheck \to \wedge^{r-1}E\spcheck \to \dots \to
  \wedge^2E\spcheck \to E \spcheck \to \rI_X \to 0,
\end{equation}
where the maps are given by contraction with the section
$s$. We recall the conormal bundle sequence
\begin{equation}
  \label{eq:NXG}
  0 \to \rI_X^2 \to \rI_X \to \iota_*N_{X/G}\spcheck \to 0,
\end{equation}
where $\iota: X \to G$ denotes the embedding of $X$ into $G$. It
follows from~\eqref{eq:resIX} and~\eqref{eq:NXG} that
\begin{equation}
  \label{eq:67}
  N_{X/G} \cong E|_X\,.
\end{equation}
Then the canonical sheaf is given in terms of the adjunction formula
\begin{equation}
  \omega_X = \omega_G \otimes \det E \otimes \rO_X.
\end{equation}
By the generalization of Bertini's theorem of~\cite{Mukai:1992ab}, the
complete intersection $X$ will be smooth if $E$ is generated by its
global sections.

Let us add a comment about twisting $F$. Note that if $L$ is a line bundle on $P$, we can consider $G' = \Gr_k(F')$
for $F'=F\otimes L$ with projection $\pi': G\to P$. Then there is an
isomorphism $f: G \xrightarrow{\sim} G'$ such that $\pi = \pi' \circ f$. The
tautological subbundle $S$, however, is not determined by the
Grassmann bundle alone, it depends on $L$. In particular, $\omega_G$ depends on the choice of $L$. If we
define the bundle $E' = E \otimes \pi^*L^{\otimes- k}$ on $G'$, and consequently
$X'=Z(s')$ for $s' \in \tH^0(G',E')$ then $\omega_G \otimes
\det E|_X \cong \omega_{G'} \otimes \det E' |_{X'}$, so that $X$ and $X'$
define isomorphic varieties, as long as both $E$ and $E'$ are globally
generated. One can normalize $E$ (or $F$) by requiring that $\ch_1(\det E)$ lies in some range.

\subsubsection*{Intersection ring}

The part of the intersection ring $\tH^*(X)$ of $X$ which is induced
from the intersection ring of $G$ is then computed in the
standard way~\cite{Fulton:1998ab}. Given $\alpha \in \tH^{2\dim X}(G)$, 
\begin{equation}
  \label{eq:42}
  \int_X \iota_*\alpha = \int_G \ch_{\Top}(E) \cup \alpha,
\end{equation}
where $\iota: X\to G$ is the embedding. By~\eqref{eq:12} the right
hand side is a polynomial in the relative Schubert classes
$\sigma_\lambda$ whose coefficients are cohomology classes in
$P$. By multiple application of the Pieri rule (see (\ref{eq:Pieri}) in Appendix~\ref{sec:calculus-with-schur}), this polynomial can be reduced to a
multiple of the class $\sigma_{(n-k)^k}$.
The actual numbers can be computed using the formula
\begin{equation}
  \label{eq:43}
  \int_{\Gr_k(\mC^n)} \sigma_{(n-k)^k} = 1.
\end{equation}

\subsubsection{Examples}
\label{sec:examples-Gr}
In this section we will be studying a number of examples that will
play an important role in the rest of the paper.

\subsubsection*{Example: elliptic curve $X_0$}
First, let $P$ be a point, $F$ a vector space of
dimension $5$ and $k=2$. Then $G$ is the Grassmannian of $2$--planes
in $F$, $\Gr_2(F) = G(2,5)$. For the bundle $E$ we take
$\rO_G(1)^{\oplus 5}$. Then $X_0=Z(s)$ is an elliptic curve of degree
$5$, also known as elliptic normal quintic, see Section~\ref{sec:normalcurves}. The GLSM realisation will be reviewed in Section~\ref{sec:glsmelliptic}.

\subsubsection*{Examples: $X_1$, $X_4$}
Next, we consider various relative versions of the previous example. We take the base of the fibrations to be $P=\mP^2$.
For the first fibration we consider the trivial bundle
\begin{equation}
  \label{eq:38}
  F =\rO_{\mP^2}^{\oplus 5} 
\end{equation}
over $\mP^2$. Let $\Gr_2(F)$ be the corresponding Grassmann bundle
over $\mP^2$. Since $F$ is the trivial rank $5$ bundle, the total
space $G$ of  $\Gr_2(F)$ is the product $\Gr_2(\mC^5) \times \mP^2$.
We compute $\omega_G = \rO_{G/\mP^2}(-5) \otimes \pi^*
\rO_{\mP^2}(-3)$. 

Now, we consider the bundle on $G$ given by
\begin{equation}
  \label{eq:40}
  E =\rO_{G/\mP^2}(1) \otimes \pi^*\left( \rO_{\mP^2}(p) \oplus  \rO_{\mP^2}(q) \oplus  \rO_{\mP^2}(r) 
    \oplus \rO_{\mP^2}^{\oplus 2} \right)\,.
\end{equation}
Then $\det E = \rO_{G/\mP^2}(5) \otimes
\pi^*\rO_{\mP^2}(p+q+r)$. Hence, the Calabi--Yau condition is
satisfied for $(p,q,r) = (1,1,1), (2,1,0)$ and $(3,0,0)$. 
Let $X_1$ be the complete intersection in $G$ given by a generic
section $s$ of $E$ for the choice $(p,q,r)=(1,1,1)$:
\begin{equation}
  X_1 = Z(s).
\end{equation}
For later purposes, we also introduce $X_4 = Z(s)$ for the choice of
$(p,q,r) = (3,0,0)$. 

We choose homogeneous coordinates $b_l\in
\tH^0(\mP^2,\rO_{\mP^2}(1))$, $l=1,2,3$, on $\mP^2$, and $x_i \in
\tH^0(G_p,S_p\spcheck)$, $i=1,\dots,5$ on the Grassmannian fiber $G_p=\Gr_2(F_{p})$ over $p\in\mP^2$. Note
that since $\rank S\spcheck=2$, $x_j = (x_j^a)_{a=1,2}$ is a 2-component
vector. Taking the determinant, $\varepsilon_{ab}x_i^ax_j^b$ is a
section of $\det S_p\spcheck = \rO_{G_p}(1)$, which is just a
Pl\"ucker coordinate. Therefore, the equations of $X_1$ can be written
as
\begin{equation}
  \label{eq:X1eqs}
  \begin{aligned}
    0 &= s_k = \sum_{i,j=1}^5\sum_{l=l}^3 \sum_{a,b=1}^2 A_{k}^{ijl} \varepsilon_{ab}x_i^ax_j^bb_l,
    & k &= 1,2,3,\\
    0 &= s_k = \sum_{i,j=1}^5 \sum_{a,b=1}^2 A_{k}^{ij} \varepsilon_{ab}x_i^ax_j^b,
    & k &= 4,5,\\
  \end{aligned}
\end{equation}
for some general complex coefficients $A_{k}^{ijl}$ and $A_{k}^{ij}$.
This is how the variety $X_1$ will appear in the GLSM in Section~\ref{sec:glsmModel1}.

Alternatively, we can think of this variety as a relative hyperplane
section 
\begin{equation}
  X_1 = (\Gr_2(\mC^5) \times \mP^2) \cap H_1 \cap \dots \cap H_5 \subset \mP^9 \times\mP^2,
  \label{eq:34}
\end{equation}
where $H_1$, $H_2$, $H_3$ are hyperplanes of degree $(1,1)$ and $H_4, H_5$ are
hyperplanes of degree $(1,0)$ in $\mP^9 \times
\mP^2$ and $\Gr_2(\mC^5)$ is embedded in $\mP^9$ via the Pl\"ucker
map.

The intersection ring of $G$ is
\begin{equation}
 \tH^*(G,\mQ) = \mQ[h,s_1,s_2]/(h^3,3\,s_{{2}}{s_{{1}}}^{2}-{s_{{1}}}^{4}-{s_{{2}}}^{2},2\,{s_{{2}}}^{2}s
_{{1}}-s_{{2}}{s_{{1}}}^{3})
  \label{eq:71}
\end{equation}
and the top Chern class of $E$ is $\ch_{5}(E) =
{s_{{1}}}^{5}+3\,{h}^{2}{s_{{1}}}^{3}+3\,h{s_{{1}}}^{4}$. Pieri's rule
yields $s_1^6 = 5s_2^3$, hence we find
\begin{equation}
  \label{eq:72}
  \begin{aligned}
    J_1^3 &=15, & J_1^2J_2&=15, & J_1J_2^2&=5, & J_2^3&=0, & \ch_2\cdot
    J_1 &=66, & \ch_2\cdot J_2 &=36.
   \end{aligned}
\end{equation}
where $J_1 = s_1|_{X_1}$ and $J_2=h|_{X_1}$. Using the criterion
of~\cite{Oguiso:1993ab}, we can read off from $ J_2^3=0$ and
$J_1J_2^2=5$ that $X_1$ admits a genus one fibration with
$5$--section, given by the projection to $\mP^2$. This argument will
be applied to most of the examples in this Section.

\subsubsection*{Example: $X_2$}
For the second fibration we consider the Grassmann bundle $\Gr_2(F)$ over $\mP^2$ with
\begin{equation}
  F = \rO_{\mP^2}^{\oplus 4} \oplus \rO_{\mP^2}(1). 
\end{equation}
The total space $G$ of $\Gr_2(F)$ is a nontrivial fibration over $\mP^2$ with general fiber $\Gr_2(\mC^5)$. 
We consider the bundle
\begin{equation}
  E = \rO_{G/\mP^2}(1) \otimes \pi^*\rO_{\mP^2}(1)^{\oplus 5},
\end{equation}
and let
\begin{equation}
  X_2 =Z(s)
  \label{eq:108}
\end{equation}
be the complete intersection in $G$ given by a generic section $s$ of $E$. Choosing homogeneous coordinates as above, the equations of $X_2$ can be written
as
\begin{equation}
  \label{eq:X2eqs}
  \begin{aligned}
    0 &= s_k = 
    \sum_{i,j=1}^4 \sum_{l=1}^3\sum_{a,b=1}^2A^{ijl}_{k}b_l\varepsilon_{ab}x_i^ax_j^b+\sum_{i=1}^4\sum_{a,b=1}^2A^{i5}_{k}\varepsilon_{ab}x_i^ax_5^b,
    & k &= 1,\dots,5.
  \end{aligned}
\end{equation}
This is how the variety $X_2$ will appear in the GLSM in
Section~\ref{sec:glsmModel2}.

The intersection ring of $G$ is
\begin{equation}
 \tH^*(G,\mQ) = \mQ[h,s_1,s_2]/(h^3,3\,s_{{2}}{s_{{1}}}^{2}-{s_{{1}}}^{4}+{s_{{1}}}^{3}h-2\,s_{{1}}s_{{2}
}h-{s_{{2}}}^{2},s_{{2}} \left( 2\,s_{{2}}s_{{1}}-s_{{2}}h-{s_{{1}}}^{
3}+{s_{{1}}}^{2}h \right))
  \label{eq:74}
\end{equation}
and the top Chern class of $E$ is $\ch_{5}(E) =
{s_{{1}}}^{5}+10\,{h}^{2}{s_{{1}}}^{3}+5\,h{s_{{1}}}^{4}$. A Groebner
basis calculation yields
\begin{equation}
  \label{eq:75}
  \begin{aligned}
    J_1^3 &=8, & J_1^2J_2&=11, & J_1J_2^2&=5, & J_2^3&=0, & \ch_2\cdot
    J_1 &=56, & \ch_2\cdot J_2 &=36.
   \end{aligned}
\end{equation}
where $J_1 = s_1|_{X_2}$ and $J_2=h|_{X_2}$. Again, we see that the
projection to $\mP^2$ induces a genus one fibration with a
5--section.

\subsection*{Example: $X_2'$}
Next, we consider an example that is not a genus one fibration but will also appear as a phase in the GLSM of Section~\ref{sec:glsmModel2}. We start with $P=\mP^4$ and consider the
Grassmann bundle $\Gr_2(F)$ over $\mP^4$ with
\begin{equation}
  \label{eq:20}
  F = \rO_{\mP^4}^{\oplus 4}.
\end{equation}
The total space $G$ of  $\Gr_2(F)$ is simply the product $\Gr_2(\mC^4) \times \mP^4$. 
We consider the bundle
\begin{equation}
  \label{eq:29}
  E = S\spcheck \otimes \pi^*\rO_{\mP^4}(1) \oplus \rO_{G/\mP^4}(1) \otimes \pi^*\rO_{\mP^4}(1)^{\oplus 3},
\end{equation}
where $S$ is the tautological rank 2 subbundle on $G$.
We let $X_2' =Z(s)$ be the subvariety in $G$ defined by the zero locus
of a generic section $s$ of $E'$. We choose homogeneous coordinates $y_l\in
\tH^0(\mP^4,\rO_{\mP^4}(1))$, $l=1,\dots,5$, on $\mP^4$, and $x_i \in
\tH^0(G_p,S_p)$, $i=1,\dots,4$ on the Grassmannian fiber
$G_p=\Gr_2(F_{p})$ over $p\in\mP^4$. Then the equations of $X_2'$ can be
written as
\begin{equation}
  \label{eq:X2peqs}
  \begin{aligned}
    0 &= s^a_{1} = \sum_{i=1}^4\sum_{l=l}^5 \sum_{b=1}^2 A^{ail}_{bk} x_i^by_l,
    & a &= 1,2,\\
    0 &= s_k = \sum_{i,j=1}^4\sum_{l=1}^5 \sum_{a,b=1}^2 A_{k}^{ijl} \varepsilon_{ab}x_i^ax_j^by_l,
    & k &= 2,\dots,4.
  \end{aligned}
\end{equation}
The intersection numbers and second Chern classes are 
\begin{equation}
  \label{eq:80}
  \begin{aligned}
    J_1^3 &=8, & J_1^2J_2&=13, & J_1J_2^2&=11, & J_2^3&=5, & \ch_2\cdot
    J_1 &=56, & \ch_2\cdot J_2 &=50,
   \end{aligned}
\end{equation}
where $J_1 = s_1|_{X_2}$ and $J_2=h|_{X_2}$. 

\subsubsection*{Example: $X_3$}
For the next example, we again take $P=\mP^2$ and $F$ as
in~\eqref{eq:38}, but consider the bundle
\begin{equation}
  E = \rO_{G/\mP^2}(1) \otimes \pi^*\left(T_{\mP^2}(-1)\oplus
    \rO_{\mP^2}(1)^{\oplus 2} \oplus \rO_{\mP^2}\right),
\end{equation}
and let $X_3 = Z(s)$ be the complete intersection in $G$ given by a
generic section $s$ of $E$. Note that $T_{\mP^2}(-1)$ is nothing but
the tautological quotient bundle $Q$ on $\mP^2$. Therefore, we choose sections
$y_l = (y^c_l)_{c=1,2} \in \tH^0(\mP^2,Q)$, $l=1,2,3$. In terms of
these sections, the homogenenous coordinates on $\mP^2$ are
sections of $\det Q \cong \rO_{\mP^2}(1)$, i.e. they are of the form
$\varepsilon_{cd}y^c_ly^d_m$. Choosing sections $x_i \in
\tH^0(G_p,S_p)$, $i=1,\dots,5$ on the Grassmannian fiber
$G_p=\Gr_2(F_{p})$ over $p\in\mP^2$ as before, the equations for $X_3$
read
\begin{equation}
  \label{eq:39}
  \begin{aligned}
    0 &= s_c = \sum_{i,j=1}^5\sum_{a,b=1}^2 \sum_{l=1}^3 \sum_{d=1}^2 A^{ijl}_{cd} y^d_l \varepsilon_{ab}x_i^ax_j^b, \qquad
    c=1,2,\\
    0 &= s_k = \sum_{i,j=1}^5\sum_{a,b=1}^2 \sum_{l,m=1}^3
    \sum_{c,d=1}^2 A^{ijlm}_{k} \varepsilon_{cd}y^c_ly^d_m
    \varepsilon_{ab}x_i^ax_j^b,\qquad k=3,4,\\
    0 &= s_5 = \sum_{i,j=1}^5\sum_{a,b=1}^2 A^{ij}_{5}
    \varepsilon_{ab}x_i^ax_j^b.\\
  \end{aligned}
\end{equation}
A GLSM construction of this example will be discussed in
Section~\ref{sec:glsmModel3}. The intersection ring is
\begin{equation}
  \label{eq:86}
  \begin{aligned}
    J_1^3 &=20, & J_1^2J_2&=15, & J_1J_2^2&=5, & J_2^3&=0, & \ch_2\cdot
    J_1 &=36, & \ch_2\cdot J_2 &=68,
   \end{aligned}
\end{equation}
where $J_1 = s_1|_{X_3}$ and $J_2=h|_{X_3}$.

\subsubsection*{Miscellaneous examples}
\label{sec:misc-exampl}

We briefly mention a few more examples over other bases than $\mP^2$.
First, let $P=\mF_0 = \mP^1\times\mP^1$. We denote the classes of the two
$\mP^1$ by $h$ and $f$, respectively. We choose $F=\rO_{\mF_0}^{\oplus 5}$
and $E=\rO_{G/\mF_0}(1)\otimes \pi^*(\rO_{\mF_0}(h+f)^{\oplus2}\oplus
\rO_{\mF_0}^{\oplus3})$. The corresponding Calabi-Yau threefold $X$ has
$h^{1,1}(X)=3, h^{2,1}(X)=43$ and nonvanishing intersection numbers\footnote{All the
  computations of intersections in Grassmann bundles have been
  performed with Schubert~\cite{Katz:1992ab}. The computation of the
  Hodge numbers can be verified using the methods in Section~\ref{sec:comp-hodge-numb}.}
\begin{equation}
  \label{eq:82}
  \begin{aligned}
    J_1^3 &=10, & J_1^2J_2 &=10, & J_1^2J_3 &=10, & J_1J_2J_3 &=5,\\
    \ch_2(X)J_1 &=64, & \ch_2(X)J_2 &= 24, & \ch_2(X)J_3 &= 24.
  \end{aligned}
\end{equation}
If we choose instead $E=\rO_{G/\mF_0}(1)\otimes \pi^*(\rO_{\mF_0}(2h+2f)\oplus
\rO_{\mF_0}^{\oplus4})$, the corresponding Calabi-Yau threefold $X$
has the same Hodge numbers $h^{1,1}(X)=3, h^{2,1}(X)=43$ and nonvanishing intersection numbers
\begin{equation}
  \label{eq:85}
  \begin{aligned}
    J_1^2J_2 &=10, & J_1^2J_3 &=10, & J_1J_2J_3 &=5,\\
    \ch_2(X)J_1 &=60, & \ch_2(X)J_2 &= 24, & \ch_2(X)J_3 &= 24.
  \end{aligned}
\end{equation}
Next, let $P=\mF_1$. We denote the classes of the base and the fiber
by $h$ and $f$, respectively.  We choose $F=\rO_{\mF_0}^{\oplus 5}$
and $E=\rO_{G/\mF_0}(1)\otimes \pi^*(\rO_{\mF_0}(2h+f)\oplus
\rO_{\mF_0}^{\oplus4})$. The corresponding Calabi-Yau threefold $X$
also has Hodge numbers $h^{1,1}(X)=3, h^{2,1}(X)=43$ and nonvanishing intersection numbers
\begin{equation}
  \label{eq:83}
  \begin{aligned}
    J_1^2J_2 &= 15, & J_1J_2^2 &= 5, & J_1^2J_3 &=10, & J_1J_2J_3 &=
    5, \\
    \ch_2(X)J_1 &=60, & \ch_2(X)J_2 &= 36, & \ch_2(X)J_3 &= 24.
  \end{aligned}
\end{equation}
Finally, let $P=\mP^1$. We choose
$F=\rO_{\mP^1}(1)\oplus\rO_{\mP^1}^{\oplus 4}$
and $E=(\rO_{G/\mP^1}(1)^{\oplus 3} \oplus \rO_{G/\mP^1}(2)) \otimes
\pi^*\rO_{\mP^1}(1)$. Note that as opposed to the cases of the genus one
fibrations, the non--trivial part comes from the $\Gr_2(\mC^5)$ fiber
and not from the base. The corresponding Calabi--Yau threefold $X$ has
Hodge numbers $h^{1,1}(X)=2, h^{2,1}(X)=53$ and intersection ring
\begin{equation}
  \label{eq:84}
  \begin{aligned}
    J_1^3 &=7, & J_1^2J_2&=10, & J_1J_2^2&=0, & J_2^3&=0, & \ch_2\cdot
    J_1 &=58, & \ch_2\cdot J_2 &=24.
   \end{aligned}
\end{equation}
where $J_1 = s_1|_{X_2}$ and $J_2=h|_{X_2}$. By~\cite{Oguiso:1993ab}, all of these examples admit
K3 fibrations. In particular, their fibers are K3 surfaces of degree
$10$. This goes beyond the K3 fibrations studied in~\cite{Klemm:2004km}, which
were all realized as complete intersection in toric varieties, and for
that reason only degrees $2$, $4$, $6$, and $8$ can be realized. This
is parallel to the discussion of the genus one fibrations in
Section~\ref{sec:normalcurves}. It would be interesting to study these
K3 fibrations further, see in particular~\cite{Klaewer:2021ab}.

\subsubsection{Remarks}
\label{sec:remarks-Gr}

Some of the constructions and examples we have discussed have appeared previously. 

A simplified version of the construction in Section~\ref{sec:general-constr-Gr} has been discussed
in~\cite{Donagi:2007hi}. In particular, a similar construction of $X_2$ has been given in~\cite[\S
4.2.3]{Donagi:2007hi}. More generally, Calabi--Yau threefolds given as
complete intersections in homogeneous vector bundles on Grassmannians have been classified
in~\cite{Inoue:2019ab}. 

The variety $X_1$ has been constructed in~\cite[\S 3.2]{Inoue:2019jle} as a complete intersection in a
projective join of $\mP^2\times \mP^2$ with $\Gr_2(\mC^5)$. There it
was denoted by $Y_1$. If we take $  (a,b,c) =(2,1,0)$ in~\eqref{eq:40}, then we
  obtain a variety that is isomorphic to $Y_2$
  in~\cite{Inoue:2019jle}. The third example of~\cite[\S
  3.2]{Inoue:2019jle}, called $Y_3$ there, is isomorphic to the first
  of the
  miscellaneous examples.
  Yet another
  construction of $X_1$ has been given in~\cite[\S 5]{Prince:2019vsu}
  in terms of a product of 2--dimensional reflexive
  polytopes.

\subsection{Pfaffian Calabi--Yau threefolds in smooth toric varieties}
\label{sec:pfaffian-calabi-yau}
In this section we discuss the construction of genus one fibrations as Pfaffian Calabi-Yaus. General references for this section are~\cite{Pragacz:1988ab},
~\cite{Fulton:1998ab},~\cite{Weyman:2003ab}, and~\cite{Fulton:1998bc}.

\subsubsection{General construction}
\label{sec:general-constr-Pf}
Let $V$ be a smooth projective variety of dimension $d$, $L \in
\Pic(V)$ a line bundle, and $F$ a vector bundle of rank $r$ on $V$.
A vector bundle morphism $\varphi : F \to F\spcheck \otimes L$ is said
to be skew-symmetric if $\varphi\spcheck \otimes \id_L = (- 1 )
\varphi$. A skew-symmetric morphism $\varphi$ corresponds to an
element $s_\varphi \in \tH^0(V, \bigwedge^2 F\spcheck  \otimes L)$; its
determinant is a section $s_{\det \varphi} \in \tH^0 (V, (\det F \spcheck)^{\otimes
  2} \otimes L^r)$. There are two different cases
depending on the parity of $r$:
If $r$ is even, then there exists a root  $s_{\Pf(\varphi)} \in
\tH^0(V,\det F \spcheck \otimes L^{\frac{r}{2}})$ of $s_{\det
  \varphi}$, the Pfaffian of the morphism $\varphi$.

Recall that for any vector bundle
map $\varphi: E_1 \to E_2$ and $k \in N$, $0 \leq k \leq \min\{\rank
E_1, \rank E_2 \}$, the $k$’th degeneracy locus of $\varphi$ is the set
\begin{equation}
  D_k(\varphi) = \{x \in X \mid \rank \varphi(x) \leq
  k\}.
  \label{eq:109}
\end{equation}
Since a linear map has rank $\leq k$ if and only if all $k +
1$--minors vanish, the set $D_k(\varphi)$ can also be described as the
zero locus of the section $s_{\wedge^{k+1}\varphi} \in
\tH^0(X,\wedge^{k+1}E_1\spcheck\otimes \wedge^{k+1}E_2)$. Therefore,
$s_{\Pf(\varphi)}$ defines the degeneracy locus $D_{r-1}(\varphi) \subset V$. 

Returning to the skew--symmetric map $\varphi: F \to F\spcheck \otimes
L$, if $r$ is odd, then $\det \varphi \equiv 0$. The first
nontrivial degeneracy locus in $V$ is therefore $D(\varphi) = D_{r-2}(\varphi) =
D_{r-3}(\varphi)$. The expected codimension of $D_k(\varphi)$ in $V$
is $\binom{r-k+1}{2}$. Therefore, the expected codimension of
$D(\varphi)$ is $3$. Moreover, the singular locus of $D(\varphi)$ is
$D_{r-4}(\varphi)$ which is of expected codimension 10.

If $\Pic(L)$ has no 2--torsion and $\codim_{V} D(\varphi) = 3$ then
there exists~\cite{Okonek:1994ab} an exact sequence 
\begin{equation}
  \label{eq:45}
  0 \to \det F\otimes L^{\otimes-\frac{r-1}{2}} \xrightarrow{\psi^T} F \xrightarrow{\varphi} 
  F\spcheck\otimes L \xrightarrow{\psi} \rI_{D(\varphi)} \otimes \det F\spcheck \otimes
  L^{\otimes\frac{r+1}{2}} \to 0,
\end{equation}
where $\rI_{D(\varphi)}$ is the ideal sheaf of $D(\varphi)$. The map
$\psi$ is given by $\psi = \frac{1}{k!} \wedge^k\varphi$. 
This is the global version of the Buchsbaum--Eisenbud complex~\cite{Buchsbaum:1977ab}. 

A codimension 3 subvariety $Y \subset V$ in a smooth projective
variety $V$ of dimension $d$ is called a Pfaffian subvariety if there exist
bundles $L \in \Pic(V)$, $F$ of odd rank $r$ over $V$, and a
skew-symmetric morphism $\varphi : F \to F\spcheck \otimes L$ such that $Y =
D(\varphi)$. In the remainder of this text, therefore, $r$ will be
assumed to be odd, and we will sometimes write $Y=\Pf(\varphi)$ for a
Pfaffian variety. 

If $Y$ a smooth Pfaffian variety then the sequence~\eqref{eq:45} yields
a resolution of $\rI_Y$ as follows:
\begin{equation}
  \label{eq:resIY}
  0 \to L_0 \to F_0 \to F_0\spcheck \otimes L_0 \to \rI_Y \to 0,
\end{equation}
where $L_0 = (\det F)^{\otimes 2} \otimes L^{\otimes -r}$ and $F_0 = F
\otimes \det F \otimes L^{\otimes -\frac{r+1}{2}}$. This is the analog
for the Pfaffian variety of the Koszul complex~\eqref{eq:resIX} for
complete intersections. In fact, for $r=3$ the
resolution~\eqref{eq:resIY} reduces to~\eqref{eq:resIX} by replacing
$F$ in~\eqref{eq:resIY} by $F\otimes \det F\spcheck \otimes L$. It
follows that every Pfaffian variety $Y$ with $r=3$ is a complete
intersection. Therefore, one gets new varieties for $r\geq 5$. Also note that if $M$ is a
line bundle on $V$, this construction is invariant under tensoring $F$
by $M$ and $L$ by $M^{\otimes 2}$ if the map $\varphi$ is modified accordingly.

Moreover, one can show that the canonical sheaf  of $Y$ is~\cite{Okonek:1994ab}
\begin{equation}
  \omega_Y = L_0\spcheck \otimes \omega_V \otimes \rO_Y.
\end{equation}
Therefore, in order for $Y$ to be a Calabi--Yau threefold, we need to
require 
\begin{equation}
  d=\dim V=6 \qquad \text{and} \qquad L_0 \cong \omega_V.
\end{equation}
More generally, in order to obtain Calabi-Yau threefolds starting from
projective varieties $V$ of dimension $\dim V > 6$, we can apply
combinations of complete intersections and Pfaffian subvarieties in order to
achieve the desired codimension. This will play a role in the
description of homological projective duality in
Section~\ref{sec:homol-proj-dual}. There
$V$ will be a projective bundle over a surface $P$ with $\dim V=11$. Then
we take a relative linear section of codimension 5 in $V$ of a
Pfaffian variety $Y' \subset V$. The resulting variety $Y$ has codimension
$5+3=8$ in $V$, hence $\dim Y=3$.

By the generalization of Bertini's theorem of~\cite{Okonek:1994ab},
the Pfaffian variety $Y$ will be smooth if $d \leq 9$ and if
$\wedge^2F\spcheck\otimes L$ is generated by its global sections.

Note that the bundles in~\eqref{eq:resIY} can be rewritten in terms of Schur functors
~\cite[\S
V.2.5]{Boffi:2006ab}, so that the exact sequence reads
\begin{equation}
  \label{eq:resIYV}
  0 \to \mS_{(2^r)}F \otimes L^{-2k-1} \to \mS_{(2,1^{r-1})}F \otimes L^{-k-1} \to \mS_{(1^{r-1})}F \otimes L^{-k} \to \rI_Y
        \to 0,
 \end{equation}
where we have set $r=2k+1$. In terms of these Schur functors, the
square of the ideal sheaf $\rI_Y^2$ admits the following minimal free resolution\footnote{In ~\cite[\S
V.2.5]{Boffi:2006ab} the
  contributions from $\det F$ and $L$ in both~\eqref{eq:resIY}
  and~\eqref{eq:resIYV} are ignored. In~\cite[\S 6.4]{Weyman:2003ab}
  there seem to be some errors in the partitions
  for~\eqref{eq:resIYV}. } ~\cite[\S 6.4]{Weyman:2003ab},~\cite[\S
V.2.5]{Boffi:2006ab} (see also~\cite{Kanazawa:2012ab} in the case $h^{1,1}(Y)=1$) 
\begin{equation}
  \label{eq:resIYV2}
  0 \to \mS_{(3^2,2^{r-2})}F\otimes L^{-2k-2}\to \mS_{(3,2^{r-2},1)}F\otimes L^{-2k-1} \to
  \mS_{(2^{{r-1}})}F\otimes L^{-2k} \to \rI_Y^2 \to 0.
\end{equation}
The knowledge of the resolutions of $\rI_Y$ and $\rI_Y^2$ allows us to
determine the push forward of the conormal bundle of $Y$ in $V$:
\begin{equation}
  \label{eq:NYV}
  0 \to \rI_Y^2 \to \rI_Y \to \iota_*N_{Y/V}\spcheck \to 0,
\end{equation}
where $\iota: Y \to V$ denotes the embedding of $Y$ into $V$.

\subsubsection*{Intersection ring and Chern classes}
To compute the intersection ring of $Y$ we need the (Chern character
of the) normal bundle $N_{Y/V}$. To this end, we apply the Riemann--Roch theorem for a closed embedding of non--singular varieties~\cite[\S
15.2]{Fulton:1998ab}. Given any holomorphic
vector bundle on $E$ on $Y$ we have
\begin{equation}
  \label{eq:RRembedding}
  \chern( i_*E) = \iota_*\left(\chern(E) \todd(N_{Y/V})^{-1}\right).
\end{equation}
In our context, we choose $E=N_{Y/V}\spcheck$. Since $\iota_*1 = [Y]$
is the fundamental class of $X$, the left hand side will be of the
form $[Y]$ times a polynomial $\alpha$ in $\tH^*(V,\mC)$ of degree
3. We can then determine the Chern classes $\ch_i(N_{Y/V}\spcheck)$ by
solving the equation $\alpha=\chern(N_{Y/V}\spcheck)
\todd(N_{Y/V})^{-1}$ in the ring $\tH^*(V,\mC)$. The fundamental class $[Y]$
will be used to compute the intersection numbers on $Y$ by the
standard formula
\begin{equation}
  \label{eq:54}
  \int_Y \iota_*\alpha = \int_V [Y] \cup \alpha.
\end{equation}
Finally, using the normal bundle sequence
\begin{equation}
  \label{eq:27}
  0 \to T_Y \to T_V|_Y \to N_{Y/V} \to 0, 
\end{equation}
we can determine the (Chern character of the) tangent bundle $T_Y$ of
$Y$ given the knowledge of $T_V$. If $Y$ is a Calabi--Yau threefold,
the remaining topological invariants of $Y$ can also be determined.

\subsubsection{Examples}
\label{sec:examples-Pf}
We will be studying a number of examples of this
construction which parallel the examples given in Section~\ref{sec:examples-Gr}.

\subsubsection*{Example: elliptic curve $Y_0$ and construction of threefolds}
First, let $V=\mP^4$,
$L=\rO_{\mP^4}(1)$, $F = \rO_{\mP^4}^{\oplus 5}$, and $\varphi:
\rO_{\mP^4}^{\oplus 5}\to  \rO_{\mP^4}^{\oplus 5}(1)$ be a generic
skew--symmetric map. Then we have $L_0 = \rO_{\mP^4}(-5) \cong \omega_{\mP^4}$, so
$Y_0=\Pf(\varphi)$ is an elliptic curve of degree $5$ in
$\mP^4$. Hence we have recovered the  Pfaffian representation of the
elliptic normal curve of Section~\ref{sec:normalcurves}.  For more
details, see see e.g.~\cite{Fisher:2010ab}.

The next examples will turn out to be relative versions of this
example whose total space is a Calabi--Yau threefold.
To construct such threefolds, we take a vector bundle $E'$ of
rank $5$ over $\mP^2$, and let $V$ be the total space
of the corresponding projective bundle $\pi: \mP(E') \to \mP^2$. Note
that this places us again into the framework of Grassmann bundles,
moreover, $\mP(E')$ is also a toric varirety. We will
denote the relative hyperplane class by $H$ and (the pullback to $V$
of) the hyperplane class on $\mP^2$ by $h$.

For various choices of a rank 5 bundle $F$ and a line bundle $L$ on
$V$, we will construct Pfaffian subvarieties $\Pf(\varphi)$ in $V$
which are Calabi--Yau. In our examples, $F$ will be the pullback of  a
homogeneous vector bundle on $\mP^2$. In fact, the roles of $E$ and
$F$ are interchanged (with $E$ replaced by $E\spcheck$) as compared to the examples in
Section~\ref{sec:examples-Gr}. We will dwell on the interpretation of
this interchange in Section~\ref{sec:homol-proj-dual}.

The choice of $L$ is related to the choice of
$E'$ as follows. Recall that $\mP(E'') \cong \mP(E')$ if $E''=E'\otimes M'$ for some line
bundle $M'$ on $\mP^2$, so that $\omega_{\mP(E'')} = \omega_V\otimes \pi^*\det
M'^{-\otimes 5}$.   Suppose $F$ and $L$ have been chosen such that the
Calabi--Yau condition $L_0 \cong \omega_V$ on $\mP(E')$ is
satisfied. If we define $L' = L\otimes \pi^*M'$ on $\mP(E'')$ and $L_0'
= (\det F)^{\otimes 2} \otimes L'{}^{-\otimes 5}$, then $L_0' =
L_0\otimes \pi^* M'{}^{-\otimes 5}$, so that the Calabi--Yau condition on
$\mP(E'')$ is automatically satisfied: $L_0' \cong
\omega_{\mP(E'')}$. Therefore, if tensoring $E'$ by $M'$ is accompanied
by tensoring $L$ by $\pi^*M'$, the two Pfaffian subvarieties $\Pf(\varphi)
\in V$ and $\Pf(\varphi') \in \mP(E'')$ will be isomorphic. 

\subsubsection*{Example: $Y_2$}
For the first example, let
\begin{equation}
  \label{eq:104}
  E'=\rO_{\mP^2}(-1)^{\oplus 5}
\end{equation}
so that $V=\mP^4\times \mP^2$. We have $\omega_{V} = \rO_{V}(-5H+2h)$. Let
$L=\rO_V(H)$ and 
\begin{equation}
  \label{eq:41}
  F =
  \pi^*\left(\rO_{\mP^2}^{\oplus 4} \oplus \rO_{\mP^2}(1)\right),
\end{equation}
and consider a skew-symmetric morphism $\varphi: F \to F\spcheck
\otimes L$. We set $Y_2 = \Pf(\varphi)$. We compute $L_0 = \det
F^{\otimes 2}\otimes L^{-\otimes 5} = \rO_{V}(-5H+2h)$, hence $Y_2$ is
a Calabi--Yau threefold.
Applying the methods from Appendix~\ref{sec:calculus-with-schur} to~\eqref{eq:resIYV} we find
that  the  ideal sheaf $\rI_{Y_2}$ of $Y_2$ admits the minimal free resolution
$0\to\rF_{-3}^1\to\rF_{-2}^1\to \rF_{-1}^1\to \rI_Y^2\to 0$ with
\begin{equation}
  \label{eq:46}
  \begin{aligned}
    \rF_{-3}^1 &= \pi^*\rO_{\mP^2}(2)\otimes \rO_{V/\mP^2}(-5), \\
    \rF_{-2}^1 &= 
    \pi^*(\rO_{\mP^2}(2)\oplus \rO_{\mP^2}(1)^{\oplus 4} )\otimes \rO_{V/\mP^2}(-3), \\
    \rF_{-1}^1 &=
    \pi^*(\rO_{\mP^2}(1)^{\oplus
      4}\oplus \rO_{\mP^2})\otimes \rO_{V/\mP^2}(-2).
  \end{aligned}
\end{equation}
From~\eqref{eq:46} we determine the fundamental class of $Y_2$ to be
\begin{equation}
  \label{eq:51}
  [Y_2] = -6\,h{H}^{2}+2\,{h}^{2}H+5\,{H}^{3} \in \tH^6(\mP^4\times \mP^2,\mZ)\,.
\end{equation}
Similarly, we find that the  ideal sheaf $\rI^2_{Y_2}$ of $Y_2$ admits the minimal free resolution
$0\to\rF_{-3}^2\to\rF_{-2}^2\to \rF_{-1}^2\to \rI_Y^2\to 0$ with
\begin{equation}
  \label{eq:49}
  \begin{aligned}
    \rF_{-3}^2 &= \pi^*(\rO_{\mP^2}(3)^{\oplus 4} \oplus
    \rO_{\mP^2}(2)^{\oplus 6} )\otimes \rO_{V/\mP^2}(-6), \\
    \rF_{-2}^2 &= 
    \pi^*(\rO_{\mP^2}(3)^{\oplus 4} \oplus \rO_{\mP^2}(2)^{\oplus
      16}\oplus \rO_{\mP^2}(1)^{\oplus 4} )\otimes \rO_{V/\mP^2}(-5),\\
    \rF_{-1}^2 &=
    \pi^*(\rO_{\mP^2}(2)^{\oplus 10} \oplus \rO_{\mP^2}(1)^{\oplus
      4}\oplus \rO_{\mP^2})\otimes \rO_{V/\mP^2}(-4).
  \end{aligned}
\end{equation}
Using~\eqref{eq:NYV} the left hand side of~\eqref{eq:RRembedding} becomes
\begin{equation}
  \label{eq:52}
  \chern(\iota_*N_{Y_2/V}\spcheck) = [Y_2] (3-{\tfrac {25}{2}}\,H+5\,h+{\tfrac {107}{4}}\,{H}^{2}-{\tfrac {107}{5}}\,
hH+{\tfrac {231}{50}}\,{h}^{2}-{\tfrac {469}{12}}\,{H}^{3}+{\tfrac {469}{
10}}\,h{H}^{2}-{\tfrac {6061}{300}}\,{h}^{2}H)
\end{equation}
while the right hand side is
\begin{equation}
  \label{eq:53}
  \chern(N_{Y_2/V}\spcheck) \todd(N_{Y_2/V})^{-1} = 3 -\tfrac{5}{2}\,c_{{1}}-\tfrac{5}{4}\,c_{{2}}+\tfrac{3}{2}\,{c_{{1}}}^{2}-\tfrac{1}{2}\,c_{{3}}+{\tfrac {29}{24}}\,c_{{1}}c_{{2}}-{\tfrac {17}{24}}\,{c_{{1}}}^{3},
\end{equation}
so that the Chern classes $c_i = \ch_i(N_{Y_2/V}), i=1,2,3$, of the
normal bundle can be read off. The Chern classes of the tangent bundle
of $V=\mP^4\times \mP^2$ are straightforward to compute. In the end,
we find, using the basis $J_1 = H|_{Y_2} - h|_{Y_2}$ and $J_2 =
h|_{Y_2}$ for $\tH^2(Y_2,\mZ)$, for the intersection numbers and the
second Chern class of $Y_2$
\begin{equation}
  \label{eq:55}
  \begin{aligned}
    J_1^3 &=5, & J_1^2J_2&=9, & J_1J_2^2&=5, & J_2^3&=0, & \ch_2\cdot
    J_1 &=50, & \ch_2\cdot J_2 &=36.
  \end{aligned}
\end{equation}
Using the criterion of~\cite{Oguiso:1993ab} we see that the projection to $\mP^2$ exhibits a genus
one fibration structure with a 5--section on $Y_2$. In
fact, the fiber is precisely the elliptic curve $Y_0$ given in the
first example.

\subsubsection*{Example: $Y_1$}
Next, let $E'=\rO_{\mP^2}^{\oplus 2}\oplus\rO_{\mP^2}(-1)^{\oplus
  3}$. Then we have $\omega_{V} = \rO_{V}(-5H)$. Let $L=\rO_{V}(H)$ and
$F=\rO_{V}^{\oplus 5}$ be the trivial bundle of rank 5 and
consider the Pfaffian variety $Y_1=\Pf(\varphi)$. 
We compute $L_0 = \det
F^{\otimes 2}\otimes L^{-\otimes 5} = \rO_{V}(-5H)$, hence $Y_1$ is
a Calabi--Yau threefold.
Since $F$ is the trivial it suffices to apply the Weyl character
formula~\eqref{eq:WeylCharacter} to~\eqref{eq:resIYV} to determine the
minimal free resolution $0\to\rF_{-3}^1\to\rF_{-2}^1\to \rF_{-1}^1\to \rI_{Y_1}^2\to 0$ of the  ideal sheaf $\rI_{Y_1}$:
\begin{equation}
  \label{eq:56}
  \begin{aligned}
    \rF_{-3}^1 &= \pi^*\rO_{\mP^2}\otimes \rO_{V/\mP^2}(-5), \\
    \rF_{-2}^1 &= 
    \pi^*\rO_{\mP^2}^{\oplus 5}\otimes \rO_{V/\mP^2}(-3), \\
    \rF_{-1}^1 &=  \pi^*\rO_{\mP^2}^{\oplus 5} \otimes \rO_{V/\mP^2}(-2).
  \end{aligned}
\end{equation}
From~\eqref{eq:46} we determine the fundamental class of $Y_1$ to be
\begin{equation}
  \label{eq:57}
  [Y_1] = 5\,{H}^{3} \in \tH^6(\mP^4\times \mP^2,\mZ).
\end{equation}
Similarly, we find that the ideal sheaf $\rI^2_{Y_1}$ of $Y_1$ admits the minimal free resolution
$0\to\rF_{-3}^2\to\rF_{-2}^2\to \rF_{-1}^2\to \rI_{Y_1}^2\to 0$ with
\begin{equation}
  \label{eq:58}
  \begin{aligned}
    \rF_{-3}^2 &= \pi^*\rO_{\mP^2}^{\oplus 10} \otimes \rO_{V/\mP^2}(-6), \\
    \rF_{-2}^2 &= \pi^*\rO_{\mP^2}^{\oplus 24} \otimes \rO_{V/\mP^2}(-5),\\
    \rF_{-1}^2 &= \pi^*\rO_{\mP^2}^{\oplus 15} \otimes \rO_{V/\mP^2}(-4).
  \end{aligned}
\end{equation}
Using~\eqref{eq:NYV} the left hand side of~\eqref{eq:RRembedding} becomes
\begin{equation}
  \label{eq:59}
  \chern(\iota_*N_{Y_1/V}\spcheck) = [Y_1] (3-{\tfrac {25}{2}}\,H+{\tfrac {107}{4}}\,{H}^{2}-{\tfrac {469}{12}}\,{H}^
{3})
\end{equation}
We proceed as in the previous example and find, using the basis $J_1 = H|_{Y_1} - h|_{Y_1}$ and $J_2 =
h|_{Y_1}$ for $\tH^2(Y_1,\mZ)$, for the intersection numbers and the second Chern class of $Y_1$
\begin{equation}
  \label{eq:73}
  \begin{aligned}
    J_1^3 &=0, & J_1^2J_2&=5, & J_1J_2^2&=5, & J_2^3&=0, & \ch_2\cdot
    J_1 &=36, & \ch_2\cdot J_2 &=36.
  \end{aligned}
\end{equation}
This shows that the projection to $\mP^2$ exhibits a genus
one fibration structure with a 5--section on $Y_1$. Again, the fiber
is the elliptic normal curve $Y_0$ given in the
first example. Moreover, $Y_1$ admits a second independent genus one
fibration with a 5--section.


\subsubsection{Remarks}
\label{sec:remarks-Pf}

Some of the constructions and examples have also appeared
previously. The example $Y_2$ is found in~\cite[\S
3.2]{Caldararu:2002ab}. An alternative construction of the variety $Y_1$ as a complete intersection in a
projective join of $\mP^2\times \mP^2$ with $\Gr_2(\mC^5)$ has been
given in~\cite[\S 3.2]{Inoue:2019jle}. There it was denoted by
$X_1$. If we replace the two summands $\rO_{\mP^2}(-1)^{\oplus
  2}$ in $E'$ by $\rO_{\mP^2}(-2)\oplus\rO_{\mP^2} $, then we
  obtain a variety that is isomorphic to $X_2$
  in~\cite{Inoue:2019jle}. An alternative construction of the variety
$Y_2$ has also been given in~\cite[Table 1,
(t.1)]{Benedetti:2020ab}.  In fact, it shown in~\cite{Benedetti:2020ab} that
the two constructions in this paper, complete intersections in Grassmann bundles and
Pfaffian subvarieties in Grassmann bundles are just special cases of the
far more general notion of orbital degeneracy loci. It would
interesting to study their realization in terms of GLSMs.

This last construction is also closely related to a very general
construction of determinantal varieties in terms of a complete
intersection of high codimension in a Grassmann bundle whose base is
the variety $V$. In the context of the GLSM for determinantal varieties this
construction has been discussed in \cite{Jockers:2012zr}. We briefly
review this construction for the special case of our Pfaffian varieties. For
details see~~\cite{Fulton:1998ab,Tjotta:2001ab,Weyman:2003ab,Benedetti:2020ab}.

Consider the map $\varphi : F \to F\spcheck\otimes L$ and its
$k$-th degeneracy locus $D_k(\varphi) \subset V$. Let $r=\rank F$ and
$k' = r-k$ and consider the Grassmann bundle $G=\Gr_{k'}(F)$ with the
projection $\pi: G \to V$, cf. Section~\ref{sec:some-title}. We can
tensor the short exact sequence~\eqref{eq:tautological} by $\pi^*(F
\otimes L)$ and obtain
\begin{equation}
  \label{eq:30}
  0 \to S\otimes \pi^*(F\otimes L) \to \pi^*(F\otimes F\otimes L) \to
  Q\otimes \pi^*(F\otimes L) \to 0. 
\end{equation}
As discussed in Section~\ref{sec:general-constr-Pf}, the map  $\varphi$ induces a section $\Pf(\varphi)
\in \tH^0(V,\wedge^2F\otimes L)$. We view $\wedge^2 F$ as a subbundle
of $F\otimes F$ and $\wedge^2S$ as subbundle of $S\otimes S$ which,
by tensoring~\eqref{eq:tautological} by $S$, is a subbundle of $S\otimes \pi^* F$. In
this way, we get an induced short exact sequence
\begin{equation}
  0 \to \wedge^2S \otimes \pi^*L \to \pi^*(\wedge^2F \otimes L) \xrightarrow{p} B
  \to 0.
  \label{eq:32}
\end{equation}
This defines the bundle $B$ as the quotient of $\pi^*(\wedge^2F
\otimes L)$ by $\wedge^2S\otimes \pi^*L$. The section $\varphi$ pulls back to a
section $\pi^*\varphi$ of $\pi^*(\wedge^2F \otimes L) $ so that we
obtain a section $s=p\circ \pi^*\varphi \in \tH^0(G,B)$. Now, one can
show that $Z(s) \subset G$ is birationally isomorphic to $D_{k}(\varphi) \subset
V$, i.e. we have a diagram
\begin{equation}
  \label{eq:31}
  \begin{tikzcd}
    Z(s) \ar[r, hook]\ar[d,"\cong"] & G\ar[d,"\pi"]\\
    D_{k}(\varphi) \ar[r,hook] & V
  \end{tikzcd}
\end{equation}
In particular, if $D_{k}(\varphi)$ is smooth, then $Z(s)$ is
isomorphic to $D_{k}(\varphi)$ and realizes it as a complete
intersection in $G$. 
For comparison with the GLSM in Section~\ref{sec:glsm1nadual}, it
is useful to write $Z(s)$ in terms of local coordinates and local
sections. For concreteness, let $V$ be a $\mP^4$ bundle over $\mP^2$
as in Section~\ref{sec:examples-Pf} with local coordinates $p_1,\dots,p_5$ in the
$\mP^4$ fiber and $b_1,b_2,b_3$ in the $\mP^2$ base. Let $F$ be a
vector bundle of rank $5$. For a point $p
\in V$ with these coordinates we choose coordinates $\widetilde x_i
\in \tH^0(G_p,S_p)$ on the Grassmann fiber $G_p$, $i=1,\dots,5$. Then,
the first degeneracy locus of the map
$\varphi$ is $D_{r-2}(\varphi)$, i.e.~we have $k'=2$. The
corresponding section is realized in a local trivalization for $F$ by an
antisymmetric $5\times 5$ matrix $A^{ij}$ that is linear in $p^k$ and
$b_l$, i.e.~the section $\pi^*\varphi$ is locally described as
$A^{ijl}_kp^kb_l$. Next, observe that $\wedge^2S$ has rank 1 and a
section of it is given by $\varepsilon_{ab} \widetilde x^a_i
\widetilde x^b_j$. By~\eqref{eq:32} we know that $\wedge^2S\otimes L$ embeds
into $\pi^*(\wedge^2F \otimes L)$, hence one of the
sections $A^{ijl}_kp^kb_l$ equals $\varepsilon_{ab} \widetilde x^a_i
\widetilde x^b_j$. Therefore, the sections of the bundle $B$ take the
form
\begin{equation}
  \label{eq:33}
  s_{ij} = A^{ijl}_kp^kb_l - \varepsilon_{ab} \widetilde x^a_i
\widetilde x^b_j, \qquad 1\leq i< j \leq 5.
\end{equation}
and one of them is identically zero.

\subsection{Homological projective duality}
\label{sec:homol-proj-dual}

In this section we reformulate the constructions of the Calabi--Yau
threefolds $X$ and $Y$ in terms of the language of homological
projective duality of Kuznetsov~\cite{Kuznetsov:2007ab}. Homological projective
duality is a vast generalization of classical projective duality for
subvarieties of projective spaces to arbitary varieties with an
algebraic morphism to projective space, and their linear
sections. This generalization often involves a noncommutative, i.e.~categorical setting. For a review, see~\cite{Kuznetsov:2014ab}.

In our case, the starting point is the classical projective duality
between $\Gr_2(\mC^5)$ with its Pl\"ucker embedding into $\mP^9$ and
again $\Gr_2(\mC^5)$, viewed as the zero locus of the Pfaffians of the
diagonal minors of a skew-symmetric $5\times5$ matrix of linear forms
on $\mC^5$. More abstractly, let $V$ be a vector space of dimension
$5$, $L \subset \wedge^2V\spcheck$ a linear subspace of dimension
$5$. An overview of the constructions of the elliptic normal curves of
degree 5 $X_0$ in
Section~\ref{sec:examples-Gr} and $Y_0$ in~\ref{sec:examples-Pf} is given in the following diagram
\begin{equation}
  \label{eq:76}
  \begin{array}{r|l}
  \begin{tikzcd}
    & \mP(\wedge^2V) \ar[dr,hookleftarrow] & \\
    \mP(L^\perp) \ar[dr,hookleftarrow] \ar[ur,hookrightarrow] & & \Gr(2,V) \\
    & X_0 \ar[ur,hookrightarrow]& 
  \end{tikzcd}&
  \begin{tikzcd}
    & \mP(\wedge^2V\spcheck) \ar[dr,hookleftarrow] & \\
    \Pf(2,V\spcheck) \ar[dr,hookleftarrow] \ar[ur,hookrightarrow] & & \mP(L) \\
    & Y_0 \ar[ur,hookrightarrow] & 
  \end{tikzcd}                
  \end{array}
\end{equation}
The classical projective duality~\cite{Gelfand:2000ab} in the middle row states that~\cite{Borisov:2009ab}
\begin{equation}
  \label{eq:77}
  \Pf(2,V\spcheck) = \{ \omega \in \mP(\wedge^2V\spcheck) \mid
  \Gr(2,V) \cap H_\omega \text{ is singular}\}. 
\end{equation}
where $H_\omega \subset \mP(\wedge^2V)$ is the hyperplane given by the
point $\omega \in \mP(\wedge^2V\spcheck)$. Furthermore, homological
projective duality in the bottom row states that~\cite{Kuznetsov:2007ac}
\begin{equation}
  \tD^{\tb}(\Coh X_0) \cong \tD^{\tb}(\Coh Y_0).
  \label{eq:78}
\end{equation}
This is, of course, obvious since $X_0$ and $Y_0$ are isomorphic.

We propose a relative version of this picture as follows: Consider the
set--up of Section~\ref{sec:some-title}: a rank 5 bundle $F$ on a
surface $P$, e.g. $P=\mP^2$. The relative Grassmannian $G=\Gr_2(F)$ embeds into the projective
bundle $\mP(\wedge^2F)$ via the Pl\"ucker embedding. Next, consider the
Pfaffian subvariety $\Pf_2(F\spcheck)$ in the dual projective bundle
$\mP(\wedge^2F\spcheck)$.  Finally, choose a subbundle
$E$~\footnote{So far we have used the notation
  $E=\rO_{G/\mP^2}(1)\otimes \pi^*E'$. Since in this subsection the
  first factor plays no role, we drop the prime on $E'$ for ease of notation.} in
$\wedge^2 F$ of rank 5 satisfying
\begin{equation}
  \label{eq:25}
  \omega_P \otimes \det(F\spcheck)^{\otimes 2} \otimes \det E = \rO_P,
\end{equation}
and consider its orthogonal complement
$E^\perp \subset \wedge^2 F$ which is also a rank 5 bundle on
$P$. Then, we consider the following intersections, or, more precisely,
fiber products (cf. e.g.~\eqref{eq:34}):
\begin{equation}
  \label{eq:24}
  \begin{aligned}
   X=X_{E,F} &= \Gr_2(F) \cap \mP(E^\perp),\\
   Y=Y_{E,F} &= \Pf_2(F\spcheck) \cap \mP(E\spcheck).\\    
  \end{aligned}
\end{equation}
The varieties $X$ and $Y$ are precisely the Calabi--Yau threefolds
that have been constructed in Sections~\ref{sec:examples-Gr}
and~\ref{sec:examples-Pf}, respectively. Note that role of $E$ and $F$
is interchanged in our constructions: On the Grassmannian side, $F$ serves for
the construction of the ambient variety and $X$ is built from a
section of $E$. On the Pfaffian side, $E\spcheck$ serves for
the construction of the ambient variety and $Y$ is built from a
section of a bundle derived from $F$, to be precise $\wedge^2 F\spcheck$. 

We therefore have the following relative analog of~\eqref{eq:76}
\begin{equation}
  \label{eq:79}
  \begin{array}{r|l}
  \begin{tikzcd}
    & \mP(\wedge^2F) \ar[dr,hookleftarrow] & \\
    \mP(E^\perp) \ar[dr,hookleftarrow] \ar[ur,hookrightarrow] & & \Gr(2,F) \\
    & X \ar[ur,hookrightarrow]& 
  \end{tikzcd}&
  \begin{tikzcd}
    & \mP(\wedge^2F\spcheck) \ar[dr,hookleftarrow] & \\
    \Pf(2,F\spcheck) \ar[dr,hookleftarrow] \ar[ur,hookrightarrow] & & \mP(E\spcheck) \\
    & Y \ar[ur,hookrightarrow] & 
  \end{tikzcd}                
  \end{array}
\end{equation}
A version of relative homological projective
duality~\cite{Kuznetsov:2007ab} now leads us to the following conjecture, similar to the
examples of the Grassmannian–Pfaffian equivalence in~\cite{Borisov:2009ab} and
the Reye congruence Calabi--Yau 3--fold in~\cite{Hosono:2011np}. The pairs $(X,Y)$
of Calabi--Yau threefolds constructed from $E$ and $F$ are derived equivalent:
\begin{equation}
  \label{eq:26}
  \tD^b(\Coh X) \cong \tD^b(\Coh Y) \,.
\end{equation}
In Section~\ref{sec:towards-class-genus} we will present a list
of pairs $(X,Y)$ obtained from pairs $(E,F)$ satisfying additional
geometric constraints (cf. Table~\ref{tab:1}). By construction, the
conjecture~\eqref{eq:26} applies to all pairs in this list. Note that
for the case of $X_1,Y_1$ a proof of homological projective duality
has been given in~\cite{Inoue:2019jle}.\footnote{To be precise, for a
  pair of varieties that is isomorphic to $X_1,Y_1$.} In particular, by explicitly
working out the birational geometry of $X_1$, $Y_1$ it was shown there
that these varieties are not birationally equivalent. We expect this
to be true for all our pairs in Table~\ref{tab:1}. However, we do not yet
have a sufficient understanding of their birational geometry. At this point, we can rely on the
GLSM. In Sections~\ref{sec:glsmModel1} and~\ref{sec:glsmModel2}  we
will argue that $X$ and $Y$ appear as neighboring phases in a GLSM
determined by the data of $E$ and $F$. This yields strong evidence for
the claim~\eqref{eq:26}. Moreover, we will observe that the phase
boundary between $X$ and $Y$ will have (at least asymptotically) two
components. We will interpret this as a sign that the two geometries
represented by these phases are not birationally equivalent (cf. the
discussion after~\eqref{eqn:glsm1disc}).

\subsection{Computation of the Hodge numbers}
\label{sec:comp-hodge-numb}
In this section we outline how to compute the Hodge numbers of the
Grassmannian and Pfaffian Calabi-Yaus we have constructed. Since
standard technologies from toric geometry do not apply, the
cohomologies are computed via spectral sequences. The Borel-Weil-Bott
theorem and the Schur functor calculus play a crucial role in
obtaining the cohomologies. We first outline the general
procedure. 
Then we move on to a selection of examples. Most of the calculations
are too involved to be done by hand. Useful software is~\cite{Katz:1992ab},~\cite{Stembridge:2005sf}, and~\cite{Schubert2Source}.

\subsubsection{The general procedure}
\label{sec:general-procedure}

We are interested in the Hodge numbers of a variety $X$
\begin{equation}
  h^{i,j}(X) = \dim \tH^j(X, \Omega^i_X).
\end{equation}
For Calabi--Yau threefolds $X$, it suffices to take $i=1$. We will
consider the Calabi--Yau threefold embedded into a Grassmann bundle
$G$ over $P$, either as a complete intersection as in Section~\ref{sec:compl-inters-calabi} or
as a Pfaffian subvariety $Y$ as in Section~\ref{sec:pfaffian-calabi-yau}. In particular, as in
the examples there, we will assume that the variety $V$ will be a
Grassmann bundle over $P$. Therefore, in this subsection we will write
$X$ for $Y$ and $G$ for $V$, in order to treat both cases in parallel.

The conormal bundle sequence allows us to relate $\Omega^1_X$ to the
conormal bundle $N_{X/G}\spcheck$ and the restriction of $\Omega^1_G$
to $X$:
\begin{equation}
  0 \to N_{X/G}\spcheck \to \Omega^1_G|_X \to \Omega^1_X \to 0.
\end{equation}
For other (not necessarily Calabi--Yau) varieties of higher dimensions one has a
relation of $\Omega^i_X$ to $\Omega^i_G|_X$ by taking wedge powers of
this sequence. The strategy to compute the cohomology of $\rE=N_{X/G}$ and
$\rE = \Omega^1_{G}|_X$ is to take a locally free resolution
\begin{equation}
  \label{eq:13}
  0 \to \rF_{-n} \to \dots \to \rF_{0} \to \rE \to 0,
\end{equation}
and to apply the hypercohomology spectral sequence~\cite{griffiths1978principles,Borcea:1983ab}
\begin{equation}
  \label{eq:28}
  E^{p,q}_1 = \tH^q( G, \rF_{p}) \Rightarrow \tH^\bullet(G,\rE)\,.
\end{equation}
If $\rE$ has only support on $X$, as is the case for  $\rE=N_{X/G}$ and
$\rE = \Omega^1_{G}|_X$, the cohomology on the right hand side is
\begin{equation}
  \label{eq:17}
  \tH^\bullet(G,\rE) \cong \tH^\bullet(X,\rE)\,.
\end{equation}
The two cases for $\rE$ and the two cases of the embeddings of $X$ into
$G$ can be treated almost in parallel. Splicing the Koszul complex~\eqref{eq:resIX} and the
Buchsbaum--Eisenbud complex~\eqref{eq:resIY}, respectively,  with the ideal sheaf
sequence
\begin{equation}
  \label{eq:14}
  0 \to \rI_X \to \rO_G \to \rO_X \to 0,
\end{equation}
yields a locally free resolution of $\rO_X$:
\begin{equation}
  \label{eq:15}
    0 \to \rF_{-n} \to \dots \to \rF_{1} \to \rO_G \to \rO_X \to 0.
\end{equation}
Tensoring this resolution with $B=\Omega^1_G$ yields the resolution
for $\rE=\Omega^1_G|_X$.

For the normal bundle we need to make a distinction between the
complete intersection and the Pfaffian subvariety. For the normal
bundle of a complete intersection in $P$, we tensor the
Koszul resolution of $\rO_X$ with $B=E\spcheck$ which yields a
resolution of $N_{X/G}$. For the normal bundle of the Pfaffian
variety, we have to proceed in two steps. The first step consists of
applying the long exact cohomology sequence to the normal bundle
sequence~\eqref{eq:NYV}
\begin{equation}
  \label{eq:16}
  0 \to \rI_X^2 \to \rI_X \to N_{X/P}\spcheck \to 0
\end{equation}
in order to express the cohomology groups $\tH^j(X, N_{X/G})$ in terms
of the cohomology groups $\tH^i(G,\rI_X^m)$, $m=1,2$.
In the second step, the latter are computed by applying the hypercohomology spectral sequence to the locally
free resolutions~\eqref{eq:resIY} and~\eqref{eq:resIYV}, respectively.  

For the computation of the cohomology groups $\tH^j(G,\rF_i)$ which
enter~\eqref{eq:28}, we observe that the $\rF_i$ are direct sums whose
summands are of the form 
\begin{equation}
  \label{eq:18}
  \rF = \rF'' \otimes \pi^*\rF',
\end{equation}
where $\rF'$ is a vector bundle on the base $P$, and
$\rF''$ is a relative homogeneous vector bundle on $G$, cf.~\eqref{eq:87}. The cotangent
bundle $\Omega^1_G$ is not of this form, but it sits in a short exact
sequence of two bundles of this form
\begin{equation}
  0 \to \pi^*\Omega^1_P \to \Omega^1_G \to \Omega^1_{G/P} \to 0.
\end{equation}
Moreover, in all our examples, $P$ will be a homogeneous variety as
well, and for the sake of exposition we assume here that $P$ is in
fact also a Grassmannian. Then, both $\rF'$ and $\rF''$ are
explicitly given in terms of a Schur functor applied to a direct sum
of homogeneous vector bundles as in~\eqref{eq:1}. Hence, these bundles can be decomposed
into irreducible components using~\eqref{eq:S_direct_sum}. Using the
fact that~\cite{Hartshorne:1983ab}
\begin{equation}
  \label{eq:3}
  \tH^q(G, E\oplus F) = \tH^q(G, E) \oplus \tH^q(G, F), 
\end{equation}
it suffices to compute the cohomology of each irreducible component
separately. So we can assume that
\begin{equation}
  \label{eq:9}
  \begin{aligned}
    \rF' &= \mS_{\lambda'} S_P\spcheck \otimes \mS_{\mu'} Q_P,\\
    \rF'' &= \mS_{\lambda''} S_G\spcheck \otimes \mS_{\mu''} Q_G,\\
  \end{aligned}
\end{equation}
where $S_P, Q_P$, $S_G, Q_G$ are the tautological bundles on $P$ and
$G$, respectively.

The cohomology of bundles on the total space $G$ of the Grassmann bundle
$\Gr_k(F) \to P$ can be determined using the Leray spectral sequence
in terms of the cohomology of bundles on the base $P$. The latter states that~\cite{griffiths1978principles}
\begin{equation}
  E^{s,t}_2 = \tH^t(P, \tR^s\pi_*\rF) \Rightarrow \tH^\bullet(G,\rF)
\end{equation}
for a sheaf (of abelian groups) $\rF$ on $G$. For $\rF$ of the
form~\eqref{eq:18} we need to determine
\begin{equation}
  \tH^t\left(P, \tR^s\pi_*\left( \rF'' \otimes \pi^*\rF'\right) \right), 
\end{equation}
and we can apply the projection formula~\cite{Fulton:1998ab}
\begin{equation}
  \label{eq:19}
  \tR^s\pi_*\left( \rF'' \otimes \pi^*\rF'\right) = \tR^s\pi_*\rF'' \otimes \rF'.
\end{equation}
Since the $\rF''$ is of the form~\eqref{eq:9}, the first factor can be
computed using the theorem of Borel--Weil--Bott. If it is nonzero,
then it is of the form $\mS_\nu F\spcheck$ for exaxtly one value of
$s$, where $F$ is the vector
bundle on $P$ underlying the Grassmann bundle $G=\Gr_k(F)$. Hence, the
Leray spectral sequence degenerates at the $E_2$-page and we have
\begin{equation}
  \label{eq:37}
  \tH^{q+s}(G, \rF ) \cong \tH^q(P, \mS_{\nu}F\spcheck
  \otimes \rF').
\end{equation}
It remains to determine the right--hand side of~\eqref{eq:37}.
Since we have assumed that $P$ is a Grassmannian and $\rF'$ is of the
form~\eqref{eq:9} as well, we can again use
the theorem of Borel--Weil--Bott to determine these cohomology groups
as representations of $\GL(V)$. In fact, in most of our examples, $P$
will be a projective space. In this case, the dimension of
$\tH^q\left(P, \mS_\nu F\spcheck \otimes \rF' \right)$ can more
easily be calculated using Bott's formula (\ref{eq:Bottformula}). For most calculations of
the spectral sequences, this is in fact sufficient. If $P$ is a toric
variety and the vector bundles $\mS_\nu F\spcheck \otimes \rF'$ are
direct sums of line bundles, then the dimensions of the cohomology
groups can be computed using e.g. CohomCalg \cite{Blumenhagen:2010pv,cohomCalg:Implementation}. 

In the following, we give the details for the examples $X_1$, $X_4$
from Section~\ref{sec:examples-Gr} and $Y_2$
from Section~\ref{sec:examples-Pf}. Further details for the examples $X_2', X_3$ and
$Y_2$ are worked out in Appendix~\ref{sec:further-examples}.

\subsubsection{Examples}
\subsubsection*{Example $X_1$}
\label{sec:genus-1-example-1}

We consider $P=\mP^2$, $F=\rO_{\mP^2}^{\oplus 5}$ as in~\eqref{eq:38}
and set $G = \Tot( \Gr_2(F) )$. We write $E$ in~\eqref{eq:40} as
\begin{equation}
  E = \rO_G(1) \otimes \pi^*E', \qquad E' = \rO_{\mP^2}(1)^{\oplus 3} \oplus \rO_{\mP^2}^{\oplus 2}\,.
\end{equation}
and $X_1=Z(s), s \in \tH^0(G,E)$. 
The Hodge numbers of $X_1$ are obtained from the Koszul spectral
sequence with $E_1^{i,j} = \tH^j(G,\wedge^{-i} E \otimes B)$ with $B =
E\spcheck, \Omega^1_{G/P}, \pi^*\Omega^1_P$. These cohomology groups are in turn
determined by the Leray spectral sequence for $\pi: G\to P$ with 
\begin{equation}
  E_2^{s,t} = \tH^t(P, \tR^s\pi_*(\wedge^{-i}E\spcheck\otimes B) ), \qquad i=-r,\dots,0.
\end{equation}
The projection formula yields
\begin{equation}
  \label{eq:21}
  \tR^s\pi_*(\wedge^{-i}E\spcheck\otimes B) =
  \begin{cases}
    \tR^s\pi_*\rO_{G/P}(i) \otimes \wedge^{-i}E'{}\spcheck \otimes
    \Omega^1_P & B = \pi^*\Omega^1_P,\\
     \tR^s\pi_*(\rO_{G/P}(i) \otimes \Omega^1_{G/P}) \otimes \wedge^{-i}E'{}\spcheck  & B = \Omega^1_{G/P},\\
     \tR^s\pi_*\rO_{G/P}(i-1) \otimes
     \wedge^{-i}E'{}\spcheck \otimes E'{}\spcheck  & B = E\spcheck.\\
  \end{cases}
\end{equation}
For the first factors in~\eqref{eq:21} we find
\begin{align}
    \tR^s\pi_*\rO_{G/P}(i) &= 
  \begin{cases}
    \mS_{(0,0,0,0,0)} F\spcheck = \rO_{\mP^2}, & (s,i)=(0,0)\\
    \mS_{(2,2,2,2,2)} F\spcheck = \rO_{\mP^2}, & (s,i)=(6,-5)\\
    0 & \text{otherwise}
  \end{cases}\\
    \tR^s\pi_*(\rO_{G/P}(i)\otimes S \otimes Q\spcheck) &= 
  \begin{cases}
    \mS_{(0,0,0,0,0)} F\spcheck = \rO_{\mP^2}, & (s,i)=(1,0)\\
    \mS_{(3,2,2,2,1)} F\spcheck = \rO_{\mP^2}^{\oplus 24}, & (s,i)=(6,-5)\\
    0 & \text{otherwise}
  \end{cases}\\
  \tR^s\pi_*\rO_{G/P}(i-1) &= 
  \begin{cases}
    \mS_{(2,2,2,2,2)} F\spcheck = \rO_{\mP^2}, & (s,i)=(6,-4)\\
    \mS_{(3,3,2,2,2)} F\spcheck = \rO_{\mP^2}^{\oplus 10}, & (s,i)=(6,-5)\\
    0 & \text{otherwise}
  \end{cases}
\end{align}
For the second factors~\eqref{eq:21} we find
\begin{equation}
  \begin{aligned}
    \wedge^{5}E'{}\spcheck &= \rO_{\mP^2}(-3)\\
    \wedge^{4}E'{}\spcheck\otimes E'{}\spcheck &= \rO_{\mP^2}(-4)^{\oplus 6} \oplus \rO_{\mP^2}(-3)^{\oplus 13} \oplus \rO_{\mP^2}(-2)^{\oplus 6}\\
    \wedge^{5}E'{}\spcheck\otimes E'{}\spcheck &= \rO_{\mP^2}(-4)^{\oplus 3} \oplus \rO_{\mP^2}(-3)^{\oplus 2}.\\
  \end{aligned}
  \label{eq:22}
\end{equation}
Therefore, the only nonvanishing contributions to the Leray spectral sequence are
\begin{equation}
  \begin{aligned}
  E_2^{0,1} &= \tH^1(\mP^2, \Omega^1_{\mP^2} ) = \mC, & i &=0\\
  E_2^{6,2} &= \tH^2(\mP^2, \Omega^1_{\mP^2}(-3) ) = \mC^8, & i &=-5,\\
  \end{aligned} 
\end{equation}
for $B=\pi^*\Omega^1_{\mP^2}$ which yields
\begin{equation}
  \tH^j(G, \wedge^{-i}E\otimes \pi^*\Omega^1_{P}) =
  \begin{cases}
    \mC & (i,j) = (0,1)\\
    \mC^8 & (i,j) = (-5,8)\\
    0 & \text{ otherwise.}
  \end{cases}
  \label{eq:105}
\end{equation}
For $B=E\spcheck$, the only nonvanishing contributions to the Leray spectral sequence are
\begin{equation}
  \begin{aligned}
  E_2^{6,2} &= \tH^2(\mP^2, \rO_{\mP^2}(-4)^{\oplus 6} \oplus \rO_{\mP^2}(-3)^{\oplus 13} ) = \mC^{31}, & i &=-4\\
  E_2^{6,2} &= \tH^2(\mP^2, \rO_{\mP^2}(-4)^{\oplus 30} \oplus \rO_{\mP^2}(-3)^{\oplus 20}) = \mC^{110}, & i &=-5.\\
\end{aligned}
\end{equation}
This yields a nontrivial differential in the Koszul spectral sequence
\begin{equation}
  \diff{}{}_1 : E_1^{-5,8} \cong \mC^{110} \to E_1^{-4,8} \cong \mC^{31}.
\end{equation}
One can show that $\coker \diff{}{}_1 = 0$, hence
\begin{equation}
  \tH^q(X,N_{X/P}\spcheck) =
  \begin{cases}
    \mC^{79} & q=3.\\
     0 & \text{otherwise.}\\
  \end{cases}
\end{equation}
This agrees with the computation of $\chi(X,N_{X/P}\spcheck) = -79$ by
the Hirzebruch--Riemann--Roch theorem. 

Finally, for $B=\Omega^1_{G/P}$ the only nonvanishing contributions to the Leray spectral sequence are
\begin{equation}
  \begin{aligned}
  E_2^{1,0} &= \tH^0(\mP^2, \rO_{\mP^2} ) = \mC, & i &=0\\
  E_2^{6,2} &= \tH^2(\mP^2, \rO_{\mP^2}(-3)^{24} ) = \mC^{24}, & i &=-5.\\
\end{aligned}
\end{equation}
Hence, we find
\begin{equation}
  \tH^j(G, \wedge^{-i}E\spcheck\otimes \Omega^1_{G/P}) =
  \begin{cases}
    \mC & (i,j) = (0,1)\\
    \mC^{24} & (i.j) = (-5,8)\\
    0, & (i,j) \text{ otherwise.}
  \end{cases}
\end{equation}
From the long exact cohomology sequences (for each $i$) associated to
\begin{equation}
  \label{eq:50}
  0 \to \wedge^{-i}E\spcheck\otimes \pi^*\Omega^1_{P} \to \wedge^{-i}E\spcheck\otimes \Omega^1_{G} \to \wedge^{-i}E\spcheck\otimes \Omega^1_{G/P} \to 0
\end{equation}
we find\footnote{One can show that the right hand sides
  of~\eqref{eq:105} and~\eqref{eq:106} hold independently of the choice of $E$.}
\begin{equation}
  \tH^j(G, \wedge^{-i}E\spcheck\otimes \Omega^1_{G}) =
  \begin{cases}
    \mC^2 & (i,j) = (0,1)\\
    \mC^{32} & (i,j) = (-5,8) \\
    0, & (i,j) \text{ otherwise.}
  \end{cases}
  \label{eq:106}
\end{equation}
Therefore we get from the Koszul spectral sequence for $\Omega^1_G$
\begin{equation}
  \tH^q(X,\Omega^1_G|_X) =
  \begin{cases}
    \mC^{2} & q=1.\\
    \mC^{32} & q=3\\
     0 & \text{otherwise.}\\
  \end{cases}
\end{equation}
Finally, the long exact cohomology sequence associated to
\begin{equation}
  0 \to N_{X/G}\spcheck \to \Omega^1_G|_X \to \Omega^1_X \to 0
\end{equation}
yields
\begin{equation}
  \begin{aligned}
    h^{1,1}(X) = h^{1}(X,\Omega^1_X) &= 2, &h^{2,1}(X) = h^{1}(X,\Omega^2_X) &= 47. 
  \end{aligned}
\end{equation}

\subsubsection*{Example $X_{4}$}
\label{sec:genus-1-example-2}

This example is special because it will turn out that $h^{1,1}(X) =
6$ instead of $2$. We will see latter that this means that the
5--section splits into five independent sections.

We consider $P=\mP^2$, $F=\rO_{\mP^2}^{\oplus 5}$ and set $G = \Tot(
\Gr_2(F) )$ as before. Now, we take $E = \rO_{G/\mP^2}(1) \otimes \pi^*E'$ with
\begin{equation}
  E' = \rO_{\mP^2}(3) \oplus \rO_{\mP^2}^{\oplus 4}.
\end{equation}
For $B=E\spcheck$, the last two equations in~\eqref{eq:22} change to
\begin{equation}
  \begin{aligned}
    \wedge^{4}E'{}\spcheck\otimes E'{}\spcheck &= \rO_{\mP^2}(-6)^{\oplus 4} \oplus \rO_{\mP^2}(-3)^{\oplus 17} \oplus \rO_{\mP^2}^{\oplus 4}\\
    \wedge^{5}E'{}\spcheck\otimes E'{}\spcheck &= \rO_{\mP^2}(-6) \oplus \rO_{\mP^2}(-3)^{\oplus 4}.\\
  \end{aligned}
\end{equation}
Note the extra term $\rO_{\mP^2}^{\oplus 4}$ in the first line. 
This leads to the following nonvanishing contributions to the Leray spectral sequence 
\begin{equation}
  \begin{aligned}
    E_2^{6,0} &= \tH^0(\mP^2, \rO_{\mP^2}(-6)^{\oplus 4} \oplus \rO_{\mP^2}(-3)^{\oplus 17} \oplus\rO_{\mP^2}^{\oplus 4}) = \mC^{4}, & i &=-4\\
    E_2^{6,2} &= \tH^2(\mP^2, \rO_{\mP^2}(-6)^{\oplus 4} \oplus \rO_{\mP^2}(-3)^{\oplus 17} \oplus\rO_{\mP^2}^{\oplus 4}) = \mC^{57}, & i &=-4\\
  E_2^{6,2} &= \tH^2(\mP^2, \rO_{\mP^2}(-6)^{\oplus 10} \oplus \rO_{\mP^2}(-3)^{\oplus 40}) = \mC^{140}, & i &=-5.\\
\end{aligned}
\end{equation}
The first line will ultimately be responsible for for $h^{1,1}(X)$
to be larger than $h^{1,1}(G) = 2$. The four summands will
correspond to four copies of the section of the genus one fibration
structure on $X$. We again have a nontrivial differential in the Koszul spectral sequence
\begin{equation}
  \diff{}{}_1 : E_1^{-5,8} \cong \mC^{140} \to E_1^{-4,8} \cong \mC^{57}.
\end{equation}
One can show that $\coker \diff{}{}_1 = 0$, hence
\begin{equation}
  \tH^q(X,N_{X/P}\spcheck) =
  \begin{cases}
    \mC^{4} & q=1,\\
    \mC^{83} & q=3,\\
     0 & \text{otherwise.}\\
  \end{cases}
\end{equation}
This agrees with the computation of $\chi(X,N_{X/P}\spcheck) = -79$ by
the Hirzebruch--Riemann--Roch theorem. 

For $B=\Omega^1_{G/P}$ and $B=\pi^*\Omega^1_{P}$ we find the same result as in the previous
example. This yields
\begin{equation}
  \begin{aligned}
    h^{1,1}(X) = h^{1}(X,\Omega^1_X) &= 2+4 = 6, &h^{2,1}(X) = h^{1}(X,\Omega^2_X) &= 51. 
  \end{aligned}
\end{equation}

\subsubsection*{Example $Y_2$}
\label{sec:genus-1-example-4}

Here we consider the Pfaffian Calabi--Yau variety $Y_2$ constructed in
Section~\ref{sec:examples-Pf}. Recall that there the ambient variety
was the projective bundle  $V=\Tot(\mP(E) \to \mP^2)$ with $E=\rO_{\mP^2}(-1)^{\oplus 3}\oplus
\rO_{\mP^2}^{\oplus 2}$. $Y_2$ is the Pfaffian
$Y=\Pf(\varphi)$ in $V$ of a general skew--symmetric morphism $\varphi: F \to
F\spcheck \otimes L$ with $F=\pi^*(\rO_{\mP^2}^{\oplus 4} \oplus
\rO_{\mP^2}(1))$.

Following the discussion in Section~\ref{sec:general-procedure} we have a minimal locally free
resolution of $\rO_{Y_2}$ given by
\begin{equation}
  \label{eq:61}
  0 \to \rF_{-3}^1 \to \rF_{-2}^1 \to \rF_{-1}^1 \to \rO_V \to \rO_{Y_2} \to 0,
\end{equation}
where the $\rF_i^1$ are given in terms of the resolution~\eqref{eq:resIYV} of
$\rI_{Y_2}$.
As in the previous examples, the cohomology of $\Omega^1_V|_{Y_2}$ is obtained from the
hypercohomology spectral
sequence of~\eqref{eq:61} with $E_1^{i,j} = \tH^j(V,\rF^1_{-i} \otimes
B)$ with $B =
\Omega^1_{V/\mP^2}, \pi^*\Omega^1_{\mP^2}$. These cohomology groups are in turn determined by the Leray spectral sequence for $\pi: V\to \mP^2$ with 
\begin{equation}
  E_2^{s,t} = \tH^t(\mP^2, \tR^s\pi_*(\rF^1_{i} \otimes B ) ), \qquad i=-3,\dots,0.
\end{equation}
For the cohomology of the normal bundle, however, we first need to  determine the cohomology groups $\tH^i(V,\rI_Y^m)$ for the
ideal sheaves\footnote{For the parallel between
  the complete intersections and the Pfaffians we temporarily renamed
  $V$ by $G$ and $Y$ by $X$ in Section~\ref{sec:general-procedure}. Here, we return to the
  notation of Section~\ref{sec:general-constr-Pf}.}  $\rI_Y^m$, $m=1,2$.
As in that Section and as in~\eqref{eq:61}, we denote the minimal free
resolutions~\eqref{eq:resIYV} and~\eqref{eq:resIYV2} of
$\rI_Y^m$ by $\rF^m_{\bullet}$, $m=1,2$, respectively. In this case,
we use the respective hypercohomology spectral
sequences with $E_1^{i,j} = \tH^j(V,\rF^m_{i})$. Note that here $i$
runs only from $-3$ to $-1$.  Again, these cohomology groups are in turn determined by the Leray spectral sequence 
\begin{equation}
  \label{eq:63}
  E_2^{s,t} = \tH^t(\mP^2, \tR^s\pi_*\rF^m_{i} ), \qquad i=-3,\dots,-1.
\end{equation}
From~\eqref{eq:resIYV} and~\eqref{eq:resIYV2} we see that each
$\rF^m_i$ is of the form
\begin{equation}
  \label{eq:60}
  \rF^m_i = L^{\otimes a}\otimes \pi^*\mS_{\mu}F, 
\end{equation}
for an integer $a$ and a partition $\mu$, both depending on $m$ and $i$. 
We first observe that by the projection formula
\begin{equation}
  \begin{aligned}
    \tR^s\pi_*\rF^m_i &= \tR^s\pi_*L^{\otimes a} \otimes \mS_{\mu}F,\\
    \tR^s\pi_*(\rF^m_i\otimes \Omega^1_{V/\mP^2}) &=
    \tR^s\pi_*(L^{\otimes a}\otimes S \otimes Q\spcheck) \otimes \mS_{\mu}F.
  \end{aligned}
\end{equation}
Then, by the theorem of Borel--Weil--Bott we find
\begin{equation}
  \label{eq:62}
  \begin{aligned}
    \tR^s\pi_*L^{\otimes a} &=
    \begin{cases}
      \mS_{(0,0,0,0,0)} E\spcheck = \rO_{\mP^2}, & (s,a)=(0,0)\\
      \mS_{(1,1,1,1,1)} E\spcheck = \rO_{\mP^2}(-5), & (s,a)=(4,-5)\\
      \mS_{(2,1,1,1,1)} E\spcheck = \rO_{\mP^2}(-6)^{\oplus 5}, & (s,a)=(4,-6)\\
      0 & \text{otherwise}.
    \end{cases}\\
      \tR^s\pi_*(L^{\otimes a}\otimes S \otimes Q\spcheck) &= 
  \begin{cases}
    \mS_{(0,0,0,0,0)} E\spcheck = \rO_{\mP^2}, & (s,a)=(1,0)\\
    \mS_{(2,1,1,1,0)} E\spcheck = \rO_{\mP^2}(-5)^{\oplus 24}, & (s,a)=(4,-5)\\
    0 & \text{otherwise}.
  \end{cases}
  \end{aligned}
\end{equation}
For $B=\pi^*\Omega^1_{\mP^2}$, the only nonvanishing contributions to the Leray spectral sequence are
\begin{equation}
  \begin{aligned}
  E_2^{0,1} &= \tH^1(\mP^2, \pi^*\Omega^1_{\mP^2}) = \mC, & i &=0\\
  E_2^{4,2} &= \tH^2(\mP^2,  \pi^*\Omega^1_{\mP^2}(-3)) = \mC^8, & i &=-3,\\
\end{aligned}
\end{equation}
which yields
\begin{equation}
  \tH^j(V, \rF_{-i}^1\otimes \pi^*\Omega^1_{\mP^2}) =
  \begin{cases}
    \mC & (i,j) = (0,1)\\
    \mC^8 & (i,j) = (-3,6)\\
    0 & \text{ otherwise.}
  \end{cases}
\end{equation}
For $B=\Omega^1_{V/\mP^2}$ the only nonvanishing contributions to
the Leray spectral sequence are
\begin{equation}
  \begin{aligned}
  E_2^{1,0} &= \tH^0(\mP^2, \rO_{\mP^2} ) = \mC, & i &=0\\
  E_2^{4,2} &= \tH^2(\mP^2, \rO_{\mP^2}(-3)^{\oplus 24} ) = \mC^{24}, & i &=-3.\\
\end{aligned}
\end{equation}
Hence, we find
\begin{equation}
  \tH^j(V, \rF^1_{-i}\otimes \Omega^1_{V/\mP^2}) =
  \begin{cases}
    \mC & (i,j) = (0,1)\\
    \mC^{24} & (i.j) = (-3,6)\\
    0, & \text{ otherwise.}
  \end{cases}
\end{equation}
At this point, it follows as in the previous examples that
\begin{equation}
  \tH^j(V, \rF^1_{-i}\otimes \Omega^1_{V}) =
  \begin{cases}
    \mC^2 & (i,j) = (0,1)\\
    \mC^{32} & (i.j) = (-3,6)\\
    0, & \text{ otherwise.}
  \end{cases}
\end{equation}
Now, we can use the hypercohomology spectral sequence for $\rO_{Y_2}$ tensored with $\Omega^1_V$
\begin{equation}
  \tH^q(Y_2,\Omega^1_V|_{Y_2}) =
  \begin{cases}
    \mC^{2} & q=1.\\
    \mC^{32} & q=3\\
     0 & \text{otherwise.}\\
  \end{cases}
\end{equation}

For the contributions from the normal bundle, we proceed as
follows. From~\eqref{eq:46} and~\eqref{eq:62} we find that the only
nonvanishing contribution to the Leray spectral sequence
in~\eqref{eq:63} is for $m=1$
\begin{equation}
  \begin{aligned}
  E_2^{4,2} &= \tH^2(\mP^2, \rO_{\mP^2}(-3) ) = \mC, & i &=-3
  \end{aligned}
\end{equation}
and from~\eqref{eq:49} for $m=2$
\begin{equation}
  \begin{aligned}
  E_2^{4,2} &= \tH^2(\mP^2, \rO_{\mP^2}(-2)^{\oplus 4} \oplus
  \rO_{\mP^2}(-3)^{\oplus 16} \oplus \rO_{\mP^2}(-4)^{\oplus 4}  ) = \mC^{28}, & i &=-2\\
  E_2^{4,2} &= \tH^2(\mP^2, \rO_{\mP^2}(-3)^{\oplus 20} \oplus \rO_{\mP^2}(-4)^{\oplus 30}  ) = \mC^{110}, & i &=-3.
  \end{aligned}
\end{equation}
This yields a nontrivial differential in the hypercohomology spectral sequence
\begin{equation}
  \diff{}{}_1 : E_1^{-3,6} \cong \mC^{110} \to E_1^{-2,6} \cong \mC^{28}.
\end{equation}
One can show that $\coker \diff{}{}_1 = 0$, hence
\begin{equation}
  \tH^j(V,\rF^2_i) =
  \begin{cases}
    \mC^{28} & (i,j) = (-2,6)\\
    \mC^{110} & (i,j) = (-3,6)\\
    0 & \text{otherwise.}
  \end{cases}
\end{equation}
Hence, we find
\begin{align}
  \tH^q(V,\rI_Y ) &=
  \begin{cases}
    \mC & q=4\\
    0 & \text{ otherwise}.
  \end{cases}\\
  \tH^q(V,\rI_Y^2 ) &=
  \begin{cases}
    \mC^{82} & q=4\\
    0 & \text{otherwise,}
  \end{cases}
\end{align}
from which we conclude that
\begin{equation}
  \tH^q(Y,N_{Y/V}\spcheck) =
  \begin{cases}
    \mC^{81} & q=3.\\
     0 & \text{otherwise.}\\
  \end{cases}
\end{equation}
This agrees with the computation of $\chi(X,N_{X/P}\spcheck) = -81$ by
Hirzebruch--Riemann--Roch. Proceeding as before, we find
\begin{equation}
  \begin{aligned}
    h^{1,1}(Y) = h^{1}(Y,\Omega^1_Y) &= 2, &h^{2,1}(Y) = h^{1}(Y,\Omega^2_Y) &= 49.
  \end{aligned}
\end{equation}

\subsection{Summary}
\label{sec:summary}

In this section we summarize our constructions with an emphasis on
comparing them and with an outlook to a detailed study of the
properties of the genus one fibrations in the later Sections.

\subsubsection{A comparison of the two constructions}
\label{sec:comp-two-constr}

In Section~\ref{sec:homol-proj-dual} we have argued that given two
vector bundles $E$, $F$ on a projective surface $P$, we have two
constructions of genus one fibered Calabi--Yau threefolds with
5--sections: Either a complete intersection determined
by $E$ in a Grassmann bundle determined by $F$, or a Pfaffian
subvariety determined by $F$ in a projective bundle determined by
$E\spcheck$. In Section~\ref{sec:comp-hodge-numb}, we have seen that the
computation of the Hodge numbers for both constructions essentially
goes in parallel, the only difference being in the computation of the
cohomology of the corresponding normal bundles. To exhibit this
parallel more clearly, we present in Table~\ref{tab:Comparison} an
overview of the two constructions and the involved quantities. We see that
the representation theory underlying the homogeneous vector bundles
$E$ and $F$ determines all the cohomology groups. 
\begin{landscape}

\begin{table}[h!]
  \centering
  \begin{tabular}{l|c|c}
    & Grassmannian & Pfaffian \\
    \hline
    base variety & $P$ & $P$ \\
    first vector bundle on $P$ & $F', \rank F'=n$ & $E', \rank E'=r$ \\
    Grassmann bundle $G$ & $\Gr_2(F') \xrightarrow{\pi} P$ & $\Gr_1(E'{}\spcheck) \xrightarrow{\pi} P$ \\
    tautological subbundle & $S$ & $S=\rO(-1)$ \\
    relative ample bundle & $\det S\spcheck = \rO(1)$ & $\det S\spcheck = \rO(1)=L$ \\ 
    canonical class of $G$ & $\pi^*\left(\omega_P \otimes \det(F'{}\spcheck)^{2} \right)\otimes \rO(-n)$ & $\pi^*\left(\omega_P \otimes \det E'\right)\otimes \rO(-r)$ \\
    \hline
    second vector bundle on $P$ & $E', \rank E'=r$ & $F', \rank F'=n=2k+1$ \\
    vector bundle on $G$ & $E = \rO(1) \otimes \pi^*E'$ & $F =\pi^*F'$ \\
    resolution of $\rI_X$ & $0\to \rF_\bullet \to \rI_X \to 0 $ & $0\to \rF_\bullet \to \rI_X \to 0 $ \\
    --- first  term & $\rF_{-1} = E\spcheck = \pi^*E'{}\spcheck \otimes L^{-1}$ & $\rF_{-1} =$ $\pi^*\mS_{(1^{2t})}F' \otimes L^{-k}$\\
    --- intermediate term(s) & $\rF_{-i} = \wedge^{i}E\spcheck = \mS_{(1^i)}E\spcheck$ & $\rF_{-2} = \pi^*\mS_{(2,1^{2t})}F' \otimes L^{-k-1}$\\
    --- last  term & $\rF_{-r} = \det E\spcheck$ & $\rF_{-3} = \pi^*\mS_{(2^{2t+1})}F' \otimes L^{-2k-1}$\\
   \hline
    Calabi--Yau condition $\rF_{\Top} = \omega_G$ & $\omega_P \otimes \det(F'{}\spcheck)^{2} = \det E'{}\spcheck$& $\omega_P \otimes \det   E' = (\det F')^2$\\
    & $n = r$ & $\rO(-r) = L^{-n}$ \\
    \hline
    spectral sequence for $\rI_X$ & $E_1^{i,j} = \tH^j(G,\rF_{-i})$ & $E_1^{i,j} = \tH^j(G,\rF_{-i})$ \\
     & $\Rightarrow  \tH^{i+j}(G,\rI_X)$   & $\Rightarrow  \tH^{i+j}(G,\rI_X)$  \\
    spectral sequence for $\rF_{-i}$ & $E_2^{s,t} = \tH^t(P, \tR^s\pi_*\rF_{-i}) $ & $E_2^{s,t} = \tH^t(P, \tR^s\pi_*\rF_{-i})  $ \\
     & $\Rightarrow  \tH^{s+t}(G,\rF_{-i})$  & $\Rightarrow  \tH^{s+t}(G,\rF_{-i})$  \\
    push forward of $\rF_{-i}$ & $\tR^s\pi_*\rF_{-i} = \tR^s\pi_*\rO(-i) \otimes \wedge^iE'{}\spcheck$   & $\tR^s\pi_*\rF_{-i} = \tR^s\pi_*L^{m_i} \otimes \mS_{\rho_i}F'{}\spcheck$  \\
     & $=\mS_{\nu_i}F'{}\spcheck \otimes \mS_{\lambda_i}E'$ & $=\mS_{\mu_i}E'{}\spcheck \otimes \mS_{\rho_i}F'{}\spcheck$ \\
  \end{tabular}
  \caption{A comparison of the various quantities involved in the
    construction of $X$ and $Y$ from the vector bundles $E$ and $F$.}
  \label{tab:Comparison}
\end{table}

\end{landscape}

\subsubsection{Towards a classification of genus one fibrations with
  5--sections}
\label{sec:towards-class-genus}

As will be reviewed in Section~\ref{sec:codim5fibers}, the Grassmannian $G(2,5)$ admits a
toric degeneration to a singular toric variety $P(2,5)$ of dimension
$6$~\cite{Batyrev:1998kx}. $P(2,5)$ admits a small crepant resolution $\widehat P(2,5)$. In
fact, $\widehat P(2,5) \cong \mP(E'')$ for
$E''=\rO_{\mP^1}^{\oplus3}\oplus \rO_{\mP^1}(1)^{\oplus2}$.
We consider a smooth Calabi--Yau complete intersection
$X_1''$ of codimension 5 in $\widehat P(2,5)\times \mP^2$ with $h^{1,1}(X_1'') = 4$ and
$h^{2,1}(X_1'')=43$. Following the discussion of
~\cite{Batyrev:1998kx}, there is
a conifold transition
\begin{equation}
  \label{eq:92}
  X_1'' \xrightarrow{\phi} \overline{X}_1 \leftrightsquigarrow X_1
\end{equation}
where $X_1$ is the Calabi--Yau variety in $G(2,5)\times \mP^2$
constructed in Section~\ref{sec:examples-Gr}. The map $\phi$ is a
birational contraction onto a singular Calabi--Yau complete
intersection $\overline{X}_1$ of codimension 5 in $P(2,5)\times
\mP^2$. The singularities of $\overline{X}_1$ are six conifold
singularities induced from the toric degeneration of $P(2,5)$. $X_1$
is a deformation of $\overline{X}_1$ smoothening these
singularities. Note that
\begin{equation}
  \label{eq:81}
  \begin{aligned}
    h^{1,1}(X_1'') &= h^{1,1}(X_1) + 2,\\
    h^{2,1}(X_1'') &= h^{2,1}(X_1) + 2 - 6\,.\\
  \end{aligned}
\end{equation}
Using the fact that $X_1$ admits a genus one fibration with a
5--section, and viewing these extremal transitions as Higgs
transitions in F--theory, we will show in
Sections~\ref{sec:codim5fibers} and~\ref{classifyU12} that, besides
$X_1$, there are nine further genus one fibrations over $\mP^2$ with
5--sections that have this behaviour. Moreover, we will see that the
same behaviour applies to genus one fibrations over $\mP^2$ with a
5--section $Y$, constructed as Pfaffian subvarieties, in accordance
with the expectations from relative homological projective
duality. The corresponding toric ambient spaces will be fibrations of
$\mP(E')$ over $\mP^2$. It is known that determinantal varieties (such
as Pfaffian varieties) admit a degeneration to a normal toric variety~\cite{Gonciulea:1996ab}. A
globalization of this degeneration should yield an explicit
description of the conifold transition of $Y$.

By the general theory of Mori (see e.g.~\cite{Rossi:2006ab}), the map $\phi$ can be
factored into two primitive contractions, $\phi = \phi'\circ\phi''$,
each of which changes $h^{1,1}$ by one, and the conifold transition~\eqref{eq:92} decomposes into two transitions
\begin{equation}
  \label{eq:94}
  X_1'' \xrightarrow{\phi''} \overline{X}''_1 \leftrightsquigarrow
  X'_1 \xrightarrow{\phi'} \overline{X}_1' \leftrightsquigarrow X_1 \,.
\end{equation}
Hence, we expect to find an intermediate smooth Calabi--Yau threefold $X_1'$
with Hodge numbers $h^{1,1}(X_1') = 3$ and $h^{2,1}(X_1') = 47 + 1 -3 =
45$. The change in $h^{2,1}$ is suggested by the fact that the
degeneration $\overline{X}_1 \leftrightsquigarrow X_1$ involves two
collections of three vanishing 3-cycles each satisfying a linear
relation.

In fact, we will show in Sections~\ref{sec:codim3fiber}
and~\ref{sec:toric-compl-inters} that $X'_1$ can be
realized as a complete intersection of codimension $3$ in a toric
variety with these Hodge numbers. We will see that the toric variety can be realized as
$\mP(E') \times \mP^2$ with $E'=\rO_{\mP^1}^{\oplus3}\oplus
\rO_{\mP^1}(1)$. The factorization of the contraction $\phi$
in~\eqref{eq:94} should have an interpretation as a two--step Higgs transition in F-theory. Independently of this
interpretation, we will show that there are conifold transitions
(single Higgs transitions)
\begin{equation}
  \label{eq:93}
  \begin{aligned}
    X' &\xrightarrow{\phi'} \overline{X}' \leftrightsquigarrow X, & Y'
    &\xrightarrow{\psi'} \overline{Y}' \leftrightsquigarrow Y, 
  \end{aligned}
\end{equation}
where $X'$, $Y'$ are Calabi--Yau
complete intersections  of codimension 3 in fibrations of $\mP(E')$
over $\mP^2$ and give a classification.

In both transitions, the conifold singularities can be understood in
terms of the genus one fibration as Kodaira fibers of type $I_2$ over
the discriminant of the fibration. 

In Table~\ref{tab:1} we list the homologically projective dual pairs
of genus one fibrations over $\mP^2$ with 5-sections that admit a
transition to a complete intersection in a toric variety either of
codimension three or five. The last example in this list is special since there the
5--section splits into 5 independent sections. This follows from
$h^{1,1}=6$ and the analysis in Section~\ref{classifyU12}, cf. the
discussion after~\eqref{eq:95}.   
\begin{table}[h!]
  \centering
  $
  \begin{small}
    \begin{array}{c|c|c|c|c|c|c|c|c|c}
      X & n & E' & F & h^{1,1}& h^{2,1} &\chi& J_1^2J_2& J_1^3
      &\ch_2J_1\\
      \hline
      & 1_a & \multirow{2}{*}{$\rO_{\mP^2}(2)\oplus \rO_{\mP^2}(1)\oplus\rO_{\mP^2}^{\oplus 3}$} &\multirow{2}{*}{$\rO_{\mP^2}^{\oplus5}$}& 2 & 47 & -90 & 15 & 10 & 64\\ 
      & 1_b & & & 2 & 47 & -90 & 5 & 5 & 38\\ 
      \hline
      X_1& 2_a & \multirow{2}{*}{$ \rO_{\mP^2}(1)^{\oplus 3}\oplus\rO_{\mP^2}^{\oplus 2}$} &\multirow{2}{*}{$\rO_{\mP^2}^{\oplus5}$}& 2 & 47 & -90 & 15 & 15 & 66\\ 
      Y_1& 2_b & & & 2 & 47 & -90 & 5 & 0 & 36\\ 
      \hline
      X_3& 3_a & \multirow{2}{*}{$ T_{\mP^2}(-1)\oplus \rO_{\mP^2}(1)^{\oplus 2}\oplus\rO_{\mP^2}$} &\multirow{2}{*}{$\rO_{\mP^2}^{\oplus5}$}& 2 & 47 & -90 & 15 & 20 & 68\\ 
      Y_3& 3_b & & & 2 & 47 & -90 & 15 & 25 & 70\\ 
      \hline
      & 4_a & \multirow{2}{*}{$ \rO_{\mP^2}(2)\oplus \rO_{\mP^2}(1)^{\oplus 3}\oplus\rO_{\mP^2}$} &\multirow{2}{*}{$\rO_{\mP^2}(1)\oplus\rO_{\mP^2}^{\oplus4}$}& 2 & 49 & -94 & 11 & 3 & 54\\ 
      & 4_b & & & 2 & 49 & -94 & 9 & 10 & 52\\ 
      \hline
      & 5_a & \multirow{2}{*}{$ \rO_{\mP^2}(3)\oplus\rO_{\mP^2}(2)^{\oplus 4}$} &\multirow{2}{*}{$\rO_{\mP^2}(1)^{\oplus 4}\oplus\rO_{\mP^2}$}& 2 & 49 & -94 & 9 & 0& 48\\ 
      & 5_b & & & 2 & 49 & -94 & 11 & 13 & 58\\ 
      \hline
      X_2& 6_a & \multirow{2}{*}{$ \rO_{\mP^2}(1)^{\oplus 5}$} &\multirow{2}{*}{$\rO_{\mP^2}(1)\oplus\rO_{\mP^2}^{\oplus4}$}& 2 & 49 & -94 & 11 & 8& 56\\ 
      Y_2& 6_b & & & 2 & 49 & -94 & 9 & 5 & 50\\ 
      \hline
      & 7_a & \multirow{2}{*}{$ \rO_{\mP^2}(3)\oplus\rO_{\mP^2}(1)^{\oplus 4}$} &\multirow{2}{*}{$\rO_{\mP^2}(1)^{\oplus 2}\oplus\rO_{\mP^2}^{\oplus3}$}& 2 & 50 & -96 & 17 & 28& 76\\ 
      & 7_b & & & 2 & 50 & -96 & 13 & 26 & 68\\ 
      \hline
      & 8_a & \multirow{2}{*}{$ \rO_{\mP^2}(3)\oplus\rO_{\mP^2}(2)^{\oplus 2}\oplus\rO_{\mP^2}(1)^{\oplus 2}$} &\multirow{2}{*}{$\rO_{\mP^2}(1)^{\oplus 3}\oplus\rO_{\mP^2}^{\oplus2}$}& 2 & 50 & -96 & 13 &11 & 62\\ 
      & 8_b & & & 2 & 50 & -96 & 7 & 7 & 46\\ 
      \hline
      & 9_a & \multirow{2}{*}{$ \rO_{\mP^2}(2)^{\oplus
           2}\oplus\rO_{\mP^2}(1)^{\oplus 3}$} &\multirow{2}{*}{$\rO_{\mP^2}(1)^{\oplus 2}\oplus\rO_{\mP^2}^{\oplus3}$}& 2 & 50 & -96 & 17 & 33& 78\\ 
      & 9_b & & & 2 & 50 & -96 & 13 & 21 & 66\\ 
      \hline
      & 10_a & \multirow{2}{*}{$ \rO_{\mP^2}(2)^{\oplus 4}\oplus\rO_{\mP^2}(1)$} &\multirow{2}{*}{$\rO_{\mP^2}(1)^{\oplus 3}\oplus\rO_{\mP^2}^{\oplus2}$}& 2 & 50 & -96 & 13 & 16& 64\\ 
      & 10_b & & & 2 & 50 & -96 & 7 & 2 & 44\\ 
      \hline
      & 11_a& \multirow{2}{*}{$ \rO_{\mP^2}(3)^{\oplus 3}\oplus\rO_{\mP^2}(2)^{\oplus 2}$} &\multirow{2}{*}{$\rO_{\mP^2}(2)\oplus\rO_{\mP^2}(1)^{\oplus 3}\oplus\rO_{\mP^2}$}& 2 & 52 & -100 & 15 & 29& 74\\ 
      & 11_b & & & 2 & 52 & -100 & 15 & 29 & 74\\ 
      \hline
      & 12_a & \multirow{2}{*}{$ \rO_{\mP^2}(3)^{\oplus 2}\oplus\rO_{\mP^2}(2)^{\oplus 2}\oplus\rO_{\mP^2}(1)$} &\multirow{2}{*}{$\rO_{\mP^2}(2)\oplus\rO_{\mP^2}(1)^{\oplus 2}\oplus\rO_{\mP^2}^{\oplus2}$}& 2 & 54 & -104 & 9 & 9& 54\\ 
      & 12_b & & & 2 & 54 & -104 & 11 & 17 & 62\\ 
      \hline
      X_4& 13_a & \multirow{2}{*}{$\rO_{\mP^2}(3)\oplus\rO_{\mP^2}^{\oplus 4}$} &\multirow{2}{*}{$\rO_{\mP^2}^{\oplus5}$}& 6 & 51 & -90 & 15 & 0 & 60\\ 
      Y_4& 13_b & & & 6 & 51 & -90 & 5 & 15 & 42\\ 
    \end{array}
  \end{small}
  $
  \caption{Genus one fibrations over $\mP^2$ with 5--sections
    admitting a conifold transition to a complete intersection in a
    toric variety.}
  \label{tab:1}
\end{table}
The label in the second column refers to the label used in
Tables~\ref{tab:codim5higgsList},~\ref{tab:codim3higgsList}
and~\ref{tab:23fibrations}. In each row, the upper entry refers to the complete
intersection 
\begin{align}
    X&\subset
    \Gr_2(F), &&E=\rO_{\Gr_2(F)/\mP^2}(1)\otimes\pi^*E',
  \label{eq:96}
\intertext{while the lower row entry refers to the Pfaffian subvariety}
    Y&\subset \mP(E'{}\spcheck), & &\varphi: \pi^*F\to \pi^*F\spcheck \otimes
    \rO_{\mP(E)/\mP^2}(1)\,.
  \label{eq:97}
\end{align}
The basis $(J_1,J_2) \in \tH^2(X,\mZ)$ is related to
$(h|_X,\sigma_1|_X)$, where $h=\pi^*\ch_1(\rO_{\mP^2}(1)), \sigma_1 = \ch_1(\rO_{\Gr_2(F)/\mP^2}(1))$, as follows: $J_2=h|_X$ and
\begin{equation}
  \label{eq:90}
  J_1 =
  \begin{cases}
    \sigma_1|_{X} & n = 1,2,3,4,6,13,\\ 
    (\sigma_1+h)|_{X} & n = 5,7,8,9,10,12,\\ 
    (\sigma_1+2h)|_{X} & n = 11.\\ 
  \end{cases}
\end{equation}
Similarly, the basis $(J_1,J_2) \in \tH^2(Y,\mZ)$ is related to
$(h|_Y,H|_Y)$, where $h=\pi^*\ch_1(\rO_{\mP^2}(1)), H = \ch_1(\rO_{\mP(E')/\mP^2}(1))$, as follows: $J_2=h|_Y$ and
\begin{equation}
  \label{eq:91}
  J_1 =
  \begin{cases}
    H|_{Y} & n = 3,\\ 
    (H-h)|_{Y} & n = 1,2,4,6,9,13,\\ 
    (H-2h)|_{Y} & n = 5,7,8,10,11,12.\\ 
  \end{cases}  
\end{equation}
All examples have
\begin{equation}
  \label{eq:98}
  \begin{aligned}
    J_2^3 &= 0, & J_1J_2^2 &=5, & \ch_2J_2 &= 36.
  \end{aligned}
\end{equation}
In the last example, the intersection numbers involving $J_1$ only hold for the class representing the
sum of the five sections.


%% file: section_2.tex
\section{GLSMs}
\label{sec:glsm}
In this section we show how the  genus one fibered Calabi-Yaus we have constructed in Section \ref{sec:geometry} arise as phases of certain non-Abelian GLSMs. After a short review of non-Abelian GLSMs, we present GLSMs whose phases correspond to the Calabi-Yaus $X_1,X_2,X_2',Y_1,Y_2$ and $X_3$ discussed in the previous section. The Pfaffian Calabi-Yaus correspond to strongly coupled phases in the GLSM. We use the non-Abelian duality \cite{Hori:2011pd} to extract information about those phases. We furthermore compute the discriminants by analyzing the Coulomb/mixed branches of the GLSMs. Finally, we compute the sphere partition function in the (weakly coupled) Grassmannian phases of these models. This will allow us to determine the fundamental periods and the Picard-Fuchs system, which will serve as an input for the subsequent sections. 
\subsection{Review of non-Abelian GLSMs}
\label{sec:glsmreview}
\subsubsection{Field content and phases}
We consider GLSMs with non-Abelian gauge group $G$. The scalar components $\phi$ of the chiral multiplets take values in a complex vector space $V$ and transform in a representation $\rho_V$ of $G$. The gauge charges $q$ are the weights of $\rho_V$. Throughout this article we will consider GLSMs describing Calabi-Yau spaces. This constrains the matter representation to $\rho_V:G\rightarrow SL(V)$. We further assume that the vector R-charges $R$ of the fields $\phi$ are between $0$ and $2$. We denote the scalar components of the vector multiplet by $\sigma$. They take values in the Lie algebra $\mathfrak{g}_{\mathbb{C}}$ of $G$. The FI-theta parameters $t=\zeta-i\frac{\theta}{2\pi}$ take values in $\mathfrak{g}_{\mathbb{C}}^{*}$ and there is a natural pairing $\langle\cdot ,\cdot\rangle:\mathfrak{g}_{\mathbb{C}}\times\mathfrak{g}_{\mathbb{C}}^{*}\rightarrow \mathbb{C}$. All our models have a non-zero superpotential $W\in\mathrm{Sym}V^*$ of vector R-charge $2$.

The classical vacua are determined by the zeroes of the scalar potential $U$ given by
\begin{equation}
  U=\frac{1}{8e^2}|[\sigma,\overline{\sigma}]|^2+\frac{1}{2}\left(|\langle q,\sigma\rangle\phi|^2+|\langle q,\overline{\sigma}\rangle\phi|^2\right)+\frac{e^2}{2}(\mu(\phi)-\zeta)^2+|dW(\phi)|^2,
  \end{equation}
where $e$ are the gauge couplings and $\mu:V\rightarrow i\mathfrak{g}^*$ is the moment map. If $\sigma=0$, the vacua are determined by the D-term and F-term equations:
\begin{equation}
  \mu(\phi)=\zeta, \qquad dW(\phi)=0.
\end{equation}
The solutions are
\begin{equation}
  X_{\zeta}=\mu^{-1}(\zeta)/G\cap dW^{-1}(0).
\end{equation}
We can write $\mu^{-1}(\zeta)/G\simeq (V-F_{\zeta})/G_{\mathbb{C}}$ where $F_{\zeta}$ is called the deleted set. In the vacua, the gauge symmetry is broken to a subgroup of $G$. If this subgroup is discrete or trivial we have a weakly coupled phase, usually referred to the Higgs branch of the theory. In the case of non-Abelian $G$ the gauge group may be broken to a continuous subgroup. If the fields $\sigma$ do not take large values (as they would on Coulomb and mixed branches), we have a strongly coupled phase where the classical analysis fails.
\subsubsection{Coulomb branch and discriminants}
Coulomb and mixed branches arise when some or all of the fields $\sigma$ can have large values and $G$ is broken to a $U(1)$-subgroup. On the Coulomb branch, the unbroken group is the maximal $U(1)$ subgroup. Then all $\sigma$ are non-zero and take values in the Lie algebra of the maximal torus of $G$. While classically unconstrained, the $\sigma$ receive a potential through one-loop corrections:
\begin{equation}
  \mathcal{W}_{eff}=-t(\sigma)-\sum_q\langle q,\sigma\rangle\left(\mathrm{log}\langle q,\sigma\rangle-1\right)+\pi i\sum_{\alpha>0}\langle\alpha,\sigma\rangle,
\end{equation}
where $\alpha>0$ denotes the positive roots of $G$. The Coulomb branch is lifted away from the critical locus of $\mathcal{W}_{eff}$.

In the context of Calabi-Yau GLSMs, the Coulomb branch encodes the principal component of the discriminant of the Calabi-Yau. Other components of the discriminant are encoded by mixed branches. In the models that we are considering all the mixed branches will eventually be lifted, but in a rather non-trivial way so that it makes sense to give more details. On a mixed branch, $G$ is broken to a non-maximal torus $T_L\subset G$ and $\sigma$ takes large values $\sigma_L$ in the complexified Lie algebra $\mathfrak{t}_{L,\mathbb{C}}$ of $T_L$. The matter fields are divided up into $\phi=(\dot{\phi},\hat{\phi})$, depending on whether or not they receive mass by $\sigma_L$. The hatted fields are charged under $T_L$ and thus receive a mass, while the uncharged fields $\dot{\phi}$ remain massless. The same holds for the $\sigma$-fields where we distinguish $\sigma=(\sigma_L,\dot{\sigma},\hat{\sigma})$, where $(\sigma_L,\dot{\sigma})$ take values in the Lie algebra $\mathfrak{c}_{L,\mathbb{C}}$ of the centraliser $C_L\subset G$ of $T_L$. These are the massless fields. The remaining fields, $\hat{\sigma}$, are massive. After integrating out the massive degrees of freedom one is left with an effective theory of the massless fields that are constrained by the the following classical potential:
\begin{align}
  U_{eff}&=\frac{1}{8e_{eff}^2}|[\dot{\sigma},\dot{\overline{\sigma}}]|^2+\frac{1}{2}\left[|\langle q,\dot{\sigma}\rangle\dot{\phi}|^2+|\langle q,\dot{\overline{\sigma}}\rangle\dot{\phi}|^2 \right]+\frac{e_{eff}^2}{2}\left(\mu^{\mathfrak{c}_L}(\dot{\phi})-\zeta_{eff}^{\mathfrak{c}_L}(\sigma_L,\dot{\sigma})\right)^2+|dW(\dot{\phi})|^2,
\end{align}
where $\zeta_{eff}^{\mathfrak{c}_L}=\mathrm{Re}\:t_{eff}^{\mathfrak{c}_L}$ with $t_{eff}^{\mathfrak{c}_L}=-d\mathcal{W}_{eff}^{\mathfrak{c}_L}(\sigma_L,\dot{\sigma})$. There is a mixed branch if there is a solution of $U_{eff}=0$ that breaks $C_L$ to $T_L$. 
\subsubsection{Non-Abelian duality}
In \cite{Hori:2011pd} Hori found a duality between two-dimensional non-Abelian models. In our case we will need the following incarnation of the duality. Given a non-Abelian GLSM with gauge group $USp(k)$ and $N$ fundamentals, the duality maps between theories with gauge groups $USp(k)$ and $USp(N-k-1)$. We will give more details on the characteristics of the dual theory once we have specialised to the class of models we will focus on. The duality maps strongly coupled phases to weakly coupled phases and vice versa. Since strongly coupled phases of GLSMs are hard to come by, we will make use of the duality to analyze them via the weakly coupled dual theory.
\subsubsection{Sphere partition function}
We will use the sphere partition function of the GLSM to extract perturbative and non-perturbative information about the Calabi-Yaus we have constructed. First constructed in \cite{Benini:2012ui,Doroud:2012xw}, it was shown in \cite{Jockers:2012dk} and later confirmed in \cite{Gomis:2012wy,Gerchkovitz:2014gta,Gomis:2015yaa} that the sphere partition function computes the exact K\"ahler potential of the superconformal field theories associated to the phases of the GLSM. In a Calabi-Yau setting, this can be used to compute Gromov-Witten invariants directly in the GLSM. The sphere partition function is defined as follows
\begin{equation}
  \label{zs2def}
  Z_{S^2}=\frac{1}{(2\pi)^{\mathfrak{h}}|\mathcal{W}|}\sum_{m\in\Lambda_m}\int d^{\mathrm{dim}\mathfrak{h}}\sigma\: Z_G(m,\sigma)Z_{\mathrm{matter}}(m,\sigma)Z_{\mathrm{cl}}(m\sigma,\zeta,\theta),
\end{equation}
where $\mathfrak{h}$ is the Cartan subalgebra of $\mathfrak{g}$ and $\mathcal{W}$ is the Weyl group. The sum over $m$ over an integer lattice $\Lambda_m\subset\mathfrak{h}$ accounts for the discrete values of the gauge fields on the sphere.  The factors in the integrand of (\ref{zs2def}) are given by
\begin{align}
  Z_G(m,\sigma)&=\prod_{\alpha>0}(-1)^{\langle\alpha,m\rangle}\left(\frac{1}{4}\langle\alpha,m\rangle^2+\langle \alpha,\sigma\rangle^2\right),\nonumber\\
  Z_{\mathrm{matter}}(m,\sigma)&=\prod_q\frac{\Gamma\left(\frac{R}{2}-i\langle q,\sigma\rangle-\frac{1}{2}\langle q,m\rangle\right)}{\Gamma\left(1-\frac{R}{2}+i\langle q,\sigma\rangle-\frac{1}{2}\langle q,m\rangle\right)},\nonumber\\
  Z_{\mathrm{cl}}(m\sigma,\zeta,\theta)&=e^{-4\pi i\langle \zeta,\sigma\rangle-i\langle\theta,m\rangle}.
\end{align}
We follow the discussion and notation of \cite{Gerhardus:2015sla} for the evaluation of the integral in (\ref{zs2def}). In the examples we consider $\mathrm{dim}{\mathfrak{h}}=3$, which makes the evaluation of the integral rather challenging, in particular in strongly coupled phases where one is faced with having to regularize divergent sums order by order. We have thus adopted a hybrid approach using the sphere partition function in combination with standard techniques from mirror symmetry and topological string theory. Concretely, we take a GLSM and evaluate the sphere partition function in a weakly coupled phase. By the results of \cite{Jockers:2012dk}, we can use the general structure of the result to read off the fundamental period of the Calabi-Yau associated to this phase. From this we can determine the Picard-Fuchs system via an ansatz. Once we have this, we can access the information on the strongly-coupled phases from changing coordinates in the Picard-Fuchs equations, rather than evaluating the sphere partition function in the strongly coupling regime.  
\subsection{Genus one fibrations via non-Abelian GLSMs}
\label{sec:glsmfibrations}
After this brief reminder on GLSMs, we can now proceed to construct the genus one fibrations. We first discuss the GLSM associated to the realizations $X_0,Y_0$ of the elliptic normal curve. Then we consider two GLSMs whose phases are $X_1,Y_1$ and $X_2,Y_2,X_2'$, respectively. Finally we outline a GLSM construction of the model $X_3$.  
\subsubsection{Elliptic curves and non-Abelian GLSMs}
\label{sec:glsmelliptic}
Let us briefly recall the GLSM description of the elliptic curve that we use to construct our genus one fibrations. The two alternative descriptions of the elliptic normal curve discussed in Section~\ref{sec:geometry} arise as phases of the same non-Abelian one-parameter GLSM.  This GLSM has already been discussed in \cite{Hori:2013gga,Caldararu:2017usq,Knapp:2019cih} to which we refer for more details. The model is the one-dimensional version of the R{\o}dland model \cite{Hori:2006dk} with $G=U(2)$. The matter content consists of five doublets $x_i^a$ ($i=1,\ldots,5$, $a=1,2$) and  five fields $p^k$ ($k=1,\ldots,5$) transforming in the following representations of $U(2)$:
\begin{equation}
  \begin{array}{c|cc|c}
    & p^{1},\ldots,p^{5}&x_{1},\ldots,x_{5}&\mathrm{FI}\\
    \hline
    U(2)&\det^{-1}&\Box&\zeta.
    \end{array}
\end{equation}
For a suitable parameterization of the maximal torus of $U(2)$ the gauge charges are
\begin{equation}
  \begin{array}{c|ccc|c}
    & p^{1},\ldots,p^{5}&x^1_{1},\ldots,x^1_{5}&x^2_{1},\ldots,x^2_{5}&\mathrm{FI}\\
    \hline
    U(1)_1&-1&1&0&\zeta\\
    U(1)_2&-1&0&1&\zeta.
    \end{array}
\end{equation}
The superpotential is
\begin{equation}
  W=\sum_{k,i,j=1}^5 \sum_{a,b=1}^2A_k^{i,j}p^kx_i^ax_j^b\varepsilon_{ab}= \sum_{i,j=1}^5A^{ij}(p)[x_ix_j],
\end{equation}
where, as in Section~\ref{sec:geometry}, $[x_ix_j]=x_i^ax_j^b\varepsilon_{ab}$ ($\varepsilon_{ab}$ being the Levi-Civit\'a symbol) and $\sum_{k=1}^5 A_k^{i,j}p^k=A^{ij}(p)$ is an antisymmetric $5\times 5$-matrix with entries linear in $p$. The coefficients $A_k^{ij}$ must satisfy the genericity condition that the two $5\times 5$-matrices $A_k^{ij}\phi_j^a$ with $a=1,2$ have a rank $5$ linear combination.

The $\zeta \gg 0$-phase is a codimension $5$ complete intersection in $G(2,5)$, defined by
\begin{equation}
  \sum_{i,j=1}^5 A_k^{i,j}[x_ix_j]=0,\qquad k=1,\ldots,5,
\end{equation}
where $[x_ix_j]$ are the Pl\"ucker coordinates. This is the model $X_0$ of Section~\ref{sec:geometry}.

The $\zeta \ll 0$-phase is a strongly coupled phase with an unbroken $SU(2)$. Following the analysis of \cite{Hori:2006dk}, one obtains a Pfaffian CY in $\mathbb{P}^4$:
\begin{equation}
  \{p\in\mathbb{P}^4|\mathrm{rk}A^{ij}(p)=2\}.
\end{equation}
As in Section~\ref{sec:geometry}, we will refer to this geometry as $Y_0$. Note that there are further equivalent descriptions of this elliptic curve \cite{Hori:2011pd,Hori:2013gga}.

The Coulomb branch analysis shows that there are two singularities at
\begin{equation}
  e^{-t}=\frac{1}{2}\left(11\pm 5\sqrt{5}\right)=\frac{1}{(1+\omega^k)^5}, \qquad \omega=e^{\frac{2\pi i}{5}}, \quad k=1,2.
\label{eqn:fiberConi}
\end{equation}
For later reference, we also recall the Picard-Fuchs operator
\begin{align}
	\mathcal{L}=\theta^2-z(11\theta^2+11\theta+2)-z^2(\theta+1)^2,
\end{align}
and the periods
\begin{align}
	\begin{split}
	\varpi_0=&1+3z+19z^2+147z^3+1251z^4+\mathcal{O}(z^5)\,,\\
	(2\pi i)\varpi_1=&\varpi_0\cdot\log(z)+5z+\frac{75}{2}z^2+\frac{1855}{6}z^3+\frac{10875}{4}z^4+\mathcal{O}(z^5)\,,
	\end{split}
\end{align}
where $\theta=z\frac{d}{dz}$ and\footnote{The extra minus sign comes from a theta-angle shift when one goes from the GLSM to the non-linear sigma model.} $z=-e^{-t}$. The inverse of the mirror map has been identified to be a modular function of $\Gamma_1(5)$ \cite{Hori:2013gga}.  Note that this result holds for both, the Grassmannian and the Pfaffian phase, because the elliptic curves corresponding to the two phases are isomorphic. Moreover, this reproduces the analysis of the periods of a family of elliptic curves with 5--torsion in~\cite{Beukers:1983ab}. 
\subsubsection{Genus $1$ fibrations}
We use the GLSM of the previous subsection to construct GLSMs whose phases are the genus $1$ fibrations constructed in Section~\ref{sec:geometry}. We consider gauge groups of the form $G=U(1)^l\times U(2)$. The elliptic fiber is characterised by the $U(2)$ GLSM above. The concrete realisation of the fibration is encoded in the $U(1)^l$-charges of the $x_i^a$ and $p^k$. Furthermore there is a set of $n$ fields $b_m$ ($m=1,\ldots,n$) that are only charged under $U(1)^l$. They determine the base manifold. 
The matter content of a GLSM of this type looks as follows:
\begin{align}
	\begin{array}{c|ccccccc|c}
		&p^{k,\,k=1,\dots,5}&x_{i,\,i=1,\dots,5}&b_{m,\,m=1,\dots,\text{dim}(B)+l}&\mathrm{FI}\\\hline
		U(2)&\det^{-1}&\square&0&\zeta\\
		U(1)_1&q^1_k&q^1_i&*&\zeta_1\\
		\dots&\dots&\dots&\dots&\dots\\
		U(1)_l&q^l_k&q^l_i&*&\zeta_{l}\\
	\end{array}
	\label{eqn:glsmFieldContentFibration}
\end{align}
The superpotential has the following form
\begin{equation}
  \label{wstructure}
  W=\sum_{ij}A^{ij}(p^k,b_m)[x_ix_j], 
\end{equation}
where $A^{ij}=-A^{ji}$ is still linear in $p$ but now has non-trivial dependence on the base coordinates $b$. The models differ in their $U(1)^{l}$-charges and the number of fields $b_m$. 
This determines the structure of the matrix $A^{ij}$. We will always choose to R-charges of the $p$-fields to be $2$ while all the other fields have charge $0$. 

It is useful to look at the non-Abelian dual of this class of models. Using $U(2)\cong\frac{U(1)\times SU(2)}{\{\pm 1,\pm{\bf 1}\}}$ and the fact that our models have $5$ fundamentals the duality maps the $SU(2)$ to an $SU(2)$ and thus the dual group $\widetilde{G}$ is $\widetilde{G}=\frac{U(1)\times SU(2)}{\{\pm1,\pm{\bf 1}\}}\cong U(2)$. The dual theory has $5$ fundamental fields $\widetilde{x}^i_a$ with gauge charges $\widetilde{q}^i_a=-q_i^a$, where $q_i^a$ are the gauge charges of the original theory. Note that this concerns {\em all} gauge charges of the fundamentals, not just those associated to the non-Abelian group. The fields $p^k$ and $b_m$ remain unaffected. In addition there are $\frac{N(N-1)}{2}=10$ singlet fields $a_{ij}=-a_{ji}$ with gauge charges $q_i^a+q_j^a$. The superpotential of the dual theory is
\begin{equation}
  \widetilde{W}=\sum_{ij}A^{ij}(p^k,b_m)a_{ij}+[\tilde{x}^i\tilde{x}^j]a_{ij}.
\end{equation}
\subsection{GLSM for $X_1$, $Y_1$}
\label{sec:glsmModel1}
In this section we discuss a GLSM that realises the geometries $X_1$ and $Y_1$ constructed in Section~\ref{sec:geometry} as different phases. We note that a different construction of these geometries using joins \cite{Inoue:2019jle} has been realised in a GLSM in \cite{Knapp:2019cih}. We discuss the connections between these two (seemingly) different GLSMs in Section~\ref{sec-match}. 
\subsubsection{Matter content and phases}
We consider a model with $G=U(1)\times U(2)$ and the following matter content:
\begin{equation}
    \begin{array}{c|c|c|c|c|c}
      &p^1,p^2,p^3&p^4,p^5&x_1^a,\ldots,x_5^a&b_1,b_2,b_3&\mathrm{FI}\\
      \hline
      U(2)&\mathrm{det}^{-1}&\mathrm{det}^{-1}&\Box&{\bf 1}&\zeta\\
      U(1)&-1&0&0&1&\zeta_1.
    \end{array}
    \end{equation}
Hence, the gauge charges are
\begin{equation}
 \begin{array}{c|c|r|cc|c|c}
    &p^1,p^2,p^3&p^4,p^5&x_1^1,\ldots,x_5^2 &x_1^2,\ldots,x_5^2&b_1,b_2,b_3&\mathrm{FI}\\
    \hline
    U(1)_1&-1&-1&1&0&0&\zeta\\
    U(1)_2&-1&-1&0&1&0&\zeta\\
    U(1)_3&-1&0&0&0&1&\zeta_1.
    \end{array}
\label{eqn:model1charges}
\end{equation}
Here $U(1)_1$ and $U(1)_2$ account for the maximal torus of the $U(2)$-factor.

The $U(2)$ D-terms have the same structure as the ones for the R{\o}dland model \cite{Hori:2006dk}:
\begin{equation}
  \label{d1}
  \sum_{i=1}^5x_i^{a}x_{i,b}^{\dagger}-\sum_{k=1}^5|p^k|\delta^a_b=\zeta\delta^a_b.
\end{equation}
The D-term associated to $U(1)_3$ is 
\begin{equation}
  \label{d2}
  -|p^1|^2-|p^2|^2-|p^3|^2+\sum_{i=1}^3 |b_i|^2=\zeta_1.
\end{equation}
The first D-term implies that for $\zeta > 0$ the $2\times 5$ matrix $x\equiv x_i^a$ must have maximal rank and for $\zeta <  0$ the point $p^1=\ldots=p^5=0$ has to be excluded, so the $p^k$ take values in a $\mathbb{P}^4$.

For $\zeta_1 > 0$ we have to exclude $b_1=b_2=b_3=0$, consistent with the base $\mathbb{P}^2$. For $\zeta_1 < 0$ the deleted set is $p^1=p^2=p^3=0$. This is also the deleted set of a $\mathbb{P}^2$, and we observe some symmetry under exchange of $p^{1,2,3}$ and $b_{1,2,3}$. The D-terms are consistent with $p^1=\ldots =p^5=0$ for $\zeta,\zeta_1 >  0$. Further note that we can subtract the $U(1)$ D-term from the diagonal components of the $U(2)$ D-term to get
\begin{equation}
  \label{dextra}
  \sum_{i=1}^5|x_i^1|^2-|p^4|^2-|p^5|^2-\sum_{i=1}^3 |b_i|^2=\sum_{i=1}^5|x_i^2|^2-|p^4|^2-|p^5|^2-\sum_{i=1}^3 |b_i|^2=\zeta-\zeta_1.
  \end{equation}
So if $\zeta-\zeta_1 > 0$ one has to exclude $x_1^1=\ldots=x_5^1=0$ and $x_1^2=\ldots=x_5^2=0$ but this is already excluded by the non-Abelian D-term anyway, because otherwise the matrix $x$ would have rank $1$. If, on the other hand, $\zeta-\zeta_1<0$ the deleted set is $p^4=p^5=b_1=b_2=b_3=0$. This constraint is again very similar to the the deleted set of the Pfaffian phase of the elliptic curve.

The superpotential has the form as indicated above where $ A_k^{i,j}(b)$ is linear in $b$ for $k=1,2,3$ and has constant entries for $k=4,5$. Concretely, we have
\begin{equation}
  W=\sum_{i,j=1}^5\left[\sum_{k,l=1}^3A^{ijl}_{k}b_lp^k+\sum_{k=4}^5A^{ij}_kp^k \right][x_ix_j].
\end{equation}
The F-term equations are
\begin{align}
  x_i^a:&\quad \sum_{j=1}^5\left[\sum_{k,l=1}^3A^{ijl}_{k}b_lp^k+\sum_{k=4,5}A^{ij}_kp^k \right]x_{j,a},\quad i=1,\ldots,5,\: a=1,2,\label{f1}\\
  p^k:&\quad \sum_{i,j=1}^5\sum_{l=1}^3A^{ijl}_{k}b_l[x_ix_j]=0,\quad k=1,2,3,\label{f2}\\
  &\quad \sum_{i,j=1}^5A^{ij}_k[x_ix_j]=0,\quad k=4,5,\label{f3}\\
  b_l:&\quad \sum_{i,j=1}^5\sum_{k=1}^3A^{ijl}_{k}p^k[x_ix_j]=0,\quad l=1,2,3.\label{f4}
\end{align}

The model has three phases. In the phase $\zeta,\zeta_1 > 0$ we can set $p^1=\ldots=p^5=0$ and the F-terms reduce to (\ref{f2}), (\ref{f3}). The D-terms (\ref{d1}) and (\ref{d2}) ensure that the $x_i^a$ ($i=1,\ldots,5$) are not allowed to vanish simultaneously. The same holds for the $b_i$ ($i=1,2,3$). There is no further information from (\ref{dextra}), consistent with the absence of a phase boundary at $\zeta=\zeta_1 > 0$. Thus we have found a weakly coupled phase that is a genus one fibration over $\mathbb{P}^2$ with the fiber being a codimension $5$ complete intersection in $G(2,5)$. We call this the Grassmannian phase. This is the geometry $X_1$ constructed in Section~\ref{sec:geometry}. 

For $\zeta < 0,\zeta_1 > 0$ there is a vacuum for $x=0$ and we expect the geometry to be non-perturbatively realised with the $U(2)$ broken to $SU(2)$. Furthermore, $b_1=b_2=b_3=0$ is disallowed, while $p^1=p^2=p^3=0$ is consistent. At $\zeta < 0,\zeta_1 < 0, \zeta-\zeta_1<0$ there is also a vacuum with $x=0$, and $b_1=b_2=b_3=0$ is now an allowed choice but $p^1=p^2=p^3=0$ is forbidden. Also this phase is expected to be strongly coupled and the deleted sets indicate that $p^{1,2,3}$ and $b_{1,2,3}$ exchange their roles. The analysis of the dual theory will show that the phase boundary between these two phases gets lifted by the F-terms and that one recovers the geometry $Y_1$ of Section~\ref{sec:geometry}. 

\begin{figure}
\begin{center}
      \begin{tikzpicture}
        \draw[->] (-3,0) -- (3,0);
        \draw[->] (0,-3) -- (0,3);
        \draw[ultra thick, blue] (0,0) -- (2.5,0);
        \draw[ultra thick, blue] (0,0) -- (0,2.5);
        \draw[ultra thick, blue] (0,0) -- (-2.5,-2.5);
        \draw[ultra thick, dashed, blue] (0,0) -- (-2.5,0);
        \draw (0.3,3) node {$\zeta_1$};
        \draw (3.3,0) node {$\zeta$};
        \draw (1.5,1.5) node {$X_1$};
        \draw (1.5,-1.5) node {$X_1$};
        \draw (-1.5,1.5) node {$Y_1$};
        \draw (2.5,0.3) node {$x_{1,\ldots,5}$};
        \draw (-0.5,2.5) node {$b_{1,2,3}$};
        \draw (-2.5,0.3) node {$p^{4,5}$};
        \draw (-2.5,-2) node {$p^{1,2,3}$};
        \end{tikzpicture}
\end{center}\caption{Classical phase diagram of the GLSM for $X_1$, $Y_1$.}\label{fig:glsm1phases}
\end{figure}
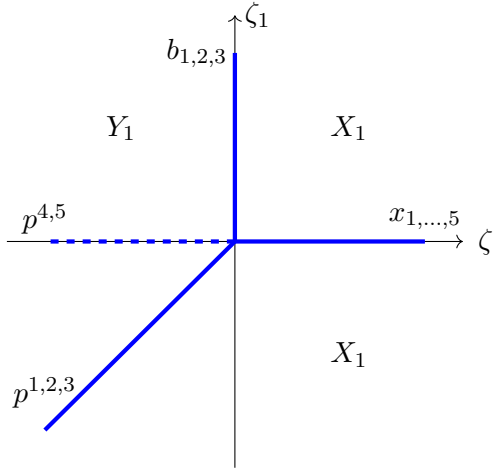

Finally there is the phase where $\zeta_1 < 0$ and $\zeta-\zeta_1 > 0$. In this phase $b_1=b_2=b_3=0$ is allowed and so is $p^4=p^5=0$, while the $x_i^a$ are not allowed to vanish simultaneously. Let us set $b_1=b_2=b_3=0$ and $p^{4}=p^5=0$. Then the F-terms reduce to
\begin{equation}
 \sum_{i,j=1}^5\sum_{k=1}^3A^{ijl}_{k}p^k[x_ix_j]=0,\quad l=1,2,3,\qquad  \sum_{i,j=1}^5A^{ij}_k[x_ix_j]=0,\quad k=4,5.
\end{equation}
This is again a complete intersection in $G(2,5)$ with $p^{1,2,3}$ taking the role of $b_{1,2,3}$. The question is if $p^{4},p^5\neq 0$ is also allowed. If this were the case, we would get $10$ more equations from (\ref{f1}). So in total there would be $15$ F-term equations constraining $15$ variables. Since we also have to satisfy the D-terms, this generically does not have a solution. We can also allow for some $b_i\neq 0$ but then all the F-terms would be non-trivial and we have too many equations for the non-zero degrees of freedom. We conclude that we have another weakly coupled phase that is realisation of the geometry $X_1$. The classical phase diagram can be found in Figure~\ref{fig:glsm1phases}. We note that the phase diagram has a structure that is compatible to the phase structures found in a recent GLSM treatment of homological projective duality \cite{Chen:2020iyo}. 
\subsubsection{Coulomb branch}
Next, we discuss the Coulomb branch and study the quantum phase structure of the GLSM.
The effective potential for the vector multiplet scalars $\sigma_i,\,i=1,\ldots,3$ is
\begin{align}
  \mathcal{W}_{eff}&=-t(\sigma_1+\sigma_2)-t_1\sigma_3-3(-\sigma_1-\sigma_2-\sigma_3)\left[\log(-\sigma_1-\sigma_2-\sigma_3)-1\right]\nonumber\\
  &\quad -2(-\sigma_1-\sigma_2)\left[\log(-\sigma_1-\sigma_2)-1\right]-5\sigma_1\left[\log\sigma_1-1\right]-5\sigma_2\left[\log\sigma_2-1\right]\nonumber\\
  &\quad -3\sigma_3\left[\log\sigma_3-1\right]+i\pi(\sigma_1-\sigma_2)\,.
\end{align}
After defining
\begin{equation}
  z=\frac{\sigma_2}{\sigma_1}, \qquad w=\frac{\sigma_3}{\sigma_1}, 
\end{equation}
the critical locus of the potential is determined by
\begin{align}
  \label{coulomb}
	\begin{split}
  e^{-t}&=\frac{1}{\left(1+\frac{1}{z}\right)^2\left(1+\frac{1}{z}+\frac{w}{z}\right)^3}\,,\quad e^{-t_1}=-\frac{1}{\left(1+\frac{1}{w}+\frac{z}{w}\right)^3}\,,\quad z^5=1\,,
	\end{split}
\end{align}
so that $z$ is restricted to the values $z=\omega^k,\,k=0,\ldots,4$ with $\omega=e^{\frac{2\pi i}{5}}$.

\begin{figure}[h!]
	\centering
	\includegraphics[width=.5\linewidth]{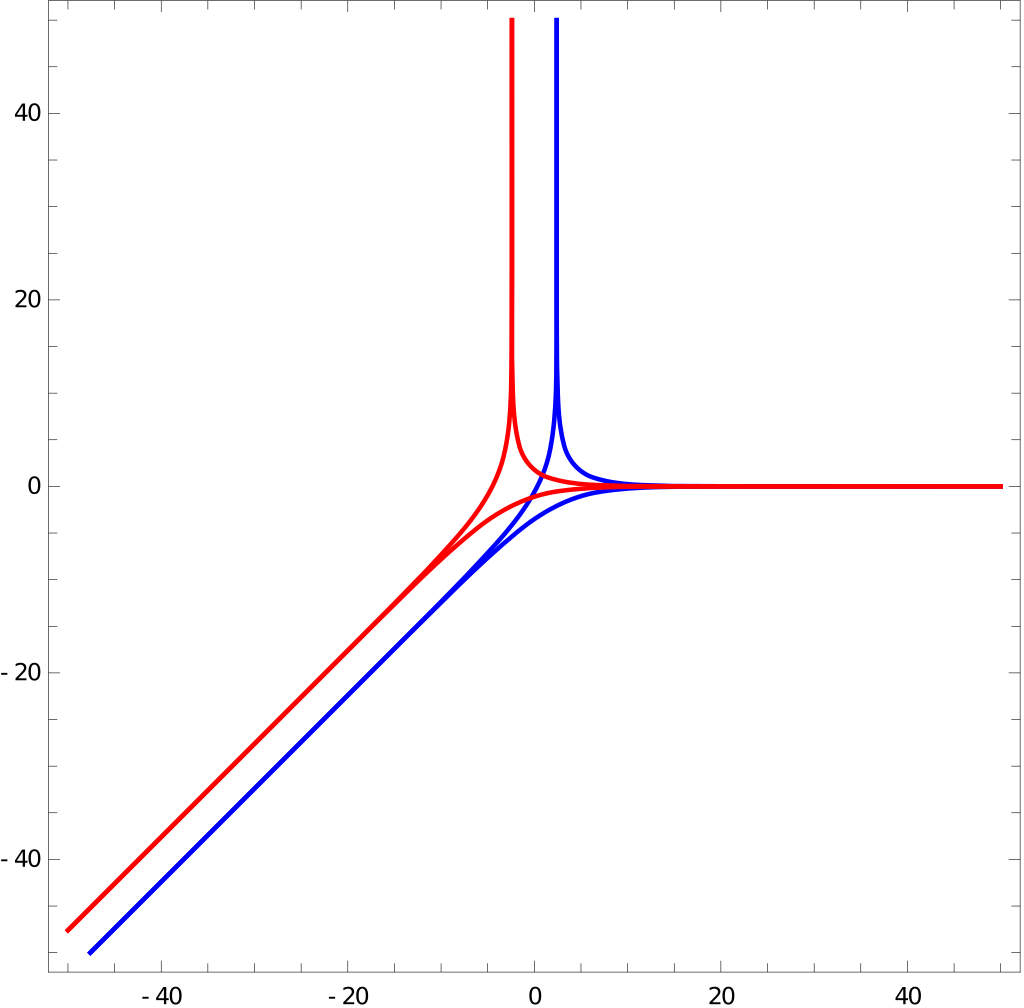}
	\caption{The two components of the discriminant in the FI parameter space, with $\{\Delta_1=0\}$ corresponding to the blue lines and $\{\Delta_2=0\}$ depicted in red. Each of the amoebas consists itself of three parts with different values for the theta angle.}
	\label{fig:splitAmoebaGLSM1}
\end{figure}

We have to remove solutions that are fixed under the Weyl group action and this eliminates $z=1$.
Moreover, the solutions for $k=3,4$ turn out to encode the same components of the discriminant as those for $k=1,2$, so we only need to consider $k=1,2$.
For each of the relevant values of $z$ we can then eliminate $w$ from the remaining system of equations.
This produces the two components of the discriminant
\begin{align}
	\begin{split}
	\Delta_1=&1-3 (5 \omega_{\pm} +8) z_1-3 z_2+3 (55 \omega_{\pm} +89) z_1^2-21 (5 \omega_{\pm} +8) z_1 z_2+3 z_2^2\\
		&+(-610 \omega_{\pm} -987) z_1^3-3 (55 \omega_{\pm} +89) z_1^2 z_2-3 (5 \omega_{\pm} +8) z_1 z_2^2-z_2^3\,,\\
	\Delta_2=&1+3 (5 \omega_{\pm} -3) z_1-3 z_2-3 (55 \omega_{\pm} -34) z_1^2+21 (5 \omega_{\pm} -3) z_1 z_2+3 z_2^2\\
		&+(610 \omega_{\pm} -377) z_1^3+3 (55 \omega_{\pm} -34) z_1^2 z_2+3 (5 \omega_{\pm} -3) z_1 z_2^2-z_2^3\,,
	\end{split}
	\label{eqn:glsm1discComps}
\end{align}
where we have introduced $z_1=-e^{-t},\,z_2=-e^{-t_1}$ and $\omega_{\pm}=\omega+\omega^{-1}$.
The corresponding amoebas are shown in Figure~\ref{fig:splitAmoebaGLSM1}.
After taking the product of $\Delta_1$ and $\Delta_2$, the powers of $\omega$ cancel and one obtains the polynomial with integer coefficients
\begin{align}
	\begin{split}
	\Delta=&1-33 z_1+360 z_1^2-1265 z_1^3-360 z_1^4-33 z_1^5-z_1^6-6 z_2-132 z_1 z_2-1602 z_1^2 z_2\\
		&+4686 z_1^3 z_2-2601 z_1^4 z_2+33 z_1^5 z_2+15 z_2^2+561 z_1 z_2^2+1755 z_1^2 z_2^2-4686 z_1^3 z_2^2\\
		&-360 z_1^4 z_2^2-20 z_2^3-561 z_1 z_2^3-1602 z_1^2 z_2^3+1265 z_1^3 z_2^3+15 z_2^4+132 z_1 z_2^4\\
		&+360 z_1^2 z_2^4-6 z_2^5+33 z_1 z_2^5+z_2^6\,.
	\end{split}
	\label{eqn:glsm1disc}
\end{align}
In the limit $z_2\rightarrow 0$ the polynomial reduces to the third power of the discriminant of the GLSM of the fiber with solutions
\begin{align}
	e^{-t}=\frac12\left(11\pm 5\sqrt{5}\right)\,.
\end{align}
Due to this structure, it is not surprising that the two geometries $X_1$ and $Y_1$ corresponding to two adjacent phases are not birational. 
On the other hand, in the limit $z_1\rightarrow 0$ the only solution to $\Delta=0$ is $e^{-t_1}=-1$.
This is also expected, because the two phases that are separated by the corresponding boundary both flow to a non-linear sigma model on $X_1$, which is trivially birational to itself.

In principle, there could also be mixed branches. One candidate is a mixed branch where $U(1)_3$ is unbroken, which could account for an extra phase boundary along the negative $\zeta$-axis. In this case $\sigma_3\equiv\sigma_L$ can take large values. We divide up the matter fields $\phi$ into $\{\dot{\phi},\hat{\phi}\}$ where $\hat{\phi}$ receive a mass for non-zero values of $\sigma_L$, which happens when they are charged under $U(1)_3$, and $\dot{\phi}$ are the massless ones. In the present example we have
\begin{equation}
  \dot\phi=\{x_i^a,p^4,p^5\}, \qquad \hat{\phi}=\{p^1,p^2,p^3,b_1,b_2,b_3\}.
\end{equation}
Similarly, we can divide up the $\sigma$-fields into $\{\dot{\sigma},\hat{\sigma},\sigma_L\}$ where the $\dot{\sigma}$ are massless and the $\hat{\sigma}$ receive mass. 
In our case we have 
\begin{equation}
  \dot{\sigma}=\{\sigma_1,\sigma_2\}, \qquad \hat{\sigma}=\{\}.
\end{equation}
These degrees of freedom generate the following effective scalar potential
\begin{align}
  U_{eff}&=\frac{1}{e^2_{eff}}|[\dot{\sigma},\dot{\bar{\sigma}}]|^2+\frac{1}{2}\left(\sum_{i=1}^5|\sigma_1x_i^1|^2+\sum_{i=1}^5|\sigma_2x_i^2|^2+\sum_{i=4}^5|(-\sigma_1-\sigma_2)p^i|^2+\sigma\leftrightarrow\bar{\sigma}\right)\nonumber\\
  &\quad+\frac{e^2_{eff}}{2}\left(\mu_{eff}(\dot{\phi})-\zeta_{eff}(\dot{\sigma},\sigma_L)\right)+|dW(\dot{\phi})|^2,
\end{align}
where (recall that there were $b$-independent entries in $A^{ij}_k(p)$)
\begin{equation}
  W(\dot{\phi})=\sum_{i,j=1}^5\sum_{k=4}^5A^{ij}_kp^k[x_ix_j]\,,
\end{equation}
and
\begin{align}
  \mu_{eff}=\sum_{i=1}^5\sum_{a=1,2}x_i^ax^{\dagger}_{i,a}-2\sum_{k=1}^3|p^k|^2\,,\quad \mu_{eff,3}=0\,,
\end{align}
and finally 
\begin{align}
  t_{eff}=2t-6\log(-\sigma_1-\sigma_2-\sigma_3)\,,\quad t_{eff,3}=t_1+\log\left(-\frac{\sigma_3}{\sigma_1+\sigma_2+\sigma_3}\right)^3\,.
\end{align}
Note that $t_{eff}:=-d\mathcal{W}_{eff}(\dot{\sigma},\sigma_L)$. We find that there is a non-trivial solution of $U_{eff}=0$ given by
\begin{equation}
  \sigma_1=\sigma_2=0, \qquad \zeta_1=0,\theta_3=\pi\mod 2\pi, \qquad dW(\dot{\phi})=0.
\end{equation}
The effective D-term associated to $U(2)$ can be satisfied for any values of $t$, implying that there is a mixed branch along $\zeta_1=0$. There seem to be two phases along this branch but the are lifted by the F-terms given by
\begin{align}
 x_i^a:&\quad \sum_{j=1}^5\sum_{k=4}^5A^{ij}_kp^kx_{j,a}=0\,, \qquad i=1,\ldots,5\,,\: b=1,2\,,\\
  p^{4,5}:&\quad \sum_{i,j=1}^5A^{ij}_k[x_ix_j]=0\,, \qquad k=4,5\,.
\end{align}
For $\zeta_{eff} > 0$ we can set $p^{4,5}=0$ and get a codimension $2$ complete intersection in $G(2,5)$.
The $\zeta_{eff} < 0$-phase seems to have a vacuum at $x=0$, indicating strong coupling phenomena.
This makes the mixed branch a candidate for a mixed Coulomb/confining branch, first observed in \cite{Hori:2016txh}.
Naively, one would expect a Pfaffian phase, but the rank $2$ locus of the matrix $A(p)$ cannot be reached if we only have two non-zero $p$-fields, and hence the branch would be lifted.
We will confirm this in the dual theory where the region in question is weakly coupled. The amoeba in Figure~\ref{fig:splitAmoebaGLSM1} therefore contains all of the phase boundaries that exist in the FI-parameter space.
\subsubsection{Sphere partition function}
\label{sec:spherePartGLSM1}
Let us now come to the sphere partition function of the GLSM and extract the fundamental period of the mirror.
This will allow us to determine the Picard-Fuchs system and to calculate the Gopakumar-Vafa invariants in Section~\ref{sec:gvandmod}.

Using the Cartan charges~\eqref{eqn:model1charges}, we can write down the localized partition function~\eqref{zs2def}, substitute $\sigma_k\rightarrow -ix_k$ and use the algebraic coordinates $ z_1=e^{-2\pi \zeta+i\theta},\,z_2=e^{-2\pi \zeta_1+i\theta_1}$ to obtain
\begin{align}
	\small
	\begin{split}
		Z=&\frac{i}{2}\sum\limits_{m_i\in\mathbb{Z}}\int\limits_{i\mathbb{R}^3}\frac{d\vec{x}^3}{(2\pi)^3}\left[\left(\frac{m_1-m_2}{2}\right)^2-\left(x_1-x_2\right)^2\right]\\
		&\cdot\left[\frac{\Gamma\left(-x_1-\frac{m_1}{2}\right)}{\Gamma\left(1+x_1-\frac{m_1}{2}\right)}\frac{\Gamma\left(-x_2-\frac{m_2}{2}\right)}{\Gamma\left(1+x_2-\frac{m_2}{2}\right)}\right]^5
		\cdot\left(\frac{\Gamma\left(1+x_1+x_2+x_3+\frac{m_1+m_2+m_3}{2}\right)}{\Gamma\left(-x_1-x_2-x_3+\frac{m_1+m_2+m_3}{2}\right)}\right)^3\\
		&\left(\frac{\Gamma\left(1+x_1+x_2+\frac{m_1+m_2}{2}\right)}{\Gamma\left(-x_1-x_2+\frac{m_1+m_2}{2}\right)}\right)^2\left[\frac{\Gamma\left(-x_3-\frac{m_3}{2}\right)}{\Gamma\left(1+x_3-\frac{m_3}{2}\right)}\right]^3\\
		&(z_1\bar{z}_1)^{x_1+x_2}(z_1/\bar{z}_1)^{-\frac{m_1+m_2}{2}}(z_2\bar{z}_2)^{x_3}(z_2/\bar{z}_2)^{-m_3}\,.
	\end{split}
\label{eqn:Z2form2}
\end{align}
To calculate the integrals, we need to evaluate the Grothendieck residues at a proper set of poles.
However, determining the relevant poles is far from trivial and we follow the algorithm that has been presented in~\cite{Gerhardus:2015sla} to solve this problem.
The algorithm is a generalization of an analogous procedure to evaluate two-dimensional Mellin-Barnes integrals, that has been developed in~\cite{Zhdanov1998} and is in turn based on a multi-dimensional generalization of the Jordan lemma.
The reader should be warned that the individual steps will seem somewhat ad-hoc and we refer to~\cite{Gerhardus:2015sla} and~\cite{Zhdanov1998} for a detailed motivation.

First we collect the divisors of poles, which can be read off from the arguments of the gamma functions in the numerator of the integrand in~\eqref{eqn:Z2form2},
\begin{align}
	\begin{split}
		D_1^{n_1}=&x_1+\frac12m_1-n_1\,,\quad D_2^{n_2}=\epsilon+x_2+\frac12m_2-n_2\,,\quad D_3^{n_3}=2\epsilon+x_3+\frac12 m_3-n_3\,,\\
D_4^{n_4}=&3\epsilon+1+n_4+\sum\limits_{i=1}^2\left(x_i+\frac12m_i\right)\,,\quad D_5^{n_5}=4\epsilon+1+n_5+\sum\limits_{i=1}^3\left(x_i+\frac12m_i\right)\,.
	\end{split}
\label{eqn:divopoles}
\end{align}
Each family $\{D_i^{n_i}=0\}$ is parameterized by an integer $n_i\in\mathbb{N}$.
A slight tilt of order $\epsilon$ has been introduced to separate poles that simultaneously lie on more than three divisors.
Considering the cancellations with zeros that arise from the poles of gamma functions in the denominator, the relevant values $n_i\in\mathbb{N}$ are restricted by
\begin{align}
\begin{split}
n_1\ge& \text{Max}(0,m_1)\,,\quad n_2\ge \text{Max}(0,m_2)\,,\quad n_3\ge \text{Max}(0,m_3)\,,\\
n_4\ge&\text{Max}\left(0,-(m_1+m_2)\right)\,,\quad n_5\ge\text{Max}\left(0,-(m_1+m_2+m_3)\right)\,.
\end{split}
\label{eqn:nrestrictions}
\end{align}

To localize the partition function in the phase that is associated to the fibration of complete intersection curves in $G(2,5)$ over $\mathbb{P}^2$ we can assume that $\zeta,\zeta_1\gg 0$.
Following the recipe from~\cite{Gerhardus:2015sla}, the poles that potentially contribute to the sphere partition function are then contained in the half-space
\begin{align}
	\begin{split}
	H=&\left\{\vec{x}\in\mathbb{R}^3\,\bigg|\,\sum\limits_{i=1}^3(1+\delta_i)x_i\ge 0\right\}\,,
	\end{split}
\end{align}
where $\delta_i,\,i=1,\ldots 3$ are again small parameters that tilt the boundary $\partial H$, which would otherwise be parallel to $\{D_5^{n_5}=0\}$. 
Without loss of generality we can assume that $\delta_1>\delta_2>\delta_3>0$.
One can now check which triple intersections of the divisors in \eqref{eqn:divopoles} are contained in $H$ and finds the relevant poles
\begin{align}
p_i=\bigcap\limits_{i\in P_i}\{D_i^{n_i}=0\}\,,
\end{align}
with the sets of divisors being
\begin{align}
	\begin{split}
		P_1=\{1,2,3\}\,,\quad P_2=\{1,3,4\}\,,\quad P_3=\{2,3,4\}\,.
	\end{split}
\end{align}

To identify the poles that actually contribute, one has to consider the intersections of each of the three corresponding divisors with $\partial H$.
A pole is relevant exactly if the triangle that is spanned by the three intersection points inside $\partial H$ contains the origin.
It is easy to show that only the poles $p_1(n_1,n_2,n_3)$ satisfy this requirement.
To see this, we define the intersections
\begin{align}
p_{m,n}=\partial H\cap\bigcap_{\tiny\begin{array}{c}i\in P_m\\i\ne n\end{array}}\{D_i^{n_i}=0\}\,,
\end{align}
and denote the corresponding third coordinates by $x_{3,m,n}$.
For $m\in\{2,3\}$ they are given by
\begin{align}
x_{3,m,1}=n_3-\frac12m_3\,,\quad x_{3,m,2}=1+n_4+\frac12(m_1+m_2)\,,\quad x_{3,m,3}=n_3-\frac12m_3\,.
\end{align}
However, the restrictions on $n_3,n_4$ in~\eqref{eqn:nrestrictions} imply that $x_{3,m,n}>0$ and therefore the origin can not be contained in the triangle spanned by $p_{m,n}$ inside $\partial H$.
Similarly one can show that the corresponding triangle associated to the poles $p_1(n_1,n_2,n_3)$ does indeed contain $0\in\mathbb{R}^3$.

We can now safely set $\epsilon$ in~\eqref{eqn:divopoles} to zero and further simplify the computation by substituting
\begin{align}
x_1\rightarrow n_1-\frac12m_1+x_1\,,\quad x_2\rightarrow n_2-\frac12m_2+x_2\,,\quad x_3\rightarrow n_3-\frac12m_3+x_3\,.
\end{align}
This moves the pole that is associated to $(n_1,n_2,n_3)$ to the origin $(x_1,x_2,x_3)=(0,0,0)$.
Expanding around this point one finds that the complex conjugate coordinates $\bar{z}_1,\bar{z}_2$ contribute with an overall factor of $\bar{z}_ 1^{n_1+n_2}\bar{z}_2^{n_3}$.

Our goal is to obtain the fundamental period and to this end we only need to consider contributions of the form
\begin{align}
f(z_1,z_2)\log(z_1)^a\log(z_2)^{3-a}\,,
\end{align}
for $a=0,\dots,3$. We can therefore assume that $n_1=n_2=n_3=0$.
Then the corresponding $z_1,z_2$-dependent factors are
\begin{align}
\begin{split}
	t(-m_1,-m_2,-m_3)=&\frac16\left[(x_1+x_2)\log(z_1)+x_3\log(z_2)\right]^3z_1^{-m_1-m_2}z_2^{-m_3}\,.\\
\end{split}
\end{align}
Finally, after substituting $m_i\rightarrow -m_i$, we obtain the contribution to the partition function
{\small
\begin{align}
\begin{split}
\tilde{Z}=&\frac{1}{12}\sum\limits_{m_1=0}^\infty\sum\limits_{m_2=0}^\infty\sum\limits_{m_3=0}^\infty z_1^{m_1+m_2}z_2^{m_3}\text{Res}_{\textbf{x}=0}\left(\left[(x_1+x_2)\log(z_1)+x_3\log(z_2)\right]^3(x_2-x_1)\phantom{\frac12}\right.\\
&(m_1-m_2+x_1-x_2)\left(\frac{\Gamma(-x_1)}{\Gamma(1+x_1+m_1)}\frac{\Gamma(-x_2)}{\Gamma(1+x_2+m_2)}\right)^5\left(\frac{\Gamma(1+x_1+x_2)}{\Gamma(-m_1-m_2-x_1-x_2)}\right)^2\\
&\left.\left(\frac{\Gamma(-x_3)}{\Gamma(1+x_3+m_3)}\frac{\Gamma(1+x_1+x_2+x_3)}{\Gamma(-m_1-m_2-m_3-x_1-x_2-x_3)}\right)^3\right)\,.
\end{split}
\label{eqn:Ztilde}
\end{align}
}
According to the general structure of $Z$, we can identify
\begin{align}
\tilde{Z}=-\tilde{c}\frac{i}{6} \varpi_0(z_1,z_2)c_{ijk}{t'}^i{t'}^j{t'}^k\,,
\end{align}
where $c_{ijk}$ are the triple intersection numbers on the Calabi-Yau, $\varpi_0$ is the fundamental period of the mirror Calabi-Yau, $\tilde{c}$ is an overall
normalization and
\begin{align}
{t'}^i=\frac{1}{2\pi i}\log(z_i)\,.
\end{align}

We expect the divisor corresponding to $z_1$ to be a $5$-section of the genus one fibration and $z_2$ to correspond to the vertical divisor that arises
from the hyperplane class in the $\mathbb{P}^2$ base.
This would fix the triple intersection numbers $c_{222}=0$ and $c_{122}=5$.
Indeed, evaluating $\tilde{Z}$ and setting $\tilde{c}=(2\pi)^3$ we find
\begin{align}
c_{111}=15\,,\quad c_{112}=15\,,\quad c_{122}=5\,,\quad c_{222}=0\,,
\label{eqn:model1_3int}
\end{align}
consistent with the results of Section~\ref{sec:geometry}. 
The leading terms of the fundamental period $\varpi_0$ read
\begin{align}
\varpi_0=1+3z_1+z_2+19z_1^2+24z_1z_2+z_2^2+147z_1^3+513z_1^2z_2+81z_1z_2^2+z_2^3+\mathcal{O}(z^4)\,.
\label{eqn:model1fundamental}
\end{align}
Anticipating the generic expression~\eqref{eqn:fundamental} from Section~\ref{sec:classification}, we can write this in closed form
\begin{align}
	\varpi_0=&\sum\limits_{\lambda_1,\lambda_2=0}^\infty\frac{\Gamma(1+\lambda_1+\lambda_2)^3}{\Gamma(1+\lambda_1)^3\Gamma(1+\lambda_2)^3}{_3}F_2(-\lambda_1,-\lambda_1,1+\lambda_1;1,1;1) z_1^{\lambda_1}z_2^{\lambda_2}\,.
	\label{eqn:model1fundClosed}
\end{align}
As a further consistency check, let us use~\eqref{eqn:Z2form2} to also calculate the Euler characteristic via the relation
\begin{align}
\begin{split}
\tilde{c}\frac{\zeta(3)}{4\pi^3}\chi=&-\frac{1}{2}\text{Res}_{\textbf{x}=0}\left((x_1-x_2)^2\left(\frac{\Gamma(-x_1)}{\Gamma(1+x_1)}\frac{\Gamma(-x_2)}{\Gamma(1+x_2)}\right)^5\left(\frac{\Gamma(1+x_1+x_2)}{\Gamma(-x_1-x_2)}\right)^2\right.\\
&\left.\left(\frac{\Gamma(-x_3)}{\Gamma(1+x_3)}\frac{\Gamma(1+x_1+x_2+x_3)}{\Gamma(-x_1-x_2-x_3)}\right)^3\right)\,.
\end{split}
\end{align}
This gives the result $\chi=-90$ which is again consistent with the calculation from Section~\ref{sec:geometry}.
\subsubsection{Non-Abelian dual}
\label{sec:glsm1nadual}
Let us now discuss the non-Abelian dual of our GLSM. This will confirm the expected Pfaffian phases and will provide a non-trivial check for the result of the discriminant.

The dual theory has gauge group $\tilde{G}=U(1)\times U(2)$. The field content is
\begin{equation}
    \begin{array}{c|c|c|c|c|c|c}
      &p^1,p^2,p^3&p^4,p^5&x_1^a,\ldots,x_5^a&b_1,b_2,b_3&a_{ij}&\mathrm{FI}\\
      \hline
      U(2)&\mathrm{det}^{-1}&\mathrm{det}^{-1}&\overline{\Box}&{\bf 1}&\mathrm{det}&\tilde{\zeta}\\
      U(1)&-1&0&0&1&0&\tilde{\zeta_1}.
    \end{array}
  \end{equation}
We thus obtain the following gauge charges:
\begin{equation}
 \begin{array}{c|c|c|cc|c|c|c}
    &p^1,p^2,p^3&p^4,p^5&\tilde{x}_1^{1},\ldots, \tilde{x}_5^1&\tilde{x}_2^{1},\ldots,\tilde{x}_5^2&b_1,b_2,b_3&a_{ij}&\mathrm{FI}\\
    \hline
    \tilde{U}(1)_1&-1&-1&-1&0&0&1&\tilde{\zeta}\\
    \tilde{U}(1)_2&-1&-1&0&-1&0&1&\tilde{\zeta}\\
    U(1)_3&-1&0&0&0&1&0&\tilde{\zeta}_1.
    \end{array}
\end{equation}
The $a_{ij}$ ($i,j\in\{1,\ldots,5\}$) are ten singlet fields satisfying $a_{ij}=-a_{ji}$. The superpotential is
\begin{equation}
  \widetilde{W}=\sum_{i,j=1}^5 A^{ij}(p,b)a_{ij}+[\tilde{x}^i\tilde{x}^j]a_{ij}\,,
\end{equation}
with the same $A^{ij}(p,b)$ as in the original theory. 
We expect that a strongly coupled geometric phase in the original theory is realised as a weakly coupled phase in the dual theory where the geometry in the dual theory is characterised by
\begin{equation}
  \label{y1dualcicy}
  A^{ij}(p,b)+[\tilde{x}^i\tilde{x}^j]=0\,.
\end{equation}
Following the arguments of~\cite{Hori:2011pd}, we can argue that this is the Pfaffian $Y_1$, realised as a complete intersection whose solutions restrict the rank of $A^{ij}(p,b)$ to be two. 

To confirm this, let us look at the phases in more detail. The D-terms are
\begin{align}
  \label{d1dual}
	\begin{split}
		-\sum_{i=1}^5\tilde{x}^i_{b}{\tilde{x}^{i,a\dagger}}-\sum_{k=1}^5|p^k|\delta^a_b+\sum_{i,j=1}^5|a_{ij}|^2\delta^a_b=&\tilde{\zeta}\delta^a_b\,,\\
		-|p^1|^2-|p^2|^2-|p^3|^2+\sum_{i=1}^3 |b_i|^2=&\tilde{\zeta}_1\,.
	\end{split}
\end{align}
There is also the linear combination
\begin{align}
  \label{dextradual}
  &-\sum_{i=1}^5|x_i^1|^2-|p^4|^2-|p^5|^2-\sum_{i=1}^3 |b_i|^2+\sum_{i,j=1}^5|a_{ij}|^2\nonumber\\
  =&-\sum_{i=1}^5|x_i^2|^2-|p^4|^2-|p^5|^2-\sum_{i=1}^3 |b_i|^2+\sum_{i,j=1}^5|a_{ij}|^2=\tilde{\zeta}-\tilde{\zeta}_1.
\end{align}
The F-term equations are
\begin{align}
  \tilde{x}^i_a:&\quad \sum_{j=1}^5a_{ij}\tilde{x}^{j,a}=0\,, \quad i=1,\ldots,5,\: a=1,2\,,\label{f1dual}\\
  p^k:&\quad \sum_{i,j=1}^5\sum_{l=1}^3A^{ijl}_{k}b_la_{ij}=0\,,\quad k=1,2,3\,,\label{f2dual}\\
  &\quad \sum_{i,j=1}^5A^{ij}_ka_{ij}=0\,,\quad k=4,5\,,\label{f3dual}\\
  b_l:&\quad \sum_{i,j=1}^5\sum_{k=1}^3A^{ijl}_{k}p^ka_{ij}=0\,,\quad l=1,2,3\,,\label{f4dual}\\
  a_{ij}:&\quad  \sum_{k=1}^5 A^{ij}(p,b)+[\tilde{x}^i\tilde{x}^j]=0\,.\label{f5dual}
\end{align}
For $\tilde{\zeta} > 0,\tilde{\zeta}_1 > 0$ the $a_{ij}$ are not allowed to vanish simultaneously, neither are the $b_{1,2,3}$. Further note that the $U(2)$ D-term still implies that if the matrix $\tilde{x}\equiv \tilde{x}^i_a$ is non-zero it has to have full rank. There is a vacuum for $\tilde{x}=0$ and$p^1=\ldots=p^5=0$ where the remaining fields are subject to (\ref{f2dual}) and (\ref{f3dual}). Hence the vacuum manifold is not a Calabi-Yau threefold, and we expect the phase to be realised via non-perturbative effects.

Next we consider $\tilde{\zeta} < 0,\tilde{\zeta}_1 > 0$. Now $\tilde{x}$ and $p$  are not allowed to vanish at the same time, nor are the $b$-fields. There is a vacuum for $a_{ij}=0$ and the non-zero fields are constrained by (\ref{f5dual}), as expected. We also have to check if there are no further solutions. Note that, since $\tilde{\zeta}_1 > 0$, $b_1=b_2=b_3=0$ is not permitted, while $p^1=p^2=p^3=0$ is allowed. However, if we set $p^1=p^2=p^3=0$ then (\ref{f4dual}) reduces to
\begin{equation}
  \sum_{k=4}^5 A_k^{ij}p^k+[\tilde{x}^i\tilde{x}^j]=0\,.
\end{equation}
This equation constrains the matrix $A(p)$ to the rank $2$-locus, giving us the Pfaffian $Y_1$ constructed in Section~\ref{sec:geometry}. If we only have two non-zero $p$-fields, this locus cannot be reached, so $p^1=p^2=p^3=0$ is actually not a solution.

If $\tilde{\zeta} < 0,\tilde{\zeta}_1 < 0, \tilde{\zeta}-\tilde{\zeta}_1 < 0$ then $b_1=b_2=b_3=0$ is allowed, while $p^1=p^2=p^3=0$ is not allowed. Again there is a vacuum for $a_{ij}=0$ that is also described by (\ref{f5dual}). So it is the same Pfaffian phase as for $\tilde{\zeta} < 0,\tilde{\zeta}_1 > 0$ with the roles of $p^{1,2,3}$ and $b_{1,2,3}$ exchanged. However, due to the symmetry between $p$ and $b$, we can argue that setting $b_1=b_2=b_3=0$ is not a solution to the F-term equations. This implies that the phase boundary between the phases $\tilde{\zeta} < 0,\tilde{\zeta}_1 > 0$ and $\tilde{\zeta} < 0,\tilde{\zeta}_1 < 0, \tilde{\zeta}-\tilde{\zeta}_1 < 0$ is lifted by the F-terms. This confirms  an earlier suspicion in the original theory that there is no phase boundary. The original theory was strongly coupled in this region of the parameter space, so the lifting of the phase boundary was hard to see, compared to the straight forward discussion in the dual theory.
In the phase $\tilde{\zeta}_1 < 0, \tilde{\zeta}-\tilde{\zeta}_1 > 0$ we are again allowed to set $b_1=b_2=b_3=0$, but are not allowed to set all $\tilde{x}$ and $p$ to zero at the same time. This is a non-perturbatively realised phase.

Let us analyze the Coulomb branch of this model to see if it matches the original model. The effective potential is 
\begin{align}
  \mathcal{W}_{eff}&=-\tilde{t}(\tilde{\sigma}_1+\tilde{\sigma}_2)-\tilde{t}_1\tilde{\sigma}_3-3(-\tilde{\sigma}_1-\tilde{\sigma}_2-\tilde{\sigma}_3)\left[\log(-\tilde{\sigma}_1-\tilde{\sigma}_2-\tilde{\sigma}_3)-1\right]\nonumber\\
  &\quad -2(-\tilde{\sigma}_1-\tilde{\sigma}_2)\left[\log(-\tilde{\sigma}_1-\tilde{\sigma}_2)-1\right]-5(-\tilde{\sigma}_1)\left[\log(-\tilde{\sigma}_1)-1\right]-5(-\tilde{\sigma}_2)\left[\log(-\tilde{\sigma}_2)-1\right]\nonumber\\
  &\quad -3\tilde{\sigma}_3\left[\log\tilde{\sigma}_3-1\right]-10(\tilde{\sigma}_1+\tilde{\sigma}_2)\left[\log(\tilde{\sigma}_1+\tilde{\sigma}_2)-1\right]+i\pi(\tilde{\sigma}_1-\tilde{\sigma}_2).
\end{align}
Defining $\tilde{z}=\frac{\tilde{\sigma}_2}{\tilde{\sigma}_1}$ and $\tilde{w}=\frac{\tilde{\sigma}_3}{\tilde{\sigma}_1}$ the critical locus is
\begin{align}
  \label{dualcoulomb}
  e^{-\tilde{t}}=-\frac{\left(1+\frac{1}{\tilde{z}}\right)^8}{\left(1+\frac{1}{\tilde{z}}+\frac{\tilde{w}}{\tilde{z}}\right)^3}\,,\quad e^{-\tilde{t}_1}=-\frac{1}{\left(1+\frac{1}{\tilde{w}}+\frac{\tilde{z}}{\tilde{w}}\right)^3}\,, \quad \frac{1}{\tilde{z}^5}=1\,,
\end{align}
so that $\tilde{z}$ is again a power of $\omega=e^{\frac{2\pi i}{5}}$ and $\omega,\omega^2$ lead to two independent solutions as in the original theory.
We aim to establish transformations that map $e^{-t_i}\leftrightarrow e^{-\tilde{t}_i}$ by giving a mapping $(\tilde{z},\tilde{w})\leftrightarrow (z,w)$.
We achieve this by showing that the mapping works for each solution of $z^5=1$. We use the identity
\begin{equation}
  \label{omegaid}
  (1+\omega)^5=-(1+\omega^2)^{-5}. 
\end{equation}
Let us assume that $\tilde{z}=\omega$. Then we have
\begin{equation}
  e^{-\tilde{t}}|_{\tilde{z}=\omega}=-\frac{(1+\omega)^8}{(1+\omega+\tilde{w})^3}\stackrel{(\ref{omegaid})}{=}\frac{(1+\omega)^3}{(1+\omega^2)^5(1+\omega+\tilde{w})^3}=\frac{1}{(1+\omega^2)\left(1+\omega^2+\tilde{w}\frac{(1+\omega^2)}{1+\omega}\right)^3}
\end{equation}
An analogous calculation for $\tilde{z}=\omega^2$ shows that
\begin{equation}
   e^{-\tilde{t}}|_{\tilde{z}=\omega^2}=\frac{1}{(1+\omega)\left(1+\omega^2+\tilde{w}\frac{(1+\omega)}{1+\omega^2}\right)^3}
\end{equation}
Comparing with (\ref{coulomb}) we find $e^{-t}=e^{-\tilde{t}}$ for 
\begin{align}
  (\tilde{z},z)=(\omega,\omega^2)&\quad& \tilde{w}\frac{1+\omega^2}{1+\omega}=w\nonumber\\
  (\tilde{z},z)=(\omega^2,\omega)&\quad& \tilde{w}\frac{1+\omega}{1+\omega^2}=w.
\end{align}
As a consistency check we show that these transformations identify $e^{-t_1}=e^{-\tilde{t}_1}$. Take $\tilde{z}=\omega$
\begin{equation}
  -\frac{1}{\left(1+\frac{1}{\tilde{w}}+\frac{\omega}{\tilde{w}}\right)^3}\mapsto -\frac{1}{\left(1+\frac{(1+\omega)}{(1+\omega^2)\tilde{w}}+\frac{\omega^2(1+\omega)}{(1+\omega^2)\tilde{w}} \right)^3}=-\frac{1}{\left(1+\frac{1}{\tilde{w}}+\frac{\omega}{\tilde{w}}\right)^3}.
\end{equation}
So the duality exchanges the two singular points.

While in the original theory it not completely obvious to see that the mixed branch with unbroken $U(1)_3$ is lifted, it is relatively straight forward to show this in the dual model. The fields are divided up as follows:
\begin{equation}
  \dot{\phi}=\{x_i^a,p^4,p^5,a_{ij}\}, \qquad \hat{\phi}=\{p^1,p^2,p^3,b_1,b_2,b_3\}, 
\end{equation}
where the dotted fields are uncharged under $U(1)_3$ and thus massless, while the hatted fields are massive. Integrating out the massive degrees of freedom, results in the following effective scalar potential:
\begin{align}
  U_{eff}&=\frac{1}{e^2_{eff}}|[\dot{\tilde{\sigma}},\dot{\bar{\tilde{\sigma}}}]|^2\nonumber\\
  &\quad +\frac{1}{2}\left(\sum_{i=1}^5|-\tilde{\sigma}_1\tilde{x}_i^1|^2+\sum_{i=1}^5|-\tilde{\sigma}_2\tilde{x}_i^2|^2+\sum_{i=4}^5|(-\sigma_1-\sigma_2)p^i|^2+\sum_{ij}|(\tilde{\sigma}_1+\tilde{\sigma}_2)a_{ij}|+\sigma\leftrightarrow\bar{\sigma}\right)\nonumber\\
  &\quad+\frac{e^2_{eff}}{2}\left(\mu(\dot{\phi})-\zeta_{eff}(\dot{\tilde{\sigma}},\tilde{\sigma}_L)\right)+|dW(\dot{\phi})|^2,
\end{align}
where
\begin{equation}
  W(\dot{\phi})=\sum_{i,j=1}^5\sum_{k=4}^5A^{ij}_kp^ka_{ij}+[\tilde{x}^i\tilde{x}^j]a_{ij},
\end{equation}
and
\begin{align}
  \mu_{eff}&=\sum_{i=1}^5\left(-x_i^ax^{\dagger}_{i,a}-2\sum_{k=1}^3|p^k|^2\right)+\sum_{ij}|a_{ij}|^2,\nonumber\\
  \mu_{eff,3}&=0,
\end{align}
and furthermore
\begin{align}
  \tilde{t}_{eff}&=2\tilde{t}-6\log(-\tilde{\sigma}_1-\tilde{\sigma}_2-\tilde{\sigma}_3)\nonumber\\
  \tilde{t}_{eff,3}&=\tilde{t}_1-3\log(-\tilde{\sigma}_1-\tilde{\sigma}_2-\tilde{\sigma}_3)+3\log\tilde{\sigma}_3=\tilde{t}_1+\log\left(-\frac{\tilde{\sigma}_3}{\tilde{\sigma}_1+\tilde{\sigma}_2+\tilde{\sigma}_3}\right)^3.
\end{align}
As usual $t_{eff}:=-d\mathcal{W}_{eff}(\dot{\sigma},\sigma_L)$. We find that there is a non-trivial solution of $U_{eff}=0$ given by
\begin{equation}
  \tilde{\sigma}_1=\tilde{\sigma}_2=0, \qquad \tilde{\zeta}_1=0,\tilde{\theta}_3=\pi\mod 2\pi, \qquad dW(\dot{\phi})=0.
\end{equation}
In the $\tilde{\zeta} < 0$-case the F-terms imply the expected determinantal phase. However, as we have argued above, the F-terms do not have a non-trivial solution if there are only two $p$-fields. This implies that the mixed branch is lifted, as expected.
\subsubsection{Relation to a GLSM constructed in \cite{Knapp:2019cih}}
\label{sec-match}
In \cite{Knapp:2019cih}, a class of non-Abelian GLSMs has been discussed that provides a physics realisation of a construction of Calabi-Yau threefolds by means of joins \cite{Inoue:2019jle} (see also \cite{Prince:2019vsu} for a different approach that also led to some of these geometries). One of these examples realised the geometries $X_1$ and $Y_1$ in a GLSM with gauge group $U(1)\times U(2)$. Here we comment on the rather non-trivial connection between this GLSM and the GLSM and its dual we have just considered. The table below gives the field content of the GLSM as presented in Section 3.2 of \cite{Knapp:2019cih}:
\begin{equation}
\begin{array}{c|ccrr|c} 
&  \phi^i_a & x_{\alpha} & y_{\beta} & q_m & \mathrm{FI}\\ \hline
SU(2) & \Box & {\bf 1} & {\bf 1} & {\bf 1}&-\\
U(1)_{\rm det} & 1 & 1 & 1 & -2&r_1 \\
U(1)_{3} & 0 & 1 & -1 & 0&r_2.
\end{array}
\end{equation}
Here we have used the isomorphism $U(2)\cong \frac{U(1)\times SU(2)}{\{\pm 1,\pm {\bf 1}\}}$. There are five doublets $\phi^i_a$ ($i=1,\ldots,5$, $a=1,2$). Furthermore, this model has $14$ singlets $x_{\alpha},y_{\beta}$ ($\alpha,\beta=1,2,3$) and $q_m$ ($m=1,\ldots,8$). This does not match with the number of singlet fields of our GLSM or its dual. It has been shown in \cite{Knapp:2019cih} that this GLSM has the following phase structure. The phase $r_1 >0,|r_2|>0$ realises the Pfaffian Calabi-Yau $Y_1$ perturbatively as a codimension $8$ complete intersection. The two phases at $r_1 <0,r_1\pm r_2<0$ both realise $X_1$ as a strongly coupled phase. This suggests that we should be able to match this model with the non-Abelian dual of our GLSM. To this end, we make the following change of basis
\begin{equation}
\begin{array}{c|ccrr|c} 
&  \phi^i_a & x_{\alpha} & y_{\beta} & q_m & \mathrm{FI}\\ \hline
  SU(2) & \Box & {\bf 1} & {\bf 1} & {\bf 1}&-\\
 \tilde{U}(1)_{\rm det} & 1 & 2 & 0 & -2&r_1+r_2 \\
U(1)_{3} & 0 & 1 & -1 & 0&r_2.
\end{array}
\end{equation}
The $U(1)\times SU(2)$-charges of our non-Abelian dual GLSM are:
\begin{equation}
    \begin{array}{c|c|c|c|c|c|c}
      &p^1,p^2,p^3&p^4,p^5&x_1^a,\ldots,x_5^a&b_1,b_2,b_3&a_{ij}&\mathrm{FI}\\
      \hline
      SU(2)&{\bf 1}&{\bf 1}&\overline{\Box}&{\bf 1}&{\bf 1}&-\\
      U(1)_{\rm det}&-2&-2&-1&0&2&2\tilde{\zeta}\\
      U(1)&-1&0&0&1&0&\tilde{\zeta_1}.
    \end{array}
  \end{equation}
Comparing the charges suggests the following identification of fields
\begin{equation}
  \label{ericidentification}
  \phi^i_a\leftrightarrow x_i^a,\quad x_{\alpha}\leftrightarrow p^{1,2,3},\quad y_{\beta}\leftrightarrow b_{1,2,3}, \quad q_m\leftrightarrow a_{ij}.
\end{equation}
Since the charges of the corresponding fields have opposite signs, we also should make the identification
\begin{equation}
  r_1+r_2\leftrightarrow -2\tilde{\zeta}, \qquad r_2\leftrightarrow -\tilde{\zeta}_1.
  \end{equation}
The matching does not quite work though: there are $10$ $a_{ij}$ but only $8$ $q_m$, and we have not found a match for $p^4,p^5$. The resolution of this puzzle is that we can actually eliminate $p^4,p^5$ and two of the $a_{ij}$ from our non-Abelian dual model. The charges of the fields are opposite, so doing this obviously does not change the Calabi-Yau condition. Furthermore, the contributions of pairs of fields with opposite charges cancel in the Coulomb branch analysis in the sense that one gets the same critical locus of the effective potential\footnote{To be precise, pairs of fields with opposite charges $\pm q_i$ would contribute a factor $(-1)^{q_i}$ to the expression $e^{-t_i}$, corresponding to a shift of $\pi\mod 2\pi$ in the theta angle. In our case, however, the charges are even, so there is indeed no contribution.}. This means that the discriminant is the same.

To see how these fields can be integrated out, we go back to the D-term and F-term equations of the non-Abelian dual, see (\ref{d1dual})--(\ref{dextradual}) and (\ref{f1dual})--(\ref{f5dual}). Since the charges of the fields we want to remove align with the charges of other fields, the phase diagrams will look the same. Now consider the F-terms (\ref{f3dual}) associated to $p^4,p^5$ which are two linear equations for the $a_{ij}$. Hence, we can use these equations to eliminate two of the ten $a_{ij}$ under suitable genericity assumptions on the coefficient matrix $A_k^{ij}$. Similarly, we can use two of the $10$ F-term equations (\ref{f5dual}) associated to $a_{ij}$ to solve explicitly for $p^4,p^5$ which are multiplied by the same coefficients $A_k^{ij}$. Now the degrees of freedom match up and the superpotential can be reorganised into the superpotential of \cite{Knapp:2019cih} using the field identification (\ref{ericidentification}). This establishes the correspondence. 
\subsection{GLSM  for $X_2$, $X_2'$, $Y_2$}
\label{sec:glsmModel2}
This model will realise the geometries $X_2,Y_2$ and $X_2'$ constructed in Section~\ref{sec:geometry}. This GLSM has also been mentioned in \cite{Donagi:2007hi} in connection with the constructions of \cite{Caldararu:2002ab}.
\subsubsection{Matter content and phases}
Let us consider another model with a more complicated fibration structure in the Grassmannian phase. The field content is
\begin{equation}
      \begin{array}{c|c|cc|c|c}
        &p^1,\ldots,p^5&x_1^a,\ldots,x_4^a&x_5^a&b_1,b_2,b_3&\mathrm{FI}\\
        \hline
        U(2)&\mathrm{det}^{-1}&\Box&\Box&{\bf 1}&\zeta\\
        U(1)&-1&0&1&1&\zeta_1
        \end{array}
      \end{equation}
and the gauge charges are
\begin{equation}
 \begin{array}{c|c|cccc|c|c}
    &p^{1},\ldots,p^{5}&x^1_{1},\ldots,x^1_{4}&x_5^1&x^2_{1},\ldots,x^2_{4}&x_5^2&b_1,b_2,b_3&\mathrm{FI}\\
    \hline
    U(1)_1&-1&1&1&0&0&0&\zeta\\
    U(1)_2&-1&0&0&1&1&0&\zeta\\
    U(1)_3&-1&0&1&0&1&1&\zeta_1.
    \end{array}
\label{eqn:model2charges}
\end{equation}
Again the D-terms for $U(2)$ are the same as for the elliptic curve. 
The extra D-term now reads
\begin{equation}
  \label{ad2}
  -\sum_{k=1}^5 |p^k|^2+\sum_{a=1}^2|x_5^a|^2+\sum_{l=1}^3 |b_l|^2=\zeta_1.
\end{equation}
We still have compatibility with $p=0$ for $\zeta_1 > 0$. Taking the difference of the D-terms, one gets
\begin{equation}
  \label{adextra}
  \sum_{i=1}^4|x_i^1|^2-\sum_{l=1}^3 |b_l|^2=\sum_{i=1}^4|x_i^2|^2-\sum_{l=1}^3 |b_l|^2 =\zeta-\zeta_1.
\end{equation}
The superpotential is now such that $A^{i,5}(p,b)$ and $A^{5,i}(p,b)$ are constant in $b$ while the remaining entries are linear. So we have
\begin{equation}
W=\sum_{i,j=1}^4\sum_{k=1}^5\sum_{l=1}^3A^{ijl}_{k}b_lp^k[x_ix_j]+\sum_{i=1}^4\sum_{k=1}^5A^{i5}_{k}p^k[x_ix_5].
\end{equation}
From this we obtain the following F-term equations:
\begin{align}
  x_i^a:&\quad \sum_{j=1}^4\sum_{k=1}^5\sum_{l=1}^3A^{ijl}_{k}b_lp^kx_{j,a}+\sum_{k=1}^5A^{i5}_{k}p^kx_5^b\epsilon_{ab}=0, \qquad i=1,\ldots,4,\: a=1,2,\label{af1}\\
  &\quad \sum_{i=1}^4\sum_{k=1}^5A^{i5}_{k}p^kx_{i,a}=0,\qquad a=1,2,\label{af2}\\
  p^k:&\quad \sum_{i,j=1}^4\sum_{l=1}^3 A^{ijl}_{k}b_l[x_ix_j]+\sum_{i=1}^4A^{i5}_{k}[x_ix_5]=0, \qquad k=1,\ldots,5,\label{af3}\\
  b_l:&\quad \sum_{i,j=1}^4\sum_{k=1}^5A^{ijl}_{k}p^k[x_ix_j]=0,\qquad l=1,2,3.\label{af4}
\end{align}
The charges of the fields now indicate that there is a phase boundary in the $\zeta>0,\zeta_1>0$-region. Experience with the previous model implies that this should be lifted by the F-terms.

We start by considering $\zeta > 0, \zeta_1 > 0, \zeta-\zeta_1 > 0$. In this case the $x$-fields are not allowed to vanish and hence the $2\times 5$-matrix $x$ has rank $2$. Furthermore $x_5$ and $b_1,b_2,b_3$ are not allowed to vanish simultaneously. Note that, in contrast to the other example, it is not completely obvious that $b_1=b_2=b_3=0$ is disallowed. The D-term (\ref{adextra}) does not give new information compared to the other D-term because $x_i^1=0$ or $x_i^2=0$ for all $i$ is not allowed since this implies $\mathrm{rk}x=1$. We can set all the $p$-fields to zero, whereupon the F-terms reduce to
\begin{equation}
  \sum_{i,j=1}^4 \sum_{l=1}^3A^{ijl}_{k}b_l[x_ix_j]+\sum_{i=1}^4A^{i5}_{k}[x_ix_5]=0, \qquad k=1,\ldots,5,
\end{equation}
which is a complete intersection in a Grassmannian. What happens if we also set $b_1=b_2=b_3=0$? Then the first term in the equation above disappears and we are left with five equations for four $[x_ix_5]$. This can only be solved for $x_i\propto x_5$ for all $i=1,\ldots,4$, but this means that $x$ has rank $1$ which is forbidden by the D-terms. Hence, $b_1=b_2=b_3=0$ is excluded by the F-terms. So, we have recovered the geometry $X_2$ of Section~\ref{sec:geometry}. 

Now consider $\zeta > 0, \zeta_1 > 0, \zeta-\zeta_1 < 0$. Then $b_1=b_2=b_3=0$ is excluded explicitly by (\ref{adextra}). Otherwise the situation is the same and we cannot distinguish this phase from the previous one. Hence, we have shown that the extra phase boundary is lifted.

The second genuine phase is $\zeta < 0, \zeta-\zeta_1 < 0$. In this phase setting all $p$-fields to zero is disallowed, and so is $b_1=b_2=b_3=0$ by (\ref{adextra}). There is a classical vacuum given by $x_1=\ldots x_5=0$. We expect this phase to be non-perturbatively realised and we will move to the dual theory to confirm that we get the geometry $Y_2$. 

The third phase is at $\zeta_1 < 0, \zeta-\zeta_1 > 0$. In this phase setting all $p$-fields to zero is disallowed and so is $x_{1}=\ldots=x_{4}=0$. We can try to set $b_1=b_2=b_3=0$ and $x_5=0$. Then the only F-terms that are left are (\ref{af2}) and (\ref{af4}). The $p$-fields, taking values in $\mathbb{P}^4$, take the role of fiber coordinates, and the remaining $x$-fields live in $G(2,4)$. What happens if we consider non-zero $b$-fields? In this case we get $8$ more equations from (\ref{af1}) and $5$ more equations from (\ref{af3}). This makes a total number of $16$ F-term conditions for $16$ variables. Together with the D-terms this will not have a solution for generic equations. Another possible solution could be to have $x_5\neq 0$ while keeping $b_1=b_2=b_3=0$. Then all the $18$ F-terms give non-trivial conditions and we again do not expect a solution. So this phase is not a genus one fibration, but rather a complete intersection in a $G(2,4)$-fibration over $\mathbb{P}^4$. This coincides with model $X_2'$ in Section~\ref{sec:geometry}.

\begin{figure}
\begin{center}
      \begin{tikzpicture}
        \draw[->] (-3,0) -- (3,0);
        \draw[->] (0,-3) -- (0,3);
        \draw[ultra thick, blue] (0,0) -- (2.5,0);
        \draw[ultra thick, blue] (0,0) -- (0,2.5);
        \draw[ultra thick, blue] (0,0) -- (-2.5,-2.5);
        \draw[ultra thick, dashed, blue] (0,0) -- (2.5,2.5);
        \draw (0.3,3) node {$\zeta_1$};
        \draw (3.3,0) node {$\zeta$};
         \draw (1.5,2) node {$X_2$};
        \draw (1.5,-1.5) node {$X_2'$};
        \draw (-1.5,1.5) node {$Y_2$};
        \draw (2.5,0.3) node {$x_{1,\ldots,4}$};
        \draw (-0.5,2.5) node {$b_{1,2,3}$};
        \draw (2.8,2.2) node {$x_5$};
        \draw (-2.5,-2) node {$p^{1,\ldots,5}$};
        \end{tikzpicture}
      \end{center}\caption{Classical phase diagram of the GLSM for $X_2$, $Y_2$.}\label{fig:glsm2phases}
\end{figure}
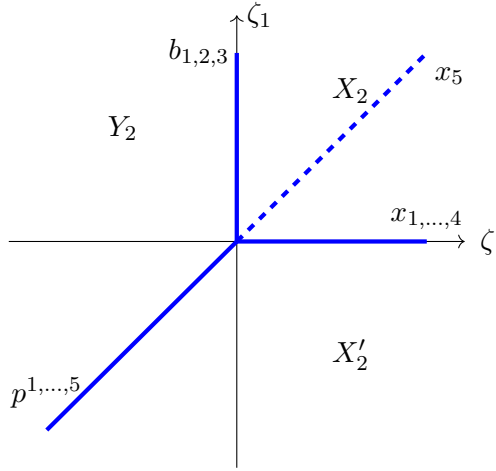

The classical phase diagram can be found in Figure~\ref{fig:glsm2phases} which again looks similar to the phase structures found in \cite{Chen:2020iyo}. 
\subsubsection{Coulomb branch}
The effective potential is
\begin{align}
  \mathcal{W}_{eff}&=-t(\sigma_1+\sigma_2)-t_1\sigma_3-5(-\sigma_1-\sigma_2-\sigma_3)\left[\log(-\sigma_1-\sigma_2-\sigma_3)-1\right]\nonumber\\
  &\quad -4\sigma_1\left[\log\sigma_1-1\right]-4\sigma_2\left[\log\sigma_2-1\right]\nonumber\\
  &\quad -(\sigma_1+\sigma_3)\left[\log(\sigma_1+\sigma_3)-1\right]-(\sigma_2+\sigma_3)\left[\log(\sigma_2+\sigma_3)-1\right]\nonumber\\
  &\quad -3\sigma_3\left[\log\sigma_3-1\right]+i\pi(\sigma_1-\sigma_2).
\end{align}
With the same definitions for $z$ and $w$ as in the previous example the critical locus is given by
\begin{align}
  e^{-t}&=\frac{1+w}{(1+z+w)^5}=\frac{1+\frac{w}{z}}{\left(1+\frac{1}{z}+\frac{w}{z}\right)^5},\nonumber\\
  e^{-t_1}&=-\frac{\left(1+\frac{1}{w}\right)\left(1+\frac{z}{w}\right)}{\left(1+\frac{1}{w}+\frac{z}{w}\right)^5}=-\frac{w^3(1+w)(1+z)}{(1+w+z)^5},\nonumber\\
  z^5\frac{\left(1+\frac{w}{z}\right)}{(1+w)}&=1,\nonumber\\
e^{-t+t_1}&=-\frac{1}{w^3(1+z)}.
\end{align}
For $w=0$ we recover what we have seen before in the GLSM for $X_1$ and $Y_1$: the two singular points of the elliptic curve, shifted to $\zeta_1\rightarrow\infty$. The third equation is linear in $w$, so we solve for it:
\begin{equation}
  w=-\frac{1-z^5}{1-z^4}.
\end{equation}
Inserting this back we get
\begin{align}
  e^{-t}&=-\frac{(1+z+z^2+z^3)^4}{z(1+z+z^2)^5},\nonumber\\
  e^{-t_1}&=\frac{(1+z+z^2+z^3+z^4)^3}{z(1+z+z^2)^5},\nonumber\\
  e^{-t+t_1}&=-\frac{(1+z+z^2+z^3)^4}{(1+z+z^2+z^3+z^4)^3}.
\end{align}
In order to find the tentacles of the amoeba we have to bear in mind that $z=1$ is not allowed as it is fixed by the Weyl group and $z=0$ is not allowed because it contradicts the third equation above. What is allowed is $z=-1$ for which we get
\begin{equation}
  e^{-t}\rightarrow 0, \qquad e^{-t_1}\rightarrow -1.
\end{equation}
We have seen this before -- it is the phase boundary along the $\zeta$-axis. Next, we can consider the roots of the $z$-polynomials. First, $1+z+z^2=0$ which is $z=\kappa^l=e^{\frac{2\pi i}{3}l}$ ($l=1,2$). Then we get
\begin{equation}
  e^{-t}\rightarrow \infty,\qquad e^{-t_1}\rightarrow \infty, \qquad  e^{-t+t_1}=1.
\end{equation}
This gives a phase boundary at $\zeta=\zeta_1 < 0$. The roots of $1 + z + z^2 + z^3$ are $z=-1,\pm i$. We have already discussed this because $z=\pm i$ yields the same behaviour as $z=-1$. Finally, the roots of $1+z+z^2+z^3+z^4$ are $z=\omega^k$ ($k=1,2$), as in the previous example. This is the same as $w=0$. This accounts for all the legs of the amoeba. One can show that the results match with topological string calculations. 
\subsubsection{Sphere partition function}
We are now going to extract the fundamental period of the mirror of $X_2$ from the sphere partition function.
The general procedure has already been discussed in Section~\ref{sec:spherePartGLSM1} and uses the algorithm presented in~\cite{Gerhardus:2015sla}.
Here we directly omit any regularization parameters, which again turn out to be irrelevant for the poles that actually contribute.
Using the Cartan charges~\eqref{eqn:model2charges}, we write down the localized partition function on $S^2$ as~\cite{Jockers:2012dk}
\begin{align}
\small
	\begin{split}
		Z=&\frac{i}{2}\sum\limits_{m_i\in\mathbb{Z}}\,\int\limits_{i\mathbb{R}^3}\frac{d\vec{x}^3}{(2\pi)^3}\left[\left(\frac{m_1-m_2}{2}\right)^2-\left(x_1-x_2\right)^2\right]\\
		&\cdot\left[\frac{\Gamma\left(-x_1-\frac{m_1}{2}\right)}{\Gamma\left(1+x_1-\frac{m_1}{2}\right)}\frac{\Gamma\left(-x_2-\frac{m_2}{2}\right)}{\Gamma\left(1+x_2-\frac{m_2}{2}\right)}\right]^4\left[\prod\limits_{i=1}^2\frac{\Gamma\left(-x_i-x_3-\frac{m_i+m_3}{2}\right)}{\Gamma\left(1+x_i+x_3-\frac{m_i+m_3}{2}\right)}\right]\\
		&\cdot\left[\frac{\Gamma\left(1+x_1+x_2+x_3+\frac{m_1+m_2+m_3}{2}\right)}{\Gamma\left(-x_1-x_2-x_3+\frac{m_1+m_2+m_3}{2}\right)}\right]^5\left[\frac{\Gamma\left(-x_3-\frac{m_3}{2}\right)}{\Gamma\left(1+x_3-\frac{m_3}{2}\right)}\right]^3\\
		&(z_1\bar{z}_1)^{x_1+x_2}(z_1/\bar{z}_1)^{-\frac{m_1+m_2}{2}}(z_2\bar{z}_z)^{x_3}(z_2/\bar{z}_2)^{-m_3}\,,
	\end{split}

\label{eqn:Z2model2form2}
\end{align}
where $z_1=e^{-2\pi\zeta +i\theta},\,z_2=e^{-2\pi\zeta_1+i\theta_1}$ and we have substituted $\sigma_i\rightarrow-ix_i$.
There are six families of divisors of poles that can be read off from the arguments of the gamma functions in the denominator
\begin{align}
	\begin{split}
		D_1^{n_1}=&x_1+\frac12m_1-n_1\,,\quad D_2^{n_2}=x_2+\frac12m_2-n_2\,,\quad D_3^{n_3}=x_3+\frac12 m_3-n_3\,,\\
		D_4^{n_4}=&1+\sum\limits_{i=1}^3\left(x_i+\frac12m_i\right)+n_4\,,\quad D_5^{n_5}=x_1+x_3+\frac12(m_1+m_3)-n_5\,,\\
		D_6^{n_6}=&x_2+x_3+\frac12(m_2+m_3)-n_6\,.
	\end{split}
\label{eqn:divopoles2}
\end{align}
Due to cancellations with zeros from the numerator, the values $n_i\in\mathbb{N}$ are restricted by
\begin{align}
\begin{split}
n_1\ge& \text{Max}(0,m_1)\,,\quad n_2\ge \text{Max}(0,m_2)\,,\quad n_3\ge \text{Max}(0,m_3)\,,\\
n_4\ge&\text{Max}\left(0,-(m_1+m_2+m_3)\right)\,,\quad n_5\ge\text{Max}\left(0,m_1+m_3\right)\,,\\
n_6\ge&\text{Max}\left(0,m_2+m_3\right)\,.
\end{split}
\label{eqn:nrestrictions2}
\end{align}
To evaluate the partition function in the geometric phase associated to $X_2$, we can assume $\zeta_1,\zeta_2\gg 0$.
Poles that potentially contribute are then contained in the critical half-space
\begin{align}
	\begin{split}
	H=&\left\{\vec{x}\in\mathbb{R}^3\,\bigg|\,\sum\limits_{i=1}^3x_i\ge 0\right\}\,.\\
	\end{split}
\end{align}
This reduces the list of candidates to
\begin{align}
p_i=\bigcap\limits_{i\in P_i}\{D_i^{n_i}=0\}\,,
\end{align}
with the sets of divisors being
\begin{align}
	\begin{split}
		P_1=&\{1,2,3\}\,,\quad P_2=\{1,2,5\}\,,\quad P_3=\{1,2,6\}\,,\quad P_4=\{1,3,6\}\,,\\
		P_5=&\{1,5,6\}\,,\quad P_6=\{2,3,5\}\,,\quad P_7=\{2,5,6\}\,,\quad P_8=\{3,5,6\}\,.
	\end{split}
\end{align}
Following~\cite{Gerhardus:2015sla}, the poles that actually contribute are those that are intersections of three divisors $D_1,\,D_2,\,D_3$ such that the triangle spanned by $D_i\cdot \partial H,\,i=1,2,3$ contains the origin.
One can check that only the poles $p_1(n_1,n_2,n_3)$ satisfy this criterion.

Considering only the contributions of the form
\begin{align}
	\tilde{Z}=\tilde{c}\left[-\frac{i}{6} \varpi_0(z_1,z_2)c_{ijk}{t'}^i{t'}^j{t'}^k+\frac{\zeta(3)}{4\pi^3}\chi(M)\right]\quad\text{with}\quad{t'}^i=\frac{1}{2\pi i}\log(z_i)\,,
\end{align}
we then obtain
{\small
\begin{align}
\begin{split}
	\tilde{Z}=&\frac{1}{2}\sum\limits_{m_1=0}^\infty\sum\limits_{m_2=0}^\infty\sum\limits_{m_3=0}^\infty z_1^{m_1+m_2}z_2^{m_3}\text{Res}_{\textbf{x}=0}\left(\frac16\left[(x_1+x_2)\log(z_1)+x_3\log(z_2)\right]^3+1\right)\\
	&\left((x_2-x_1)(m_1-m_2+x_1-x_2)\left[\frac{\Gamma(-x_1)}{\Gamma(1+x_1+m_1)}\frac{\Gamma(-x_2)}{\Gamma(1+x_2+m_2)}\right]^4\left[\frac{\Gamma(-x_3)}{\Gamma(1+x_3+m_3)}\right]^3\right.\\
	&\left.\left[\prod\limits_{i=1}^2\frac{\Gamma\left(-x_i-x_3\right)}{\Gamma\left(-x_i-x_3-m_i-m_3\right)}\right]\left[\frac{\Gamma(1+x_1+x_2+x_3)}{\Gamma(-m_1-m_2-m_3-x_1-x_2-x_3)}\right]^5\right)\,.
\end{split}
\label{eqn:Ztildemodel2}
\end{align}
}
Evaluating this expression, we can extract the fundamental period $\varpi_0$ with leading terms
\begin{align}
	\varpi_0=1 + 3 z_1 + 19 z_1^2 + 147 z_1^3 + z_2 + 24 z_1 z_2 + 513 z_1^2 z_2 + z_2^2 + 81 z_1 z_2^2 + z_2^3 + \mathcal{O}(z^4)\,,
\end{align}
where we have substituted $z_2\rightarrow -z_2$.
Again, we can use the generic expression~\eqref{eqn:fundamental} that we obtain in Section~\ref{sec:classification} and write this in closed form
\begin{align}
	\varpi_0=&\sum\limits_{\lambda_1,\lambda_2=0}^\infty\frac{\Gamma(1+\lambda_1)}{\Gamma(1-\lambda_1+\lambda_2)}\frac{{_3}F_2(-\lambda_1,-\lambda_1,1-\lambda_2+\lambda_1;1-\lambda_2,1-\lambda_2;1)}{\Gamma(1-\lambda_2)\Gamma(1-\lambda_2)\Gamma(1+\lambda_2)^3} z_1^{\lambda_1}z_2^{\lambda_2}\,.
\end{align}
We can also extract the triple intersection numbers
\begin{align}
	c_{111}=8\,,\quad c_{112}=11\,,\quad c_{122}=5\,,\quad c_{222}=0\,,
\end{align}
where we have performed the normalization using $c_{122}=5$. This is consistent with the results of Section~\ref{sec:geometry}. 
\subsubsection{Non-Abelian dual}
Again, we use the dual theory to recover the Pfaffian phase. The field content of the dual theory is
\begin{equation}
      \begin{array}{c|c|cc|c|cc|c}
        &p^1,\ldots,p^5&x_1^a,\ldots,x_4^a&x_5^a&b_1,b_2,b_3&a_{ij}&a_{i5}&\mathrm{FI}\\
        \hline
        U(2)&\mathrm{det}^{-1}&\overline{\Box}&\overline{\Box}&{\bf 1}&\mathrm{det}&\mathrm{det}&\tilde{\zeta}\\
        U(1)&-1&0&-1&1&0&1&\tilde{\zeta}_1,
        \end{array}
      \end{equation}
where $i,j\in\{1,\ldots,4\}$. The gauge charges are
\begin{equation}
 \begin{array}{c|c|cccc|c|cc|c}
    &p^{1},\ldots,p^{5}&\tilde{x}_1^{1},\ldots,\tilde{x}_1^{4}&\tilde{x}^5_1&\tilde{x}_2^{1},\ldots,\tilde{x}_2^{4}&\tilde{x}^5_2&b_1,b_2,b_3&a_{ij}&a_{i5}&\mathrm{FI}\\
    \hline
    \tilde{U}(1)_1&-1&-1&-1&0&0&0&1&1&\tilde{\zeta}\\
    \tilde{U}(1)_2&-1&0&0&-1&-1&0&1&1&\tilde{\zeta}\\
    U(1)_3&-1&0&-1&0&-1&1&0&1&\tilde{\zeta}_1,
    \end{array}
\end{equation}
   where $a_{ij}$ has $i,j\in\{1,\ldots,4\}$. The superpotential is 
   \begin{equation}
     \widetilde{W}=\sum_{i,j=1}^4\sum_{k=1}^5\sum_{l=1}^3A_k^{ijl}b_lp^ka_{ij}+\sum_{i=1}^4\sum_{k=1}^5A_k^{i5}p^ka_{i5}+\sum_{i,j=1}^4[\tilde{x}^i\tilde{x}^j]a_{ij}+\sum_{i=1}^4[\tilde{x}^i\tilde{x}^5]a_{i5}.
   \end{equation}
   The D-term equations are
   \begin{equation}
  \label{ad1dual}
  -\sum_{i=1}^5\tilde{x}^i_{a}\tilde{x}^{i,b\dagger}-\sum_{k=1}^5|p^k|\delta_a^b+\sum_{i,j=1}^5|a_{ij}|^2\delta_a^b=\tilde{\zeta}\delta_a^b,
   \end{equation}
   and
   \begin{equation}
  \label{ad2dual}
  -\sum_{k=1}^5 |p^k|^2-\sum_{a=1}^2|\tilde{x}^5_a|^2+\sum_{l=1}^3 |b_l|^2+\sum_{i=1}^4|a_{i5}|^2 = \tilde{\zeta}_1.
   \end{equation}
   We also have 
   \begin{equation}
  \label{adextradual}
  -\sum_{i=1}^4|\tilde{x}^i_1|^2+|\tilde{x}^5_2|^2-\sum_{l=1}^3 |b_l|^2+\sum_{i,j=1}^4|a_{ij}|^2=\tilde{\zeta}-\tilde{\zeta}_1,
   \end{equation}
   and similarly with the $1,2$-components of the $\tilde{x}$ exchanged. The F-term equations are
   \begin{align}
     \tilde{x}^i_a:&\quad \sum_{j=1}^4\tilde{x}^{j,a}a_{ij}=0,\quad i=1,\ldots,4,\: a=1,2,\label{af1dual}\\
     &\quad \sum_{j=1}^4\tilde{x}^{j,a}a_{j5}=0, \quad a=1,2,\label{af2dual}\\
     p^k:&\quad \sum_{i,j=1}^4\sum_{l=1}^3A_k^{ijl}b_la_{ij}+\sum_{i=1}^4A_k^{i5}a_{i5}=0, \quad k=1,\ldots,5,\label{af3dual}\\
     b_l:&\quad \sum_{i,j=1}^4\sum_{k=1}^5A_k^{ijl}p^ka_{ij}=0,\quad l=1,2,3,\label{af4dual}\\
     a_{ij}:&\quad \sum_{k=1}^5\sum_{l=1}^3A_k^{ijl}b_lp^k+[\tilde{x}^i\tilde{x}^j]=0,\quad i,j\in\{1,\ldots,4\},\label{af5dual}\\
       &\quad \sum_{k=1}^5A_k^{i5}p^k+[\tilde{x}^i\tilde{x}^5]=0,\quad i=1,\ldots,4.\label{af6dual}
   \end{align}
   The phase diagram has the same structure as for the original model.

   For $\tilde{\zeta} > 0,\tilde{\zeta}_1 > 0,\tilde{\zeta}-\tilde{\zeta}_1 > 0$ the $a_{ij}$ are not allowed to vanish simultaneously, neither are $\{b_l,a_{i5}\}$. The same holds for $\{a_{ij},\tilde{x}^5\}$ with $i,j\in\{1,\ldots,4\}$. There is a vacuum for $\tilde{x}^1=\ldots=\tilde{x}^5=0$ and ${p}^1=\ldots=p^5=0$. The same also holds for $\tilde{\zeta} > 0,\tilde{\zeta}_1 > 0,\tilde{\zeta}-\tilde{\zeta}_1 < 0$ but then $\{\tilde{x}^{1},\ldots,\tilde{x}^4,b_1,b_2,b_3\}$ are in the deleted set. The question is whether the F-terms distinguish these two cases. To see this, we check if the deleted sets coming from the different signs of $\tilde{\zeta}-\tilde{\zeta}_1$ lead to sensible solutions of the F-term equations. If we set $\{a_{ij},\tilde{x}^5\}=0$ we are left with $2$ equations from (\ref{af2dual}), $5$ F-terms from (\ref{af3dual}), $6$ conditions from (\ref{af5dual}) and $4$ conditions from (\ref{af6dual}), so a total of $17$ F-terms. The non-zero degrees of freedom are $4$ $a_{i5}$, $5$ $p^k$, $3$ $b_l$, and $8$ $\tilde{x}^i$. That makes $20$ degrees of freedom that have to satisfy the D-terms in addition. On the other hand, we can consider $\{\tilde{x}^{1},\ldots,\tilde{x}^4,b_1,b_2,b_3\}=0$. Then we have $5+3+4=12$ F-terms left for $2$ $\tilde{x}^5$, $10$ $a_{ij}$ and $5$ $p^k$. It looks like this can have a solution, but setting $\tilde{x}^{1}=\ldots=\tilde{x}^4=0$ is only allowed by the non-Abelian D-term if $p\neq 0$. However, since we are not allowed to set all $a_{ij}$ to zero in either of the two phases, (\ref{af4dual}) and (\ref{af6dual}) are can only be solved if $p=0$. Therefore we expect the phase boundary in the $\zeta,\zeta_1>0$-region to be lifted. 

   We expect the Pfaffian phase to be where $\tilde{\zeta} < 0,\tilde{\zeta}-\tilde{\zeta}_1 < 0$. The classical phase diagram indicates a phase boundary along the negative $\zeta$-axis. The deleted set for $\zeta_1>0$ is $\{b_l,a_{i5}\}\cup \{\tilde{x}^1,\ldots,\tilde{x}^5,p^1,\ldots p^5\}\cup \{\tilde{x}^1,\ldots,\tilde{x}^4,b_1,b_2,b_3\}$ whereas in the $\tilde{\zeta}_1<0,\tilde{\zeta}-\tilde{\zeta}_1 < 0$-region we have $\{\tilde{x}^1,\ldots\tilde{x}^5,p^1,\ldots,p^5\}\cup \{\tilde{x}^5,p^1,\ldots,p^5\}\cup\{\tilde{x}^{1},\ldots,\tilde{x}^4,b_1,b_2,b_3\}$. Again, the phase boundary will be lifted if the differing deleted sets are not solutions to the classical equations of motion. Let us start with $\{b_1,b_2,b_3,a_{i5}\}=0$. Then we are left with $21$ non-trivial F-term conditions, plus the D-terms, for $21$ non-zero fields, so we do not expect these equations to have a non-trivial solution. Next we consider $\{\tilde{x}^5,p^1,\ldots,p^5\}=0$. Also in this case one gets $21$ non-trivial F-terms and $21$ unknowns, so we are again in the same situation. Hence, we conclude that the phase boundary is lifted. We note that the D-terms would allow for $b_1=b_2=b_3=0$ in this phase. However, when $a_{ij}=0$, (\ref{af5dual}) implies that $\tilde{x}^i\sim\tilde{x}^j$ for $i,j=1,\ldots,4$. This means that we can find a basis of solutions such that e.g.~$\tilde{x}^i_1=0$ for all $i$, but this is in contradiction with (\ref{adextradual}). 
   
   Finally, we can consider $\tilde{\zeta}_1< 0,\tilde{\zeta}-\tilde{\zeta}_1 < 0 > 0$, which we also expect to be strongly coupled. The deleted set is $\{p^1,\ldots,p^5,\tilde{x}^5\}\cup \{\tilde{x}^5,a_{ij}\}$. We observe that $\tilde{x}^1=\ldots\tilde{x}^4=b_1=b_2=b_3=p^1=\ldots=p^5=a_{i5}=0$ solves the F-terms and is consistent with the D-terms. We will not go into the details of the non-perturbative realisation of this phase, but we note that the $a_{ij}$ ($i,j=1,\ldots,4$) are the Pl\"ucker coordinates of the Grassmannian we have encountered in the weakly coupled phase of the original theory.

Let us also analyze the Coulomb branch. The effective potential is
   \begin{align}
     \mathcal{W}_{eff}=&-\tilde{t}(\tilde{\sigma}_1+\tilde{\sigma}_2)-\tilde{t}_1\tilde{\sigma}_3-5(-\tilde{\sigma_1}-\tilde{\sigma}_2-\tilde{\sigma}_3)[\log(-\tilde{\sigma_1}-\tilde{\sigma}_2-\tilde{\sigma}_3)-1]\nonumber\\
     &-4(-\tilde{\sigma}_1)[\log(-\tilde{\sigma}_1)-1]-(-\tilde{\sigma}_1-\tilde{\sigma}_3)[\log(-\tilde{\sigma}_1-\tilde{\sigma}_3)-1]\nonumber\\
     &-4(-\tilde{\sigma}_2)[\log(-\tilde{\sigma}_2)-1]-(-\tilde{\sigma}_2-\tilde{\sigma}_3)[\log(-\tilde{\sigma}_2-\tilde{\sigma}_3)-1]-3\tilde{\sigma}_3[\log\tilde{\sigma}_3-1]\nonumber\\
     &-6(\tilde{\sigma}_1+\tilde{\sigma}_2)[\log(\tilde{\sigma}_1+\tilde{\sigma}_2)-1]-4(\tilde{\sigma}_1+\tilde{\sigma}_2+\tilde{\sigma}_3)[\log(\tilde{\sigma}_1+\tilde{\sigma}_2+\tilde{\sigma}_3)-1]\nonumber\\
     &+i\pi(\tilde{\sigma}_1+\tilde{\sigma}_2).
   \end{align}
   Defining $\tilde{z}=\frac{\tilde{\sigma}_2}{\tilde{\sigma}_1}$ and $\tilde{w}=\frac{\tilde{\sigma}_3}{\tilde{\sigma}_1}$ the critical locus is given by
   \begin{align}
     e^{-\tilde{t}}&=-\frac{(1+\tilde{z})^6}{(1+\tilde{w})(1+\tilde{z}+\tilde{w})}=-\frac{\left(1+\frac{1}{\tilde{z}}\right)^6}{\left(1+\frac{\tilde{w}}{\tilde{z}}\right)\left(1+\frac{1}{\tilde{z}}+\frac{\tilde{w}}{\tilde{z}}\right)},\nonumber\\
     e^{-\tilde{t}_1}&=-\frac{1}{\left(1+\frac{1}{\tilde{w}}\right)\left(1+\frac{\tilde{z}}{\tilde{w}}\right)\left(1+\frac{1}{\tilde{w}}+\frac{\tilde{z}}{\tilde{w}}\right)}=-\frac{\tilde{w}^3}{(1+\tilde{w})(\tilde{z}+\tilde{w})(1+\tilde{z}+\tilde{w})},\nonumber\\
     1&=\tilde{z}^5\frac{\left(1+\frac{\tilde{w}}{\tilde{z}}\right)}{(1+\tilde{w})},\nonumber\\
     e^{-\tilde{t}+\tilde{t}_1}&=\frac{(1+\tilde{z})^6(1+\tilde{w})}{\tilde{z}^4\tilde{w}^3}=\left(1+\frac{1}{\tilde{z}}\right)^6\left(1+\frac{\tilde{w}}{\tilde{z}}\right)\frac{\tilde{z}^3}{\tilde{w}^3}.
   \end{align}
   As in the original theory, we observe that for $\tilde{w}=0$ we recover the conditions for the discriminant of the elliptic curve. The two spikes get pushed towards $\tilde{t}_1\rightarrow\infty$. The third equation is the same as in the original theory and we have
   \begin{equation}
     \tilde{w}=-\frac{1-\tilde{z}^5}{1-\tilde{z}^4}.
   \end{equation}
   Inserting this back into the other equations we get
   \begin{align}
     e^{-\tilde{t}}&=\frac{(1+\tilde{z})^8(1+\tilde{z}^2)^2}{\tilde{z}^5(1+\tilde{z}+\tilde{z}^2)},\nonumber\\
     e^{-\tilde{t}_1}&=\frac{(1+\tilde{z}+\tilde{z}^2+\tilde{z}^3+\tilde{z}^4)^3}{\tilde{z}^5(1+\tilde{z}+\tilde{z}^2)},\nonumber\\
     e^{-\tilde{t}+\tilde{t}_1}&=\frac{(1+\tilde{z})^8(1+\tilde{z}^2)^2}{(1+\tilde{z}+\tilde{z}^2+\tilde{z}^3+\tilde{z}^4)^3}.
   \end{align}
   The discussion is similar to the original model. We exclude $\tilde{z}=0$ and $\tilde{z}=1$. For $\tilde{z}=-1,\pm i$, i.e.~the roots of $(1+\tilde{z})^8(1+\tilde{z}^2)^2$ we get
   \begin{equation}
     e^{-\tilde{t}}=0, \quad e^{-\tilde{t}_1}=-1,\quad e^{-\tilde{t}+\tilde{t}_1}=0.
   \end{equation}
   This is the $\tilde{\zeta}\geq0$-axis. Inserting the roots of $1+\tilde{z}+\tilde{z}^2$ we get
   \begin{equation}
     e^{-\tilde{t}}\rightarrow\infty, \quad e^{-\tilde{t}_1}=\infty,\quad e^{-\tilde{t}+\tilde{t}_1}=-1.
   \end{equation}
   This gives the phase boundary at $\tilde{\zeta}=\tilde{\zeta}_2 < 0$. 
   \begin{equation}
       e^{-\tilde{t}}=const., \quad e^{-\tilde{t}_1}=0,\quad e^{-\tilde{t}+\tilde{t}_1}\rightarrow\infty,
   \end{equation}
   where $const.$ stands for the locus of the two singularities of the elliptic fiber. This gives the two spikes along the $\zeta_1$-axis. So we have recovered the amoeba of the original theory. 
   \subsection{GLSM  for $X_3$}
   \label{sec:glsmModel3}
In this section we propose a GLSM for the genus $1$ fibration involving the tangent bundle of $\mathbb{P}^2$ discussed in Section~\ref{sec:compl-inters-calabi}. The Calabi-Yau also makes an appearance in the list of genus 1 fibrations over $\mathbb{P}^2$ in Section \ref{sec:examples-Gr} where it is labeled as model $X_3$. Since the corresponding GLSM is rather complicated we only identify the Grassmannian phase and match the Coulomb branch with the discriminant obtained via the Picard-Fuchs operator. 
\subsubsection{Matter content and Grassmannian phase}
We consider a GLSM with gauge group $U(2)\times U(2)$ with the following matter content
\begin{equation}
\label{eqn:glsmU2U2matter}
  \begin{array}{c|cccc|cc|c}
    &p^1 & p^2 & p^3 & p^4 & x_{1},\ldots,x_{5} & y_{1},y_2,y_3&\mathrm{FI}\\
    \hline
    U(2)_1&\det^{-1}&\det^{-1}&\det^{-1}&\det^{-1}&\Box&0&\zeta_1\\
    U(2)_2&\overline{\rule{0ex}{1.3ex}\Box}&\det^{-1}&\det^{-1}&0&0&\Box&\zeta_2.\\
  \end{array}
\end{equation}
The corresponding gauge charges are thus
\begin{equation}
  \begin{array}{c|rrrrr|cccc|c}
   & p^{1,1} & p^{1,2} & p^2 & p^3 & p^4 & x^1_{1,\ldots,5} & x^2_{1,\ldots,5} & y^1_{1,2,3} & y^2_{1,2,3} &\mathrm{FI}\\
    \hline
   U(1)_1& -1 & -1 & -1 & -1 & -1 & 1 & 0 & 0 & 0 &\zeta_1 \\
   U(1)_2& -1 & -1 & -1 & -1 & -1 & 0 & 1 & 0 & 0 &\zeta_1\\
   U(1)_3& -1 &  0 & -1 & -1 &  0 & 0 & 0 &1 & 0 &\zeta_2 \\  
   U(1)_4&  0 & -1 & -1 & -1 &  0 & 0 & 0 & 0 & 1 &\zeta_2.   
  \end{array}
\end{equation}
The superpotential is
\begin{equation}
  W = \sum_{i,j=1}^5 \left( \sum_{\ell=1}^3 \sum_{a=1}^2  p^{1,a} A_{1a}^{ij\ell} y_{\ell}^a +\sum_{\ell,m=1}^3 \sum_{k=2}^3 p^k A_{k}^{ij\ell m} [y_\ell y_m] + p^4 A^{ij}_4 \right) [x_ix_j]. 
\end{equation}
The three terms correspond to sections of $Q=T_{\mP^2}(-1)$, $\rO_{\mP^2}(1)^{\oplus 2}$, and $\rO_{\mP^2}$, respectively. The D-term equations read
\begin{align}
  \left(  -|p^{1,1}|^2-|p^{1,2}|^2-\sum_{k=2}^4|p^k|^2\right)\delta_a^b+\sum_{i=1}^5x_{i,a}x^{\dagger,b}_i&=\zeta_1\delta_a^b,\nonumber\\
  -p^1_ap^{-\dagger,b}+\left(-\sum_{k=2}^4|p^k|^2\right)\delta_a^b+\sum_{i=1}^3y_{i,a}y^{\dagger,b}_i=\zeta_2\delta_a^b.
\end{align}
In the phase $\zeta_1,\zeta_2>0$ the ranks of the matrices $x_i^a$ and $y_i^a$ have to be two. Setting $p=0$, the geometry is defined by the following F-terms
\begin{align}
  \sum_{i,j=1}^5\sum_{\ell=1}^3\sum_{a=1}^2A_{1a}^{ij\ell} y_{\ell}^a[x_ix_j]&=0,\qquad a=1,2,\nonumber\\
  \sum_{i,j=1}^5  \sum_{\ell=1}^3 \sum_{\ell,m=1}^3A_{k}^{ij\ell m} [y_\ell y_m][x_ix_j] &=0,\qquad k=2,3,\nonumber\\
   \sum_{i,j=1}^5A^{ij}_4[x_ix_j]&=0. 
\end{align}
This coincides with the Calabi-Yau $X_3$ constructed in Section~\ref{sec:geometry}. 
\subsubsection{Coulomb branch}
The effective potential of the Coulomb branch can be deduced from the gauge charges. Parametrising the four-dimensional maximal torus of the gauge group by $\sigma_i$ ($i=1,\ldots,4$) and further defining $z=\frac{\sigma_2}{\sigma_1}, w=\frac{\sigma_3}{\sigma_1}, u=\frac{\sigma_4}{\sigma_1}$ the critical locus of $\mathcal{W}_{eff}$ is at
\begin{align}
  e^{-t_1}&=\frac{1}{(1+z)(1+z+w)(1+z+u)(1+z+w+u)^2},\nonumber\\
  e^{-t_2}&=\frac{1}{\left(1+\frac{1}{w}+\frac{z}{w}\right)\left(1+\frac{1}{w}+\frac{z}{w}+\frac{u}{w}\right)^2},\nonumber\\
  z^5&=1\qquad \frac{u^3}{w^3}\frac{1+z+w}{1+z+u}=1
\end{align}
Using the Picard--Fuchs operators for $3_a$ in Appendix~\ref{sec:fibrationsOverP2}, this matches the corresponding discriminant upon identifying $z_1=-e^{-t_2}, z_2=-e^{-t_1}$. 


%% file: section_3.tex
\section{Gopakumar-Vafa invariants and modular forms}
\label{sec:gvandmod}
It was first noted in~\cite{Candelas:1994hw} that topological string amplitudes on elliptically fibered Calabi-Yau threefolds exhibit modular properties.
This can be seen as a consequence of the covariant transformation of the topological string partition function under certain monodromies in the quantum K\"ahler moduli space~\cite{Candelas:1994hw,Aganagic:2006wq,Gunaydin:2006bz}.
More precisely, for genus one fibered Calabi-Yau threefolds with $N$-sections, the large volume monodromies and the relative conifold monodromies generate an action of $\Gamma_1(N)$ on the category of topological B-branes and the corresponding quantum periods.
This has been shown for elliptic fibrations without reducible fibers in~\cite{Andreas:2000sj,Andreas:2001ve}.
The calculation was extended to fibrations that have reducible fibers, and to genus one fibrations that are not elliptic, in~\cite{Schimannek:2019ijf,Cota:2019cjx}.
Together with the holomorphic anomaly equations, this seems to imply that the topological string partition function can be expanded in terms of weak Jacobi forms~\cite{Klemm:2012sx,Alim:2012ss,Huang:2015sta,DelZotto:2017mee,Lee:2018urn,Lee:2018spm,Cota:2019cjx}.
A physical derivation for an elliptic fibration without reducible fibers has been performed in~\cite{Huang:2015sta} while a general, although still physical, derivation of the elliptic transformation law was given in~\cite{Cota:2019cjx}.
A partial mathematical proof of the elliptic transformation law for fibrations without reducible fibers was obtained in~\cite{Oberdieck:2016nvt}.
In many cases the expansion in weak Jacobi forms can also be seen as a consequence of the properties of the elliptic genera of certain non-critical strings~\cite{Haghighat:2013gba,Haghighat:2014vxa,DelZotto:2017mee,Lee:2018urn,Lee:2018spm}.
However, a complete derivation for fibrations with reducible fibers is, even to physicists' standards, still missing.
Nevertheless, there is by now a very precise conjecture about the modular structure of the topological string partition function on general genus one fibrations~\cite{Cota:2019cjx}.

The modular structure, together with the knowledge of some Gopakumar-Vafa invariants then fixes the coefficients, in an expansion of the topological string partition function in the complexified volumes of the base, up to some degree.
To this end, one can use genus zero and genus one invariants, that can be calculated using mirror symmetry or localization, as well as certain vanishing conditions that, however, are only known in few restricted cases~\cite{Huang:2015sta}.
The coefficients of the topological string partition function then encode the Gopakumar-Vafa invariants of the corresponding base degree and for all worldsheet genera and fiber degrees.
This procedure is generally referred to as the \textit{modular bootstrap}.
In this section we apply the modular bootstrap to genus one fibrations with $5$-sections.
This will lead to expressions for the Gopakumar-Vafa invariants for low base degrees in terms of Jacobi forms for $\Gamma_1(5)$.
Morever, we observe that the corresponding expressions for the fibrations of Pfaffian curves and the complete intersections in fibrations of Grassmannians can be transformed into each other by acting on the modular parameter with a certain element in $\Gamma_0(5)$.
We interpret this as a modular manifestation of relative homological projective duality which therefore enhances the $\Gamma_1(5)$ action on the category of B-branes to $\Gamma_0(5)$.

\subsection{Gromov-Witten and Gopakumar-Vafa invariants}
Gromov-Witten invariants are perhaps more familiar to mathematicians than Gopakumar-Vafa invariants and we will briefly review their relation.
This also serves as an opportunity to fix some notation.

We denote by $\overline{M}_{g,n}(X,\beta)$ the moduli stack of genus $g$, $n$-pointed stable maps into a Calabi-Yau $d$-fold $X$ that represent the class $\beta\in H_2(X,\mathbb{Z})$.
Its virtual dimension is given by
\begin{align}
	\text{vdim}\,\overline{M}_{g,n}(X,\beta)=n\,.
\end{align}
For $n=0$ we have a well-defined enumerative problem and the \textit{Gromov-Witten invariants} $N_{g,\beta}^X$ ``count'' the points in the virtual fundamental class
\begin{align}
	N_{g,\beta}^X=\text{deg}\,[\overline{M}_{g,0}(X,\beta)]^{\text{vir.}}\,.
\end{align}
The corresponding generating functions
\begin{align}
	F_g(Q)=\sum\limits_{\beta\in H_2(X,\mathbb{Z})}N_{g,\beta}^X \cdot Q^\beta\,,
\end{align}
are called the \textit{topological string free energies}\footnote{To be precise, this is only correct up to classical terms that contribute a polynomial in $\log(Q)$ to the free energies at genus zero and genus one.}.
The free energies can be combined into the \textit{topological string partition function}
\begin{align}
	Z_{\text{top.}}(Q,\lambda)=\exp\left(\sum\limits_{g=0}^\infty \lambda^{2g-2}F_g(Q)\right)\,,
	\label{eqn:ztopf}
\end{align}
where the formal parameter $\lambda$ is called the \textit{topological string coupling}.
Both, the free energies and the partition function, can also be defined via the path integral of topological string theory with target space $X$.
For complete intersections in toric ambient spaces, the Gromov-Witten invariants for genus zero and genus one can then be calculated using mirror symmetry~\cite{Candelas:1990rm,Hosono:1993qy,Hosono:1994ax}.
If a GLSM realization is available, the same information can also be extracted directly from the sphere partition function~\eqref{zs2def}~\cite{Jockers:2012dk}.

Unfortunately, the moduli stacks $\overline{M}_{g,0}(X,\beta)$ in general have orbifold singularities and the Gromov-Witten invariants are therefore rational numbers.
This is remedied by the \textit{Gopakumar-Vafa invariants} $n_\beta^{(g)}$~\cite{Gopakumar:1998ii,Gopakumar:1998jq} that are conjecturally integral but still lack a general mathematical definition.
They encode a certain trace over multiplicities of BPS states in the five-dimensional effective theory that arises from M-theory on the Calabi-Yau $X$ and are encoded in the topological string partition function via
\begin{align}
	\log\left(Z_{\text{top.}}\right)=\sum\limits_{g=0}^\infty\sum\limits_{\beta\in H_2(X,\mathbb{Z})}\sum\limits_{m=1}^\infty n_\beta^{(g)}\frac{1}{m}\left(2\sin\frac{m\lambda}{2}\right)^{2g-2}Q^{m\beta}\,.
	\label{eqn:gvexp}
\end{align}
To be precise, the five-dimensional little group of massive particles is given by
\begin{align}
	SO(4)=SU(2)_1\times SU(2)_2\,.
\end{align}
The BPS particles arise from M2-branes that wrap curves in $X$ and the charges are correspondingly labelled by $\beta\in H_2(X,\mathbb{Z})$.
Now let $N^\beta_{j_1,j_2}$ denote the multiplicity of BPS states of charge $\beta$ in the representation
\begin{align}
	\left[\left(\frac12,0\right)\oplus2(0,0)\right]\otimes (j_1,j_2)\,.
\end{align}
The Gopakumar-Vafa invariants $n_\beta^{(g)}$ are then defined via the relation
\begin{align}
	\sum\limits_{g=0}^\infty n_\beta^{(g)}I_g=\sum\limits_{j_1,j_2}(-1)^{2j_2}(2j_2+1)N_{j_1,j_2}^\beta\cdot[j_1]\,,
\end{align}
where $I_g=([\frac12]+2[0])^g$.
In general it is very difficult to match these invariants with geometric quantities.
However, if $X$ is elliptically or genus one fibered some of these multiplicities are directly related to the multiplicities of reducible fibers~\cite{Paul-KonstantinOehlmann:2019jgr,Kashani-Poor:2019jyo}.
We will use this relation in Section~\ref{sec:ftheory} to obtain general expressions for the multiplicities and structures of the reducible fibers of generic genus one fibered Calabi-Yau threefolds with $5$-sections.

\subsection{The modular bootstrap for genus one fibrations with 5-sections}
\label{sec:bootstrapreview}
Let us now review the modular bootstrap for genus one fibrations with $N$-sections that has been developed in~\cite{Cota:2019cjx}.
While the derivation in~\cite{Cota:2019cjx} only covered cases with $N\le 4$, the ansatz for the topological string partition function immediately extends to $N=5$.
We assume that the fibration is generic in that it does not exhibit any fibral divisors or additional $N$-sections.

To perform the modular bootstrap, one first needs to choose suitable coordinates on the complexified K\"ahler moduli space.
This choice is somewhat intricate and can be derived by explicitly calculating the action of the Fourier-Mukai transform with the ideal sheaf of the relative diagonal in $M\times_B M$ on the complexified volumes of curves in $M$~\cite{Schimannek:2019ijf,Cota:2019cjx}.
The basis is then singled out by requiring that the complexified volume $\tau$, of the fiber component that intersects the $N$-section once, transforms like a modular parameter with $\tau\rightarrow\tau/(1+N\tau)$ and, up to subleading contributions, the volumes of base curves transform like Jacobi forms.
We will discuss the Fourier-Mukai transform in more detail in Section~\ref{sec:monmod}.

Let us assume that $M$ is a Calabi-Yau threefold that is genus one fibered over a base $B$ such that $h^{1,1}(M)=h^{1,1}(B)+1$ and $N\in\mathbb{N}$ is the lowest value for which $M$ has an $N$-section.
We denote the class of the $N$-section by $E_0$ and a basis of the K\"ahler cone of the base by $\tilde{D}_i,\,i=1,\dots,h^{1,1}(B)$.
As this is true for all geometries in this paper we will assume that the K\"ahler cone is simplicial.
The dual basis of the Mori cone will be denoted by $\tilde{C}^i,\,i=1,\dots,h^{1,1}(B)$.
On the fibration itself we then have the vertical divisors and curves
\begin{align}
	D_i=\pi^{-1}(\tilde{D}_i)\,,\quad C^i=E_0\cdot\pi^{-1}(\tilde{C}^i)\,,\quad i=1,\dots,h^{1,1}(B)\,.
	\label{eqn:fibrationsdivcur}
\end{align}
Note that due to $E_0$ being an $N$-section they intersect as
\begin{align}
	C^i\cdot D_j=N\cdot \delta_j^i\,.
\end{align}
We also need to define the special vertical divisor
\begin{align}
	D=-\pi^{-1}\pi_*(E_0\cdot E_0)=\tilde{a}^iD_i\,,\quad \tilde{a}^i=-\int_ME_0\cdot C^i\,,
	\label{eqn:bootstrapA}
\end{align}
which is the height pairing of the $N$-section with itself.
The modular parametrization of the complexified K\"ahler form $\omega$ then reads
\begin{align}
	\omega=\tau\cdot\left(E_0+\frac{1}{2N} D\right)+\sum\limits_{i=1}^{h^{1,1}(B)}t^i\cdot D_i\,.
	\label{eqn:bootstrapKaehlerForm}
\end{align}

The topological string partition function depends on the complexified K\"ahler parameters $\tau, t^i$ and the topological string coupling $\lambda$.
To perform the modular bootstrap we expand in the exponentiated K\"ahler parameters $Q^i=\exp(2\pi i t^i)$
\begin{align}
	Z_{\text{top.}}(\tau,t^i,\lambda)=Z_0(\tau,\lambda)\left(1+\sum\limits_{\beta\in H_2(B)}Z_\beta(\tau,\lambda)Q^\beta\right)\,,
\end{align}
where we use
\begin{align}
	Q^\beta=\exp\left(2\pi i t^i\beta_i\right)\quad\text{with}\quad \beta_i=D_i\cdot\beta.
\end{align}
The coefficients $Z_\beta(\tau,\lambda)$ are conjectured to take the form
\begin{align}
	Z_\beta(\tau,\lambda)=\frac{1}{\eta(N\tau)^{12\cdot c_1(B)\cdot \beta}}\frac{\phi_\beta(\tau,\lambda)}{\prod_{i=1}^{b_2(B)}\prod_{s=1}^{\beta_i}\phi_{-2,1}(N\tau,s\lambda)}\,,
	\label{eqn:bootstrapAnsatz}
\end{align}
where $\eta(\tau)$ is the Dedekind $\eta$-function and $\phi_{-2,1}(\tau,\lambda)$ is a certain Jacobi form of weight $-2$ and index $1$ that can be found in Appendix~\ref{sec:appendixJacobiForms}.
The numerator $\phi_\beta(\tau,\lambda)$ depends on the particular fibration and is a Jacobi form
\begin{align}
	\phi_\beta(\tau,\lambda)\in M_\bullet(N)\left[\phi_{-2,1}(N\tau,\lambda),\phi_{0,1}(N\tau,\lambda)\right]\cdot \Delta_{2N}(\tau)^{1-\frac{r_\beta}{N}\,\text{mod}\,1}\,,
	\label{eqn:bootstrapRing}
\end{align}
with $M_\bullet(N)$ being the ring of modular forms for $\Gamma_1(N)$.
Moreover, the exponent of $\Delta_{2N}$ is determined by the congruence relation
\begin{align}
	1-\frac{r_\beta}{N}\equiv\frac12\left[Nc_1(B)-\frac{1}{N}D\right]\cdot\beta\,\text{ mod }\,1\,.
	\label{eqn:bootstrapCongruence}
\end{align}
The fact that a rational power of $\Delta_{2N}$ can appear in $\phi_\beta(\tau,\lambda)$, as well as the corresponding congruence relation, has been derived in~\cite{Cota:2019cjx} by considering Higgs transitions in F-theory that relate elliptic and genus one fibrations.
The weight $w$ and index $r$ of $\phi_\beta(\tau,\lambda)$ are determined by the fact that $Z_\beta(\tau,\lambda)$ has weight $0$ and index
\begin{align}
	r=\frac{1}{2N}\beta\cdot(\beta-c_1(B))\,.
\end{align}
The modular forms that generate the ring $M_\bullet(5)$ are described in Appendix~\ref{sec:appModularForms}.

For any given geometry and curve degree $\beta$ one can then, using the generators of the ring of weak Jacobi forms, make an ansatz for $\phi_\beta(\tau,\lambda)$ of the correct weight and index.
At least for low curve degrees the coefficients can then be fixed using e.g. the knowledge of some Gopakumar-Vafa invariants or certain vanishing conditions.

Before we consider examples, it is useful to specialize some of the expressions for geometries with base $B=\mathbb{P}^2$.
Let us introduce $J_1=E_0,\,J_2=D_1$, with $E_0$ being the $N$-section and $D_1$ the hyperplane class, and denote the corresponding triple intersection numbers by
\begin{align}
	c_{ijk}=\int_M J_iJ_jJ_k\,.
\end{align}
The height pairing of the $N$-sections is then given by
\begin{align}
	D=-c_{112}J_2\,.
	\label{eqn:heightpairingP2}
\end{align}
The exponent of $\Delta_{2N}$ in~\eqref{eqn:bootstrapRing} then satisfies the congruence relation
\begin{align}
	1-\frac{r_{n\cdot D_1}}{N}\equiv\frac{n}{2}\left[N+\frac{c_{112}}{N}\right]\,\text{ mod }\,1\,.
	\label{eqn:bootstrapCongruence2}
\end{align}
With these expressions we can now apply the modular bootstrap to a genus one fibered Calabi-Yau threefold with a $5$-section.

\subsection{Examples: The dual fibrations $X_1$ and $Y_1$}
Let us now apply the modular bootstrap to the Calabi-Yau threefolds $X_1$ and $Y_1$ that have been constructed in Section~\ref{sec:geometry}.
Recall that $X_1$ is a fibration of complete intersection curves in $G(2,5)$ over $\mathbb{P}^2$ while $Y_1$ is a fibration of Pfaffian curves over the same base.
The corresponding GLSM has been discussed in Section~\ref{sec:glsmModel1}, where we found that non-linear sigma models on $X_1$ and $Y_1$ respectively arise as the infrared limit of a weakly and a strongly coupled phase.
By localizing the sphere partition function we obtained the fundamental period of the mirror of $X_1$ and we will use it now to determine the corresponding Picard-Fuchs system.
The Picard-Fuchs system that is associated to the mirror of $Y_1$ can then be obtained by a coordinate transformation\footnote{Although not discussed in this paper, we have also performed the localization calculation in the weakly coupled Hori dual GLSM of the strongly coupled phase. The result for the fundamental period matches with the regular solution of the Picard-Fuchs system that one obtains from a coordinate transformation.}.
Having the generators of the Picard-Fuchs system allows us to calculate the Gopakumar-Vafa invariants up to genus one and use them to perform the modular bootstrap.
Recall that the realization as two phases of the same GLSM implies that the geometries $X_1$ and $Y_1$ are derived equivalent, as an example of relative homological projective duality, and we will see that this manifests itself in a modular relation between the topological string partition functions.

\subsubsection{The Grassmannian fibration $X_1$}
\label{sec:gvinvariantsModel1Weak}
The Calabi-Yau threefold $X_1$ was constructed in Section~\ref{sec:examples-Gr} as a fibration of codimension five complete intersection curves in $G(2,5)$ over $\mathbb{P}^2$ and realized as a phase of a GLSM in Section~\ref{sec:glsmModel1}.
By construction, the genus one fibration has a $5$-section and we have calculated the Hodge numbers
\begin{align}
	h^{1,1}=2\,,\quad h^{2,1}=47\,,
\end{align}
which also fix the Euler characteristic $\chi=-90$.
As a basis of the K\"ahler cone we can take the 5-section $J_1$ and the vertical divisor $J_2$ that is associated to the hyperplane class in the base $\mathbb{P}^2$.
For the sake of convenience let us reproduce the intersection numbers $c_{ijk}=J_i\cdot J_j\cdot J_k$ from~\eqref{eq:72} and~\eqref{eqn:model1_3int}
\begin{align}
c_{111}=15\,,\quad c_{112}=15\,,\quad c_{122}=5\,,\quad c_{222}=0\,,
	\label{eqn:gvmod1triple}
\end{align}
as well as the intersections with the second Chern class
\begin{align}
	b_1=c_2(X_1)\cdot J_1=66\,,\quad b_2=c_2(X_1)\cdot J_2=36\,.
	\label{eqn:secchernx1}
\end{align}
Note that $c_{222}=0$ and $b_2=36$ also imply that $X_1$ is a genus one fibration over $\mathbb{P}^2$.

To perform the modular bootstrap, we first calculate the genus zero and genus one Gopakumar-Vafa invariants, which we can then use to fix the coefficients of the modular ansatz.
By localizing the sphere partition function we obtained the fundamental period~\eqref{eqn:model1fundClosed} of the mirror Calabi-Yau, which takes the form
\begin{align}
	\varpi_0=&\sum\limits_{\lambda_1,\lambda_2=0}^\infty\frac{\Gamma(1+\lambda_1+\lambda_2)^3}{\Gamma(1+\lambda_1)^3\Gamma(1+\lambda_2)^3}{_3}F_2(-\lambda_1,-\lambda_1,1+\lambda_1;1,1;1) z_1^{\lambda_1}z_2^{\lambda_2}\,.
\end{align}
For two parameter families of Calabi-Yau threefolds one can always choose a basis for the Picard-Fuchs system that consists of one operator of order $2$ and another operator of order $3$~\cite{Hosono:1993qy}.
Making an ansatz we find
{
\begin{align}
\begin{split}
\mathcal{D}_1=&\theta_1^2-3\theta_1\theta_2+6\theta_2^2+z_2^2(10\theta_1^2+15\theta_1\theta_2+6\theta_2^2+15\theta_1+12\theta_2+6)\\
&+z_1(z_2-1)(11\theta_1^2+11\theta_1+3)-z_2(11\theta_1^2+12\theta_1\theta_2+12\theta_2^2+15\theta_1+12\theta_2+6)\\
&-z_1^2(3\theta_1^2+3\theta_1\theta_2+3\theta_2+2\theta_1+3\theta_2+1)\,,\\
\mathcal{D}_2=&\theta_2^3-z_2(\theta_1+\theta_2+1)^3\,.
\end{split}
\label{eqn:model1pf}
\end{align}
}
As expected, the point $(z_1,z_2)=(0,0)$ is of maximally unipotent monodromy.
The two single logarithmic periods can be choosen such that
\begin{align}
	\varpi^1=\varpi_0\log(z_1)+\mathcal{O}(z_1,z_2)\,,\quad \varpi^2=\varpi_0\log(z_2)+\mathcal{O}(z_1,z_2)\,,
\end{align}
and the flat coordinates $t^i=\varpi^i/\varpi_0$ relate the algebraic complex structure coordinates $z_1,z_2$ of the mirror with the complexified K\"ahler moduli of $X_1$.
Inverting this mirror map we obtain
\begin{align}
\begin{split}
	z_1(q_1,q_2)=&q_1 - 9 q_1 q_2 + 27 q_1 q_2^2+\mathcal{O}(q^4)\,,\\
	z_2(q_1,q_2)=&q_2 - 3 q_1 q_2 + 3 q_1^2 q_2 - 5 q_2^2 - 5 q_1 q_2^2 + 15 q_2^3+\mathcal{O}(q^4)\,,
\end{split}
\end{align}
where $q^i(z_1,z_2)=\exp(2\pi i t^i)$.
The intersection numbers~\eqref{eqn:gvmod1triple} can be used to normalize the triple logarithmic solution which takes the form
\begin{align}
	\varpi_3=\varpi_0\frac{1}{3!}\sum\limits_{i,j,k=1}^2c_{ijk}\log(z_i)\log(z_j)\log(z_k)+\mathcal{O}(z_1,z_2)\,.
\end{align}
Up to classical terms, the genus zero free energy is then given by
\begin{align}
	F_0=\frac12{\varpi_3}/{\varpi_0}\big|_{z_i\rightarrow z_i(q_1,q_2)}+\mathcal{O}(t^1,t^2)\,.
\end{align}
Together with~\eqref{eqn:ztopf} and~\eqref{eqn:gvexp} this is sufficient to obtain the Gopakumar-Vafa invariants at genus zero that are shown in Table~\ref{tab:gv0model1}.
Here and in the following we denote the Gopakumar-Vafa invariant at genus $g$ that corresponds to a curve class $\beta$ with
\begin{align}
	d_1=J_1\cdot\beta\,,\quad d_2=J_2\cdot\beta\,,
\end{align}
by $n^{(g)}_{d_1,d_2}$.
We refer to $d_1$ as the fiber degree and to $d_2$ as the base degree.
\begin{table}[h!]
{\tiny
\begin{align*}
\begin{array}{c|cccccc}
	n^{(0)}_{d_1,d_2}&d_2=0&1&2&3&4&5\\\hline
d_1=0& 0 & 30 & 0 & 0 & 0 & 0 \\
1& 105 & 330 & 105 & 0 & 0 & 0 \\
2& 120 & 2865 & 6585 & 2865 & 120 & 0 \\
3& 120 & 17400 & 151260 & 283755 & 151260 & 17400 \\
4& 105 & 87150 & 2141265 & 11044335 & 18347055 & 11044335 \\
5& 90 & 368670 & 22279830 & 256967580 & 974066175 & 1488072900 \\
6& 105 & 1377840 & 186120810 & 4267143150 & 31595446320 & 97322962410 \\
7& 120 & 4644030 & 1311908070 & 55405726800 & 729262582320 & 4007703642030 \\
8& 120 & 14441100 & 8065898475 & 594374999280 & 13050194338080 & 118409369639565 \\
9& 105 & 42003450 & 44272540830 & 5463083502630 & 191094069663765 & 2712537543756540 \\
10& 90 & 115593255 & 220759120890 & 44140588111590 & 2375090868607470 & 50686607599977960
\end{array}
\end{align*}
}
	\caption{Genus zero Gopakumar-Vafa invariants for the genus one fibration $X_1$ over $\mathbb{P}^2$.}
	\label{tab:gv0model1}
\end{table}
Note that the invariants exhibit five-periodicity at base degree zero which is expected for a genus one fibration with $5$-sections (see Section~\ref{eqn:gvspecReview} for a discussion).

The genus one topological string free energy takes the generic form~\cite{Bershadsky:1993ta,Bershadsky:1993cx,Hosono:1994ax}
\begin{align}
	\begin{split}
	F_1=&-\frac12\left(3+h^{1,1}-\frac{\chi}{12}\right)K-\frac12\log\,\det\,G\\
		&-\frac{1}{24}\sum\limits_{i=1}^{h^{1,1}}(b_i+12)\log\,z^i-\frac{1}{12}\sum\limits_ic_i\log\,\Delta_i\,,\\
	\end{split}
	\label{eqn:f1ansatz}
\end{align}
where the last sum is over the components $\Delta_i$ of the discriminant.
The coefficient $c_i$ is the difference of the number of hypermultiplets and vector multiplets that become massless on the sublocus $\{\Delta_i=0\}$ of the moduli space~\cite{Vafa:1995ta}.
Moreover, $b_i$ are the intersections with the second Chern class~\eqref{eqn:secchernx1}.
In the holomorphic limit, the K\"ahler potential of the moduli space metric is $K=-\log(\varpi_0)$, while, up to an irrelevant factor, the determinant of the Weil-Petersson metric becomes $\det(G_{ij})\rightarrow\det(\partial_{z_i}t^j)$.

With the help of the GLSM we have determined that the discriminant consists of the two components~\eqref{eqn:glsm1discComps}.
We can use the base degree one topological string partition function, which we obtain from the modular bootstrap below, in order to fix the coefficients $c_1=c_2=1$.
This is possible because the former can be fixed already using the genus zero Gopakumar-Vafa invariants.
Note that the intersections with the second Chern class~\eqref{eqn:secchernx1} can therefore also be determined using the modular bootstrap.
The resulting invariants for low degrees are listed in Table~\ref{tab:gv1model1}.
\begin{table}[h!]
{\tiny
\begin{align*}
\begin{array}{c|ccccccc}
	n^{(1)}_{d_1,d_2}& d_2=0 & 1 & 2 & 3 & 4 & 5 & 6 \\\hline
 d_1=0 & 0 & 0 & 0 & 0 & 0 & 0 & 0 \\
 1 & 0 & 0 & 0 & 0 & 0 & 0 & 0 \\
 2 & 0 & 0 & 0 & 0 & 0 & 0 & 0 \\
 3 & 0 & 0 & 0 & -1 & 0 & 0 & 0 \\
 4 & 0 & 0 & 120 & 2970 & 6915 & 2970 & 120 \\
 5 & 3 & -60 & 16125 & 703230 & 4208475 & 7375071 & 4208475 \\
 6 & 0 & -660 & 372540 & 33573185 & 413584710 & 1620769755 & 2512209775 \\
 7 & 0 & -5730 & 4756425 & 849268260 & 19357530555 & 142122194100 & 437958571425 \\
 8 & 0 & -34800 & 43430730 & 14645587326 & 575578641975 & 7201911667665 & 38548073582460 \\
 9 & 0 & -174300 & 316804320 & 193187602800 & 12444000331665 & 249450953710980 & 2142984186155265 \\
 10 & 3 & -737520 & 1958504520 & 2080508838285 & 211337363714955 & 6492780643180803 & 84886235042252880 \\
\end{array}
\end{align*}
}
	\caption{Genus one Gopakumar-Vafa invariants genus one fibration $X_1$ over $\mathbb{P}^2$.}
	\label{tab:gv1model1}
\end{table}

Let us now apply the modular bootstrap that we reviewed in Section~\ref{sec:bootstrapreview}.
To obtain the modular parametrization of the K\"ahler form we identify $E_0=J_1,\,D_1=J_2$ and $C^i=E_0\cdot J_2$.
The triple intersection numbers~\eqref{eqn:gvmod1triple} then determine the height pairing~\eqref{eqn:heightpairingP2}
\begin{align}
	D=-\pi^{-1}\pi_*(E_0\cdot E_0)=-15J_2\,.
\end{align}
The modular expansion~\eqref{eqn:bootstrapKaehlerForm} of the K\"ahler form $\omega$ on $X_1$ is then given by
\begin{align}
	\omega=\tau\cdot\left(J_1-\frac32\cdot J_2\right)+t\cdot J_2\,.
\end{align}
We denote the corresponding exponentiated parameters by $q=e^{2\pi i \tau},\,Q=e^{2\pi i t}$ and expand the topological string partition function as
\begin{align}
	Z_{\text{top}}(q,Q,\lambda)=Z_0(q)\left(1+\sum\limits_{d=1}^\infty Z_d(q,\lambda)Q^d\right)\,,
	\label{eqn:ztopoex}
\end{align}
where $\lambda$ is the topological string coupling.
We also calculate the fractional part of the exponent of $\Delta_{10}$ in~\eqref{eqn:bootstrapRing}, which is determined by the congruence
\begin{align}
	1-\frac{r_\beta}{5}\equiv 9 J_2\cdot\beta\text{ mod }1\equiv 0\text{ mod } 1\,.
\end{align}
The genus zero Gopakumar-Vafa invariants from Table~\ref{tab:gv0model1} can then be used to fix the coefficients in the general ansatz~\eqref{eqn:bootstrapAnsatz} and one obtains
\begin{align}
	Z_1(q,\lambda)=\frac{-15}{\eta(5\tau)^{36}\phi_{-2,1}(5\tau,\lambda)}E_{1,1}^{\,7}E_{1,2}^{\,6}\left(2E_{1,1}^{\,3}-14E_{1,1}^{\,2}E_{1,2}+35E_{1,1}E_{1,2}^{\,2}-8E_{1,2}^{\,3}\right)\,.
	\label{eqn:zbd1model1}
\end{align}
From this we can extract the Gopakumar-Vafa invariants $n^g_{d_1,d_2=1}$ for arbitrary genera.
The invariants up to genus $5$ and fiber degree $20$ are shown in Table~\ref{tab:gvBD1model1}.
\begin{table}[h!]
{\small
\begin{align*}
\begin{array}{c|ccccccc}
	n^{(g)}_{d_1,1}& g=0 & 1 & 2 & 3 & 4 & 5 \\\hline
 d_1=0 & 30 & 0 & 0 & 0 & 0 & 0 \\
 1 & 330 & 0 & 0 & 0 & 0 & 0 \\
 2 & 2865 & 0 & 0 & 0 & 0 & 0 \\
 3 & 17400 & 0 & 0 & 0 & 0 & 0 \\
 4 & 87150 & 0 & 0 & 0 & 0 & 0 \\
 5 & 368670 & -60 & 0 & 0 & 0 & 0 \\
 6 & 1377840 & -660 & 0 & 0 & 0 & 0 \\
 7 & 4644030 & -5730 & 0 & 0 & 0 & 0 \\
 8 & 14441100 & -34800 & 0 & 0 & 0 & 0 \\
 9 & 42003450 & -174300 & 0 & 0 & 0 & 0 \\
 10 & 115593255 & -737520 & 90 & 0 & 0 & 0 \\
 11 & 303353190 & -2757660 & 990 & 0 & 0 & 0 \\
 12 & 764067540 & -9305250 & 8595 & 0 & 0 & 0 \\
 13 & 1856177550 & -28986600 & 52200 & 0 & 0 & 0 \\
 14 & 4367282100 & -84529800 & 261450 & 0 & 0 & 0 \\
 15 & 9985141200 & -233398770 & 1106490 & -120 & 0 & 0 \\
 16 & 22247210190 & -614976060 & 4138800 & -1320 & 0 & 0 \\
 17 & 48416987520 & -1556022180 & 13977930 & -11460 & 0 & 0 \\
 18 & 103134292500 & -3799140900 & 43601700 & -69600 & 0 & 0 \\
 19 & 215400960300 & -8987282100 & 127404750 & -348600 & 0 & 0 \\
 20 & 441766481520 & -20666791710 & 352679835 & -1475580 & 150 & 0 \\
\end{array}
\end{align*}
}
	\caption{Gopakumar-Vafa invariants $n^{(g)}_{d_1,1}$ of the genus one fibration $X_1$ over $\mathbb{P}^2$.}
	\label{tab:gvBD1model1}
\end{table}

The information that we get from the genus zero and genus one invariants is not sufficient to fix the modular bootstrap ansatz for base degree two.
In~\cite{Huang:2015sta} the authors derived a Castelnuovo-like bound for elliptic fibrations over $\mathbb{P}^2$ that requires
\begin{align}
	n^{(g)}_{d_1,d_2}=0\,,\quad\text{for}\quad d_1\ge 3d_2-1,\,g\ge d_1d_2-\frac12(3d_2^2-d_2-4)\,,
\end{align}
and this allows to fix the numerator, at least in principle, up to base degree $20$.
But the bound does not hold for genus one fibrations that are not elliptic.
However, at least for base degree $2$ we can experimentally determine an analogous restriction
\begin{align}
	n^{(g)}_{d_1,2}=0\,,\quad\text{for}\quad g\ge 1+\left\lfloor\frac{d_1+1}{5}\right\rfloor+\left\lfloor\frac{d_1+3}{5}\right\rfloor\,.
	\label{eqn:gvrestriction}
\end{align}
This holds for fiber degrees smaller than $24$, which is the highest degree for that we have calculated the invariants.
For the geometry at hand, this bound can be tightened, but~\eqref{eqn:gvrestriction} seems to be satisfied by all of the genus one fibrations over $\mathbb{P}^2$ for which we have tested the modular bootstrap. 
Note that the existence of such a bound is implied by the Gopakumar-Vafa finiteness conjecture by Bryan and Pandharipande~\cite{10.2140/gt.2001.5.287} which has recently been proven by Doan, Ionel and Walpuski~\cite{doan2021gopakumarvafa}.

Based on the information that we obtain from genus zero and genus one, we further assume that the numerator exhibits an overall factor of $E_{1,1}^{\,14}E_{1,2}^{\,12}$.
This fixes the $185$ coefficients in the ansatz completely and we obtain
\begin{align}
	Z_2(q,\lambda)=\frac{E_{1,1}^{\,14}E_{1,2}^{\,12}\left(\phi_{6}\cdot A^4+\phi_{8}\cdot A^3B+\phi_{10}\cdot A^2B^2+\phi_{12}\cdot AB^3+\phi_{14}\cdot B^4\right)}{\eta(5\tau)^{72}\phi_{-2,1}(5\tau,\lambda)\phi_{-2,1}(5\tau,2\lambda)}\,,
	\label{eqn:Z2model1a}
\end{align}
with $A\equiv\phi_{0,1}(5\tau,\lambda),\,B\equiv\phi_{-2,1}(5\tau,\lambda)$ and
\begin{align}
	\begin{split}
		\phi_{6}(\tau)=&\frac{25}{1152}\left(2 a^3 - 14 a^2 b + 35 a b^2 - 8 b^3\right)^2\,,\\
		\phi_{8}(\tau)=&-\frac{5}{576}\left(11 a^8 - 214 a^7 b + 2960 a^6 b^2 - 13874 a^5 b^3 + 28595 a^4 b^4\right. \\
		&\left.- 37378 a^3 b^5 + 43112 a^2 b^6 - 3600 a b^7 + 192 b^8\right)\,,\\
		\phi_{10}(\tau)=&-\frac{5}{192}\left(a^2 - 11 a b - b^2\right) \left(9 a^8 - 113 a^7 b + 515 a^6 b^2 - 340 a^5 b^3\right.\\
		&\left.+ 429 a^4 b^4 + 3545 a^3 b^5 - 1565 a^2 b^6 + 1348 a b^7 - 156 b^8\right)\,,\\
		\phi_{12}(\tau)=&\frac{5}{576}\left(47 a^{12} - 1354 a^{11} b + 18138 a^{10} b^2 - 131738 a^9 b^3 + 423642 a^8 b^4\right.\\
		&- 506248 a^7 b^5 + 586990 a^6 b^6 - 2759698 a^5 b^7 + 569379 a^4 b^8 + 1030654 a^3 b^9 \\
		&\left.- 28284 a^2 b^{10} - 1984 a b^{11} + 872 b^{12}\right)\,,\\
		\phi_{14}(\tau)=&-\frac{5}{1152}\left(38 a^{14} - 1288 a^{13} b + 21542 a^{12} b^2 - 211916 a^{11} b^3\right.\\
		&+ 1075151 a^{10} b^4 - 2239468 a^9 b^5 + 1648420 a^8 b^6 - 11424676 a^7 b^7 \\
		&+ 7569260 a^6 b^8 - 8748236 a^5 b^9 + 8478662 a^4 b^{10} + 1576352 a^3 b^{11}\\
		&\left.- 48985 a^2 b^{12} - 1776 a b^{13} + 744 b^{14}\right)\,,
	\end{split}
	\label{eqn:Z2model1b}
\end{align}
where we further introduced $a=E_{1,1},\,b=E_{1,2}$.
From this we can extract the base degree two Gopakumar-Vafa invariants up to arbitrary fiber degree and genus.
The first $15$ fiber degrees up to genus $6$ are shown in Table~\ref{tab:gvBD2model1}.
\begin{table}[h!]
	\centering
	{\tiny
	\begin{align*}
\begin{array}{c|ccccccc}
	n^{(g)}_{d_1,2}& g=0 & 1 & 2 & 3 & 4 & 5 & 6 \\\hline
 d_1=0 & 0 & 0 & 0 & 0 & 0 & 0 & 0 \\
 1 & 105 & 0 & 0 & 0 & 0 & 0 & 0 \\
 2 & 6585 & 0 & 0 & 0 & 0 & 0 & 0 \\
 3 & 151260 & 0 & 0 & 0 & 0 & 0 & 0 \\
 4 & 2141265 & 120 & 0 & 0 & 0 & 0 & 0 \\
 5 & 22279830 & 16125 & 0 & 0 & 0 & 0 & 0 \\
 6 & 186120810 & 372540 & 0 & 0 & 0 & 0 & 0 \\
 7 & 1311908070 & 4756425 & -4170 & 0 & 0 & 0 & 0 \\
 8 & 8065898475 & 43430730 & -166800 & 0 & 0 & 0 & 0 \\
 9 & 44272540830 & 316804320 & -2731560 & 360 & 0 & 0 & 0 \\
 10 & 220759120890 & 1958504520 & -28989060 & 47070 & 0 & 0 & 0 \\
 11 & 1013572957380 & 10640032350 & -235363635 & 1088550 & 0 & 0 & 0 \\
 12 & 4331159754645 & 52040873400 & -1583604465 & 13913100 & -8340 & 0 & 0 \\
 13 & 17375814175500 & 233115403215 & -9244669020 & 127798200 & -333600 & 0 & 0 \\
 14 & 65916243157860 & 968623984560 & -48221564355 & 942273570 & -5465040 & 600 & 0 \\
 15 & 237871858142880 & 3770188504875 & -229340754150 & 5914257480 & -58225560 & 78450 & 0 \\
\end{array}
	\end{align*}
	}
	\caption{Gopakumar-Vafa invariants $n^{(g)}_{d_1,2}$ of the genus one fibration $X_1$ over $\mathbb{P}^2$.}
	\label{tab:gvBD2model1}
\end{table}

\subsubsection{The Pfaffian fibration $Y_1$}
Let us now apply the modular bootstrap to the Calabi-Yau threefold $Y_1$ that is a fibration of Pfaffian curves over $\mathbb{P}^2$ and related by relative homological projective duality to $X_1$.
Recall that the algebraic geometry of this Calabi-Yau has been studied in Section~\ref{sec:examples-Pf} and the associated GLSM phase was discussed in Section~\ref{sec:glsm1nadual}.
The Hodge numbers and the Euler characteristic are
\begin{align}
	h^{1,1}=2\,,\quad h^{1,2}=47\,,\quad \chi=-90\,.
	\label{eqn:mody1hodge}
\end{align}

We can transform the Picard-Fuchs system~\eqref{eqn:model1pf} to the large volume point associated to $Y_1$ by performing the coordinate change
\begin{align}
	z_1\rightarrow -v_1^{-1}\,,\quad z_2\rightarrow -v_2 v_1^{-1}\,.
\end{align}
Here the signs have been choosen such that all of the terms in the expansion of the fundamental period around $(v_1,v_2)=(0,0)$ are positive.
This leads to the differential operators
\begin{align}
	\begin{split}
		\mathcal{D}_1=&\theta_1^2-\theta_1 \theta_2+\theta_2^2-z_1 \left(3+11 \theta_1+11 \theta_1^2+11 \theta_2+22 \theta_1 \theta_2+11 \theta_2^2\right)\\
		&-z_2 \left(3+11 \theta_1+11 \theta_1^2+11 \theta_2+22 \theta_1 \theta_2+11 \theta_2^2\right)-z_1^2 \left(1+2 \theta_1+\theta_1^2+5 \theta_2\right.\\
		&\left.+5 \theta_1 \theta_2+10 \theta_2^2\right)-z_1 z_2 \left(2+7 \theta_1+11 \theta_1^2+7 \theta_2+10 \theta_1 \theta_2+11 \theta_2^2\right)\\
		&-z_2^2 \left(1+5 \theta_1+10 \theta_1^2+2 \theta_2+5 \theta_1 \theta _2+\theta_2^2\right)\,,\\
		\mathcal{D}_2=&\theta_2^3-z_2 \left(1+\theta_1+\theta_2\right) \left(3+11 \theta_1+11 \theta_1^2+11 \theta_2+22 \theta_1 \theta_2+11 \theta_2^2\right)\\
		&-z_1 z_2 \left(2+5 \theta _1+4 \theta_1^2+\theta_1^3+11 \theta_2+17 \theta_1 \theta_2+6 \theta_1^2 \theta_2+25 \theta_2^2+15 \theta_1 \theta_2^2+20 \theta_2^3\right)\\
		&-z_2^2 \left(2+11 \theta_1+25 \theta_1^2+5 \theta_2+17 \theta_1 \theta_2+15 \theta_1^2 \theta_2+4 \theta_2^2+6 \theta_1 \theta_2^2+\theta_2^3\right)\,.
	\end{split}
	\label{eqn:model1pfscp}
\end{align}
The fundamental period reads
\begin{align}
	\begin{split}
	\varpi_0=&1+3 v_1+3 v_2+19 v_1^2+152 v_1 v_2+19 v_2^2\\
		&+147 v_1^3+3969 v_1^2 v_2+3969 v_1 v_2^2+147 v_2^3+\mathcal{O}(v^4)\\
		=&\sum\limits_{\lambda_1,\lambda_2=0}^\infty\frac{\Gamma(1+\lambda_+)^3}{\Gamma(1+\lambda_1)^3\Gamma(1+\lambda_2)^3}{_3}F_2(-\lambda_+,-\lambda_+,1+\lambda_+;1,1;1) v_1^{\lambda_1}v_2^{\lambda_2}\,,
	\end{split}
\end{align}
with $\lambda_+=\lambda_1+\lambda_2$ and, as expected, the point $(v_1,v_2)=(0,0)$ is of maximally unipotent monodromy.

Let us again denote the basis of the K\"ahler cone by $J_1,\,J_2$, where $J_2$ is the vertical divisor and $J_1=E_0-J_2$ with $E_0$ being the $5$-section.
The triple intersection numbers $c_{ijk}=J_i\cdot J_j\cdot J_k$ have been calculated in~\eqref{eq:73} and are given by
\begin{align}
	c_{111}=0\,,\quad c_{112}=5\,,\quad c_{122}=5\,,\quad c_{222}=0\,,
\end{align}
while the intersections with the second Chern class are
\begin{align}
	b_1=c_2(Y_1)\cdot J_1=36\,,\quad b_2=c_2(Y_1)\cdot J_2=36\,.
	\label{eqn:mody1int}
\end{align}
Note that $c_{111}$ differs from the correpsonding intersection number that we obtained in the weakly coupled phase.
Following the procedure outlined in Section~\ref{sec:gvinvariantsModel1Weak} we can again extract the genus zero Gopakumar-Vafa invariants.
The invariants for low degrees are listed in Table~\ref{tab:gv0model1scp}.
\begin{table}[h!]
	\centering
	{\tiny
	\begin{align*}
\begin{array}{c|cccccc}
	n^{(g)}_{d_1,d_2}& d_2=0 & 1 & 2 & 3 & 4 & 5 \\\hline
 d_1=0 & 0 & 120 & 105 & 105 & 120 & 90 \\
 1 & 120 & 2085 & 15690 & 83400 & 362850 & 1365060 \\
 2 & 105 & 15690 & 569475 & 9690270 & 107459880 & 901887570 \\
 3 & 105 & 83400 & 9690270 & 418812780 & 10086474180 & 164859436335 \\
 4 & 120 & 362850 & 107459880 & 10086474180 & 472152998265 & 13800385325580 \\
 5 & 90 & 1365060 & 901887570 & 164859436335 & 13800385325580 & 675995017391805 \\
 6 & 120 & 4621020 & 6204484125 & 2041590595410 & 286700834960805 & 22351196770131870 \\
 7 & 105 & 14399490 & 36701125005 & 20496053409240 & 4593254607725475 & 546563929916334210 \\
 8 & 105 & 41932200 & 192593575110 & 174405931797135 & 59937858896889555 & 10518492857890739820 \\
 9 & 120 & 115485075 & 916315955820 & 1297448843314125 & 661998422042833065 & 166511015537610566130 \\
 10 & 90 & 303166710 & 4015843886955 & 8630138044756890 & 6364684023911207415 & 2240097475662256021890 \\
\end{array}
	\end{align*}
	}
	\caption{Genus zero Gopakumar-Vafa invariants for the genus one fibration $Y_1$ over $\mathbb{P}^2$.}
	\label{tab:gv0model1scp}
\end{table}
The contribution of the discriminant to the genus one free energy has to be the same as for $X_1$ and, together with~\eqref{eqn:mody1hodge} and~\eqref{eqn:mody1int}, this again fixes the genus one free energy~\eqref{eqn:f1ansatz}.
Some of the genus one invariants are listed in Table~\ref{tab:gv1model1scp}.
\begin{table}[h!]
\centering
	{\tiny
	\begin{align*}
\begin{array}{c|cccccc}
	n^{(1)}_{d_1,d_2}& d_2=0 & 1 & 2 & 3 & 4 & 5 \\\hline
 d_1=0 & 0 & 0 & 0 & 0 & 0 & 3 \\
 1 & 0 & 0 & 0 & 0 & 0 & -240 \\
 2 & 0 & 0 & 330 & 17400 & 368790 & 4662420 \\
 3 & 0 & 0 & 17400 & 2967033 & 152464260 & 4141021440 \\
 4 & 0 & 0 & 368790 & 152464260 & 15549957525 & 762487576830 \\
 5 & 3 & -240 & 4662420 & 4141021440 & 762487576830 & 62771612774814 \\
 6 & 0 & -4170 & 42570870 & 74376148860 & 23171036639520 & 3041228666778570 \\
 7 & 0 & -31380 & 311782950 & 996296178705 & 499448622404940 & 100290841470936765 \\
 8 & 0 & -166800 & 1937387460 & 10705096100433 & 8293309463942745 & 2462636031682103055 \\
 9 & 0 & -725700 & 10575317925 & 96755745250020 & 112056887343462270 & 47824236177306164880 \\
 10 & 3 & -2730840 & 51926307780 & 760141098535650 & 1280236061312374950 & 766594359928331774304 \\
\end{array}
	\end{align*}
	}
	\caption{Genus one Gopakumar-Vafa invariants for the genus one fibration $Y_1$ over $\mathbb{P}^2$.}
	\label{tab:gv1model1scp}
\end{table}

%
The modular parametrization of the K\"ahler form $\omega$ on $Y_1$ is given by 
\begin{align}
	\omega=\tau\cdot\left(J_1-\frac12\cdot J_2\right)+t\cdot J_2\,,
\end{align}
and we denote the exponentiated parameters by $q=e^{2\pi i \tau},\,Q=e^{2\pi i t}$.
Using the expansion of the topological string partition function~\eqref{eqn:ztopoex} as well as the genus zero Gopakumar-Vafa invariants we can fix
\begin{align}
	Z_1(q,\lambda)=\frac{-15}{\eta(5\tau)^{36}\phi_{-2,1}(5\tau,\lambda)}E_{1,1}^{\,6}E_{1,2}^{\,7}\left(2E_{1,2}^{\,3}+14E_{1,2}^{\,2}E_{1,1}+35E_{1,2}E_{1,1}^{\,2}+8E_{1,1}^{\,3}\right)\,.
	\label{eqn:zbd1model1scp}
\end{align}
Some of the corresponding higher genus invariants are listed in Table~\ref{tab:gvBD1model1scp}.
\begin{table}[h!]
\centering
	{
	\begin{align*}
	\begin{array}{c|ccccc}
	n^{(g)}_{d_1,1}&g=0&1&2&3&4\\\hline
	d_1=0& 120 & 0 & 0 & 0 & 0 \\
	1& 2085 & 0 & 0 & 0 & 0 \\
	2& 15690 & 0 & 0 & 0 & 0 \\
	3& 83400 & 0 & 0 & 0 & 0 \\
	4& 362850 & 0 & 0 & 0 & 0 \\
	5& 1365060 & -240 & 0 & 0 & 0 \\
	6& 4621020 & -4170 & 0 & 0 & 0 \\
	7& 14399490 & -31380 & 0 & 0 & 0 \\
	8& 41932200 & -166800 & 0 & 0 & 0 \\
	9& 115485075 & -725700 & 0 & 0 & 0 \\
	10& 303166710 & -2730840 & 360 & 0 & 0 \\
	11& 763767660 & -9254550 & 6255 & 0 & 0 \\
	12& 1855698750 & -28893120 & 47070 & 0 & 0 \\
	13& 4366532100 & -84364800 & 250200 & 0 & 0 \\
	14& 9984045450 & -233147250 & 1088550 & 0 & 0 \\
	15& 22245496710 & -614524740 & 4097100 & -480 & 0 \\
	16& 48414415230 & -1555278120 & 13896420 & -8340 & 0 \\
	17& 103130437560 & -3797919960 & 43449510 & -62760 & 0 \\
	18& 215395264050 & -8985324600 & 127131000 & -333600 & 0 \\
	19& 441758392350 & -20663904150 & 352260825 & -1451400 & 0 \\
	\end{array}
	\end{align*}
	}
	\caption{Gopakumar-Vafa invariants $n^{(g)}_{d_1,1}$ of the genus one fibration $Y_1$ over $\mathbb{P}^2$.}
	\label{tab:gvBD1model1scp}
\end{table}

To fix the base degree two partition function we need to again employ a vanishing condition and for this geometry the experimental bound~\eqref{eqn:gvrestriction},
\begin{align}
	n^{(g)}_{d_1,2}=0\,,\quad\text{for}\quad g\ge 1+\left\lfloor\frac{d_1+1}{5}\right\rfloor+\left\lfloor\frac{d_1+3}{5}\right\rfloor\,,
\end{align}
also leads to integer invariants for $d_1\le 24$.
Based on the information that we obtain from genus zero and genus one, we also assume that the numerator exhibits an overall factor of $E_{1,1}^{\,12}E_{1,2}^{\,14}$.
Again, this fixes the ansatz completely and we obtain
\begin{align}
	Z_2(q,\lambda)=\frac{E_{1,1}^{\,12}E_{1,2}^{\,14}\left(\phi_{6}\cdot A^4+\phi_{8}\cdot A^3B+\phi_{10}\cdot A^2B^2+\phi_{12}\cdot AB^3+\phi_{14}\cdot B^4\right)}{\eta(5\tau)^{72}\phi_{-2,1}(5\tau,\lambda)\phi_{-2,1}(5\tau,2\lambda)}\,,
	\label{eqn:Z2model1ascp}
\end{align}
with $A\equiv\phi_{0,1}(5\tau,\lambda),\,B\equiv\phi_{-2,1}(5\tau,\lambda)$ and
\begin{align}
	\begin{split}
		\phi_{6}(\tau)=&\frac{25}{1152}\left(35 a^2 b+8 a^3+14 a b^2+2 b^3\right)^2\,,\\
		\phi_{8}(\tau)=&-\frac{5}{576}\left(192 a^8+3600 a^7 b+43112 a^6 b^2+37378 a^5 b^3\right.\\
		&\left.+28595 a^4 b^4+13874 a^3 b^5+2960 a^2 b^6+214 a b^7+11 b^8\right)\,,\\
		\phi_{10}(\tau)=&-\frac{5}{192}\left(-a^2 + 11 a b + b^2\right)\cdot\left(-156 a^8 - 1348 a^7 b - 1565 a^6 b^2  \right.\\
		&\left.- 3545 a^5 b^3+ 429 a^4 b^4 + 340 a^3 b^5 + 515 a^2 b^6 + 113 a b^7 + 9 b^8\right)\,,\\
		\phi_{12}(\tau)=&\frac{5}{576}\left(-28284 a^{10} b^2-1030654 a^9 b^3+569379 a^8 b^4+2759698 a^7 b^5\right.\\
		&+586990 a^6 b^6+506248 a^5 b^7+423642 a^4 b^8+131738 a^3 b^9\\
		&\left.+18138 a^2 b^{10}+1984 a^{11} b+872 a^{12}+1354 a b^{11}+47 b^{12}\right)\,,\\
		\phi_{14}(\tau)=&-\frac{5}{1152}\left(744 a^{14}+1776 a^{13} b-48985 a^{12} b^2-1576352 a^{11} b^3\right.\\
		&+8478662 a^{10} b^4+8748236 a^9 b^5+7569260 a^8 b^6+11424676 a^7 b^7\\
		&+1648420 a^6 b^8+2239468 a^5 b^9+1075151 a^4 b^{10}+211916 a^3 b^{11}\\
		&\left.+21542 a^2 b^{12}+1288 a b^{13}+38 b^{14}\right)\,,
	\end{split}
	\label{eqn:Z2model1bscp}
\end{align}
where we use $a=E_{1,1},\,b=E_{1,2}$.
Some of the corresponding invariants are listed in Table~\ref{tab:gvBD2model1scp}.
\begin{table}[h!]
	\centering
	{\tiny
	\begin{align*}
\begin{array}{c|cccccc}
	n^{(g)}_{d_1,2}& g=0 & 1 & 2 & 3 & 4 & 5 \\\hline
 d_1=0 & 105 & 0 & 0 & 0 & 0 & 0 \\
 1 & 15690 & 0 & 0 & 0 & 0 & 0 \\
 2 & 569475 & 330 & 0 & 0 & 0 & 0 \\
 3 & 9690270 & 17400 & 0 & 0 & 0 & 0 \\
 4 & 107459880 & 368790 & -60 & 0 & 0 & 0 \\
 5 & 901887570 & 4662420 & -5730 & 0 & 0 & 0 \\
 6 & 6204484125 & 42570870 & -174300 & 0 & 0 & 0 \\
 7 & 36701125005 & 311782950 & -2759640 & 990 & 0 & 0 \\
 8 & 192593575110 & 1937387460 & -29091000 & 52200 & 0 & 0 \\
 9 & 916315955820 & 10575317925 & -235611630 & 1106730 & -120 & 0 \\
 10 & 4015843886955 & 51926307780 & -1583995065 & 14000850 & -11460 & 0 \\
 11 & 16407512285100 & 233286772710 & -9242696730 & 128101950 & -348600 & 0 \\
 12 & 63079928506830 & 971353243680 & -48198804525 & 943160730 & -5524560 & 1650 \\
 13 & 229902653975760 & 3786044994885 & -229205802990 & 5916859050 & -58460400 & 87000 \\
 14 & 799125914198880 & 13924247889780 & -1008585209280 & 32766537720 & -477126120 & 1845150 \\
 15 & 2662321293079905 & 48634484327985 & -4152265143450 & 164053917480 & -3242633010 & 23392050 \\
 16 & 8536545947114475 & 162188834827680 & -16130826253485 & 755246580540 & -19167567900 & 215246250 \\
 17 & 26436605275331250 & 518726285663670 & -59536640053755 & 3237337250760 & -101412573480 & 1599561090 \\
 18 & 79311367370246610 & 1597098769044060 & -209928404589150 & 13045602784170 & -489830544600 & 10153930950 \\
\end{array}
	\end{align*}
	}
	\caption{Gopakumar-Vafa invariants $n^{(g)}_{d_1,2}$ of the genus one fibration $Y_1$ over $\mathbb{P}^2$.}
	\label{tab:gvBD2model1scp}
\end{table}

Comparing the modular expressions~\eqref{eqn:zbd1model1scp} and~\eqref{eqn:Z2model1ascp} for the base degree one and two topological string partition functions of $Y_1$ to the corresponding results~\eqref{eqn:zbd1model1} and~\eqref{eqn:Z2model1a} for $X_1$ leads to a striking observation.
It turns out that we can obtain the results for either geometry by applying the transformation
\begin{align}
	E_{1,1}\rightarrow E_{1,2}\,,\quad E_{1,2}\rightarrow -E_{1,1}\,,
\end{align}
to the partition function of the dual Calabi-Yau.
We will explain this phenomenon in Section~\ref{sec:transferrelation} and show that the transfer matrix that relates the modular parameters of the two geometries is an element of $\Gamma_0(5)$.
The topological string partition functions, being expressed in terms of $\Gamma_1(5)$ Jacobi forms, transform vector-valued under this action and this provides a modular manifestation of relative homological projective duality.
However, before we come to this, let us first recall the general structure of the complexified K\"ahler moduli space and how the monodromies relate to the modular properties of the topological string.

\subsection{Monodromies, Fourier-Mukai transforms and $\Gamma_1(5)$ modularity}
\label{sec:monmod}
The modular properties of the topological string partition function can be seen as a consequence of the monodromies in the stringy K\"ahler moduli space.
This has been shown in~\cite{Schimannek:2019ijf,Cota:2019cjx} by calculating the effect of B-field shifts and relative conifold monodromies on the K\"ahler parameters using Fourier-Mukai transforms on the derived category of coherent sheaves.
At least for genus one fibrations that have $N$-sections with $N\le4$ these monodromies generate a $\Gamma_1(N)$ action and, together with general automorphic properties of the topological string partition function and the holomorphic anomaly equations, imply the expansion in terms of Jacobi forms that underlies the modular bootstrap.
However, if the fibration only has $5$-sections, a third monodromy is needed to generate $\Gamma_1(5)$.
To understand the nature of this additional generator, it is instructive to first review the action of the B-field shifts and the relative conifold monodromy.
This will also motivate the parametrization of the K\"ahler form~\eqref{eqn:bootstrapKaehlerForm} that was necessary to perform the modular bootstrap and has been derived in~\cite{Cota:2019cjx}.
We will assume a basic familiarity with the derived category of coherent sheaves and Fourier-Mukai transforms and accessible introductions can be found in~\cite{Aspinwall:2004jr,Andreas:2004uf}.

First, let us recall the connection between the monodromies in the stringy K\"ahler moduli space and Fourier-Mukai transforms.
The homological mirror symmetry conjecture implies that the monodromies act as auto-equivalences on the category of topological B-branes, which is the derived category of coherent sheaves $D^b(X)$, and thus also on the associated K-theory and central charges.
From this perspective, the K\"ahler parameters can be interpreted as central charges of 2-branes which are normalized by dividing with the central charge of a 0-brane.
A Fourier-Mukai transform $\Phi_{\mathcal{E}}:\,D^b(X)\rightarrow D^b(Y)$ relates the derived categories of two varieties $X,Y$ and is determined by a kernel $\mathcal{E}\in D^b(X\times Y)$.
Denoting the projection on the $i$-th factor of $X\times Y$ by $\pi_i$, the transform acts on a complex $\mathcal{F}^\bullet\in D^b(X)$ as
\begin{align}
	\phi_{\mathcal{E}}:\,\mathcal{F}^\bullet\mapsto R\pi_{2*}\left(\mathcal{E}\otimes_L L\pi_{1}^*\mathcal{F}^\bullet\right)\,.
	\label{eqn:fmtrans}
\end{align}
For some kernels $\mathcal{E}\in D^b(X\times X)$ this is an auto-equivalence of $D^b(X)$ and Orlov proved that for every auto-equivalence, and thus for every monodromy, there exists a kernel that realizes it as a Fourier-Mukai transform.

Perhaps the simplest example is given by the large volume monodromies.
To obtain the corresponding Fourier-Mukai kernel let us denote the embedding of the diagonal in $X\times X$ by
\begin{align}
	j:\Delta\xhookrightarrow{}X\times X\,.
\end{align}
One can show that for any Cartier divisor $D$ on $X\simeq \Delta$, the Fourier-Mukai transform associated to the kernel $j_*\mathcal{O}_\Delta(-D)$ is an auto-equivalence of $D^b(X)$ that acts as
\begin{align}
	\Phi_{j_*\mathcal{O}_\Delta(-D)}:\,\mathcal{F}^\bullet\mapsto\mathcal{F}^\bullet\otimes\mathcal{O}_X(-D)\,.
	\label{eqn:fmtransbfield}
\end{align}
The quantum periods of the Calabi-Yau correspond to the central charges of topological B-branes and the asymptotic terms can be calculated using the $\Gamma$-class formula
\begin{align}
	\Pi_{\text{asy}}(\mathcal{F}^\bullet)=\int_X e^{\omega}\Gamma_{\mathbb{C}}(X)\left(\text{ch}\,\mathcal{F}^\bullet\right)^\vee\,,
\end{align}
where $\omega$ is the complexified K\"ahler class and the $\Gamma$-class of a Calabi-Yau threefold can be expressed in terms of the Chern classes $c_2,c_3$ as
\begin{align}
	\Gamma_{\mathbb{C}}(X)=1+\frac{1}{24}c_2+\frac{\zeta(3)}{(2\pi i)^3}c_3\,.
\end{align}
The linear involution $\vee:\,\oplus_k H^{k,k}(X)\rightarrow H^{k,k}(X)$ is determined by $\delta^\vee\rightarrow (-1)^i\delta$ for $\delta\in H^{i,i}(X)$.
Applying this to the right hand side of~\eqref{eqn:fmtransbfield} and using $\text{ch}(V\otimes V')=\text{ch}(V)\cdot\text{ch}(V')$ as well as $\text{ch}\,\mathcal{O}_X(D)=e^D$ one can see that the Fourier-Mukai transform indeed corresponds to the large volume monodromy that shifts the $B$-field by $D$.

Another standard example is the monodromy around the conifold locus where the 6-brane becomes massless.
The corresponding Fourier-Mukai kernel is the ideal sheaf $\mathcal{I}_\Delta$ of the diagonal $\Delta$.
One can show that the corresponding transformation~\eqref{eqn:fmtrans} acts non-trivially only on the 0-brane, which maps into a bound state of a 0-brane and an anti-6-brane.
If the Calabi-Yau is a fibration over a base $B$ then one can also consider relative Fourier-Mukai transforms with kernel in $D^b(X\times_B X)$ by performing all operations in~\eqref{eqn:fmtrans} relative to $B$.
In particular, at least in cases where the generic fiber is itself Calabi-Yau, we can use the ideal sheaf $\mathcal{I}_{\Delta_B}$ of the relative diagonal $\Delta_B$ in $X\times_B X$ to perform a relative conifold transformation. 
Using the GLSM one can show that this corresponds to a monodromy around the wall in the stringy K\"ahler moduli space where the volume of the generic fiber goes to zero~\cite{Schimannek:2019ijf}.
It has also been found that the relative conifold transformation can be expressed as a combination of large volume monodromies and ordinary conifold monodromies~\cite{Cota:2019cjx}.
The action of the relative conifold transformation on the brane charges can be calculated using the singular Grothendieck-Riemann-Roch formula and we refer to~\cite{Andreas:2004uf,Schimannek:2019ijf,Cota:2019cjx} for details.

Let us now review the action of the B-field shift and the relative conifold monodromy on the K\"ahler parameters.
To this end, we assume that $X$ is a genus one fibered Calabi-Yau manifold over a base $B$ with one $N$-section and no fibral divisors.
A basis of divisors and curves on the fibration has already been introduced in~\eqref{eqn:fibrationsdivcur} and we use the modular parametrization of the K\"ahler form~\eqref{eqn:bootstrapKaehlerForm}
Let us denote the Fourier-Mukai transform with kernel $j_*\mathcal{O}_{\Delta}(-E_0)$ by $T$ and the relative conifold transform with kernel $\mathcal{I}_{\Delta_B}$ by $U$.
Interpreting the parameters $\tau$ and $t^i$ as central charges of 2-branes one obtains the actions
\begin{align}
	\def\arraystretch{1.2}
	T:\,\left\{\begin{array}{rcl}
		\tau&\mapsto&\tau+1\\
		t^i&\mapsto&t^i+\frac{\tilde{a}^i}{2N}
	\end{array}\right.\,,\quad
	U:\,\left\{\begin{array}{rcl}
		\tau&\mapsto&\frac{\tau}{N\tau+1}\\
		t^i&\mapsto&t^i+a^i\,,
	\end{array}\right.\,,
	\label{eqn:UTtrans}
\end{align}
with $a^i=c_1(B)\cdot \tilde{C}^i$.
The images of the base parameters $t^i$ under relative conifold transformations also receive subleading corrections that are exponentially surpressed in the large base limit and can be neglected.
For $N\le 4$ the actions of $T$ and $U$ on $\tau$ generate $\Gamma_1(N)$ and can be used to derive the modular properties of the topological string partition function~\cite{Cota:2019cjx}.

\begin{figure}[h!]
	\centering
	\begin{tikzpicture}[node distance=4mm, >=latex',block/.style = {draw, rectangle, minimum height=35mm, minimum width=43mm,align=center},
		yblock/.style = {draw=none, rectangle, rounded corners=0.5em,align=center,minimum width=4.3cm,fill=white},
		]
		\node[] at (5.5,4.9) {\includegraphics[width=.5\linewidth]{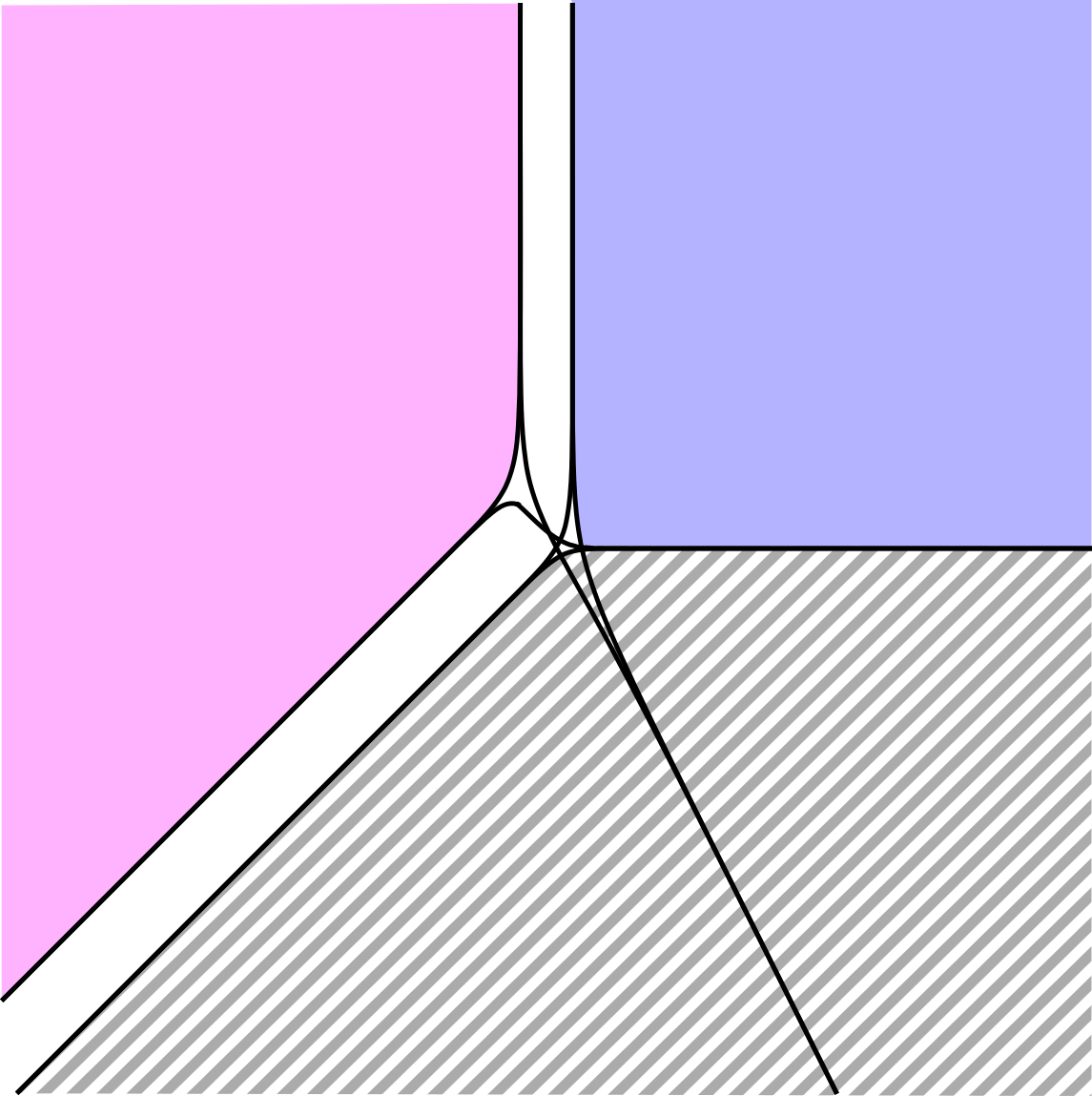}};
		\node[align=center] at (7.7,6.5) {Fibration of CI\\in Grassmannian};
		\node[align=center] at (3.5,6.5) {Fibration of\\Pfaffian curves};
		\draw[->,black,thick,overlay] (1,4) -- (1,7.5);
		\draw[->,black,thick,overlay] (5.5,.5) -- (9.5,.5);
		\draw[->,black,thick,overlay] (5.0,.5) -- (1.0,.5);
		\node[align=center] at (0,6) {large\\base};
		\node[align=center] at (11,6) {\phantom{large}};
		\node[align=center] at (7.3,.2) {large fiber};
		\node[align=center] at (3.0,.2) {large fiber};
		\node[yblock, align=center, fill opacity=.7, text opacity=1.] at (7.0,2.3) {non generic\\phases};
	\end{tikzpicture}
	\caption{The phase structure of genus one fibered Calabi-Yau manifolds with 5-sections.}
	\label{fig:amoebasketch}
\end{figure}
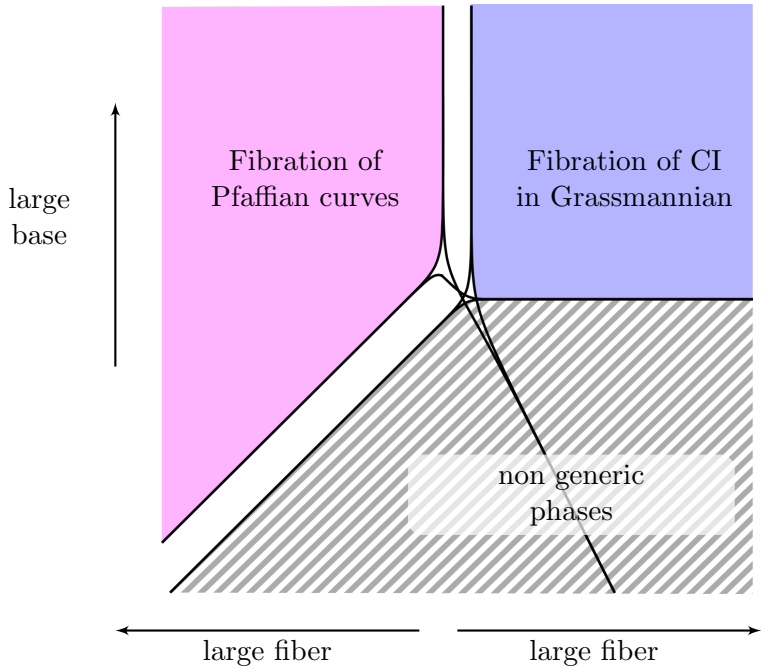
However, to generate $\Gamma_1(5)$ a third monodromy is needed and this is provided by the second component of the conifold that we observed in the amoeba of the geometry $X_1$ in Figure~\ref{fig:splitAmoebaGLSM1}.
This structure appears for all of the geometries that we discuss in this paper and is induced by the splitting of the conifold point in the moduli space of the fiber.
Based on this, the generic phases in the stringy K\"ahler moduli space of a genus one fibered Calabi-Yau with $5$-sections is shown in Figure~\ref{fig:amoebasketch}.
In the large base limit there are two walls that separate the two geometric phases from each other, with a non-geometric region in between.
From the perspective of either of the geometric phases, the monodromy around the component at the boundary of the phase corresponds to the relative conifold transform.
However, the Fourier-Mukai kernel that corresponds to the monodromy around the respective component that is separated by the non-geometric region is not so clear.

To see that it provides the necessary third generator of $\Gamma_1(5)$ it is fortunately sufficient to obtain the action on the $\tau$-parameter of the complexified K\"ahler form of the fibration.
We will calculate this action in the next section by analytically continuing the periods of the generic fiber.
In that way we also obtain the transfer matrix that connects the periods of the fiber in one geometric limit to the other.

\subsection{Relative homological projective duality enhances $\Gamma_1(5)$ to $\Gamma_0(5)$}
\label{sec:transferrelation}
Let us now try to understand the action of the monodromy around the second conifold component on the modular parameter as well as the striking relation between the topological string partition functions on dual fibrations.
To this end, we restrict to the moduli space of the generic fiber and analytically continue the corresponding periods.
The restriction is possible because the complexified K\"ahler moduli space of the fiber embeds into the moduli space of the fibration and the periods of the fiber are recovered in the large base limit.

As we have already discussed before, the generic family of degree $5$ curves can be realized as an intersection of five Pfaffians in $\mathbb{P}^4$ or, equivalently, as the intersection of five hyperplanes in the Grassmannian $\text{Gr}(2,5)$.
The corresponding GLSM has been constructed in~\cite{Hori:2013gga} and was reviewed in Section~\ref{sec:glsmelliptic}.
By localizing the sphere partition function and extracting the fundamental period, the Picard-Fuchs operator in the Grassmannian phase was found to be
\begin{align}
	\mathcal{L}=\theta^2-z(11\theta^2+11\theta+2)-z^2(\theta+1)^2\,.
	\label{eqn:pfrod}
\end{align}
Note that this operator is just the $z_2\rightarrow 0$ limit of $\mathcal{D}_1$ in~\eqref{eqn:model1pf}.

From~\eqref{eqn:pfrod} one can easily derive the discriminant locus $\Delta=0$ at which the periods that are annihilated by $\mathcal{L}$ develop singularities and finds
\begin{align}
	\Delta=z\cdot (1-11z+z^2)\,.
\end{align}
The vanishing locus of this polynomial consists of three singular points $z=z_i,\,i=1,...,3$, namely the large volume limit $z_0=0$ and two other points $z_1,z_2$ with
\begin{align}
	z_1=-\frac12\left(11-5\sqrt{5}\right)\sim 0.09\,,\quad z_2=-\frac12\left(11+5\sqrt{5}\right)\sim-11.09\,.
\end{align}
The latter two coincide with the points~\eqref{eqn:fiberConi} in the FI-theta parameter space of the associated GLSM where there is a Coulomb branch.
An additional singularity exists at $z\rightarrow\infty$ and corresponds to the large volume limit of the Pfaffian curve.
Note that the component of the discriminant which is closer to a large volume limit is expected to be the point where the 2-brane on the respective curve becomes massless.

Let us choose a basis $\vec{\Pi}=(\omega_1,\omega_0)$ of periods at the large volume point with
\begin{align}
	\begin{split}
	\omega_0=&1+3z+19z^2+147z^3+1251z^4+\mathcal{O}(z^5)\,,\\
	(2\pi i)\omega^1=&\omega_0\cdot\log(z)+5z+\frac{75}{2}z^2+\frac{1855}{6}z^3+\frac{10875}{4}z^4+\mathcal{O}(z^5)\,.
	\end{split}
	\label{eqn:fiberbasis}
\end{align}
Numerical analytic continuation\footnote{An alternative approach is to transport D-branes from the Grassmannian phase to the Pfaffian phase and back using the GLSM. To do so, one requires the ``grade restriction rule'' \cite{Herbst:2008jq} which encodes the information on the path along which the branes are transported. The grade-restriction rule can be obtained from the asymptotic behaviour of the hemisphere partition function \cite{Hori:2013ika,never,beijing}, see also \cite{donovan2020stringy} for recent results in mathematics. Applying this to this particular model, one obtains equivalent results.} produces the corresponding monodromy matrices
\begin{align}
	M_0=\left(\begin{array}{rr}
		1&1\\0&1
	\end{array}\right)\,,\quad
	M_1=\left(\begin{array}{rr}
		1&0\\-5&1
	\end{array}\right)\,,\quad
	M_2=\left(\begin{array}{rr}
		-9&5\\-20&11
	\end{array}\right)\,.
\end{align}
Note that $M_0$ is the large volume monodromy while the monodromy $M_1$ around $z=z_1$ corresponds to a Seidel-Thomas twist with respect to the $2$-brane.
The matrices act via M\"obius transformations on the modular parameter $\tau=\omega^1/\omega_0$ and $M_0$ and $M_1^{-1}$ reproduce the action of the $T$ and $U$ monodromy~\eqref{eqn:UTtrans}. 
One can immediately see that all of the matrices are contained in $\Gamma_1(5)$.

We can now use SageMath~\cite{sagemath} to obtain a set of generators for $\Gamma_1(5)$
\begin{align}
	g_1=\left(\begin{array}{rr}
		1&1\\0&1
	\end{array}\right)\,,\quad
	g_2=\left(\begin{array}{rr}
		-5&1\\-6&1
	\end{array}\right)\,,\quad
	g_3=\left(\begin{array}{rr}
		7&-3\\12&-5
	\end{array}\right)\,,
\end{align}
and one easily checks the relations
\begin{align}
	g_1=M_0\,,\quad g_2=M_0\cdot M_1\,,\quad g_3=M_2\cdot M_0\cdot M_1\,.
\end{align}
The additional monodromy from the second conifold component therefore indeed provides the required third generator of $\Gamma_1(5)$. Note that the splitting of the conifold so far has only been observed for Calabi-Yaus that are not torically realized and for the associated non-Abelian GLSMs. The non-toric nature of the fiber is thus closely connected to the modular properties. 

Comparing the modular expressions~\eqref{eqn:zbd1model1scp},~\eqref{eqn:Z2model1ascp},~\eqref{eqn:Z2model1bscp} for the topological string partition function of $Y_1$ to the corresponding expressions~\eqref{eqn:zbd1model1},~\eqref{eqn:Z2model1a},~\eqref{eqn:Z2model1b} for $X_1$, we observed that one can obtain the expressions for either geometry from that of the dual geometry by replacing
\begin{align}
	E_{1,1}\rightarrow E_{1,2}\,,\quad E_{1,2}\rightarrow -E_{1,1}\,.
	\label{eqn:e11e12exchange}
\end{align}
To understand this phenomen, we transform the Picard-Fuchs operator~\eqref{eqn:pfrod} to the point at infinity by using $z\rightarrow -v^{-1}$ and obtain
\begin{align}
	\mathcal{D}=(\theta-1)^2-v(11\theta^2-11\theta+3)-v^2\theta^2\,.
\end{align}
We can now analytically continue the basis~\eqref{eqn:fiberbasis} from $z=0$ to the analogous basis at $v=0$ and obtain the transformation matrix
\begin{align}
	T=\left(\begin{array}{cc}
		2&-1\\
		5&-2
	\end{array}\right)\in\Gamma_0(5)\,.
	\label{eqn:fibertransfer}
\end{align}
This satisfies $T^2=-1$ and is in particular not an element of $\Gamma_1(5)$.
We also observe that
\begin{align}
	T^{-1}\cdot M_1\cdot T=M_2\,,
\end{align}
which reflects the fact that the second singularity behaves from the perspective of the Pfaffian curve like an ordinary conifold point.

The transfer matrix~\eqref{eqn:fibertransfer} explains the relation between the topological string partition functions as a result of the vector valued transformation behaviour of $\Gamma_1(5)$ modular forms under $\Gamma_0(5)$.
To see this, we can express the modular forms $E_{1,1}$ and $E_{1,2}$ in terms of the Eisenstein series with character $E_1^{\chi_1},E_1^{\chi_2}$, that are defined in Appendix~\ref{sec:basis-modular-forms}.
The latter are $\Gamma_1(5)$ modular forms but can also be interpreted as $\Gamma_0(5)$ modular forms with character.
The replacement~\eqref{eqn:e11e12exchange} acts on the Eisenstein series as
\begin{align}
	E_1^{\chi_1}\rightarrow i E_1^{\chi_1}\,,\quad E_1^{\chi_2}\rightarrow -iE_1^{\chi_2}\,,
\end{align}
and this is indeed the transformation behaviour under $\tau\rightarrow\gamma\cdot\tau$, with
\begin{align}
	\gamma=\left(\begin{array}{cc}a&b\\c&d\end{array}\right)\in\Gamma_0(5)\,,
\end{align}
and $d=3\,\text{mod}\,5$.
Let us point out again that the action of the transfer matrix on the modular parameter is dictated by properties of the generic fiber.
We thus find that relative homological projective duality between genus one fibrations with five sections always manifests as a $\Gamma_0(5)$ action on the modular parameter and in this way relates the topological string partition functions of dual fibrations.


%% file: section_4.tex
\section{M/F-theory, Higgs transitions and anomaly cancellation}
\label{sec:ftheory}

Let us briefly summarize the picture that has emerged until now.
A genus one fibered Calabi-Yau manifold with 5-sections can be realized as a fibration of Pfaffian curves or as a fibration of complete intersection curves in Grassmannians.
For every such fibration we have found a relative homologically projective dual Calabi-Yau, that is also genus one fibered over the base with $5$-sections.
Both fibrations can be realized as phases of the same GLSM and are then separated by two walls, corresponding to the respective limits of zero fiber volume, with a non-geometric phase in between.

In this section we will study the fibrations from the perspective of M- and F-theory.
After reviewing the basics, we will start by describing an extremal transition from the family of Grassmannian genus one curve to a family of elliptic curves.
The latter are realized as codimension five complete intersections in a toric ambient space and can be used to construct elliptically fibered Calabi-Yau manifolds.
The elliptic fibrations exhibit three independent sections, leading to a $U(1)\times U(1)$ gauge group in F-theory, and are connected, again via an extremal transition, to genus one fibrations with 5-sections.
We will first study M- and F-theory on the elliptic fibrations and then, using these results, on genus one fibrations with $5$-sections.

To determine the field content of the effective theories, we apply GV-spectroscopy~\cite{Paul-KonstantinOehlmann:2019jgr} to the elliptic fibrations.
This provides us with base independent expressions for the multiplicities of reducible fibers over isolated points in the base\footnote{For Calabi-Yau manifolds with complex dimension larger than three, the reducible fibers are part of a family and the base independent expressions give the class of the base of this family. The actual multiplicities then also depend on a choice of flux.}, which in turn correspond to hypermultiplets in the six-dimensional effective supergravity.
The resulting spectrum satisfies all of the six-dimensional anomaly constraints in a completely base independent manner.
In particular, it contains hypermultiplets with $U(1)\times U(1)$ charges $\pm(4,3)$ and $\pm(1,2)$.
This allows us to relate the extremal transition, from the elliptic fibration to the Grassmannian genus one fibration, with a Higgs transition that breaks the gauge group
\begin{align*}
	U(1)^2\rightarrow\mathbb{Z}_5\,.
\end{align*}

Using this transition we relate the base independent expressions for the multiplicities of reducible fibers and the Euler characteristic of the elliptic fibration into corresponding expressions for the Grassmannian genus one fibration.
It turns out that they can be written as relatively simple polynomials in terms of the first Chern class of the base and the Chern classes of the bundles that determine the fibration.
Since the base degree zero Gopakumar-Vafa invariants of the Pfaffian fibration are directly encoded in those of the Grassmannian fibration we thus obtain a complete unterstanding of the structure of all of the fibrations.
Again, the six-dimensional anomalies generically cancel.

The Higgs transition associated to the codimension five complete intersection fibers has been a two step process and the intermediate geometry does not have a manifest realization as a complete intersection in a toric ambient space.
To remedy this situation, we will also construct a family of elliptic curves that are codimension three complete intersections in a toric ambient space.
Using GV-spectroscopy we show that F-theory on the corresponding fibrations exhibits a $U(1)$ gauge group and charge $5$ matter.
To our knowledge, this is the first construction of elliptic fibrations that realize such vacua.
The corresponding Higgs transition $U(1)\rightarrow\mathbb{Z}_5$ implies that the fibrations again admit an extremal transition to genus one fibrations with $5$-sections\footnote{Matching the bundles is less straightforward, which is the reason why we derive the base independent matter multiplicities using the codimension five complete intersection fibers.}.

Finally, combining our knowledge about the fibrations, the spectra and the stringy moduli spaces, we are in a position to discuss the relation between dual pairs of Pfaffian and Grassmannian fibrations from the perspective of F-theory.
We argue that the corresponding effective physics are identical und thus provide strong evidence for the conjecture that they are elements of the same Tate-Shafarevich group.
This also relates the relative homological projective duality to a class of derived equivalences between genus one fibrations that have been proposed in~\cite{Caldararu:2002ab}.

\subsection{F-theory and GV-spectroscopy}
Since we aim to address a relatively diverse audience, we will first spend some time to introduce the relevant aspects of F-theory.
In particular, the details of how extremal transitions between elliptic and genus one fibrations manifest themselves in the F-theory compactifications as Higgs transitions with Wilson lines will be reviewed.
We then discuss the relation between the reducible fibers of the fibrations and certain Gopakumar-Vafa invariants.
Following~\cite{Paul-KonstantinOehlmann:2019jgr} this allows us to determine the structures and multiplicities of the reducible fibers in the elliptic fibrations and genus one fibrations with $5$-sections in a base independent manner.

\subsubsection{F-theory on genus one fibrations and Higgs transitions}
F-theory encodes the varying axio-dilaton profile of Type IIB string theory, compactified on a variety $B$, as the complex structure of the fibers in an elliptic or genus one fibration $M$ over $B$~\cite{Vafa:1996xn,Morrison:1996na,Morrison:1996pp}.
This leads to a rich dictionary between geometry and physics that has been uncovered over the last two and a half decades.
Here we will only review the aspects that are most relevant to our discussion.
For a thorough introduction we refer to the recent reviews~\cite{Weigand:2018rez,Cvetic:2018bni}.

If the non-compact directions of the Type IIB compactification form a Minkowski space, the fibration $M$ has to be Calabi-Yau.
We will be interested in the case where $M$ is a Calabi-Yau threefold and then the F-theory compactification leads to a six-dimensional $N=1$ effective supergravity~\cite{Bonetti:2011mw}.
In particular, the gauge group and matter spectrum of this theory are completely determined by the structure of the fibration.
In the following we will discuss the geometric manifestations of Abelian gauge symmetries, discrete gauge symmetries and the relation among them via Higgs transitions in six and, after circle compactification, in five dimensions.

Let us first recall the general field content of the six-dimensional effective supergravities with eight supercharges.
The fields arrange themselves into gravitational-, tensor-, vector- and hypermultiplets:
\begin{itemize}
	\item There is only one gravitational multiplet and it contains the graviton and the Weyl left-handed gravitino, as well as a self-dual tensor field of rank two.
	\item Each tensor multiplet contains an anti-self-dual tensor field of rank two, as well as one Weyl right-handed tensorino and one real scalar.
The number of tensor multiplets is given by $n_T=h^{1,1}(B)-1$.
	\item Each vector multiplet contains one vector field and one Weyl left-handed gaugino.
We will assume that the fibration $M$ over $B$ is flat, meaning that the complex dimension of every fiber is one, and then the number of vector multiplets is given by $n_V=h^{1,1}(M)-h^{1,1}(B)-2$.
	\item Finally, each hypermultiplet contains two complex scalars and one Weyl right-handed hyperino.
Gravitational anomaly cancellation fixes the number of hypermultiplets to be $n_H=273+n_V-29\cdot n_T$~\cite{Morrison:1996pp}.
As we will describe below, the hypermultiplets are geometrically encoded in the complex structure deformations as well as in isolated or enhanced singular fibers over points in the base.
The cancellation of gravitational anomalies therefore imposes a highly non-trivial constraint on the structure of elliptic~\cite{Grassi:2000we,Grassi:2011hq} and genus one fibered~\cite{Braun:2014oya} Calabi-Yau threefolds.
\end{itemize}
In addition to the purely gravitational anomaly cancellation condition, the spectrum of six-dimensional supergravities satisfies other constraints to ensure that the
gauge-gravitational and pure gauge anomalies can be cancelled via a generalized Green-Schwarz mechanism~\cite{Sadov:1996zm}.
We will list the anomalies that are relevant to our models as well as the corresponding constraints in Section~\ref{sec:anomCanc}.

From a IIB perspective the non-trivial axio-dilaton profile is sourced by stacks of D7-branes and O7-orientifold planes that wrap curves in $B$ and backreact on the compact space metric which is thus no longer Ricci flat.
The 7-brane stacks correspond to singular fibers in $M$ and, depending on the Kodaira type of the fibers as well as potential monodromies, exhibit non-Abelian gauge theories on their worldvolume.
A non-trivial gauge group $G$ on the worldvolume corresponds to a curve $C\subset B$ of singular fibers, that need to be resolved in order for the total space $M$ to be smooth~\cite{Katz:1996ht,Witten:1996qb,Morrison:1996pp,Bershadsky:1996nh}.
Physically, the resolution corresponds to a non-vanishing Wilson line that can be turned on after compactfiying on a circle to five-dimensions.
This generates a mass for some of the gauge bosons and charged hypermultiplets and thus breaks the gauge symmetry while preserving the rank of the gauge group.
The resolved fibers are then reducible with the components being $\mathbb{P}^1s$ that intersect like the affine Dynkin diagram of $G$.
In particular, the fibration of each $\mathbb{P}^1$ over $C$ leads to a so-called \textit{fibral divisor} in $M$.
We will only be interested in Calabi-Yau manifolds without fibral divisors, where the gauge groups of the corresponding effective theories are Abelian.

The origin of $U(1)$ gauge symmetries is somewhat obscured from the 7-brane perspective and can be best understood using the duality with M-theory.
To this end one compactifies the F-theory vacuum again on an additional circle, with the total compact space then being $M\times S^1$, and the resulting theory is dual to M-theory on the Calabi-Yau $M$.
The M-theory 3-form $C_3$ can then be expanded along harmonic 2-forms in $M$ as
\begin{align}
	C_3=\dots + A^0\wedge [S^{(n)}_0]+\sum\limits_{i=1}^{n_V} A^i\wedge [\phi(S^{(n)}_i)]+\sum\limits_{i=1}^{h^{1,1}(B)}\tilde{A}^i\wedge [\pi^{-1}(D^b_i)]\,.
	\label{eqn:c3expansion}
\end{align}
Here $D^b_i,\,i=1,\dots,h^{1,1}(B)$ is a basis of divisors on $B$, $n\in\mathbb{N}$ is the minimal value for which $M$ exhibits \textit{$n$-sections}, i.e. divisors that intersect the generic fiber $n$ times, and $[D]$ is the harmonic 2-form that is associated to a divisor $D$.
Moreover, $S^{(n)}_0$ is some $n$-section that can be choosen freely and $S^{(n)}_i$ are additional $n$-sections that together with $S^{(n)}_0$ and the \textit{vertical divisors} $\pi^{-1}(D^b_i)$ form a basis of $H^2(M)$.
To diagonalize the gauge coupling matrix and to guarantee that the corresponding gauge fields lift to six dimensions, the $S^{(n)}_0$ have to be orthogonalized via the \textit{generalized Shioda map}
\begin{align}
	\phi(S^{(n)})=S^{(n)}-S^{(n)}_0+D^b\,,
	\label{eqn:genshioda}
\end{align}
where $\pi^{-1}(D^b)$ is a vertical divisor that is fixed by demanding that
\begin{align}
	\phi(S^{(n)})\cdot S_0^{(n)}\cdot \pi^{-1}(\tilde{D}^b)=0\,,
\end{align}
for every divisor $\tilde{D}^b$ on $B$~\cite{Park:2011ji,Klevers:2014bqa,Grimm:2015wda}.
If $M$ is an elliptic fibration, and therefore $n=1$, the map $\phi$ coincides with the Shioda map that provides an isomorphism between the Mordell-Weil group and a subset of the N\'{e}ron-Severi lattice of $M$~\cite{Park:2011ji,Morrison:2012ei}.
Note that $\phi(S^{(n)}_0)=0$ and we therefore refer to $S^{(n)}_0$ as the \textit{zero-$n$-section}.
The five-dimensional gauge field $A^0$ in~\eqref{eqn:c3expansion} is the Kaluza-Klein vector that descends from the six-dimensional metric with one leg along the circle.
Similarily, the $\tilde{A}^i$ arise from Kaluza-Klein reduction of the tensor fields.
Only the $n_V$ 1-forms $A^i$ lift to six-dimensional Abelian gauge fields.
Therefore $U(1)$ gauge symmetries of the F-theory effective action correspond to additional sections or multi-sections in the elliptic or genus one fibration $M$.

F-theory on genus one fibrations with $n$-sections also exhibits a discrete gauge symmetry~\cite{morrisonLecture,deBoer:2001wca,Morrison:2014era,Anderson:2014yva,Klevers:2014bqa,Mayrhofer:2014laa} and the discrete gauge group is, at least in the absence of multiple fibers, conjectured to be identical to the Tate-Shafaverich (TS) group of the genus one fibration~\cite{morrisonLecture,deBoer:2001wca,Morrison:2014era}.
If the fibration is flat and the total space is smooth, the latter is known to be an extension of a certain \'Etale cohomology group by $\mathbb{Z}_n$, where we have again used the lowest value $n\in\mathbb{N}$ for which $M$ exhibits an $n$-section~\cite{dolgachev1992elliptic,Morrison:2014era,Braun:2014oya}.
We will discuss the Tate-Shafarevich group in some more detail in Section~\ref{sec:tsgroup}.
At this point we only note that, assuming $\text{Br}(M)=0$ or equivalently $\text{Tors}(H^3(M,\mathbb{Z}))=0$, the TS-group is for the genus one fibrations with $5$-sections that we consider in this paper exactly given by $\mathbb{Z}_5$.

Discrete symmetries are conjectured to be absent in consistent theories of quantum gravity~\cite{Banks:2010zn,Hellerman:2010fv} and therefore a $\mathbb{Z}_n$ gauge symmetry must arise as a remnant of a broken $U(1)$ gauge symmetry.
The breaking is achieved by turning on a non-vanishing vacuum expectation value for the scalar fields in massless hypermultiplets of charge $n$~\footnote{This is assuming that the hypermultiplet is not charged under any other factor of the gauge group and otherwise potential mixing has to be taken into account.}.
Charged hypermultiplets in F-theory compactifications on elliptic and genus one fibrations without fibral divisors are encoded in $I_2$ singular fibers over isolated points of the base where the discriminant locus itself is singular.
From an M-theory perspective the corresponding particles arise from M2-branes that wrap the $\mathbb{P}^1$s that resolve the singular fibers and the corresponding mass is proportional to the volume of these exceptional curves.
Geometrically, the Higgs transition $U(1)\rightarrow\mathbb{Z}_n$ therefore manifests itself as an extremal transition where tuning the hypermultipet mass to zero amounts to a deformation of the K\"ahler structure, such that the volume of the corresponding exceptional curve is zero, while the vacuum expectation values of the scalar fields correspond to subsequent complex structure deformations~\cite{Morrison:2014era,Mayrhofer:2014laa,Cvetic:2015moa}.
Moreover, the charge of the particles under a $U(1)$ gauge symmetry is given by the intersection of the exceptional curve with the image under the generalized Shioda map~\eqref{eqn:genshioda} of the $n$-section that is associated to the $U(1)$ in~\eqref{eqn:c3expansion}.
The $\mathbb{Z}_n$ charge corresponds to the intersection of the curve with the zero-$n$-section.
In five dimensions, this intersection can be identified with the Kaluza-Klein charge $q_{KK}$.

As we mentioned in the beginning, the genus one fibration $M$ encodes the varying axio-dilaton profile of a Type IIB compactification on the base $B$.
All the information about the six-dimensional physics is determined by this profile.
To every genus one fibration $M$, that does not exhibit a section but only $n$-sections, there exists an associated Jacobian fibration that is elliptic and encodes the same axio-dilaton profile.
The corresponding F-theory vacua of the genus one fibration and the associated Jacobian fibration are therefore identical.
This changes after we compactify to five dimensions where we have the additional degree of freedom of turning on a Wilson line along the circle.

Let us denote the radius of the circle by $R$ and the value of the $U(1)$ gauge field along the circle by $\xi$.
The mass of a five-dimensional particle of $U(1)_{KK}\times U(1)_{6D}$ charge $(q_{KK},q_{6D})$ is then given by
\begin{align}
	m=R\cdot |q_{KK}+\xi\cdot q_{6D}|\,.
\end{align}
We therefore have to choose $\xi=-q_{KK}/q_{6D}$ in order for the particle to become massless.
However, since every particle comes with an entire tower of Kaluza-Klein excitations, we can in fact choose any value $\xi=k/q_{6D}$ with $k\in\mathbb{Z}$ to ensure that some state of the tower is massless.
A shift $\xi\rightarrow \xi+1$ then reshuffles the states in the tower and we only need to consider equivalence classes $[\xi]\in\mathbb{Z}_{q_{6D}}$.
In this way one finds $q_{6D}$ inequivalent vacua and these are conjecturally dual to M-theory compactified on the $q_{6D}$ elements of the Tate-Shafarevich group~\cite{deBoer:2001wca,Cvetic:2015moa,Paul-KonstantinOehlmann:2019jgr}.

\subsubsection{Fibration structures from GV-spectroscopy}
\label{eqn:gvspecReview}
To determine the non-Abelian part of the gauge group and the matter spectrum of an F-theory compactification on a smooth genus one fibration, it is necessary to understand the structure of the reducible fibers.
The locus in the base over which the fiber becomes singular or reducible is determined by the vanishing of the discriminant $\Delta$.
For an elliptic curve in Weierstrass form
\begin{align}
	y^2=x^3+fxz^4+gz^6\,,
\end{align}
with $[x:y:z]$ being homogeneous coordinates in $\mathbb{P}_{231}$, this is given by
\begin{align}
	\Delta=4f^3+27g^2\,.
\end{align}
Every genus one fibration has an associated elliptic fibration, namely the Jacobian fibration, which can be brought into Weierstrass form and has the same discriminant $\Delta$.
The corresponding coefficients $f,g$ are then sections of line bundles on the base $B$.
For genus one fibrations that are realized as hypersurfaces in toric ambient spaces it is usually possible to explicitly calculate the associated fibration of Weierstrass curves~\cite{Klevers:2014bqa}.
At least for simple cases one can then explicitly determine the loci of the reducible fibers and, if the fibral divisors and $n$-sections are realized as toric divisors, calculate the intersection with the irreducible components and thus obtain the charges of the associated representations.

However, a particular problem is posed by the isolated reducible fibers, which occur over points where the discriminant locus $\{\Delta=0\}\subset B$ itself is singular.
They lead to hypermultiplets that are only charged under the Abelian part of the gauge group.
Even for hypersurfaces it is highly non-trivial to determine the number of such fibers and the intersections of the irreducible components with the (multi-)sections~\cite{Klevers:2014bqa}.
For complete intersections in toric ambient spaces the situation is even more involved and the difficulty increases with the codimension~\cite{Cvetic:2013qsa,Braun:2014qka,Oehlmann:2016wsb}.
In the case of genus one fibrations with $5$-sections the discriminant is known to be a polynomial of degree $50$ in the base coordinates and the direct approach seems essentially hopeless~\cite{fisher1}.
To circumvent these difficulties, an alternative technique to determine the structure of the fibrations has been developed in~\cite{Paul-KonstantinOehlmann:2019jgr}, which uses the information encoded in certain Gopakumar-Vafa invariants.

The definition of Gopakumar-Vafa invariants in terms of M-theory on a Calabi-Yau threefold $M$, as well as the relation to Gromov-Witten invariants, has already been summarized in Section~\ref{sec:gvandmod}.
However, to understand the geometric information that is contained in the Gopakumar-Vafa invariants, it is useful to compactify M-theory on an additional circle.
The resulting theory is dual to Type IIA string theory on $M$ and from this perspective the invariants encode bound states of D2 and D0 branes.
This leads to the relation~\cite{Gopakumar:1998jq}
\begin{align}
	n^0_\beta=(-1)^d\chi(\widehat{\mathcal{M}}_\beta)\,,\quad d=\text{dim}\left(\widehat{\mathcal{M}}_\beta\right)\,,
	\label{eqn:gvD2D0}
\end{align}
where $\widehat{\mathcal{M}}_\beta$ is the moduli space of curves in the class $\beta$ together with a choice of a flat $U(1)$ connection.

Let us assume that $\beta$ is the class of a union of fiber components and denote the curves in such a class as \textit{fibral curves}.
This is the case if $\beta\cdot \pi^{-1}(D)=0$ for every divisor $D\in H_2(B)$.
We denote the restriction of the Gopakumar-Vafa invariants to such classes by $\tilde{n}^0_\beta$ and refer to them as \textit{fiber Gopakumar-Vafa invariants}.
The reducible fibers are unions of rational curves and therefore the flat $U(1)$ connections are trivial.
We can thus replace $\widehat{\mathcal{M}}_\beta=\mathcal{M}_\beta$ in~\eqref{eqn:gvD2D0}, where $\mathcal{M}_\beta$ is the moduli space of curves in the class $\beta$.
The information in the fiber GV-invariants can then be reduced to three different cases~\cite{Paul-KonstantinOehlmann:2019jgr}:
\begin{enumerate}[label=\arabic*)]
	\item If $\beta$ is the class of the component of an isolated reducible fiber, or one that only exists over an isolated point of a fibral divisor, the invariant $\tilde{n}^0_\beta$ is just the number of curves in this class.
	\item If $\beta$ is the class of a component of the generic fiber in a fibral divisor, the moduli space is the base of the fibral divisor and $-\tilde{n}^0_\beta$ is the Euler characteristic of that base.
	\item If $\beta$ is the class of the generic fiber, the invariants $\tilde{n}_\beta$ actually correspond to the $I_1$ singular fibers of the fibration.
		It turns out that in this case $-\tilde{n}_\beta$ is the Euler characteristic of the Calabi-Yau threefold.
		This was discussed in~\cite{Candelas:1994hw} for the generic elliptic fibration over $\mathbb{P}^2$ but is not proven in general.
		However, it has been checked for a large number of geometries~\cite{Cota:2019cjx,Paul-KonstantinOehlmann:2019jgr}.
\end{enumerate}
We refer to the corresponding classes as \textit{elementary classes} of type one, two and three\footnote{Non-flat fibrations can have additional types of curves, that appear in the higher dimensional fibers and lead to so-called superconformal matter. We will not discuss such geometries in this paper and refer to~\cite{Paul-KonstantinOehlmann:2019jgr} for an example of how to treat these cases.}.
All other invariants $\tilde{n}^0_\beta$ are then fixed by the periodicity
\begin{align}
	\tilde{n}^0_\beta=\tilde{n}^0_{\beta+F},
	\label{eqn:periodicity}
\end{align}
where $F$ is the class of the generic fiber.
From the perspective of F-theory, the periodicity accounts for the Kaluza-Klein tower that is associated to every state.
We are not aware of any mathematical proof of the periodicity property but the three elementary cases are sufficient to determine the spectrum completely.
Another useful property is the reflection identity
\begin{align}
	\tilde{n}^0_{\beta}=\tilde{n}^0_{F-\beta}\,,
\end{align}
which holds when $\beta$ is an elementary class of type one or two.

Let us illustrate this at the hand of an example.
In Section~\ref{sec:gvandmod} we used a localization calculation in the associated GLSM to obtain the genus zero and genus one Gopakumar-Vafa invariants for generic genus one fibrations with $5$-sections.
For those fibrations we can denote the class of a fibral curve $C$ by its intersection with the $5$-section $S_0^{(5)}$, i.e. $\beta=C\cdot S_0^{(5)}$.
Moreover, we found that the fiber GV-invariants take the form
\begin{align}
	\tilde{n}^0_1=n_{\pm1}\,,\quad \tilde{n}^0_2=n_{\pm2}\,,\quad \tilde{n}^0_3=n_{\pm2}\,,\quad\tilde{n}^0_4=n_{\pm1}\,,\quad \tilde{n}^0_5=-\chi\,,
\end{align}
where $n_{\pm 1},n_{\pm 2}\in\mathbb{N}$ and $\chi$ is the Euler characteristic of the fibration.
Since there are no fibral divisors we can conclude that $n_{\pm 1}$ and $n_{\pm 2}$ are multiplicities of $I_2$ fibers, which consist of two rational curves that intersect in two points.
We also know the intersection of those curves with the $5$-section.
There are $n_{\pm1}$ fibers with one component intersecting the $5$-section once while the other component intersects it four times.
The components of the other $n_{\pm2}$ reducible fibers intersect the $5$-section respectively two and three times.

In the previous section we explained that the F-theory compactification on the genus one fibration exhibits a $\mathbb{Z}_5$ gauge symmetry.
Moreover, the isolated reducible fibers lead to hypermultiplets and the $\mathbb{Z}_5$ charge is determined by the intersection with the $5$-section.
Note that a hypermultiplet consists of two half-hyper multiplets that transform in conjugate representations.
The F-theory spectrum therefore contains $n_{\pm1}$ hypermultiplets that contain half-hypermultiplets of $\mathbb{Z}_5$ charge $1$ and $-1$ as well as $n_{\pm 2}$ hypermultiplets that contain half-hypermultiplets of charge $2$ and $-2$.
In this way the fiber GV-invariants allow us to deduce the structure of the fibration and thus also the corresponding F-theory spectrun.

Note that for complete intersections in toric varieties the genus zero Gopakumar-Vafa invariants can be calculated using mirror symmetry~\cite{Batyrev:1994pg,Hosono:1994ax,Cox:2000vi}.
It turns out that the fiber GV-invariants can then be determined in a base independent manner, which produces quadratic polynomials in terms of Chern classes of line bundles on a generic base, that parametrize the structure of the fibration (see below)~\cite{Paul-KonstantinOehlmann:2019jgr}.
In the following we will apply the technique from~\cite{Paul-KonstantinOehlmann:2019jgr} to calculate base independent expressions for the relevant Gopakumar-Vafa invariants in elliptic fibrations, that are connected via an extremal transition to the genus one fibrations.
We can then use the relation of the spectrum under Higgs transitions to obtain expressions for $n_{\pm1}$ and $n_{\pm2}$ in terms of the Chern classes of the bundles $E'$ and $F^*$, that were introduced in Section~\ref{sec:geometry}, or, equivalently, in terms of the charges of an associated GLSM.

Recently, the connection between the Gopakumar-Vafa invariants and the F-theory spectrum has also been used in~\cite{Grassi:2021wii} to study fibrations with large Mordell-Weil rank.

\subsection{Toric codimension 5 complete intersection fibers with $G=U(1)^2$}
\label{sec:codim5fibers}
In this section we are constructing elliptic curves that are codimension five complete intersections in a toric ambient space and can be used to engineer elliptically fibered Calabi-Yau threefolds.
The corresponding F-theory vacua will exhibit a $U(1)^2$ gauge symmetry and, using the techniques recently developed in~\cite{Paul-KonstantinOehlmann:2019jgr}, we determine base independent expressions for the spectrum of charged hypermultiplets.
We then check that all six-dimensional anomalies cancel and that the spectrum allows us to perform a Higgs transition $U(1)^2\rightarrow \mathbb{Z}_5$.
The base independent expressions for the spectrum will then be used in Section~\ref{sec:baseIndependentSpecZ5} to obtain corresponding expressions for generic genus one fibrations with 5-sections.
\subsubsection{Determining the base independent spectrum}
In~\cite{sturmfels1996grobner} Sturmfels has described an extremal transition from the Grassmannian $G(2,5)$ to a six-dimensional toric variety $P(2,5)$.
An extension of this construction to more general Grassmannians $G(k,n)$ and, in particular, to corresponding complete intersection Calabi-Yau threefolds was applied in~\cite{Batyrev:1998kx} to obtain mirror partners for complete intersections in Grassmannians.
Here we will use the construction by Sturmfels to obtain an extremal transition from the family of genus one curves, that are codimension five complete intersection in $G(2,5)$, to a family of elliptic curves, that are again codimension five complete intersections but in a toric ambient space.
The latter can then be fibered over two-dimensional bases to obtain elliptically fibered Calabi-Yau threefolds with three sections, that in turn exhibit an extremal transition to genus one fibrations with 5-sections.

Let us start with the six-dimensional toric variety $P(2,5)$ constructed in~\cite{sturmfels1996grobner} that is determined by the data
\begin{align}
	\begin{blockarray}{rrrrrrrrrrl}
		\begin{block}{r(rrrrrr|rrr)l}
			v_1& 1& 0& 0& 0& 0& 0& 1& 0& 0&\\
			v_2&-1& 0& 0& 1& 0& 0& 1& 0&-1&\\
			v_3& 0&-1& 0& 0& 1& 0& 0&-1& 1&\\
			v_4& 0& 0&-1& 0& 0& 1& 0& 1& 0&\\
			v_5& 0&-1& 1& 0& 0& 0& 0& 1& 0&\\
			v_6& 0& 0& 0& 0& 0&-1& 1& 0& 0&\\
			v_7& 0& 0& 0&-1& 1& 0& 1& 0&-1&\\
			v_8& 0& 0& 0& 0&-1& 1& 1&-1& 0&\\
			v_9&-1& 1& 0& 0& 0& 0& 0& 0& 1&\\
			& 0& 0& 0& 0& 0& 0&-5&0& 0&\\
		\end{block}
	\end{blockarray}\,.
	\label{eqn:toricdata}
\end{align}
The last three columns contain the linear relations that correspond to the generators of the Mori cone and $v_i,\,i=1,\dots,9$ are the homogeneous coordinates.
We use $[v_i]$ to denote the class that is associated to the divisor $\{v_i=0\}\subset P(2,5)$ and determine the relations
\begin{align}
\begin{split}
[v_1]=&[v_9]+[v_7]\,,\quad [v_2]=[v_7]\,,\quad [v_3]=[v_8]-[v_7]\,,\\
[v_4]=&[v_9]+[v_7]-[v_8]\,,\quad [v_5]=[v_9]+[v_7]-[v_8]\,,\quad [v_6]=[v_9]+[v_7]\,.
\end{split}
\end{align}
It is easy to check using PALP~\cite{Kreuzer:2002uu,Braun:2012vh} that
\begin{align}
	-K_{P(2,5)}=D_{\nabla_1}+D_{\nabla_2}+D_{\nabla_3}+D_{\nabla_4}+D_{\nabla_5}\,,
	\label{eqn:u1u1nef}
\end{align}
with
\begin{align}
\begin{split}
D_{\nabla_1}=&[v_1]\,,\quad D_{\nabla_2}=[v_2]+[v_3]+[v_4]\,,\quad D_{\nabla_3}=[v_9]+[v_7]\,,\\
D_{\nabla_4}=&[v_5]+[v_8]\,,\quad D_{\nabla_5}=[v_6]\,,
\end{split}
\end{align}
is a nef-partition and therefore determines a smooth genus one curve $C$ in codimension five.
Using the linear equivalence relations among the divisors we observe that in fact
\begin{align}
D_{\nabla_1}=\dots=D_{\nabla_5}\,,
\end{align}
and the corresponding equations read
\begin{align}
\begin{split}
p_i=&c_{i,1}\cdot v_1+c_{i,2}\cdot v_6+c_{i,3}\cdot v_2v_9+c_{i,4}\cdot v_4v_8+c_{i,5}\cdot v_9v_7+c_{i,6}\cdot v_5v_8\\
&c_{i,7}\cdot v_2v_3v_4 + c_{i,8}\cdot v_2v_3v_5 + c_{i,9}\cdot v_3v_4v_7 + c_{i,10}\cdot v_3v_5v_7 \,,
\end{split}
\label{eqn:codim5equations}
\end{align}
where $c_{i,j}$ are $50$ complex coefficients.
Moreover, using SageMath~\cite{sagemath} we calculate that the intersections of the divisors $[v_i]$ with this curve are
\begin{align}
C\cdot [v_i]=(5,\,2,\,1,\,2,\,2,\,5,\,2,\,3,\,3)\,.
\label{eqn:codim5inter}
\end{align}

One can now choose a two-dimensional base $B$ and promote the coefficients $c_{i,j}$ to sections of line bundles on $B$ such that the resulting fibration is a Calabi-Yau threefold.
The classes of the $50$ bundles on $B$ are then equivalently encoded in the modified linear relations among $[v_i],\,i=1,\dots,9$ together with modified nef divisors.
Without loss of generality we can choose the relations to be
\begin{align}
\begin{split}
[v_1]=&[v_9]+[v_7]+d_1\,,\quad [v_2]=[v_7]+d_2\,,\quad [v_3]=[v_8]-[v_7]+d_3\,,\\
[v_4]=&[v_9]+[v_7]-[v_8]+d_4\,,\quad [v_5]=[v_9]+[v_7]-[v_8]+d_5\,,\\
[v_6]=&[v_9]+[v_7]+d_6\,,
\label{eqn:shiftedx}
\end{split}
\end{align}
with $d_i,\,i=1,\dots,6$ being classes of vertical divisors $d_i=\pi^{*}(\tilde{d}_i)$ that are associated to Cartier divisors $\tilde{d}_i$ on $B$.
Moreover, we assume that the divisors
\begin{align}
D_{\nabla_i}=[v_9]+[v_7]+s_{i}\,,
\label{eqn:shiftedd}
\end{align}
where $s_i,\,i=1,\dots 5$ are again associated to vertical divisors, form a nef partition of the anti-canonical class of the corresponding fibration of $P(2,5)$ over $B$.
In order for the fibration to be Calabi-Yau one needs
\begin{align}
s_5=c_1+\sum_{i=1}^6d_i-\sum_{i=1}^{4}s_i\,,
\label{eqn:cycondition}
\end{align}
where $c_1\equiv \pi^*c_1(B)$ is the first Chern class of the base.

We can now apply the GV-spectroscopy developed in~\cite{Paul-KonstantinOehlmann:2019jgr} to obtain the cohomology classes of the degeneration loci of the fiber in codimension one and two of the base as polynomials in $c_1,d_{i=1,\dots,6},s_{i=1,\dots,4}$.
To this end we parametrize the K\"ahler class on the fiber as $\omega=\sum_{i=1}^3t^iJ_i$, with a basis of the K\"ahler cone on the toric ambient space given by
\begin{align}
	J_1=[v_1]\,,\quad J_2=[v_4]\,,\quad J_3=[v_9]\,.
\end{align}
Using the techniques from~\cite{Paul-KonstantinOehlmann:2019jgr} we can then calculate $s_i,d_i$ dependent expressions for the fiber GV-invariants.
The calculation is utilizing mirror symmetry~\cite{Hosono:1994ax,Batyrev:1994pg} but, due to the base independent nature, somewhat non-trivial to conceptualize.
For a detailed discussion we therefore refer to~\cite{Paul-KonstantinOehlmann:2019jgr} and here only state the result.
The class of a fibral curve $\beta$ is determined by the intersections with $J_i,\,i=1,\dots,3$ and we therefore use $\tilde{n}^0_{k_1,k_2,k_3}$ to denote the invariant $\tilde{n}^0_\beta$ with
\begin{align}
	\beta\cdot J_i=k_i\,,\quad i=1,\dots 3\,.
\end{align}
Elementary invariants are listed in the Tables~\ref{tab:codim5gv_1} and~\ref{tab:codim5gv_2}.
All fiber GV-invariants that are not listed can be obtained using the periodicity and symmetry relations
\begin{align}
\tilde{n}^0_{k_1,k_2,k_3}=\tilde{n}^0_{k_1+5,k_2+2,k_3+3}\,,\quad \tilde{n}^0_{k_1,k_2,k_3}=\tilde{n}^0_{5-k_1,2-k_2,3-k_3}\,.
\end{align}
From these invariants we can now determine the structure of the fibration as well as the corresponding F-theory spectrum and check that all of the anomalies are generically cancelled.
\begin{table}
	\centering
\begin{tabular}{c|l}
	$(k_1,k_2,k_3)$&$\tilde{n}^0_{k_1,k_2,k_3}$\\\hline
	\multirow{7}{*}{$(1,0,0)$}&$3 c_1^2 - 2 c_1 d_1 - c_1 d_2 - 2 d_1 d_2 - 2 d_2^2 - 2 c_1 d_3 - 2 d_1 d_3 - 2 d_2 d_3 - 2 c_1 d_4$\\
&$- 2 d_1 d_4 - d_2 d_4 + d_4^2 - 2 c_1 d_5 - 2 d_1 d_5 - d_2 d_5 - 4 d_4 d_5 + d_5^2 - 2 c_1 d_6 - 2 d_1 d_6$\\
&$- 2 d_2 d_6 - 2 d_3 d_6 - 2 d_4 d_6 - 2 d_5 d_6 + 2 c_1 s_1 + 2 d_1 s_1 + 2 d_2 s_1 + 2 d_3 s_1 + 2 d_4 s_1$\\
&$+ 2 d_5 s_1 + 2 d_6 s_1 - 2 s_1^2 + 2 c_1 s_2 + 2 d_1 s_2 + 2 d_2 s_2 + 2 d_3 s_2 + 2 d_4 s_2 + 2 d_5 s_2$\\
&$+ 2 d_6 s_2 - 2 s_1 s_2 - 2 s_2^2 + 2 c_1 s_3 + 2 d_1 s_3 + 2 d_2 s_3 + 2 d_3 s_3 + 2 d_4 s_3 + 2 d_5 s_3$\\
&$+ 2 d_6 s_3 - 2 s_1 s_3 - 2 s_2 s_3 - 2 s_3^2 + 2 c_1 s_4 + 2 d_1 s_4 + 2 d_2 s_4 + 2 d_3 s_4 + 2 d_4 s_4$\\
&$+ 2 d_5 s_4 + 2 d_6 s_4 - 2 s_1 s_4 - 2 s_2 s_4 - 2 s_3 s_4 - 2 s_4^2$\\\hline
\multirow{5}{*}{$(0,1,0)$}&$c_1 d_1 - c_1 d_2 - d_1 d_2 - d_1 d_3 - d_2 d_3 - d_1 d_4 - d_2 d_4 - d_1 d_5 - d_2 d_5 - c_1 d_6 - d_1 d_6$\\
&$- d_2 d_6 - d_3 d_6 - d_4 d_6 - d_5 d_6 + c_1 s_1 + d_1 s_1 + d_2 s_1 + d_3 s_1 + d_4 s_1 + d_5 s_1 + d_6 s_1$\\
&$- s_1^2 + c_1 s_2 + d_1 s_2 + d_2 s_2 + d_3 s_2 + d_4 s_2 + d_5 s_2 + d_6 s_2 - s_1 s_2 - s_2^2 + c_1 s_3$\\
&$+ d_1 s_3 + d_2 s_3 + d_3 s_3 + d_4 s_3 + d_5 s_3 + d_6 s_3 - s_1 s_3 - s_2 s_3 - s_3^2 + c_1 s_4 + d_1 s_4$\\
&$+ d_2 s_4 + d_3 s_4 + d_4 s_4 + d_5 s_4 + d_6 s_4 - s_1 s_4 - s_2 s_4 - s_3 s_4 - s_4^2$\\\hline
\multirow{6}{*}{$(0,0,1)$}&$c_1 d_1 - d_1 d_2 - d_1 d_3 - c_1 d_4 - d_1 d_4 - d_2 d_4 - d_3 d_4 - c_1 d_5 - d_1 d_5 - d_2 d_5 - d_3 d_5$\\
&$- d_4 d_5 - c_1 d_6 - d_1 d_6 - d_2 d_6 - d_3 d_6 - d_4 d_6 - d_5 d_6 + c_1 s_1 + d_1 s_1 + d_2 s_1 + d_3 s_1$\\
&$+ d_4 s_1 + d_5 s_1 + d_6 s_1 - s_1^2 + c_1 s_2 + d_1 s_2 + d_2 s_2 + d_3 s_2 + d_4 s_2 + d_5 s_2 + d_6 s_2$\\
&$- s_1 s_2 - s_2^2 + c_1 s_3 + d_1 s_3 + d_2 s_3 + d_3 s_3 + d_4 s_3 + d_5 s_3 + d_6 s_3 - s_1 s_3 - s_2 s_3$\\
&$- s_3^2 + c_1 s_4 + d_1 s_4 + d_2 s_4 + d_3 s_4 + d_4 s_4 + d_5 s_4 + d_6 s_4$\\
&$- s_1 s_4 - s_2 s_4 - s_3 s_4 - s_4^2$\\\hline
\multirow{7}{*}{$(1,0,1)$}&$3 c_1^2 - 2 c_1 d_1 - c_1 d_2 - 2 d_1 d_2 - 2 d_1 d_3 - 2 d_2 d_3 - 2 d_3^2 - c_1 d_4 - 2 d_1 d_4 - d_2 d_4$\\
&$- 2 d_3 d_4 - c_1 d_5 - 2 d_1 d_5 - d_2 d_5 - 2 d_3 d_5 - 4 d_4 d_5 - 2 c_1 d_6 - 2 d_1 d_6 - 2 d_2 d_6$\\
&$- 2 d_3 d_6 - 2 d_4 d_6 - 2 d_5 d_6 + 2 c_1 s_1 + 2 d_1 s_1 + 2 d_2 s_1 + 2 d_3 s_1 + 2 d_4 s_1 + 2 d_5 s_1$\\
&$+ 2 d_6 s_1 - 2 s_1^2 + 2 c_1 s_2 + 2 d_1 s_2 + 2 d_2 s_2 + 2 d_3 s_2 + 2 d_4 s_2 + 2 d_5 s_2 + 2 d_6 s_2$\\
&$- 2 s_1 s_2 - 2 s_2^2 + 2 c_1 s_3 + 2 d_1 s_3 + 2 d_2 s_3 + 2 d_3 s_3 + 2 d_4 s_3 + 2 d_5 s_3 + 2 d_6 s_3$\\
&$- 2 s_1 s_3 - 2 s_2 s_3 - 2 s_3^2 + 2 c_1 s_4 + 2 d_1 s_4 + 2 d_2 s_4 + 2 d_3 s_4 + 2 d_4 s_4$\\
&$+ 2 d_5 s_4 + 2 d_6 s_4 - 2 s_1 s_4 - 2 s_2 s_4 - 2 s_3 s_4 - 2 s_4^2$\\\hline
\multirow{6}{*}{$(1,1,0)$}&$c_1^2 + c_1 d_1 + c_1 d_2 + d_1 d_2 + 2 c_1 d_3 + d_1 d_3 + d_2 d_3 + d_3^2 + c_1 d_4 + d_1 d_4 + d_2 d_4$\\
&$+ d_3 d_4 + c_1 d_5 + d_1 d_5 + d_2 d_5 + d_3 d_5 + d_4 d_5 + c_1 d_6 + d_1 d_6 + d_2 d_6 + d_3 d_6 + d_4 d_6$\\
&$+ d_5 d_6 - c_1 s_1 - d_1 s_1 - d_2 s_1 - d_3 s_1 - d_4 s_1 - d_5 s_1 - d_6 s_1 + s_1^2 - c_1 s_2 - d_1 s_2$\\
&$- d_2 s_2 - d_3 s_2 - d_4 s_2 - d_5 s_2 - d_6 s_2 + s_1 s_2 + s_2^2 - c_1 s_3 - d_1 s_3 - d_2 s_3 - d_3 s_3$\\
&$- d_4 s_3 - d_5 s_3 - d_6 s_3 + s_1 s_3 + s_2 s_3 + s_3^2 - c_1 s_4 - d_1 s_4 - d_2 s_4 - d_3 s_4$\\
&$- d_4 s_4 - d_5 s_4 - d_6 s_4 + s_1 s_4 + s_2 s_4 + s_3 s_4 + s_4^2$\\\hline
\end{tabular}
	\caption{Part one of the non-vanishing fiber Gopakumar-Vafa invariants for an elliptic fibration with generic fiber given by a codimension five complete intersection in $\mathbb{P}_\Delta$ with toric data~\eqref{eqn:toricdata}.}
	\label{tab:codim5gv_1}
\end{table}
\begin{table}
	\centering
\begin{tabular}{c|l}
	$(k_1,k_2,k_3)$&$n^0_{k_1,k_2,k_3}$\\\hline
\multirow{7}{*}{$(1,1,1)$}&$3 c_1^2 - 2 c_1 d_1 - 2 c_1 d_2 - 2 d_1 d_2 + d_2^2 - 2 c_1 d_3 - 2 d_1 d_3 - c_1 d_4 - 2 d_1 d_4 - d_2 d_4$\\
&$- 2 d_3 d_4 - 2 d_4^2 - c_1 d_5 - 2 d_1 d_5 - d_2 d_5 - 2 d_3 d_5 - 2 d_5^2 - 2 c_1 d_6 - 2 d_1 d_6 - 2 d_2 d_6$\\
&$- 2 d_3 d_6 - 2 d_4 d_6 - 2 d_5 d_6 + 2 c_1 s_1 + 2 d_1 s_1 + 2 d_2 s_1 + 2 d_3 s_1 + 2 d_4 s_1 + 2 d_5 s_1$\\
&$+ 2 d_6 s_1 - 2 s_1^2 + 2 c_1 s_2 + 2 d_1 s_2 + 2 d_2 s_2 + 2 d_3 s_2 + 2 d_4 s_2 + 2 d_5 s_2 + 2 d_6 s_2$\\
&$- 2 s_1 s_2 - 2 s_2^2 + 2 c_1 s_3 + 2 d_1 s_3 + 2 d_2 s_3 + 2 d_3 s_3 + 2 d_4 s_3 + 2 d_5 s_3 + 2 d_6 s_3$\\
&$- 2 s_1 s_3 - 2 s_2 s_3 - 2 s_3^2 + 2 c_1 s_4 + 2 d_1 s_4 + 2 d_2 s_4 + 2 d_3 s_4 + 2 d_4 s_4 + 2 d_5 s_4$\\
&$+ 2 d_6 s_4 - 2 s_1 s_4 - 2 s_2 s_4 - 2 s_3 s_4 - 2 s_4^2$\\\hline
\multirow{7}{*}{$(2,0,1)$}&$3 c_1^2 + 3 c_1 d_1 + c_1 d_2 + 3 d_1 d_2 + 2 c_1 d_3 + 3 d_1 d_3 + 2 d_2 d_3 + 3 c_1 d_4 + 3 d_1 d_4$\\
&$+ 2 d_2 d_4 + d_3 d_4 + d_4^2 + 3 c_1 d_5 + 3 d_1 d_5 + 2 d_2 d_5 + d_3 d_5 + d_4 d_5 + d_5^2 + 3 c_1 d_6$\\
&$+ 3 d_1 d_6 + 3 d_2 d_6 + 3 d_3 d_6 + 3 d_4 d_6 + 3 d_5 d_6 - 3 c_1 s_1 - 3 d_1 s_1 - 3 d_2 s_1 - 3 d_3 s_1$\\
&$- 3 d_4 s_1 - 3 d_5 s_1 - 3 d_6 s_1 + 3 s_1^2 - 3 c_1 s_2 - 3 d_1 s_2 - 3 d_2 s_2 - 3 d_3 s_2 - 3 d_4 s_2$\\
&$- 3 d_5 s_2 - 3 d_6 s_2 + 3 s_1 s_2 + 3 s_2^2 - 3 c_1 s_3 - 3 d_1 s_3 - 3 d_2 s_3 - 3 d_3 s_3 - 3 d_4 s_3$\\
&$- 3 d_5 s_3 - 3 d_6 s_3 + 3 s_1 s_3 + 3 s_2 s_3 + 3 s_3^2 - 3 c_1 s_4 - 3 d_1 s_4 - 3 d_2 s_4 - 3 d_3 s_4$\\
&$- 3 d_4 s_4 - 3 d_5 s_4 - 3 d_6 s_4 + 3 s_1 s_4 + 3 s_2 s_4 + 3 s_3 s_4 + 3 s_4^2$\\\hline
\multirow{6}{*}{$(2,1,1)$}&$9 c_1^2 - c_1 d_1 - c_1 d_2 - d_1 d_2 - 2 d_2^2 - 2 c_1 d_3 - d_1 d_3 - d_2 d_3 - d_3^2 - c_1 d_4 - d_1 d_4$\\
&$- d_2 d_4 - d_3 d_4 - 2 d_4^2 - c_1 d_5 - d_1 d_5 - d_2 d_5 - d_3 d_5 + 3 d_4 d_5 - 2 d_5^2 - c_1 d_6 - d_1 d_6$\\
&$- d_2 d_6 - d_3 d_6 - d_4 d_6 - d_5 d_6 + c_1 s_1 + d_1 s_1 + d_2 s_1 + d_3 s_1 + d_4 s_1 + d_5 s_1 + d_6 s_1$\\
&$- s_1^2 + c_1 s_2 + d_1 s_2 + d_2 s_2 + d_3 s_2 + d_4 s_2 + d_5 s_2 + d_6 s_2 - s_1 s_2 - s_2^2 + c_1 s_3$\\
&$+ d_1 s_3 + d_2 s_3 + d_3 s_3 + d_4 s_3 + d_5 s_3 + d_6 s_3 - s_1 s_3 - s_2 s_3 - s_3^2 + c_1 s_4 + d_1 s_4$\\
&$+ d_2 s_4 + d_3 s_4 + d_4 s_4 + d_5 s_4 + d_6 s_4 - s_1 s_4 - s_2 s_4 - s_3 s_4 - s_4^2$\\\hline
\multirow{7}{*}{$(2,1,2)$}&$3 c_1^2 + 3 c_1 d_1 + 3 c_1 d_2 + 3 d_1 d_2 + d_2^2 + 2 c_1 d_3 + 3 d_1 d_3 + d_2 d_3$\\
&$+ c_1 d_4 + 3 d_1 d_4 + 2 d_2 d_4 + 2 d_3 d_4 + c_1 d_5 + 3 d_1 d_5 + 2 d_2 d_5 + 2 d_3 d_5 + 4 d_4 d_5$\\
&$+ 3 c_1 d_6 + 3 d_1 d_6 + 3 d_2 d_6 + 3 d_3 d_6 + 3 d_4 d_6 + 3 d_5 d_6 - 3 c_1 s_1 - 3 d_1 s_1 - 3 d_2 s_1$\\
&$- 3 d_3 s_1 - 3 d_4 s_1 - 3 d_5 s_1 - 3 d_6 s_1 + 3 s_1^2 - 3 c_1 s_2 - 3 d_1 s_2 - 3 d_2 s_2 - 3 d_3 s_2$\\
&$- 3 d_4 s_2 - 3 d_5 s_2 - 3 d_6 s_2 + 3 s_1 s_2 + 3 s_2^2 - 3 c_1 s_3 - 3 d_1 s_3 - 3 d_2 s_3 - 3 d_3 s_3$\\
&$- 3 d_4 s_3 - 3 d_5 s_3 - 3 d_6 s_3 + 3 s_1 s_3 + 3 s_2 s_3 + 3 s_3^2 - 3 c_1 s_4 - 3 d_1 s_4 - 3 d_2 s_4$\\
&$- 3 d_3 s_4 - 3 d_4 s_4 - 3 d_5 s_4 - 3 d_6 s_4 + 3 s_1 s_4 + 3 s_2 s_4 + 3 s_3 s_4 + 3 s_4^2$\\\hline
\multirow{1}{*}{$(5,2,3)$}&$2 (5 c_1^2 + 2 c_1 d_1 + c_1 d_2 + 2 d_1 d_2 + 2 d_2^2 + 2 d_1 d_3 + 2 d_2 d_3 + 2 d_3^2 + c_1 d_4$\\
&$+ 2 d_1 d_4 + d_2 d_4 + 2 d_3 d_4 + 2 d_4^2 + c_1 d_5 + 2 d_1 d_5 + d_2 d_5 + 2 d_3 d_5 + 2 d_5^2 + 2 c_1 d_6$\\
&$+ 2 d_1 d_6 + 2 d_2 d_6 + 2 d_3 d_6 + 2 d_4 d_6 + 2 d_5 d_6 - 2 c_1 s_1 - 2 d_1 s_1 - 2 d_2 s_1 - 2 d_3 s_1$\\
&$- 2 d_4 s_1 - 2 d_5 s_1 - 2 d_6 s_1 + 2 s_1^2 - 2 c_1 s_2 - 2 d_1 s_2 - 2 d_2 s_2 - 2 d_3 s_2 - 2 d_4 s_2$\\
&$- 2 d_5 s_2 - 2 d_6 s_2 + 2 s_1 s_2 + 2 s_2^2 - 2 c_1 s_3 - 2 d_1 s_3 - 2 d_2 s_3 - 2 d_3 s_3 - 2 d_4 s_3$\\
&$- 2 d_5 s_3 - 2 d_6 s_3 + 2 s_1 s_3 + 2 s_2 s_3 + 2 s_3^2 - 2 c_1 s_4 - 2 d_1 s_4 - 2 d_2 s_4 - 2 d_3 s_4$\\
&$- 2 d_4 s_4 - 2 d_5 s_4 - 2 d_6 s_4 + 2 s_1 s_4 + 2 s_2 s_4 + 2 s_3 s_4 + 2 s_4^2)$\\\hline
\end{tabular}
	\caption{Part two of the non-vanishing fiber Gopakumar-Vafa invariants for an elliptic fibration with generic fiber given by a codimension five complete intersection in $\mathbb{P}_\Delta$ with toric data~\eqref{eqn:toricdata}.}
	\label{tab:codim5gv_2}
\end{table}

All intersection numbers~\eqref{eqn:codim5inter} are non-zero and this implies that the fibration does not exhibit any fibral divisors that descend from the ambient space of the generic fiber.
Moreover, the classification in Section~\ref{sec:classification} shows that a generic fibration over $\mathbb{P}^2$ has $h^{1,1}(M)=4$, and therefore the fiber does not induce non-toric divisors either.
We thus conclude that the fiber generically leads to fibrations with three independent sections.
For a suitable choice of sections the corresponding divisor classes are
\begin{align}
E_1=2J_2-J_3\,,\quad E_2=J_1-2J_2\,,\quad E_3=J_3-J_2\,.
\end{align}
Choosing $E_1$ as the zero-section we find that the images of $E_2$ and $E_3$ under the Shioda map are, up to base divisors, given by
\begin{align}
	\sigma(E_2)=E_2-E_1=J_1-4J_2+J_3\,,\quad \sigma(E_3)=E_3-E_1=2J_3-3J_2\,.
\end{align}
The vertical contributions do not affect the intersections with fibral curves and can therefore be neglected.
Expanding the M-theory three-form along the harmonic forms that are associated to these classes leads to a $U(1)^2$ gauge symmetry in the corresponding effective action.

From the absence of fibral divisors it follows that the fibration exhibits only isolated reducible fibers.
Furthermore, the fiber GV-invariants tell us that all of those fibers are of $I_2$ type, i.e. they consist of two rational curves that intersect in two points.
The fibral curves counted by $\tilde{n}^0_{k_1,k_2,k_3}$ correspond to half-hypermultiplets with KK-charge $q_{KK}$ and $U(1)$ charges $(q_1,q_2)$ given by
\begin{align}
	q_1=k_1-4k_2+k_3\,,\quad q_2=2k_3-3k_2\,,\quad q_{KK}=2k_2-k_3\,.
\end{align}
We therefore find the representations listed in Table~\ref{tab:u1u1spec}.
\begin{table}[h!]
\begin{align*}
\renewcommand{\arraystretch}{1.2}
\begin{array}{ccc|c}
q_{KK}&q_1&q_2&\text{Multiplicity $n_{\pm(q_1,q_2)}$}\\\hline
-1&1&2&\tilde{n}^0_{0,0,1}\\
-1&2&2&\tilde{n}^0_{1,0,1}\\
-1&3&2&\tilde{n}^0_{2,0,1}\\
-1&3&3&\tilde{n}^0_{4,1,3}=\tilde{n}^0_{1,1,0}\\
-1&4&3&\tilde{n}^0_{5,1,3}=\tilde{n}^0_{0,1,0}\\\hline
0&1&0&\tilde{n}^0_{1,0,0}\\
0&0&1&\tilde{n}^0_{2,1,2}\\
0&1&1&\tilde{n}^0_{3,1,2}=\tilde{n}^0_{2,1,1}\\
0&2&1&\tilde{n}^0_{4,1,2}=\tilde{n}^0_{1,1,1}
\end{array}
\hspace{1cm}
\begin{array}{ccc|c}
q_{KK}&q_1&q_2&\text{Multiplicity $n_{\pm(q_1,q_2)}$}\\\hline
1&0&0&\tilde{n}^0_{5,2,3}\\
1&-2&-1&\tilde{n}^0_{1,1,1}\\
1&0&-1&\tilde{n}^0_{3,1,1}=\tilde{n}^0_{2,1,2}\\
1&-1&-1&\tilde{n}^0_{2,1,1}\\
1&-1&0&\tilde{n}^0_{4,2,3}=\tilde{n}^0_{1,0,0}\\\hline
2&-4&-3&\tilde{n}^0_{0,1,0}\\
2&-3&-3&\tilde{n}^0_{1,1,0}\\
2&-1&-2&\tilde{n}^0_{5,2,2}=\tilde{n}^0_{0,0,1}\\
2&-3&-2&\tilde{n}^0_{3,2,2}=\tilde{n}^0_{2,0,1}\\
2&-2&-2&\tilde{n}^0_{4,2,2}=\tilde{n}^0_{1,0,1}
\end{array}
\end{align*}
	\caption{The $U(1)\times U(1)$ charged half-hypermultiplet spectrum of M-theory on elliptically fibered Calabi-Yau threefolds that are constructed using the fiber that is a codimension five complete intersection in the toric ambient space~\eqref{eqn:toricdata} with nef-partition~\eqref{eqn:u1u1nef}.}
\label{tab:u1u1spec}
\end{table}
Recall that the half-hypermultiplets with charges $(q_{KK},q_1,q_2)$ and $(1-q_{KK},-q_1,-q_2)$ combine into one hypermultiplet.
We denote the multiplicity of hypermultiplets with a given set of conjugate charges by $n_{\pm(q_1,q_2)}$.
Each one is of course also accompanied by a tower of higher Kaluza-Klein excitations.

\subsubsection{Cancellation of six-dimensional anomalies}
\label{sec:anomCanc}
Let us now check that this spectrum indeed leads to an anomaly free six-dimensional supergravity.
There are three types of anomalies that can potentially affect this theory, namely the pure gravitational, the Abelian-gravitational and the pure Abelian anomalies.
They can be cancelled by a generalized Green-Schwarz mechanism if the following conditions are satisfied~\cite{Park:2011wv}:
\begin{align}
\begin{split}
\renewcommand{\arraystretch}{1.5}
\begin{array}{rl}
\text{Pure gravitational: }&H-V+29T=273\,,\quad 9-T=a\cdot a\\
\text{Abelian-gravitational: }&-\frac16\sum_{\vec{q}}x_{q_i,q_j}q_iq_j=a\cdot b_{ij}\\
\text{Pure Abelian: }&\sum_{\vec{q}}x_{q_i,q_j,q_k,q_l}q_mq_nq_kq_l=b_{(ij}\cdot b_{kl)}
\end{array}
\end{split}
	\label{eqn:6danomalies}
\end{align}
Here $H,V$ and $T$ respectively denote the number of hyper-, vector- and tensormultiplets.
On the other hand, $x_{q_{m_1},\dots,q_{m_k}}$ denotes the number of hypermultiplets that carry charge $(q_{m_1},\dots,q_{m_k})$ under some set of $U(1)$ gauge symmetries $m_1,\dots,m_k$ while $b_{ij}$ denotes the symmetric \textit{height pairing} that is associated
to the $n$-sections which give rise to the $i$-th and $j$-th $U(1)$.
For $n$-sections $S_i^{(n)}$ and $S_j^{(n)}$ this is defined as
\begin{align}
	b_{ij}=-\pi\left(\sigma(S_i^{(n)})\cdot\sigma(S_j^{(n)})\right)\,.
	\label{eqn:heightpairing}
\end{align}
Note that the same $U(1)$ factor can appear multiple times. Finally, $a$ is the canonical class of the base $B$.
The intersections on the right-hand sides are evaluated in $B$.

We will first consider the pure gravitational anomalies.
Since we are working base independently, the second constraint can be used to express the number of tensor multiplets that arise from curves in the base as $T=9-c_1^2$.
The number of neutral hypermultiplets is given by
\begin{align}
H_{\text{neut.}}=T+3+\text{rk}(G)-\frac12\chi(M)\,,
\end{align}
where $\chi(M)$ is the Euler characteristic of the fibration.
Together with the number of charged hypermultiplets
\begin{align}
	H_{\text{char.}}=\frac12\sum\limits_{k_1=0}^5\sum\limits_{k_2=0}^2\sum\limits_{k_3=0}^3\tilde{n}^0_{k_1,k_2,k_3}-\frac12\tilde{n}^0_{5,2,3}\,,
\end{align}
and using $\chi(M)=-\tilde{n}^0_{5,2,3}$ this leads to a total number of hypermultiplets
\begin{align}
	H=14-c_1^2+\frac12\sum\limits_{k_1=0}^5\sum\limits_{k_2=0}^2\sum\limits_{k_3=0}^3\tilde{n}^0_{k_1,k_2,k_3}=14+29c_1^2\,.
\end{align}
The gauge group is $G=U(1)^2$ which implies $V=2$ and we find that indeed all of the pure gravitational anomalies cancel.
Note that for smooth fibrations without non-flat fibers or fibral divisors the gravitational anomaly cancellation condition can always be written in terms of the fiber GV-invariants as
\begin{align}
	\sum\limits_{\tiny\begin{array}{c}\beta\in H_2(M)\\\beta\le F\end{array}}\tilde{n}^0_\beta=60c_1^2\,,
	\label{eqn:gravangv}
\end{align}
where $F$ is the class of the generic fiber.

Next we are looking at the Abelian-gravitational anomalies which will also enable us to obtain the height pairings.
Note that even without knowing the height pairing beforehand, the required factorization on the right hand side provides another highly non-trivial consistency check on the spectrum.
Recall the we denote the multiplicity of six-dimensional hypermultiplets of $U(1)\times U(1)$ charge $(q_1,q_2)$ by $n_{\pm(q_1,q_2)}$.
Setting for example $m=1,n=2$ we obtain
\begin{align}
\begin{split}
	&-\frac16\left(2n_{\pm(1,2)}+4n_{\pm(2,2)}+6n_{\pm(3,2)}+9n_{\pm(3,3)}+12n_{\pm(4,3)}+n_{\pm(1,1)}+2n_{\pm(2,1)}\right)\\
=&c_1(-9c_1+d_2-4d_3-3d_4-3d_5)\,,
\end{split}
\end{align}
and deduce that $b_{12}=9c_1-d_2+4d_3+3d_4+3d_5$.
In an analogous fashion we also determine
\begin{align}
b_{11}=2(6c_1-d_2+2d_3+2d_4+2d_5)\,,\quad b_{22}=2(4c_1+2d_3+d_4+d_5)\,.
\end{align}
Finally we can use these height pairings to check the pure Abelian anomalies and find that they are indeed generically cancelled.
Taken together this provides an extremely strong consistency check of the identification of the gauge group, the classes of the sections and in particular the corresponding spectrum that we obtained from the fiber GV-invariants.

\subsubsection{The Higgs transition  $U(1)^2\rightarrow\mathbb{Z}_5$}
\label{sec:higgsu12}
Let us now perform a Higgs transition using the hypermultiplets of charge $(q_{KK},q_1,q_2)=(-1,1,2)$ that appear with multiplicity $n_{\pm(1,2)}=\tilde{n}^0_{0,0,1}$.
These are massless after taking the limit $t^3\rightarrow 0$ in the K\"ahler moduli space.
After a subsequent complex structure deformation, which amounts to turning on a vacuum expectation value for the corresponding scalar fields, the gauge group is broken into a $U(1)$ and the remaining particles carry charges $q=q_2-2q_1$.
The resulting spectrum is summarized in Table~\ref{tab:higgsspec} and, as expected, contains half-hypermultiplets of charge $5$.
\begin{table}[h!]
\begin{align*}
\begin{array}{cc|c}
	q_{KK}&q&\text{Multiplicity $n'_{\pm q}$}\\\hline
-1&-2&\tilde{n}^0_{1,0,1}\\
-1&-3&\tilde{n}^0_{4,1,3}=\tilde{n}^0_{1,1,0}\\
-1&-4&\tilde{n}^0_{2,0,1}\\
-1&-5&\tilde{n}^0_{5,1,3}=\tilde{n}^0_{0,1,0}\\\hline
0&-3&\tilde{n}^0_{4,1,2}=\tilde{n}^0_{1,1,1}\\
0&-2&\tilde{n}^0_{1,0,0}\\
0&-1&\tilde{n}^0_{3,1,2}=\tilde{n}^0_{2,1,1}\\
0&1&\tilde{n}^0_{2,1,2}
\end{array}
\hspace{1cm}
\begin{array}{cc|c}
	q_{KK}&q&\text{Multiplicity $n'_{\pm q}$}\\\hline
1&-1&\tilde{n}^0_{3,1,1}=\tilde{n}^0_{2,1,2}\\
1&0&\tilde{n}^0_{5,2,3}+\tilde{n}^0_{0,0,1}+\tilde{n}^0_{5,2,2}\\
1&1&\tilde{n}^0_{2,1,1}\\
1&2&\tilde{n}^0_{4,2,3}=\tilde{n}^0_{1,0,0}\\
1&3&\tilde{n}^0_{1,1,1}\\\hline
2&2&\tilde{n}^0_{4,2,2}=\tilde{n}^0_{1,0,1}\\
2&3&\tilde{n}^0_{1,1,0}\\
2&4&\tilde{n}^0_{3,2,2}=\tilde{n}^0_{2,0,1}\\
2&5&\tilde{n}^0_{0,1,0}
\end{array}
\end{align*}
	\caption{The $U(1)$ charged half-hypermultiplet spectrum that arises after Higgsing the spectrum in Table~\ref{tab:u1u1spec} by giving a vev to the scalars with charge $(q_{KK},q_1,q_2)=(-1,1,2)$. }
\label{tab:higgsspec}
\end{table}
The anomalies are automatically cancelled, since we have started with an anomaly free spectrum before the Higgs transition.

Recall that the Kaluza-Klein charge only exists in the five-dimensional theory, that arises from M-theory on the elliptically fibered Calabi-Yau or from F-theory after compactifying on an additional circle.
Therefore the multiplicity of six-dimensional hypermultiplets with charge $q$ can decompose into five-dimensional multiplicities of half-hypermultiplets with different Kaluza-Klein charges.
In order to distinguish them from the corresponding multiplicities in the compactifications with $G=\mathbb{Z}_5$, let us denote the multiplicities of six-dimensional hypermultiplets that contain half-hypermultiplets of charge $\pm q$ by $n'_{\pm q}$.
We then find
\begin{align}
	\begin{split}
		n'_{\pm 1}=&\tilde{n}^0_{2,1,1}+\tilde{n}^0_{2,1,2}\,,\quad n'_{\pm 2}=\tilde{n}^0_{1,0,0}+\tilde{n}^0_{1,0,1}\,\quad n'_{\pm3}=\tilde{n}^0_{1,1,0}+\tilde{n}^0_{1,1,1}\,,\\
		n'_{\pm4}=&\tilde{n}^0_{2,0,1}\,,\quad n'_{\pm5}=\tilde{n}^0_{0,1,0}\,.
	\end{split}
\end{align}

To perform the second Higgs transition, which breaks $U(1)\rightarrow\mathbb{Z}_5$, we note that in five dimensions there are two types of charge five half-hypermultiplets, with $(q_{KK},q)$ respectively given by $(-1,-5)$ and $(2,5)$.
We can make the particles with charge $(2,5)$ massless by setting $t^2\rightarrow 0$.
Giving the corresponding scalar fields a non-zero vacuum expectation value, which geometrically amounts to a complex structure deformation, then breaks the gauge group such that in six dimensions there is a discrete remnant $G=\mathbb{Z}_5$.
On the other hand, in order to perform a Higgs transition with the $(-1,-5)$ modes we would need to take the limit $t^2\rightarrow -5t^1$.
This is clearly not contained in a boundary of the K\"ahler cone, as some curves acquire negative volume, and we would first need to find a birational phase that allows us to take this limit.

Note that it is also possible start the Higgs transition using the multiplets with six-dimensional $U(1)\times U(1)$ charge $(q_1,q_2)=(4,3)$.
Geometrically this just exchanges the order in which the limits for the K\"ahler parameters are taken.

\subsection{Base independent spectra for Grassmannian and Pfaffian fibrations}
\label{sec:baseIndependentSpecZ5}
Using the Higgs transition we can write down base independent expressions for the spectrum and the Euler characteristic of the generic genus one fibrations with 5-sections.
The gauge group of the associated F-theory vacuum is then $G=\mathbb{Z}_5$ and the spectrum of charged hypermultiplets contains the charges $\pm1$ and $\pm2$.
The corresponding multiplicities $n_{q_{\mathbb{Z}_5}}$ are
\begin{align}
\begin{split}
	n_{\pm1}=&n'_{\pm2}+n'_{\pm3}=\tilde{n}^0_{1,0,0}+\tilde{n}^0_{1,1,0}+\tilde{n}^0_{1,0,1}+\tilde{n}^0_{1,1,1}\,,\\
	n_{\pm2}=&n'_{\pm 1}+n'_{\pm 4}=\tilde{n}^0_{2,1,2}+\tilde{n}^0_{2,1,1}+\tilde{n}^0_{2,0,1}\,,
\end{split}
\end{align}
where we have choosen the charges with respect to $e^{4\pi i/5}\in U(1)$ as a generator for the unbroken $\mathbb{Z}_5$ subgroup.
Note that the 5-section corresponding to $[v_1]$ and $[v_2]$ is not affected by the extremal transition and therefore the corresponding intersection numbers with the components of the reducible fibers directly translate into those of the genus one fibration.
The change of the Euler characteristic can be calculated from the change of hypermultiplets and vectormultiplets~\cite{Klevers:2014bqa}.
This gives us the Euler characteristic for the genus one fibration with five section,
\begin{align}
\begin{split}
\chi_{\mathbb{Z}_5}=&\chi_{U(1)^2}-2(\tilde{n}^0_{0,1,0}+\tilde{n}^0_{0,0,1})\\
=&-2 (5 c_1^2 + 2 d_2^2 + d_2 d_3 + 2 d_3^2 - d_2 d_4 + d_3 d_4 + 2 d_4^2 - d_2 d_5 + d_3 d_5 - d_4 d_5 + 2 d_5^2)\,.
\end{split}
\end{align}
that one obtains after performing the extremal transition.
Compared to $\chi_{U(1)^2}=-n^0_{5,2,3}$ this is a surprisingly simple expression as it only depends on a small subset of the parameters.

However, in order to use these expressions directly, let us match the parameters $d_{1,\dots,6},\,s_{1,\dots,5}$ with the charges of the fields in the GLSM~\eqref{eqn:glsmFieldContentFibration}.
To this end we denote the $U(1)^k$ charges of the generic field content as follows:
\begin{align}
	\begin{array}{c|ccccccc|c}
		&p_{i,\,i=1,\dots,5}&x_{i,\,i=1,\dots,5}&b_{1,\dots,\text{dim}(B)+k}&\textrm{FI}\\\hline
		U(2)&\det^{-1}&\square&0&\zeta_1\\
		U(1)_1&q^1(p_i)&q^1(x_i)&*&\zeta_2\\
		\dots&\dots&\dots&\dots&\dots\\
		U(1)_k&q^k(p_i)&q^k(x_i)&*&\zeta_{1+k}\\
	\end{array}
	\label{eqn:glsmFieldContentSec4}
\end{align}
Here we assume that the base $B$ is a toric variety that is realized by a $U(1)^k$ GLSM and that the $k$ charge vectors correspond to a basis of the Mori cone.
For simplicity we will also assume that the Mori cone is simplicial.
Then one can denote a dual basis of the K\"ahler cone by $J_1,\dots,J_k$ and the charges $q_i(p_j),q_i(x_j)$ determine classes $Q(\bullet)=q^i(\bullet)J_i$.

Let us first note that the identifications~\eqref{eqn:shiftedx},~\eqref{eqn:shiftedd} only fix the charges of the fields $v_1,\dots,v_6$ in a corresponding GLSM with respect to the base $U(1)^k$ up to twists with the torus action on the ambient space of the fiber.
Using the limit where $q_i(x_j)=0$ allows us to determine the $Q(p_i)$ dependence of $s_{1,\dots,5}$.
However, due to the twist, this result will in general be shifted by a linear function $\alpha$ of $Q(x_i)$ and this leads us to identify
\begin{align}
\begin{split}
s_1=&-Q(p_1)-\alpha\,,\quad s_2=-Q(p_2)-\alpha\,,\quad s_3=-Q(p_3)-\alpha\,,\\
s_4=&-Q(p_4)-\alpha\,,\quad s_5=-Q(p_5)-\alpha\,.
\end{split}
\end{align}
On the other hand, Theorem 3.2.3 in~\cite{Batyrev:1998kx} implies that the sections
{\small
\begin{align}
\{v_1, v_2 v_9, v_2 v_3 v_4, v_2 v_3 v_5, v_7 v_9, v_7 v_3 v_4, v_7 v_3 v_5, v_8 v_4, v_8  v_5, v_6\}\in \Gamma\left(P(2,5),\mathcal{O}([v_1])\right)\,,
\label{eqn:pluckersections}
\end{align}
}
embed $P(2,5)$ into a $\mathbb{P}^9$ that is in turn parametrized by the Plücker coordinates on the Grassmannian $G(2,5)$.
Identifying the vertical parts of the classes of the Pl\"ucker coordinates with those of the sections~\eqref{eqn:pluckersections} leads to the system of equations
\begin{align}
\begin{split}
d_1+\alpha=&Q(x_1)+Q(x_2)\,,\quad d_2+\alpha=Q(x_1)+Q(x_3)\,,\\
d_2+d_3+d_4+\alpha=&Q(x_1)+Q(x_4)\,,\quad d_2+d_3+d_5+\alpha=Q(x_1)+Q(x_5)\,,\\
\alpha=&Q(x_2)+Q(x_3)\,,\quad d_3+d_4+\alpha=Q(x_2)+Q(x_4)\,,\\
d_3+d_5+\alpha=&Q(x_2)+Q(x_5)\,,\quad d_4+\alpha=Q(x_3)+Q(x_4)\,,\\
d_5+\alpha=&Q(x_3)+Q(x_5)\,,\quad d_6+\alpha=Q(x_4)+Q(x_5)\,.
\end{split}
\end{align}
This has a unique solution and we find
\begin{align}
\begin{split}
d_1=&Q(x_1)-Q(x_3)\,,\quad d_2=Q(x_1)-Q(x_2)\,,\quad d_3=Q(x_2)-Q(x_3)\,,\\
d_4=&Q(x_4)-Q(x_2)\,,\quad d_5=Q(x_5)-Q(x_2)\,,\\
d_6=&Q(x_4)+Q(x_5)-Q(x_2)-Q(x_3)\,,\quad s_1=-Q(x_2)-Q(x_3)-Q(p_1)\,,\\
s_2=&-Q(x_2)-Q(x_3)-Q(p_2)\,,\quad s_3=-Q(x_2)-Q(x_3)-Q(p_3)\,,\\
s_4=&-Q(x_2)-Q(x_3)-Q(p_4)\,,\quad s_5=-Q(x_2)-Q(x_3)-Q(p_5)\,.
\end{split}
\label{eqn:xptods}
\end{align}
We can then rewrite
\begin{align}
\begin{split}
\chi=&-2\left(5c_1^2+2S_{1,x}^2-5S_{2,x}\right)\,,\\
n_{\pm1}=&10c_1^2-c_1\left(8S_{1,x}+5S_{1,p}\right)-6S_{1,x}^2-5S_{2,x}-10S_{1,x}S_{1,p}-5S_{1,p}^2+5S_{2,p}\,,\\
n_{\pm2}=&15c_1^2+c_1\left(8S_{1,x}+5S_{1,p}\right)+4S_{1,x}^2+10S_{2,x}+10S_{1,x}S_{1,p}+5S_{1,p}^2-5S_{2,p}\,,
\end{split}
\label{eqn:genericMultis}
\end{align}
where $S_{i,x/p}$ are elementary symmetric polynomials of degree $i$ in $Q(x_{1,\dots,5})/Q(p_{1,\dots,5})$.

This brings us very close to the most general expression for $n_{\pm1}$, $n_{\pm2}$ and $\chi$.
For the final ingredient, let us first note that the identification~\eqref{eqn:xptods} necessarily implies $d_1=d_2+d_3$ and $d_6=d_3+d_4+d_5$.
As we will see in the following section, by classifying fibrations over $\mathbb{P}^2$, these relations are in general not satisfied.
Although the corresponding genus one fibrations are still related to the toric complete intersections via an extremal transition, in most cases this is not just a fiberwise application of the process described in~\cite{Batyrev:1998kx}.
Nevertheless, the expressions~\eqref{eqn:genericMultis} remain valid as long as the corresponding GLSM is of the form~\eqref{eqn:glsmFieldContentSec4}.
However, the classification leads us to also consider at least one class of genus one fibrations with 5-sections that require a different form of the GLSM.
The corresponding fibration of curves in Grassmannians involves the tangent bundle of the base $\mathbb{P}^2$.
To cover these cases as well, it is necessary to identify the polynomials $S_{i,x/p}$ with quantities that are intrinsic to the geometries.

Using the generic construction from Section~\ref{sec:geometry}, we can express them in terms of Chern classes of the bundles that determine the fibration of Grassmannian curves,
\begin{align}
	S_{1,x} = c_1({E'}^\vee)\,,\quad
S_{1,p} = c_1(F)\,,\quad
	S_{2,x} = c_2({E'}^\vee)\,,\quad
S_{2,p} = c_2(F)\,.
\end{align}
Inserting this into~\eqref{eqn:genericMultis} and using $c_{i,\bullet}\equiv c_i(\bullet)$ leads us to the final result
\begin{align}
\begin{split}
	\chi=&-2\left(5c_{1,B}^2+2c_{1,{E'}^\vee}^2-5c_{2,{E'}^\vee}\right)\,,\\
	n_{\pm1}=&10c_{1,B}^2-c_{1,B}\left(8c_{1,{E'}^\vee}+5c_{1,F}\right)-6c_{1,{E'}^\vee}^2 -5c_{2,{E'}^\vee} \\
	&-10c_{1,{E'}^\vee}c_{1,F}-5c_{1,F}^2+5c_{2,F}\,,\\
	n_{\pm2}=&15c_{1,B}^2+c_{1,B}\left(8c_{1,{E'}^\vee}+5c_{1,F}\right)+4c_{1,{E'}^\vee}^2 +10c_{2,{E'}^\vee} \\
	&+10c_{1,{E'}^\vee}c_{1,F}+5c_{1,F}^2-5c_{2,F}\,.
\end{split}
\label{eqn:fullyGenericMultis}
\end{align}
The corresponding expressions for fibrations of Pfaffian curves can be obtained by exchanging $n_{\pm1}$ and $n_{\pm2}$.

\paragraph{Examples} Let us now check these expressions for two of our examples.
The Calabi-Yau $X_1$ corresponds to the parameters
\begin{align}
Q(p_{1,\dots,3})=-H\,,\quad Q(p_{4,5})=Q(x_{1,\dots,5})=0\,,
\end{align}
where $H$ is the hyperplane class of $B=\mathbb{P}^2$.
This determines the symmetric polynomials
\begin{align}
S_{1,x}=S_{2,x}=0\,,\quad S_{1,p}=-3H\,,\quad S_{2,p}=3H^2\,.
\end{align}
Together with $c_1=3H$ we then find
\begin{align}
\chi=-90\,,\quad n_{\pm1}=105\,,\quad n_{\pm2}=120\,,
\end{align}
which matches the result obtained by calculating the genus zero Gopakumar-Vafa invariants in Section~\ref{sec:gvinvariantsModel1Weak}.

On the other hand, for the Calabi-Yau $Y_1$ the GLSM charges are such that
\begin{align}
Q(p_{1,\dots,5})=-H\,,\quad Q(x_{1,\dots,4})=0\,,\quad Q(x_5)=H\,.
\label{eqn:model2Qcharges}
\end{align}
The corresponding symmetric polynomials are
\begin{align}
S_{1,x}=H\,,\quad S_{2,x}=0\,,\quad S_{1,p}=-5H\,,\quad S_{2,p}=10H^2\,,
\end{align}
and this leads to the invariants
\begin{align}
\chi=-94\,,\quad n_{\pm1}=110\,,\quad n_{\pm2}=113\,.
\end{align}
The corresponding genus zero Gopakumar-Vafa invariants are listed in Appendix~\ref{sec:fibrationsOverP2} and verify this result.

\subsection{Toric codimension 3 complete intersection fibers with $G=U(1)$}
\label{sec:codim3fiber}
In the last section we studied an extremal transition between the elliptic fibrations of codimension 5 complete intersections in a toric ambient space and genus one fibrations with 5-sections.
However, this transition was a two step process, without a clear geometric realization of the intermediate geometry.
To remedy this situation, we are now constructing a class of codimension 3 complete intersection fibers in a four dimensional toric ambient space.
The corresponding fibrations generically have two independent sections and directly lead to F-theory vacua with $U(1)$ gauge symmetry and massless hypermultiplets with charges $q=1,\dots,5$.

To construct the polytope and nef partition let us first recall that the fundamental period of the degree $5$ genus one curve can be written as
\begin{align}
	\begin{split}
		\varpi_0=&\sum_{n=0}^\infty\sum_{k=0}^n\left(\begin{array}{c}n\\k\end{array}\right)^2\left(\begin{array}{c}n+k\\k\end{array}\right)z^n\\
			=&\sum_{n=0}^\infty\sum_{k=0}^n\frac{\Gamma(1+n)^2\Gamma(1+n+k)}{\Gamma(1+n-k)^2\Gamma(1+k)^3\Gamma(1+n)}z^n\,.
	\end{split}
\end{align}
It is natural to interpret the second sum as a result of setting the second complex structure modulus in a two parameter geometry to one and this leads us to consider geometries with fundamental period
\begin{align}
	\varpi_0'=\sum\limits_{n_1,n_2=0}^\infty\frac{\Gamma(1+n_1)^2\Gamma(1+n_1+n_2)}{\Gamma(1+n_1-n_2)^2\Gamma(1+n_2)^3\Gamma(1+n_1)}z_1^{n_1}z_2^{n_2}\,.
\end{align}
The arguments of the Gamma functions determine the charge matrix of an Abelian GLSM
\begin{align}
	\begin{split}
		\begin{array}{c|rrr||rrrrrr}
			&p^1&p^2&p^3&v_1&v_2&v_3&v_4&v_5&v_6\\\hline
			U(1)_1&-1&-1&-1  & 1& 1& 0& 0& 0& 1\\
			U(1)_2& 0& 0&-1  &-1&-1& 1& 1& 1& 0\\
		\end{array}\,.
	\end{split}
	\label{eqn:cicy3glsm}
\end{align}
Taking the kernel of the charge matrix of the fields $v_i,\,i=1,\dots,6$ we obtain a four-dimensional reflexive polytope with points
\begin{align}
	\begin{blockarray}{rrrrrrrl}
		\begin{block}{r(rrrr|rr)l}
			v_1& 0& 0& 0& 1& 1&-1&\\
			v_2& 1& 1& 1&-1& 1&-1&\\
			v_3& 1& 0& 0& 0& 0& 1&\\
			v_4& 0& 1& 0& 0& 0& 1&\\
			v_5& 0& 0& 1& 0& 0& 1&\\
			v_6&-1&-1&-1& 0& 1& 0&\\
		\end{block}
	\end{blockarray}\,.
	\label{eqn:toricdata3}
\end{align}
The points admit a unique fine regular star triangulation and the relations correspond to the Mori cone of the associated toric variety.
We will denote the homogeneous coordinates also by $v_i,\,i=1,\dots,6$.
The charges of the $p$-fields in~\eqref{eqn:cicy3glsm} determine a codimension 3 nef-partition
\begin{align}
	D_{\nabla_1}=[v_1]+[v_4]\,,\quad D_{\nabla_2}=[v_2]+[v_5]\,,\quad D_{\nabla_3}=[v_3]+[v_6]\,,
\end{align}
and the generic sections of the associated bundles take the form
\begin{align}
	\begin{split}
		p_1=&c_{1,1} v_6+c_{1,2} v_3 v_1+c_{1,3} v_1 v_4+c_{1,4} v_1 v_5+c_{1,5} v_3 v_2+c_{1,6} v_4 v_2+c_{1,7} v_5 v_2\,,\\
		p_2=&c_{2,1} v_6+c_{2,2} v_3 v_1+c_{2,3} v_1 v_4+c_{2,4} v_1 v_5+c_{2,5} v_3 v_2+c_{2,6} v_4 v_2+c_{2,7} v_5 v_2\,,\\
		p_3=&c_{3,1} v_1 v_3^2+c_{3,2} v_2 v_3^2+c_{3,3} v_1 v_3 v_4+c_{3,4} v_2 v_3 v_4+c_{3,5} v_1 v_4^2+c_{3,6} v_2 v_4^2\\
		&+c_{3,7} v_1 v_3 v_5+c_{3,8} v_2 v_3 v_5+c_{3,9} v_1 v_4 v_5+c_{3,10} v_2 v_4 v_5+c_{3,11} v_1 v_5^2\\
		&+c_{3,12} v_2 v_5^2+c_{3,13} v_3 v_6+c_{3,14} v_4 v_6+c_{3,15} v_5 v_6\,.
	\end{split}
	\label{eqn:codim3equations}
\end{align}
The intersections of the toric divisors with the complete intersection curve $C$ can be calculated using SageMath~\cite{sagemath} and are given by
\begin{align}
	C\cdot[v_i]=(2,2,3,3,3,5)\,.
\end{align}
One can use CohomCalg to check that the bundles associated to the divisors
\begin{align}
	E_1=2[v_5]-[v_6]\,,\quad E_2=2[v_6]-3[v_5]\,,
	\label{eqn:codim3sections}
\end{align}
contain non-trivial sections and therefore the divisors themselves correspond to classes of sections.
We choose $E_1$ to be the class of the zero section. Then the image of $E_2$ under the Shioda map is
\begin{align}
	\sigma(E_2)=E_2-E_1=3[v_6]-5[v_5]\,.
	\label{eqn:codim3shioda}
\end{align}
For fibrations over toric bases that are constructed with this fiber we can again use CohomCalg to calculate the Hodge numbers and in general find $h^{1,1}=2+h^{1,1}(B)$.
This implies that there are no additional non-toric sections and the F-theory effective action associated to generic fibrations will therefore exhibit a $U(1)$ gauge symmetry.

We now introduce fibration parameters such that
\begin{align}
	[v_1]=[v_6]-[v_5]+d_1\,,\quad[v_2]=[v_6]-[v_5]+d_2\,,\quad [v_3]=[v_5]+d_3\,,\quad [v_4]=[v_5]+d_4\,,
\end{align}
as well as
\begin{align}
	D_{\nabla_1}=[v_1]+[v_4]+s_1\,,\quad D_{\nabla_2}=[v_2]+[v_5]+s_2\,,\quad D_{\nabla_3}=[v_3]+[v_6]+s_3\,,
\end{align}
where the Calabi-Yau condition imposes
\begin{align}
	s_3=c_1+\sum\limits_{i=1}^4d_i-(s_1+s_2)\,.
\end{align}
A basis for the K\"ahler cone is given by
\begin{align}
	J_1=[v_6]\,,\quad J_2=[v_5]\,,
	\label{eqn:codim3kaehler}
\end{align}
and the K\"ahler class can then be parametrized as $\omega= t^iJ_i$.
We apply the technique developed in~\cite{Paul-KonstantinOehlmann:2019jgr} to obtain the fiber Gopakumar-Vafa invariants listed in Table~\ref{tab:codim3gv}.
\begin{table}[h!]
	\begin{align*}
		\begin{array}{c|l}
			(k_1,k_2)&\tilde{n}^0_{k_1,k_2}\\\hline
			(0,1)&\multirow{2}{*}{$s_1s_2$}\\
			(5,2)&\\\hline
			\multirow{3}{*}{$\begin{array}{c}(1,0)\\(4,3)\end{array}$}&-3 c_1 d_1-3 c_1 d_2-2 c_1 d_3-2 c_1 d_4+6 c_1 s_1+6 c_1 s_2+c_1^2+3 d_1 s_1+3 d_2 s_1\\
				&+2 d_3 s_1+2 d_4 s_1+3 d_1 s_2+3 d_2 s_2+2 d_3 s_2+2 d_4 s_2-2 d_1^2-2 d_2^2+d_3^2+d_4^2\\
				&+d_1 d_2-d_1 d_3-d_2 d_3-d_1 d_4-d_2 d_4-2 d_3 d_4-3 s_1^2-3 s_2^2-5 s_1 s_2\\\hline
			\multirow{4}{*}{$\begin{array}{c}(1,1)\\(4,2)\end{array}$}&-3 c_1 d_1-3 c_1 d_2-2 c_1 d_3-2 c_1 d_4+6 c_1 s_1+6 c_1 s_2+6 c_1^2\\
				&+6 d_1 s_1+6 d_2 s_1+4 d_3 s_1+4 d_4 s_1+6 d_1 s_2+6 d_2 s_2+4 d_3 s_2\\
				&+4 d_4 s_2+d_1^2+d_2^2-2 d_3^2-2 d_4^2-8 d_1 d_2-2 d_1 d_3-2 d_2 d_3\\
				&-2 d_1 d_4-2 d_2 d_4-6 s_1^2-6 s_2^2-8 s_1 s_2\\\hline
			\multirow{4}{*}{$\begin{array}{c}(2,1)\\(3,2)\end{array}$}&3 c_1 d_1+3 c_1 d_2+2 c_1 d_3+2 c_1 d_4-6 c_1 s_1-6 c_1 s_2+15 c_1^2\\
				&-3 d_1 s_1-3 d_2 s_1-2 d_3 s_1-2 d_4 s_1-3 d_1 s_2-3 d_2 s_2-2 d_3 s_2\\
				&-2 d_4 s_2-2 d_1^2-2 d_2^2-d_3^2-d_4^2+7 d_1 d_2+d_1 d_3+d_2 d_3\\
				&+d_1 d_4+d_2 d_4+2 d_3 d_4+3 s_1^2+3 s_2^2+4 s_1 s_2\\\hline
			\multirow{2}{*}{$\begin{array}{c}(2,2)\\(3,1)\end{array}$}&-3 d_1 s_1-3 d_1 s_2-3 d_2 s_1-2 d_3 s_1-2 d_4 s_1-3 d_2 s_2-2 d_3 s_2-2 d_4 s_2+d_1^2\\
				&+d_2 d_1+d_3 d_1+d_4 d_1+d_2^2+d_2 d_3+d_2 d_4+d_3 d_4+3 s_1^2+3 s_2^2+3 s_1 s_2\\\hline
			\multirow{4}{*}{$(5,3)$}&2 \left(3 c_1 d_1+3 c_1 d_2+2 c_1 d_3+2 c_1 d_4-6 c_1 s_1-6 c_1 s_2\right.\\
			&+8 c_1^2-3 d_1 s_1-3 d_2 s_1-2 d_3 s_1-2 d_4 s_1-3 d_1 s_2-3 d_2 s_2\\
			&-2 d_3 s_2-2 d_4 s_2+2 d_1^2+2 d_2^2+2 d_3^2+2 d_4^2-d_1 d_2+d_1 d_3\\
			&\left.+d_2 d_3+d_1 d_4+d_2 d_4-d_3 d_4+3 s_1^2+3 s_2^2+5 s_1 s_2\right)\\\hline
		\end{array}
	\end{align*}
	\caption{The independent non-vanishing fiber Gopakumar-Vafa invariants for an elliptic fibration with generic fiber given by a codimension three complete intersection in the toric ambient space with data~\eqref{eqn:toricdata3}.}
	\label{tab:codim3gv}
\end{table}
All non-vanishing fiber GV-invariants are related to those in Table~\ref{tab:codim3gv} by the periodicty $\tilde{n}_{k_1,k_2}=\tilde{n}_{k_1+5,k_2+3}$.

Using~\eqref{eqn:codim3sections},~\eqref{eqn:codim3shioda} and~\eqref{eqn:codim3kaehler} we find that the $U(1)_{KK}\times U(1)_{6d}$ charge $(q_{KK},q_{6d})$ of half-hypermultiplets with multiplicity $\tilde{n}^0_{k_1,k_2}$ is given by
\begin{align}
	q_{KK}=2k_2-k_1\,,\quad q_{6d}=3k_1-5k_2\,.
\end{align}
The number of uncharged hypermultiplets are encoded in the Euler characteristic of the fibration which is related to the fiber GV-invariants via
\begin{align}
	\chi=-\tilde{n}^0_{5,3}\,.
\end{align}
From the remaining fiber GV-invariants we can then deduce the five-dimensional spectrum of charged half-hypermultiplets that is listed in Table~\ref{tab:codim3spec}.
\begin{table}[h!]
\begin{align*}
\begin{array}{rr|c}
	q_{KK}&q_{6d}&\text{Multiplicity}\\\hline
	 0&1&\multirow{2}{*}{$\tilde{n}^0_{2,1}$}\\
	 1&-1&\\\hline
	 0&2&\multirow{2}{*}{$\tilde{n}^0_{1,1}$}\\
	 1&-2&\\\hline
	-1&3&\multirow{2}{*}{$\tilde{n}^0_{1,0}$}\\
	 2&-3&\\
\end{array}
\hspace{1cm}
\begin{array}{rr|c}
	q_{KK}&q_{6d}&\text{Multiplicity}\\\hline
	-1&4&\multirow{2}{*}{$\tilde{n}^0_{2,2}$}\\
	2&-4&\\\hline
	-1&5&\multirow{2}{*}{$\tilde{n}^0_{0,1}$}\\
	 2&-5&\\
\end{array}
\end{align*}
	\caption{The charged half-hypermultiplet spectrum, that arises from the codimension 3 complete intersection fiber in the ambient space with toric data~\eqref{eqn:toricdata3}.}
\label{tab:codim3spec}
\end{table}
Recall that each pair of half-hypermultiplets with opposite $q_{6d}$ charges combines into one six-dimensional hypermultiplet.

\subsubsection{Cancellation of six-dimensional anomalies}
The five-dimensional spectrum of half-hypermultiplets in Table~\ref{tab:codim3spec} determines the six-dimensional multiplicities $n_q$ of hypermultiplets with $U(1)$ charge $q$
\begin{align}
	n_{\pm1}=\tilde{n}^0_{2,1}\,,\quad n_{\pm2}=\tilde{n}^0_{1,1}\,,\quad n_{\pm3}=\tilde{n}^0_{1,0}\,,\quad n_{\pm4}=\tilde{n}^0_{2,2}\,,\quad n_{\pm5}=\tilde{n}^0_{0,1}\,.
	\label{eqn:codim6fthspc}
\end{align}
Let us now check that the six-dimensional anomalies~\eqref{eqn:6danomalies} are generically cancelled.

As discussed in Section~\ref{sec:anomCanc}, the gravitational anomaly cancellation condition amounts to the identity
\begin{align}
	\sum\limits_{k_1=0}^5\sum\limits_{k_2=0}^3\tilde{n}^0_{k_1,k_2}=60c_1^2\,,
\end{align}
which is satisfied by the fiber GV-invariants invariants in Table~\ref{tab:codim3spec}.
We will again use the Abelian-gravitational anomalies to obtain the height pairing~\eqref{eqn:heightpairing} and then check that the pure Abelian anomaly is cancelled.
Note that with the base independent intersection calculus that has been developed in~\cite{Paul-KonstantinOehlmann:2019jgr} it would be possible, although tedious, to calculate the height pairing directly.
The Abelian-gravitational anomaly fixes the height pairing
\begin{align}
	\begin{split}
		b_{11}=&\frac{1}{6c_1}(n_{\pm1}+2^2n_{\pm2}+3^2n_{\pm3}+4^2n_{\pm4}+5^2n_{\pm5})\\
	=&2 \left(4 c_1-3 d_1-3 d_2-2 d_3-2 d_4+6 s_1+6 s_2\right)\,.
	\end{split}
\end{align}
The pure Abelian anomaly then amounts to the identity
\begin{align}
	n_{\pm1}+2^4n_{\pm2}+3^4n_{\pm3}+4^4n_{\pm4}+5^4n_{\pm5}=\frac{4!}{2!\cdot 2!}b_{11}^2\,,
\end{align}
which is indeed satisfied.

\subsubsection{The Higgs transition $U(1)\rightarrow\mathbb{Z}_5$}
\label{sec:higgsU1Z5}
The six-dimensional F-theory spectrum~\eqref{eqn:codim6fthspc} contains the expected hypermultiplets of charge five, which allow for a Higgs transition $U(1)\rightarrow\mathbb{Z}_5$.
Again, the corresponding extremal transition is best understood in terms of the five-dimensional theory which contains two half-hypermultiplets of $U(1)_{6d}$ charge five that have different Kaluza-Klein charges.
The multiplet with charge $(q_{KK},q_{6d})=(2,-5)$ corresponds to curves in the class $(k_1,k_2)=(0,1)$ which become massless in the limit $t^2\rightarrow 0$.
This lies at the boundary of the K\"ahler moduli space and indeed allows us to perform an extremal transition.
On the other hand, the multiplet with charge $(q_{KK},q_{6d})=(2,-5)$ arises from curves in the class $(k_1,k_2)=(5,2)$.
To make the corresponding states massless we have to take the limit $t^2\rightarrow-\frac{5}{2}t^1$.
This would cause the curves in the class $(k_1,k_2)=(0,1)$ to acquire negative volume and thus takes us outside the K\"ahler cone.
We therefore find that again only one extremal transition is possible.

\subsubsection{A map to the cubic and factorization over the charge 5 locus}
The multiplicity of charge $5$ matter takes the particularly simple form $n_{\pm5}=\tilde{n}^0_{0,1}=s_1\cdot s_2$, with $s_1$ and $s_2$ being the classes of the coefficients $c_{1,1},c_{2,1}$ in~\eqref{eqn:codim3equations}.
This suggests that the corresponding $I_2$ singular fibers arise over points in the base where $c_{1,1}=c_{2,1}=0$.
We can check this assumption by first mapping the curve into a cubic hypersurface in $\mathbb{P}^2$.
To this end we take the line bundle associated to $[v_3]=[v_4]=[v_5]$ which is of degree $3$ on the curve and use the sections $v_3,v_4,v_5$ to embed the curve into the projective space.

In order to obtain the defining equation we first restrict to the patch $v_1=v_3=1$ and later homogenize the result.
We can use $p_1$ in~\eqref{eqn:codim3equations} to solve for $v_6$.
Taking the resultant of $p_2$ and $p_3$ with respect to $v_2$ then leaves us with a degree $3$ polynomial in $v_4$ and $v_5$.
After restoring the dependence on $v_3$ and introducing $u\equiv v_3,\,v\equiv v_4,\,w\equiv v_5$ we are left with a cubic polynomial $p(u,v,w)$ that
defines the image of the embedding as a hypersurface.
Due to a lack of space we do not provide the result but it can be easily reproduced by following the previously described steps.
Note that for fixed values $v_3,v_4,v_5$ the three polynomials $p_i,\,i=1,\dots,3$ in~\eqref{eqn:codim3equations} become linear in $v_1,v_2,v_6$ and can be used
to solve for these coordinates to define an inverse map.

We can now restrict to the locus $c_{1,1}=c_{2,1}=0$ and find that the polynomial $p(u,v,w)$ indeed factorizes into
\begin{align}
	p(u,v,w)\big|_{c_{1,1}=c_{2,1}=0}=(u\cdot c_{3,13}+v\cdot c_{3,14}+w\cdot c_{3,15})\cdot p'(u,v,w)\,,
	\label{eqn:factorization}
\end{align}
with the quadratic factor given by
\begin{align}
	\begin{split}
	p'(u,v,w)=&u^2 \left(c_{1,5} c_{2,2}-c_{1,2} c_{2,5}\right)+u v \left(c_{1,6} c_{2,2}+c_{1,5} c_{2,3}-c_{1,3} c_{2,5}-c_{1,2} c_{2,6}\right)\\
	&+u w \left(c_{1,7} c_{2,2}+c_{1,5} c_{2,4}-c_{1,4} c_{2,5}-c_{1,2} c_{2,7}\right)+v^2 \left(c_{1,6} c_{2,3}-c_{1,3} c_{2,6}\right)\\
		&+v w \left(c_{1,7} c_{2,3}+c_{1,6} c_{2,4}-c_{1,4} c_{2,6}-c_{1,3} c_{2,7}\right)+w^2 \left(c_{1,7} c_{2,4}-c_{1,4} c_{2,7}\right)\,.
	\end{split}
	\label{eqn:quadfac}
\end{align}
Let us denote the components of the fiber over $c_{1,1}=c_{2,1}=0$ that respectively correspond to the linear and quadratic factor in~\eqref{eqn:factorization} by $C_l$ and $C_q$.
The homogeneous coordinate $v_5$ can be identified with one of the coordinates on $\mathbb{P}^2$ and this implies the intersections
\begin{align}
	[v_5]\cdot C_l=1\,,\quad [v_5]\cdot C_q=2\,.
\end{align}
To obtain the intersections with $[v_6]$ note that the only equation in~\eqref{eqn:codim3equations} that still depends on $v_6$ is $p_3$.
Setting $v_6\rightarrow 0$ and demanding that the linear factor in~\eqref{eqn:factorization} vanishes simultaneously with $p_1,p_2$ and $p_3$ leads to an overdetermined system.
On the other hand, the quadratic factor~\eqref{eqn:quadfac} is not independent.
The homogeneous coordinates $v_6$, $v_1$ and $v_2$ are not allowed to vanish simultaneously and without loss of generality we can assume $v_1\ne 0$ and use the $U(1)$ action to set $v_1=1$.
The factor $p'(u,v,w)$ then arises by solving $p_1$ for $v_2$ and inserting the result into $p_2$.
Together with the fact that $[v_6]\cdot (C_l+C_q)=5$ this leads to the intersections
\begin{align}
	[v_6]\cdot C_l=0\,,\quad [v_6]\cdot C_q=5\,.
\end{align}
We have thus verified that the $I_2$ singular fibers counted by the fiber GV-invariants $\tilde{n}^0_{0,1}$ and $\tilde{n}^0_{5,2}$ indeed correspond to the codimension two locus $c_{1,1}=c_{2,1}=0$ in the base of the fibration.

\subsection{The Tate-Shafarevich group, derived equivalences and F-theory}
\label{sec:tsgroup}
Given a smooth genus one fibration $X$ one can construct the associated Jacobian fibration $J\equiv X^{(0)}$, at least away from the discrimant locus, by replacing each fiber with the corresponding moduli space of degree zero line bundles.
A more refined construction is necessary to incorporate the singular and reducible fibers and can be found e.g. in~\cite{Caldararu:2002ab} but for our purpose we can ignore the associated subtleties.
The result of this construction is an elliptic fibration that leads to the same axio-dilaton profile in an F-theory compactification.
Again up to technical issues related to the singular fibers, one can consider the relative moduli space $X^{(k)}$ of degree one line bundles on the fibers of the genus one fibration and there is an isomorphism $X=X^{(1)}$.

Given a section of $X^{(0)}$ we can consider the tensor product of the corresponding family of degree $0$ sheaves with the degree $1$ sheaves in $X^{(1)}$ and thus define an action of $X^{(0)}$ on $X$.
This turns $X$ into an $X^{(0)}$ torsor and the set of $X^{(0)}$ torsors can in turn be equipped with a group structure.
It is then called the Weil-Ch\^{a}telet group $WC(X^{(0)})$.
The genus one fibrations $X'\in WC(X^{(0)})$ that are ``locally elliptic'', i.e. they do not have multiple fibers, form a subgroup $\Sh(X^{(0)})$ that is called the Tate-Shafarevich group.
The theory of the Tate-Shafarevich group for genus one fibered threefolds was developed in~\cite{dolgachev1992elliptic} and for an explicit description of the group structure on $\Sh(X^{(0)})$ we refer the reader to Appendix A in~\cite{Cvetic:2015moa}.
A precise definition of multiple fibers can be found in~\cite{griffiths1978principles} and here we only note that they are absent in all of the geometries that we consider.

In~\cite{dolgachev1992elliptic} it was proven that the Tate-Shafarevich group associated to a generic fibration of cubics in $\mathbb{P}^2$ over $\mathbb{P}^2$ is given by $\mathbb{Z}_3$.
An analogous argument holds for generic genus one fibered threefold with five-sections and implies that the Tate-Shaferevich group is $\mathbb{Z}_5$, if we assume that the Brauer group of the fibration is trivial.
For Calabi-Yau $d$-folds $X$ with $d\ge 3$ there is an isomorphism~\cite{Batyrev:2005jc}
\begin{align}
	\text{Br}(X)\equiv \text{Tors}(H^3(X,\mathbb{Z}))\,.
\end{align}
Since the genus one fibered Calabi-Yau threefolds that we consider in this paper are smooth and arise via extremal transition from torsion free elliptic fibrations, we thus assume that the corresponding Tate-Shafarevich group is isomorphic to $\mathbb{Z}_5$.

It follows from the arguments in~\cite{caldararuThesis,Caldararu:2002ab} that if $k$ is co-prime to the order of $X$ in $\Sh(J)$ then the element $X'=(X)^k\in\Sh(J)$ is derived equivalent to $X$, i.e. $D^b(X)=D^b(X')$.
Moreover, for any element $X'\in \Sh(J)$ there is an isomorphism $X'=(X')^{-1}$.
Note that $X'$ and $(X')^{-1}$ as elements of $\Sh(J)$ are equipped with an action of the Jacobian fibration and thus the same geometry corresponds to two different elements of the Tate-Shafarevich group.
This has been discussed in the context of genus one fibered Calabi-Yau threefolds with 3-sections in~\cite{Cvetic:2015moa}. 

For smooth genus one fibrations $X$ with $5$-sections and Jacobian fibration $J$ such that $\Sh(J)=\mathbb{Z}_5$ this implies that the Tate-Shafarevich group contains three different geometries, namely the Jacobian elliptic fibration and two smooth genus one fibrations that are not elliptic.
Physically, the smoothness of both genus one fibrations is a consequence of $5$ being a prime number and thus Higgs transitions with non-trivial Wilson line do not produce localized uncharged matter which would geometrically correspond to terminal singularities.
Indeed we found that generic genus one fibered Calabi-Yau threefolds with $5$-sections come in pairs $(X,X^\vee)$ that are derived equivalent and share the same stringy moduli space.
It is natural to assume that these two geometries correspond to the non-trivial elements of the Tate-Shafarevich group.

This would imply that they lead to identical F-theory vacua.
We have seen that for each pair the number of massless hypermultiplets with $\mathbb{Z}_5$ charge $\pm 1$ that arises from $X$ is identical to the number of charge $\pm2$ hypermultiplets that arise from $X^\vee$ and vice versa.
However, the charges can also be exchanged by choosing $e^{4\pi i/5}$ instead of $e^{2\pi i/5}$ as a generator for the $\mathbb{Z}_5$ gauge symmetry.
The massless spectra associated to both geometries are therefore physically equivalent.

Furthermore, the non geometric region between the two conifold loci that separate both geometries shrinks to a flat wall in M-theory~\cite{Witten:1996qb} and approaching this wall from either of the geometric regions amounts to sending the volume of the corresponding generic fiber to zero.
This means that the two genus one fibrations $X$ and $X^\vee$ indeed share the same F-theory limit.
We therefore find strong evidence that they correspond to different elements of the same Tate-Shafarevich group.

Note that in this example the derived equivalence among different elements of the Tate-Shafarevich group proven in~\cite{Caldararu:2002ab} is realized as relative homological projective duality. 
A similar observation has been made in the context of a hybrid phase in the GLSM of the intersection of two quadrics in $\mathbb{P}^3$~\cite{Caldararu:2010ljp}.
This was interpreted by the authors of~\cite{Caldararu:2010ljp} as a homological projective duality involving a non-commutative resolution and related to the twisted derived equivalences among genus one fibrations from~\cite{Caldararu:2002ab}.


%% file: section_5.tex
\section{Classification of fibrations over $B=\mathbb{P}^2$}
\label{sec:classification}
We now want to use the two toric complete intersection fibers that we discussed in the last section and attempt to construct all generic genus one fibrations with 5-sections over $\mathbb{P}^2$~\footnote{Recall that generic means that there are no additional fibral divisors or independent 5-sections, such that the Calabi-Yau $M$ has $h^{1,1}(M)=2$.}.
To this end we first classify all fibrations of the toric complete intersection fibers over $\mathbb{P}^2$ with $h^{1,1}=k+2,\,k=1,2$ and then restrict to those geometries that allow for the Higgs transition $U(1)^k\rightarrow\mathbb{Z}_5$ in the corresponding F-theory spectrum.
In this way we obtain a list of $23$ different genus one fibrations that are contained in $12$ moduli spaces (in one case the same geometry appears both in the Grassmannian and the Pfaffian phase).
For each of the geometries we find a GLSM that reproduces the corresponding GV invariants and fundamental periods via localization.
Moreover, in each case we check that the base independent expressions for the Euler characteristic and the multiplicities of $I_2$ fibers in~\eqref{eqn:genericMultis} hold.
The data will be summarized in Appendix~\ref{sec:fibrationsOverP2}.

As an additional result, we find a generic expression for the fundamental periods of the mirrors.
By scanning over parameters of the generic form for the fundamental period and trying to find associated Picard-Fuchs systems we can reproduce the list of models.
It is of course possible that additional geometries exist with parameters outside the range of our scan, and that do not exhibit an extremal transition to either of the two types of elliptic fibrations.
In particular, there are probably many additional complete intersection fibers in toric varieties that lead to fibrations with Higgs transitions to genus one fibrations with 5-sections.
However, we take the agreement between the list obtained from the two types of fibers and the explicit scan over fundamental periods as evidence, that those transitions will not lead to genus one fibrations outside of our list.

Let us point out that this classification supplements the geometric construction, in that we can not check directly if a given choice of bundles leads to a smooth fibration.
By constructing elliptic fibrations as complete intersections in toric ambient spaces, such that smoothness follows from the results in~\cite{Batyrev:1994pg}, and using Higgs transitions in F-theory we can nevertheless obtain the data of smooth genus one fibrations.
This is possible, because in the transitions that we consider singularities in the fiber could only arise from uncharged localized hypermultiplets~\cite{Arras:2016evy}.
Having base independent expressions for the complete spectrum from Section~\ref{sec:ftheory}, we can easily exclude this possiblity.

\subsection{Warm-up: Classification of cubic hypersurfaces}
Before considering the classification of genus one fibrations with 5-sections, let us first illustrate the procedure for the easier case of genus one fibrations with 3-sections.
Every genus one fibration with a 3-section can be mapped into a fibration of cubics~\cite{Braun:2014oya}.
\begin{figure*}[t]
\centering
\vspace{3cm}
\begin{tikzpicture}[remember picture,overlay,node distance=4mm, >=latex',block/.style = {draw, rectangle, minimum height=65mm, minimum width=83mm,align=center},]
\begin{scope}[shift={(3,0)},scale=1.2]
\draw[help lines, overlay, lightgray!30] (-8.5,-0.5) grid (-3.5,4.5);
\node[overlay] at (-8.5,4) {a), $M$};
\draw [line width=1] (-7,3) -- (-5,2);
\draw [line width=1] (-7,3) -- (-6,1);
\draw [line width=1] (-6,1) -- (-5,2);
\draw [line width=1] (-6,1) -- (-6,2);
\draw [line width=1] (-7,3) -- (-6,2);
\draw [line width=1] (-5,2) -- (-6,2);

\node[overlay] at (-7.3, 3.3) {$w$};
\node[overlay] at (-6.3, .7) {$v$};
\node[overlay] at (-4.7, 2) {$u$};

\node at (-6,2) {\textbullet};
\end{scope}

\begin{scope}[shift={(3,0)},scale=1.2]
\draw[help lines, overlay, lightgray!30] (-3.5,-0.5) grid (2.5,4.5);
\node[overlay] at (-2.5,4) {b), $N$};
\draw [line width=1, lightgray!100] (-1,3) -- ( 2,3);
\draw [line width=1, lightgray!100] ( 2,3) -- (-1,0);
\draw [line width=1, lightgray!100] (-1,0) -- (-1,3);

\node[overlay] at (-1, 3) {\small$w^3$};
\node[overlay] at (-1, 2) {\small$w^2v$};
\node[overlay] at (-1, 1) {\small$w v^2$};
\node[overlay] at (-1, 0) {\small$v^3$};
\node[overlay] at ( 0, 1) {\small$v^2u$};
\node[overlay] at ( 1, 2) {\small$vu^2$};
\node[overlay] at ( 2, 3) {\small$u^3$};
\node[overlay] at ( 1, 3) {\small$wu^2$};
\node[overlay] at ( 0, 3) {\small$w^2u$};
\node[overlay] at ( 0, 2) {\small$uvw$};
\end{scope}
\end{tikzpicture}
\caption{
The dual pair of polytopes $F_1$ and $F_{16}$ is shown in a) and b).
	We also indicate the toric fan obtained from a complete star triangulation of $F_1$ and labelled the homogeneous coordinates that parametrize $\mathbb{P}^2$.
The points of the newton polytope $F_{16}$ correspond to monomials in the homogeneous coordinates which are in turn sections of the anti-canonical bundle on $\mathbb{P}^2$.
}
\label{fig:f1f16}
\end{figure*}
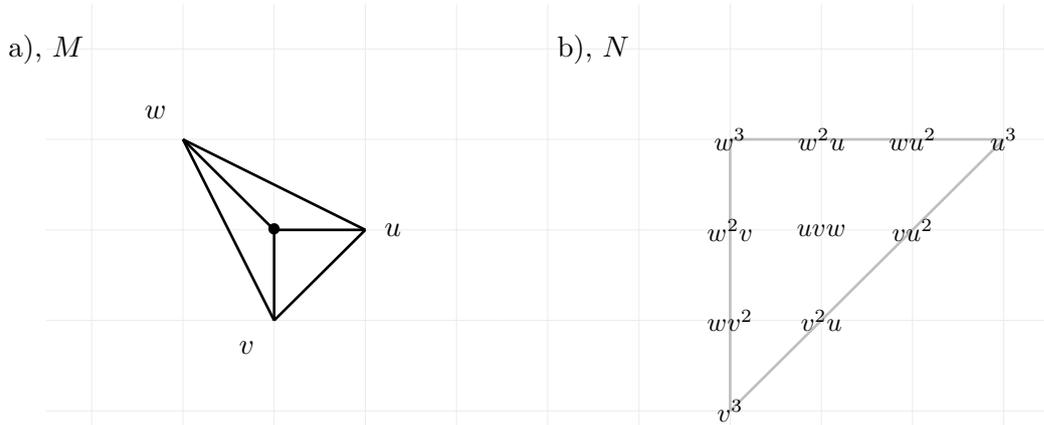
A generic cubic polynomial in the homogenous coordinates on $\mathbb{P}^2$ takes the form
\begin{align}
	\small
	\begin{split}
	p=s_1 u^3+s_2u^2v+s_3uv^2+s_4v^3+s_5u^2w+s_6uvw+s_7v^2w+s_8uw^2+s_9vw^2+s_{10}w^3\,,
	\end{split}
	\label{eqn:cubic}
\end{align}
and by adjunction the vanishing locus $\{p=0\}\subset\mathbb{P}^2$ is an elliptic curve.
For generic choices of coefficients $s_i,\,i=1,...,10$ this curve is smooth.
To construct a family of genus one fibered Calabi-Yau threefolds $M$ we can promote the coefficients to sections of line bundles on a base $B$.
Let us denote the class of the divisor on $M$ that is associated to the homogeneous coordinate $w$ by $[w]=H$.
Then there are vertical divisors $D_{u/v}=\pi^{-1}(D_{u/v}')$ on $M$ that correspond to Cartier divisors $D_u',D_v'$ on $B$ such that
\begin{align}
	[u]=H+D_u\,,\quad[v]=H+D_v\,.
\end{align}
The Calabi-Yau condition $[p]=\pi^*c_1(B)+[u]+[v]+[w]$ fixes 
\begin{align}
	\begin{split}
		[s_1]=&c_1-2D_u+D_v\,,\quad [s_2]=c_1-D_u\,,\quad [s_3]=c_1-D_v\,,\quad [s_4]=c_1+D_u-2D_v\,,\\
		[s_5]=&c_1-D_u+D_v\,,\quad [s_6]=c_1\,,\quad [s_7]=c_1+D_u-D_v\,,\quad [s_8]=c_1+D_v\,,\\
		[s_9]=&c_1+D_u\,,\quad [s_{10}]=c_1+D_u+D_v\,,
	\end{split}
	\label{eqn:sbundles}
\end{align}
where we introduced $c_1=\pi^*c_1(B)$.
The Euler characteristic of $M$ can be expressed as
\begin{align}
	\chi=-6(3c_1^2+{D'}_u^2-D'_uD'_v+{D'}_v^2)\,,
\end{align}
where the right hand side is the intersection of the corresponding divisors on $B$~\cite{Klevers:2014bqa}.
Another important quantity is the number of isolated $I_2$ fibers which is given by
\begin{align}
	n_{I_2}=\frac12\chi+240\,.
\end{align}

Let us now specialize to the case $B=\mathbb{P}^2$ and introduce the homogeneous coordinates
\begin{align}
	[b_1:b_2:b_3]\sim[\lambda b_1:\lambda b_2:\lambda b_3]\quad\text{for all }\lambda\in\mathbb{C}^*\,.
\end{align}
We denote the hyperplane class of $\mathbb{P}^2$ by $H$ and the class of the associated vertical divisor by $D_b=\pi^{-1}H$.
One can then expand the classes of $D_u',D_v'$ as
\begin{align}
	D_u'=p\cdot H\,,\quad D_v'=q\cdot H\,,
	\label{eqn:duvexpansion}
\end{align}
and also replace $c_1=3H$.
In terms of $p,q$ the Euler characteristic is then given by
\begin{align}
	\chi(p,q)=-6(27+p^2-pq+q^2)\,.
\end{align}

To obtain a genus one fibration that does not exhibit a section, or curves of singularities in the fiber that need to be resolved with fibral divisors, we need to ensure that the coefficients $s_1$, $s_4$ and $s_{10}$ are non-vanishing over a generic point of the base $B$.
This is equivalent to demanding that the bundles on $B$ of which those coefficients are supposed to be sections actually admit any non-trivial sections.
In particular it ensures that all of the other coefficients are also generically non-vanishing.

\begin{figure}[h]
\centering
\begin{tikzpicture}
\begin{scope}[shift={(6,0)},scale=1]
\draw[fill=gray, ultra thick] ( 3, 3) -- (0,-3) -- (-3,0) -- cycle; 
\draw [->,line width=1, opacity=.5] (0,-3) -- (0,3.7);
\draw [->,line width=1, opacity=.5] (-3,0) -- (3.7,0);
	\draw[help lines, overlay, black, opacity=.3] (-3,-3) grid[step={(1,1)}] (3,3);
	\node at (.3,3.9) {$q$};
	\node at (-3.5,.5) {$(-3,0)$};
	\node at (.7,-3.4) {$(0,-3)$};
	\node at (3.5,3.5) {$(3,3)$};
	\node at (3.9,.3) {$p$};
	\node at (-3, 0) {\color{blue}$\bullet$};
	\node at (-2,-1) {$\bullet$};
	\node at (-1,-2) {$\bullet$};
	\node at ( 0,-3) {\color{blue}$\bullet$};
	\node at ( 1,-1) {$\bullet$};
	\node at ( 2, 1) {$\bullet$};
	\node at ( 3, 3) {\color{blue}$\bullet$};
	\node at (-1, 1) {$\bullet$};
	\node at ( 1, 2) {$\bullet$};
	\node at ( 2, 2) {$\bullet$};
	\node at (-1, 0) {$\bullet$};
	\node at ( 0,-2) {$\bullet$};
	\node at ( 0,-1) {$\bullet$};
	\node at (-2, 0) {$\bullet$};
	\node at ( 1, 1) {$\bullet$};
	\node at ( 0, 1) {$\bullet$};
	\node at ( 1, 0) {$\bullet$};
	\node at ( 0, 0) {$\bullet$};

	\draw (0,0) circle (.15cm);
	\draw (0,1) circle (.15cm);
	\draw (1,1) circle (.15cm);
	\draw (1,2) circle (.15cm);
	\draw (2,2) circle (.15cm);
	\draw (3,3) circle (.15cm);
	\node at (-1,-1) {$\bullet$};
\end{scope}
\end{tikzpicture}
\caption{The points $(p,q)$ that parametrize the fibrations of $\mathbb{P}^2$ over $B=\mathbb{P}^2$ and the corresponding genus one fibered Calabi-Yau threefolds are restricted to lie
inside a triangle $\Delta_{pq}$ that is congruent to $3\cdot F_1$. A representative set of $6$ inequivalent choices is marked with a circle. The corners of the triangle are highlighted in blue and correspond
to elliptic fibrations where the 3-section splits into three sections}
\label{fig:f1overF0charges}
\end{figure}
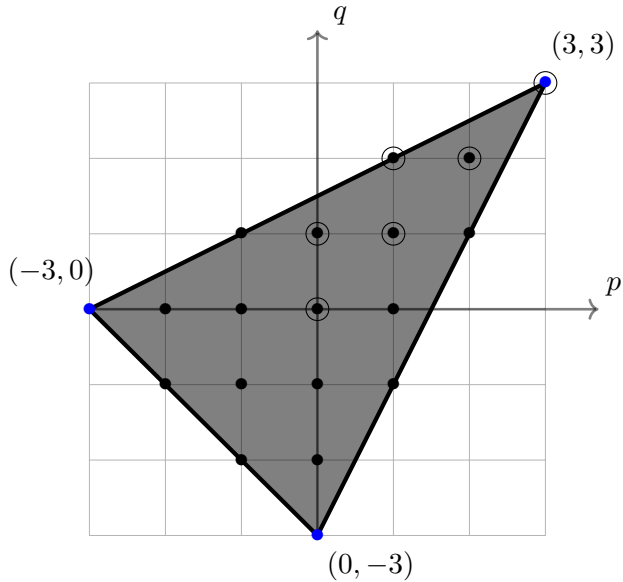

The number of sections of a bundle $\mathcal{O}(a H)$ on $\mathbb{P}^2$ is greater than zero iff $a\ge 0$.
Using~\eqref{eqn:sbundles} together with~\eqref{eqn:duvexpansion} we can therefore translate the requirement that for generic choice of complex structure on $M$ the coefficients $s_1,s_4,s_{10}$ are non-trivial sections into the set of inequalities
\begin{align}
	q-2p\ge-3\,,\quad p-2q\ge -3\,,\quad p+q\ge-3\,.
\end{align}
It is easy to see that this restricts the points $(p,q)$ to lie inside the triangle $\Delta_{pq}$ depicted in Figure~\ref{fig:f1overF0charges} that is congruent to $3\cdot F_1$.
This procedure for obtaining a complete set of fibrations of a hypersurface in a toric ambient space over a given base was first described in~\cite{Cvetic:2013uta} at the hand of the generic elliptic curves in the Hirzebruch surface $\mathbb{F}_1$.
We will also treat that case below.
In the case of fibrations of cubics over $\mathbb{P}^2$ we find $19$ potential choices for $p,q$ and from all of these choices we can recover a reflexive polytope by considering generators of the kernel of the charge matrix
\begin{align}
	Q=\left(\begin{array}{rrrrrr}
	1&1&1&0&0&0\\
	p&q&0&1&1&1\\
	\end{array}\right)\,.
\end{align}

The corresponding toric data reads as follows:
\begin{align}
\begin{blockarray}{crrrrl}
	&&&&\\
\begin{block}{c(rrrr)l}
	u& 1& 0& 0& 0&\leftarrow\text{3-section }H+D_u\\
	v& 0& 1& 0& 0&\leftarrow\text{3-section }H+D_v\\
	w&-1&-1& 0& 0&\leftarrow\text{3-section }H\\
	b_1& 0& 0& 1& 0&\leftarrow\text{vertical divisor }D_b\\
	b_2& 0& 0& 0& 1&\phantom{x}\hspace{1.5cm}\text{\ditto}\\
	b_3&-p&-q&-1&-1&\phantom{x}\hspace{1.5cm}\text{\ditto}\\
\end{block}
\end{blockarray}
\label{eqn:pqpoints}
\end{align}
The rows correspond to generators of the 1-dimensional cones of a toric fan and we indicate the corresponding homogeneous coordinates and divisor classes.
Moreover, the Stanley-Reiner ideal is always given by
\begin{align}
	\mathcal{SRI}=\langle b_1b_2b_3,\,uvw\rangle\,.
\end{align}
Using the normal form for reflexive polytopes~\cite{Kreuzer:2002uu} we find that there are only $6$ inequivalent choices for $(p,q)$ and a representative set is given by
\begin{align}
	(p,q)\in\{(0,0),\,(0,1),\,(1,1),\,(1,2),\,(2,2),\,(3,3)\}\,.
\end{align}
The choice $(p,q)=(3,3)$ is special in that the ramification locus of the 3-section becomes trivial and it splits into three sections.

Let us try to understand this directly from the cubic equation~\eqref{eqn:cubic}.
We have three different $3$-sections that are induced from the ambient space, namely those corresponding to $\{u=0\},\,\{v=0\}$ and $\{w=0\}$.
Setting $u=0$ and choosing the patch $w=1$ we obtain the cubic polynomial
\begin{align}
	s_4v^3+s_7v^2+s_9v+s_{10}=0\,,
\end{align}
with discriminant
\begin{align}
	\Delta_u=-27 s_{10}^2 s_4^2 - 4 s_{10} s_7^3 + 18 s_{10} s_4 s_7 s_9 + s_7^2 s_9^2 - 4 s_4 s_9^3\,.
\end{align}
The corresponding divisor class is $[\Delta_u]=2(2c_1+2D_u-D_v)$.
For the other sections we can perform an analogous calculation and obtain $[\Delta_v]=2(2c_1-D_u+2D_v)$ as well as $[\Delta_w]=2(2c_1-D_u-D_v)$.
It turns out that the corners of the polytope on Figure~\ref{fig:f1overF0charges} correspond to those fibrations where one of the three classes is trivial.

\begin{figure}[h!]
\centering
\begin{tikzpicture}
\begin{scope}[shift={(6,0)},scale=2]
	\draw[fill=gray, ultra thick] ( 3, 3) -- ( 3, 0) -- (0,-3) -- (-3,0) -- cycle; 
\draw [dashed, -,line width=1, opacity=.5] (0,-3) -- (3,3);
\draw [->,line width=1, opacity=.5] (0,-3) -- (0,3.7);
\draw [->,line width=1, opacity=.5] (-3,0) -- (3.7,0);
	\draw[help lines, overlay, black, opacity=.3] (-3,-3) grid[step={(1,1)}] (3,3);
	\node at (.3,3.9) {$q$};
	\node at (-3.5,.5) {$(-3,0)$};
	\node at (3.5,.5) {$(3,0)$};
	\node at (.7,-3.4) {$(0,-3)$};
	\node at (3.5,3.5) {$(3,3)$};
	\node at (3.9,.3) {$p$};

	\node at (-3, 0) {\color{blue}$\bullet$};
	\draw[opacity=.5, white, fill] (-3,0) circle (.4);
	\node at (-3,0) {$\begin{array}{c}h^{1,1}=5\\n_3=36\end{array}$};

	\node at (-2,-1) {$\bullet$};
	\draw[opacity=.5, white, fill] (-2,-1) circle (.4);
	\node at (-2,-1) {$\begin{array}{c}h^{1,1}=3\\n_3=20\end{array}$};

	\node at (-1,-2) {$\bullet$};
	\draw[opacity=.5, white, fill] (-1,-2) circle (.4);
	\node at (-1,-2) {$\begin{array}{c}h^{1,1}=3\\n_3=8\end{array}$};

	\node at ( 0,-3) {\color{blue}$\bullet$};
	\draw[opacity=.5, white, fill] (0,-3) circle (.4);
	\node at (0,-3) {$\begin{array}{c}h^{1,1}=4\\n_3=0\end{array}$};

	\node at ( 1,-1) {$\bullet$};
	\draw[opacity=.5, white, fill] (1,-1) circle (.4);
	\node at (1,-1) {$\begin{array}{c}h^{1,1}=3\\n_3=2\end{array}$};

	\node at ( 2, 1) {$\bullet$};
	\draw[opacity=.5, white, fill] (2,1) circle (.4);
	\node at (2,1) {$\begin{array}{c}h^{1,1}=3\\n_3=2\end{array}$};

	\node at ( 3, 3) {\color{blue}$\bullet$};
	\draw[opacity=.5, white, fill] (3,3) circle (.4);
	\node at (3,3) {$\begin{array}{c}h^{1,1}=4\\n_3=0\end{array}$};

	\node at ( 3, 2) {$\bullet$};
	\draw[opacity=.5, white, fill] (3,2) circle (.4);
	\node at (3,2) {$\begin{array}{c}h^{1,1}=3\\n_3=0\end{array}$};

	\node at ( 3, 1) {$\bullet$};
	\draw[opacity=.5, white, fill] (3,1) circle (.4);
	\node at (3,1) {$\begin{array}{c}h^{1,1}=3\\n_3=0\end{array}$};

	\node at ( 3, 0) {\color{blue}$\bullet$};
	\draw[opacity=.5, white, fill] (3,0) circle (.4);
	\node at (3,0) {$\begin{array}{c}h^{1,1}=3\\n_3=0\end{array}$};

	\node at ( 2, 0) {$\bullet$};
	\draw[opacity=.5, white, fill] (2,0) circle (.4);
	\node at (2,0) {$\begin{array}{c}h^{1,1}=3\\n_3=1\end{array}$};

	\node at ( 2,-1) {$\bullet$};
	\draw[opacity=.5, white, fill] (2,-1) circle (.4);
	\node at (2,-1) {$\begin{array}{c}h^{1,1}=3\\n_3=0\end{array}$};

	\node at (-1, 1) {$\bullet$};
	\draw[opacity=.5, white, fill] (-1,1) circle (.4);
	\node at (-1,1) {$\begin{array}{c}h^{1,1}=3\\n_3=20\end{array}$};

	\node at ( 1, 2) {$\bullet$};
	\draw[opacity=.5, white, fill] (1,2) circle (.4);
	\node at (1,2) {$\begin{array}{c}h^{1,1}=3\\n_3=8\end{array}$};

	\node at ( 2, 2) {$\bullet$};
	\draw[opacity=.5, white, fill] (2,2) circle (.4);
	\node at (2,2) {$\begin{array}{c}h^{1,1}=3\\n_3=3\end{array}$};

	\node at (-1, 0) {$\bullet$};
	\draw[opacity=.5, white, fill] (-1,0) circle (.4);
	\node at (-1,0) {$\begin{array}{c}h^{1,1}=3\\n_3=16\end{array}$};

	\node at ( 0,-2) {$\bullet$};
	\draw[opacity=.5, white, fill] (0,-2) circle (.4);
	\node at (0,-2) {$\begin{array}{c}h^{1,1}=3\\n_3=3\end{array}$};

	\node at ( 0,-1) {$\bullet$};
	\draw[opacity=.5, white, fill] (0,-1) circle (.4);
	\node at (0,-1) {$\begin{array}{c}h^{1,1}=3\\n_3=6\end{array}$};

	\node at (-2, 0) {$\bullet$};
	\draw[opacity=.5, white, fill] (-2,0) circle (.4);
	\node at (-2,0) {$\begin{array}{c}h^{1,1}=3\\n_3=25\end{array}$};
	
	\node at ( 1, 1) {$\bullet$};
	\draw[opacity=.5, white, fill] (1,1) circle (.4);
	\node at (1,1) {$\begin{array}{c}h^{1,1}=3\\n_3=6\end{array}$};

	\node at ( 0, 1) {$\bullet$};
	\draw[opacity=.5, white, fill] (0,1) circle (.4);
	\node at (0,1) {$\begin{array}{c}h^{1,1}=3\\n_3=12\end{array}$};

	\node at ( 1, 0) {$\bullet$};
	\draw[opacity=.5, white, fill] (1,0) circle (.4);
	\node at (1,0) {$\begin{array}{c}h^{1,1}=3\\n_3=4\end{array}$};

	\node at ( 1,-2) {$\bullet$};
	\draw[opacity=.5, white, fill] (1,-2) circle (.4);
	\node at (1,-2) {$\begin{array}{c}h^{1,1}=3\\n_3=0\end{array}$};

	\node at ( 0, 0) {$\bullet$};
	\draw[opacity=.5, white, fill] (0,0) circle (.4);
	\node at (0,0) {$\begin{array}{c}h^{1,1}=3\\n_3=9\end{array}$};

	\node at (-1,-1) {$\bullet$};
	\draw[opacity=.5, white, fill] (-1,-1) circle (.4);
	\node at (-1,-1) {$\begin{array}{c}h^{1,1}=3\\n_3=12\end{array}$};
\end{scope}
\end{tikzpicture}
\caption{The points $(p,q)$ that parametrize the fibrations of Hirzebruch surface $\mathbb{F}_1$ over $B=\mathbb{P}^2$, as well as the corresponding genus one fibered Calabi-Yau threefolds, are restricted to lie
inside a polytope that is congruent to $3\cdot F_3$. At each point we indicate the Hodge number $h^{1,1}$ of the Calabi-Yau as well as the number $n_3$ of isolated reducible fibers that lead to charge $3$ hyper multiplets.}
\label{fig:f3overF0charges}
\end{figure}
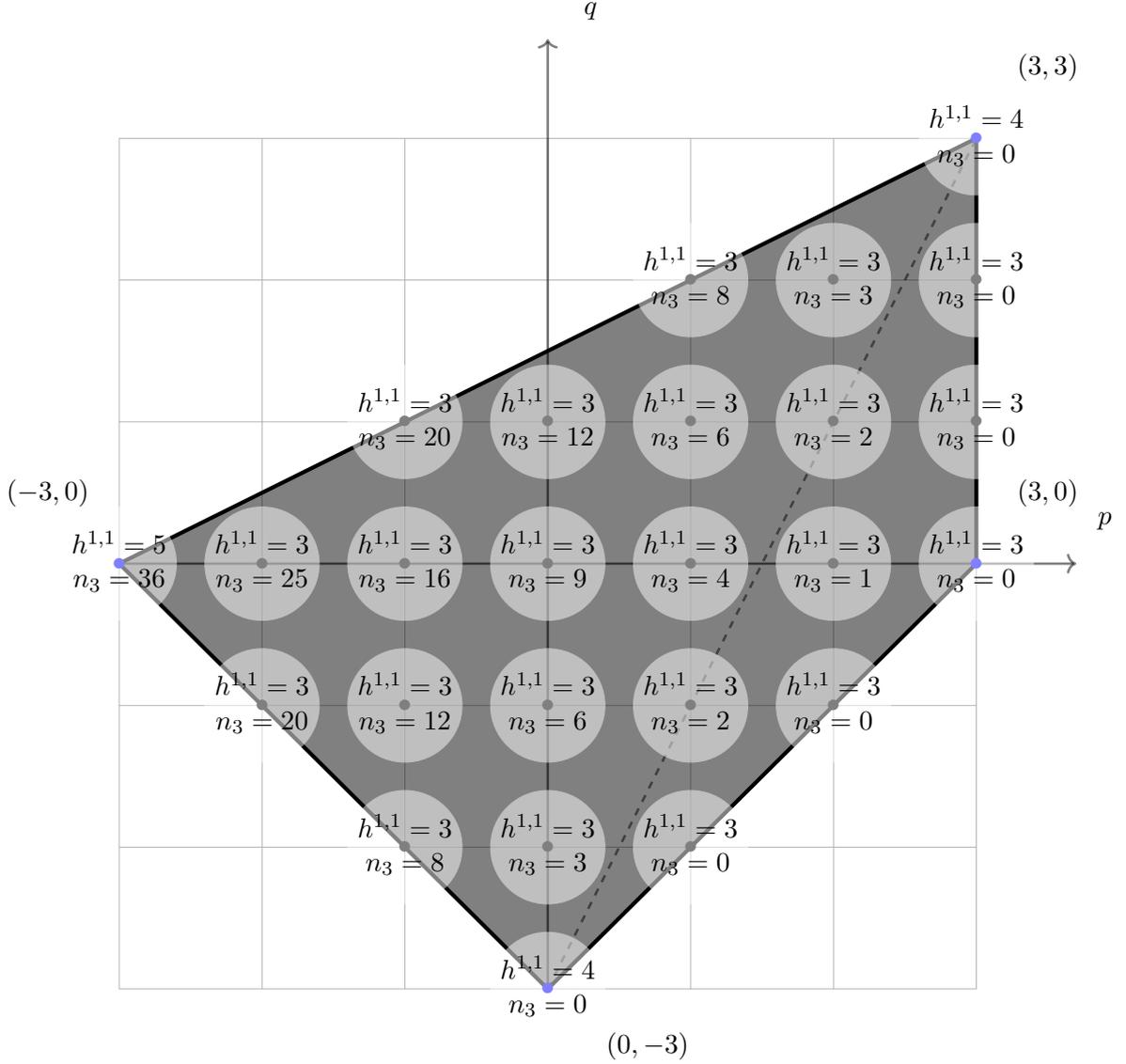

From a fibration of cubics we can, via an extremal transition that physically amounts to unhiggsing $\mathbb{Z}_3\rightarrow U(1)$, go to a fibration of hypersurfaces in the Hirzebruch surface $\mathbb{F}_1$.
Concretely one performs a complex structure deformation such that in the defining polynomial~\eqref{eqn:cubic} the coefficient $s_{1}$ becomes zero and then resolves the resulting singularity using one toric blow-up.
If on the other hand we directly start with fibrations of hypersurfaces in $\mathbb{F}_1$ over $\mathbb{P}^2$ there are additional choices for $(p,q)$ that lead to smooth Calabi-Yau threefolds.
We indicate the corresponding Hodge numbers $h^{1,1}$ as well as the number of charge $3$ hyper multiplets in Figure~\ref{fig:f3overF0charges}.
The latter can be obtained using GV-spectroscopy~\cite{Paul-KonstantinOehlmann:2019jgr} or the results from~\cite{Klevers:2014bqa}.

It turns out that Higgs transitions to fibrations of cubics are only possible for values of $(p,q)$ that are inside the admissible region shown in Figure~\ref{fig:f1overF0charges}.
The reason is that, to form a D-flat direction in the corresponding six-dimensional supergravity effective action, one needs at least two hypermultiplets with identical quantum numbers~\cite{Honecker:2006qz}.
From a geometrical perspective this manifests itself such that the polytope that is obtained by removing the exceptional divisor is not reflexive for choices of $(p,q)$ outside of that region.
Let us further note that via lattice automorphisms the values $(p,q)$ are equivalent to $(p-q,-q)$ such that we can restrict ourselves to $q\ge 0$.
Moreover, all of the values in the admissible part of the upper half-plane in Figure~\ref{fig:f3overF0charges} lead to different topological invariants (for $p=3$ they differ by the number of reducible fibers that lead to charge $2$ matter).

We therefore found a way to classify all inequivalent fibrations of cubics over $\mathbb{P}^2$ by studying fibrations of generic hypersurfaces in $\mathbb{F}_1$ over the same base.
One just needs to construct all possible fibrations of the latter type, which can be done systematically, and then check which of those have $h^{1,1}=3$ and lead to enough hyper multiplets in the associated effective action to perform the Higgs transition.

This criterion can also be applied for different toric bases and, most important for us, to the codimension three and five complete intersection fibers that exhibit extremal transitions to genus one fibrations with five-sections.
To this end, for a given set of GLSM charges, we need to check that the complete intersection fibration $M$ does not contain non-toric divisors and the multiplicities of the hyper multiplets that are involved in the Higgs transition do not vanish.
We then expect that the corresponding fibration of Pfaffian genus one curves exhibits a genuine $5$-section and has $h^{1,1}=1+h^{1,1}(B)$.

\subsection{Classification of toric complete intersections}
\label{sec:cicyclassification}
The classification of admissible line bundles can be extended to complete interesections in toric ambient spaces.
This has been demonstrated for certain codimension two complete intersections in~\cite{Cvetic:2013qsa} and we will start by extending the procdure to more general fibers.

For ease of exposition we will again restrict to the base $B=\mathbb{P}^2$ and can then choose the parametrization of the fibration structure such that only the first $r+1$ homogeneous coordinates $v_1,\dots,v_{r+1}$ of the $r+1$-dimensional ambient space of the fiber transform non-trivially under the associated $U(1)$ action.
We denote the corresponding weights by $d_1,\dots,d_{r+1}$ and the remaining weights $d_{r+2},\dots,d_{r+k+1}$ are zero.
Let us then write the $r$ sections that define the complete intersection in the form
\begin{align}
p_m=\sum\limits_{y_j\in\Delta_m}a_{m,j}\prod\limits_{n=1}^r\prod\limits_{x_i\in\nabla_n}v_i^{\langle y_j,x_i\rangle +\delta_{n,m}}\,,\quad m=1,\dots,r\,,
\end{align}
with the coefficients $a_{m,j}$ being sections of line bundles on $\mathbb{P}^2$ such that $p_m$ are sections
\begin{align}
p_m\in\Gamma\left(\mathbb{P}^2,\mathcal{O}(s_m H)\right)\,,\quad s_m\ge 0\,,
\end{align}
and the Calabi-Yau condition imposes
\begin{align}
\sum\limits_{i=1}^r s_i=\sum\limits_{i=1}^{r+1}d_i+3\,.
\label{eqn:classificationCYcondition}
\end{align}

In order for the coefficients $a_{m,j}$ to be non-vanishing, the associated divisors on $\mathbb{P}^2$ need to be effective.
This is equivalent to the conditions
\begin{align}
s_m-\sum\limits_{n=1}^r\sum\limits_{x_i\in\nabla_n}\left(\langle y,x_i\rangle+\delta_{m,n}\right)d_i\ge 0\,,\quad\forall m=1,\dots,r\,,\quad y\in\Delta_m\,.
	\label{eqn:classIneq}
\end{align}
Together with the conditions $s_m\ge 0$ this defines a reflexive Gorenstein cone in $\mathbb{Z}^{r}\times\mathbb{Z}^{r+1}$ with coordinates $(s_1,\dots,s_r,d_1,\dots,d_{r+1})$.
It is actually equivalent to the reflexive Gorenstein cone that defines the complete intersection in the toric ambient space of the fiber~\cite{Batyrev:1997tv}.
In order to obtain the admissible values for $s_i,d_j$ we just have to intersect this cone with the Calabi-Yau condition~\eqref{eqn:classificationCYcondition}.

\subsubsection{Toric complete intersections with $G=U(1)^2$}
\label{classifyU12}
Let us apply this machinery to the codimension five complete intersection fibers from Section~\ref{sec:codim5fibers}.
The toric data~\eqref{eqn:toricdata} and sections~\eqref{eqn:codim5equations} determine $50$ inequalities~\eqref{eqn:classIneq}, which due to spatial constraints we are not able to list here.
The inequalities determine a reflexive Gorenstein cone and, after intersecting with the Calabi-Yau condition~\eqref{eqn:cycondition},
we obtain the polytope $\Delta=3\cdot \Delta'$ in $\mathbb{Z}^{r}\times\mathbb{Z}^{r+1}$, with $\Delta'$ being the convex hull of the rows of the matrix
\begin{align}
	M=\left(
\begin{array}{ccccc|cccccc}
 0 & 0 & 0 & 0 & 0 & -1 & 0 & 0 & 0 & 0 & 0 \\
 1 & 1 & 1 & 1 & 1 & 1 & 1 & -1 & 1 & 1 & 1 \\
 1 & 1 & 1 & 1 & 1 & 1 & 0 & 0 & 1 & 1 & 1 \\
 1 & 0 & 0 & 0 & 0 & 0 & 0 & 0 & 0 & 0 & 0 \\
 0 & 1 & 0 & 0 & 0 & 0 & 0 & 0 & 0 & 0 & 0 \\
 0 & 0 & 1 & 0 & 0 & 0 & 0 & 0 & 0 & 0 & 0 \\
 0 & 0 & 0 & 1 & 0 & 0 & 0 & 0 & 0 & 0 & 0 \\
 0 & 0 & 0 & 0 & 1 & 0 & 0 & 0 & 0 & 0 & 0 \\
 0 & 0 & 0 & 0 & 0 & 0 & 0 & 1 & -1 & -1 & 0 \\
 0 & 0 & 0 & 0 & 0 & 0 & 0 & 0 & 0 & 0 & -1 \\
 0 & 0 & 0 & 0 & 0 & 0 & 0 & 0 & 0 & -1 & 0 \\
 0 & 0 & 0 & 0 & 0 & 0 & 0 & 0 & -1 & 0 & 0 \\
 0 & 0 & 0 & 0 & 0 & 0 & 0 & -1 & 0 & 0 & 0 \\
 0 & 0 & 0 & 0 & 0 & 0 & -1 & 0 & 0 & 0 & 0 \\
\end{array}
\right)\,.
\end{align}
The parameters~\eqref{eqn:shiftedx} and~\eqref{eqn:shiftedd} are constrained to take values in this polytope, i.e.
\begin{align}
	(s_1,s_2,s_3,s_4,s_5,d_1,d_2,d_3,d_4,d_5,d_6)\in\Delta\,.
\end{align}
Moreover, each of the $532$ points corresponds to a valid choice for the parameters and thus to an elliptic fibration over $\mathbb{P}^2$.
As for the fibrations of cubics we do not expect all of those elliptic fibrations to be inequivalent.
However, we are mainly interested in the inequivalent genus one fibrations with five sections, that we can obtain via an extremal transition and that we are now going to discuss.

Let us recall that the matching between the fibration parameters of the codimension five fibers with the non-Abelian GLSMs that engineer generic 5-section fibrations performed in~\eqref{eqn:xptods} implies the non-trivial relations
\begin{align}
	d_1=d_2+d_3\,,\quad d_6=d_3+d_4+d_5\,.
	\label{eqn:drelation}
\end{align}
They are only satisfied for $65$ points in $\Delta'$ and these correspond to the five inequivalent GLSM charges
\begin{align}
  \label{eq:95}
	\begin{array}{c|rrrrr|rrrrr|ccc}
		\#&\multicolumn{5}{c}{q(p_1)\dots q(p_5)}&\multicolumn{5}{|c|}{q(x_1)\dots q(x_5)}&n_{\pm1}&n_{\pm2}&\chi\\\hline
		1&-2&-1&0&0&0&0&0&0&0&0&100&125&-90\\
		2&-1&-1&-1&0&0&0&0&0&0&0&105&120&-90\\
		3&-1&0&0&0&0&-1&0&0&0&0&108&115&-94\\
		4&-1&-1&-1&-1&-1&1&0&0&0&0&110&113&-94\\
		5&-3&0&0&0&0&0&0&0&0&0&90&135&-90\\
	\end{array}\,,
\end{align}
where the complete field content is given by
\begin{align}
	\begin{array}{c|ccccc|c}
		&p_{i,\,i=1,\dots,5}&x_{i,\,i=1,\dots,5}&b_1&b_2&b_3&\textrm{FI}\\\hline
		U(2)&\det^{-1}&\square&0&0&0&\zeta_1\\
		U(1)&q(p_i)&q(x_i)&1&1&1&\zeta_2
	\end{array}\,.
	\label{eqn:p2fieldcontent}
\end{align}
Note that we have used the action of the determinant of $U(2)$ to set $q(x_5)=0$.

For the first four sets of charges we have used CohomCalg~\cite{Blumenhagen:2010pv,cohomCalg:Implementation} to check that the associated codimension five complete intersections, that are connected to the genus one fibrations with $5$-sections via extremal transition, satisfy $h^{1,1}=4$.
The codimension five complete intersection that is associated to the fifth set of charges corresponds to a vertex of $\Delta$.
In particular, the corresponding F-theory effective action does not exhibit any hypermultiplets that would induce a higgs transition $U(1)^2\rightarrow \mathbb{Z}_5$.
This can be checked by plugging the GLSM charges into~\eqref{eqn:xptods} and the resulting values for $d_i,s_i$ into the generic expressions in Table~\ref{tab:codim5gv_2} for $n_{\pm(4,3)}=\tilde{n}^0_{0,1,0}$ and $n_{\pm(1,2)}=\tilde{n}^0_{0,0,1}$.
Analogous to the example of the cubic fibrations, we therefore expect for this choice of parameters that the $5$-section of the genus one fibration is actually the union of multiple independent sections.
This is in line with the geometric calculation of the associated Hodge number $h^{1,1}=6$ in Table~\ref{tab:1}.

Nevertheless, many of the geometries that do not satisfy the relations~\eqref{eqn:drelation} still lead to F-theory vacua with a number of Higgs multiplets that is sufficient to perform the transition $U(1)^2\rightarrow\mathbb{Z}_5$ in supergravity.
For all of those geometries we construct a polytope and nef-partition from the charge vectors, triangulate the polytope and determine the Mori cone with SageMath~\cite{sagemath}.
With CohomCalg~\cite{Blumenhagen:2010pv,cohomCalg:Implementation} we also check that the fibrations have $h^{1,1}=4$.
Using the generic formula from~\cite{Hosono:1994ax}, we then write down the fundamental period for the mirror of the elliptic fibration, and take the limit that is dual to the extremal transition discussed in Section~\ref{sec:higgsu12}.
In this way we find a total of $20$ inequivalent two-parameter fundamental periods for mirrors of genus one fibrations, that all take the form
\begin{align}
	\begin{split}
	w_0=&\sum\limits_{\lambda_1,\lambda_2=0}^\infty\frac{{_3}F_2(d_1\lambda_2-\lambda_1,-\lambda_1,1-d_2\lambda_2+\lambda_1;1-e_1\lambda_2,1-e_2\lambda_2;1)}{\Gamma(1-e_1\lambda_2)\Gamma(1-e_2\lambda_2)\Gamma(1+\lambda_2)^3}\\
		&\cdot\prod\limits_{i=1}^3\frac{\Gamma(1+a_i\lambda_2+\lambda_1)}{\Gamma(1+b_i\lambda_2+\lambda_1)}z_1^{\lambda_1}z_2^{\lambda_2}\,.
	\end{split}
	\label{eqn:fundamental}
\end{align}
The parameters for the fundamental period~\eqref{eqn:fundamental}, as well as representative values for the parameters of the associated elliptic fibration, are listed in Table~\ref{tab:codim5higgsList}.
\begin{table}[h!]
{\small
\begin{align*}
	\arraycolsep=2.5pt
	\begin{array}{c|ccc||rrr|rrr|rr|rr||rrrrr|rrrrrr}
		\#&n_{\pm1}&n_{\pm2}&\chi&a_1&a_2&a_3&b_1&b_2&b_3&d_1&d_2&e_1&e_2&s_1&s_2&s_3&s_4&s_5&d_1&d_2&d_3&d_4&d_5&d_6\\\hline
		 1_a&100 & 125 & -90 &  0 & 1 & 2 &  0 &  0 &  0 & 0 & 0 & 0 & 0 & 2 & 1 & 0 & 0 & 0 & 0 & 0 & 0 & 0 & 0 & 0 \\
		 1_b&125 & 100 & -90 &  0 & 0 & 0 & -2 & -1 &  0 & 0 & 0 & 0 & 0 & 0 & 0 & 0 & 0 & 0 &-1 & 0 & 0 & 0 & 0 &-2 \\\hline
		 2_a&105 & 120 & -90 &  0 & 0 & 2 & -1 &  0 &  0 & 0 & 0 & 0 & 0 & 2 & 0 & 0 & 0 & 0 &-1 & 0 & 0 & 0 & 0 & 0 \\
		2'_a&105 & 120 & -90 &  1 & 1 & 1 &  0 &  0 &  0 & 0 & 0 & 0 & 0 & 1 & 1 & 1 & 0 & 0 & 0 & 0 & 0 & 0 & 0 & 0 \\
		 2_b&120 & 105 & -90 &  0 & 0 & 1 & -2 &  0 &  0 & 0 & 0 & 0 & 0 & 1 & 0 & 0 & 0 & 0 &-2 & 0 & 0 & 0 & 0 & 0 \\\hline
		 3_a&110 & 115 & -90 &  0 & 1 & 1 & -1 &  0 &  0 & 0 & 0 & 0 & 0 & 1 & 1 & 0 & 0 & 0 & 0 & 0 & 0 & 0 & 0 &-1 \\
		 3_b&115 & 110 & -90 &  0 & 0 & 1 & -1 & -1 &  0 & 0 & 0 & 0 & 0 & 1 & 0 & 0 & 0 & 0 &-1 & 0 & 0 & 0 & 0 &-1 \\\hline
		 4_a&105 & 118 & -94 &  0 & 0 & 1 &  0 &  0 &  0 & 0 & 1 & 1 & 1 & 2 & 1 & 1 & 1 & 1 & 1 & 0 & 0 & 0 & 1 & 1 \\
		 4_b&118 & 105 & -94 &  0 & 0 & 0 & -2 & -1 &  0 & 1 & 0 & 0 & 0 & 1 & 1 & 1 & 1 & 1 & 1 & 1 &-1 & 1 & 1 &-1 \\\hline
		 5_a&108 & 115 & -94 &  0 & 0 & 1 & -1 &  0 &  0 & 0 & 1 & 0 & 1 & 2 & 1 & 1 & 1 & 1 & 1 & 0 & 1 & 0 & 0 & 1 \\
		 5_b&115 & 108 & -94 & -1 & 0 & 0 & -2 &  0 &  0 & 0 & 1 & 1 & 1 & 0 & 0 & 0 & 0 & 0 & 0 & 0 &-1 & 0 & 0 &-2 \\\hline
		 6_a&110 & 113 & -94 &  0 & 0 & 0 & -1 &  0 &  0 & 0 & 1 & 1 & 1 & 0 & 0 & 0 & 0 & 0 &-1 & 0 &-1 & 0 & 0 &-1 \\
		 6_b&113 & 110 & -94 &  0 & 0 & 0 & -1 & -1 &  0 & 0 & 1 & 0 & 1 & 1 & 1 & 1 & 1 & 1 & 1 & 0 & 1 & 0 & 0 & 0 \\\hline
		 8_a&104 & 118 & -96 &  0 & 0 & 1 & -1 &  0 &  0 & 1 & 1 & 0 & 1 & 1 & 0 & 0 & 0 & 0 & 0 &-1 & 0 &-1 & 0 & 0 \\
		 8_b&118 & 104 & -96 &  0 & 0 & 0 & -2 & -1 &  0 & 0 & 1 & 0 & 0 & 0 & 0 & 0 & 0 & 0 &-2 & 0 & 1 &-1 &-1 & 0 \\\hline
		 9_a&108 & 114 & -96 &  0 & 0 & 1 & -1 &  0 &  0 & 0 & 0 & 0 & 1 & 2 & 1 & 1 & 1 & 1 & 0 & 0 & 0 & 1 & 1 & 1 \\\hline
		 10_a&109 & 113 & -96 &  0 & 0 & 0 & -1 & -1 &  0 & 1 & 1 & 0 & 1 & 1 & 1 & 1 & 1 & 1 & 1 &-1 & 1 & 0 & 0 & 1 \\
		 10_b&113 & 109 & -96 &  0 & 0 & 0 & -1 & -1 & -1 & 0 & 1 & 0 & 0 & 0 & 0 & 0 & 0 & 0 &-1 & 0 & 1 &-1 &-1 &-1 \\\hline
		11_{ab}&110 & 110 & -100 & 0 & 0 & 0 &  0 &  0 &  0 & 0 & 2 & 2 & 1 & 1 & 1 & 1 & 1 & 1 & 1 & 0 & 0 &-1 & 1 & 1 \\\hline
		12_b&110 & 108 & -104 & 0 & 0 & 0 & -1 &  0 &  0 & 0 & 1 & 0 & 2 & 2 & 2 & 2 & 2 & 2 & 2 & 0 & 1 & 1 & 1 & 2 \\
	\end{array}
\end{align*}
}
	\caption{The data of 19 genus one fibrations with $5$-sections, that are connected via extremal transitions to fibrations of the codimension five complete intersection fiber from Section~\ref{sec:codim5fibers} over $\mathbb{P}^2$.
	The values for $s_i,\,i=1,\dots 5$ and $d_i,\,i=1,\dots,6$ are representative and there are in general multiple elliptic fibrations that admit a transition to a given genus one fibration..
	}
	\label{tab:codim5higgsList}
\end{table}
Two of those fundamental periods, number $2$ and $2'$, correspond to genus one fibrations that are Wall equivalent and their intersection numbers, as well as the Gopakumar-Vafa invariants, are identical after a change of basis for the K\"ahler cone.
We therefore find $19$ inequivalent genus one fibrations over $\mathbb{P}^2$.

In Table~\ref{tab:codim5higgsList} we have grouped pairs of geometries for which the values $n_{\pm1}$ and $n_{\pm2}$ are exchanged.
This suggests that they appear in the same moduli space with one being realized as a Grassmannian fibration and the other as a Pfaffian fibration.
Transforming the variables in the Picard-Fuchs systems we have checked that the first assumption indeed holds true.
Later we are going to construct a GLSM for each of the geometries that reproduces the same invariants and, using localization, the correct fundamental period.
This verifies also the second assumption.

\subsubsection{Toric complete intersections with $G=U(1)$}
\label{sec:toric-compl-inters}
Let us now apply the same reasoning to the codimension three complete intersection fibers from Section~\ref{sec:codim3fiber} and construct the corresponding fibrations over $B=\mathbb{P}^2$.
We find that $(s_1,\dots,s_3,d_1\dots,d_4)$ are constrained to take values in the polytope $\Delta=3\cdot \Delta'\subset \mathbb{Z}^3\times\mathbb{Z}^4$ with $\Delta'$ being the convex hull of the rows of the matrix
\begin{align}
	M=\left(\begin{array}{rrr|rrrr}
 1 & 0 & 0 & 0 & 0 & 0 & 0 \\
 0 & 1 & 0 & 0 & 0 & 0 & 0 \\
 0 & 0 & 1 & 0 & 0 & 0 & 0 \\
 1 & 1 & 1 & 1 & 1 & 0 & 0 \\
 0 & 0 & 1 & -1 & -1 & 1 & 1 \\
 0 & 0 & 0 & 0 & 0 & 0 & -1 \\
 0 & 0 & 0 & 0 & 0 & -1 & 0 \\
 0 & 0 & 0 & 0 & -1 & 0 & 0 \\
 0 & 0 & 0 & -1 & 0 & 0 & 0 \\
	\end{array}\right).
\end{align}
Following the same procedure as in Section~\ref{sec:cicyclassification} we obtain $10$ distinct genus one fibrations.
The most important topological invariants as well, as the parameters for the fundamental period of the mirror of the genus one fibration~\eqref{eqn:fundamental}, are listed in Table~\ref{tab:codim3higgsList}.
\begin{table}[h!]
	{\small
\begin{align*}
	\begin{array}{c|ccc||rrr|rrr|rr|rr||rrr|rrrr}
		\#&n_{\pm1}&n_{\pm2}&\chi&a_1&a_2&a_3&b_1&b_2&b_3&d_1&d_2&e_1&e_2&s_1&s_2&s_3&d_1&d_2&d_3&d_4\\\hline
 1_a & 100 & 125 &-90 & 0 & 1 & 2 & 0 & 0 & 0 & 0 & 0 & 0 & 0 & 1 & 2 & 0 & 0 & 0 & 0 & 0 \\\hline
 2_a & 120 & 105 &-90 & 0 & 0 & 1 & -2 & 0 & 0 & 0 & 0 & 0 & 0 & 2 & 3 & 2 & 2 & 2 & 0 & 0 \\
 2_b & 105 & 120 &-90 & 0 & 0 & 2 & -1 & 0 & 0 & 0 & 0 & 0 & 0 & 1 & 3 & 1 & 1 & 1 & 0 & 0 \\\hline
 3_a & 110 & 115 &-90 & 0 & 1 & 1 & -1 & 0 & 0 & 0 & 0 & 0 & 0 & 2 & 2 & 1 & 1 & 1 & 0 & 0 \\\hline
 4_b & 118 & 105 &-94 & 0 & 0 & 0 & -2 & -1 & 0 & 1 & 0 & 0 & 0 & 2 & 2 & 2 & 1 & 2 & 0 & 0 \\\hline
 5_a & 108 & 115 &-94 & 0 & 0 & 1 & -1 & -1 & 0 & 1 & 0 & 0 & 0 & 1 & 2 & 1 & 0 & 1 & 0 & 0 \\\hline
 7_b & 119 & 103 &-96 & 0 & 0 & 1 & -2 & 0 & 0 & 0 & -1 & 0 & 0 & 2 & 2 & 3 & 2 & 2 & 0 & 0 \\\hline
 8_b & 118 & 104 &-96 & 0 & 0 & 0 & -2 & 0 & 0 & 0 & 0 & 0 & 1 & 2 & 2 & 3 & 1 & 1 & 1 & 1 \\\hline
 9_a & 108 & 114 &-96 & 0 & 0 & 1 & -1 & 0 & 0 & 0 & 0 & 0 & 1 & 1 & 2 & 2 & 0 & 0 & 1 & 1 \\\hline
10_a & 109 & 113 &-96 & 0 & 1 & 1 & -1 & 0 & 0 & 0 & -1 & 0 & 0 & 1 & 2 & 2 & 1 & 1 & 0 & 0 \\
	\end{array}
\end{align*}
	}
	\caption{The data of 10 genus one fibrations with $5$-sections, that are connected via extremal transitions to fibrations of the codimension three complete intersection fiber from Section~\ref{sec:codim3fiber} over $\mathbb{P}^2$.
	The values for $s_i,\,i=1,\dots 3$ and $d_i,\,i=1,\dots,4$ are representative and there are in general multiple elliptic fibrations that admit a transition to a given genus one fibration.
}
	\label{tab:codim3higgsList}
\end{table}
Again, we also provide the parameters of a representative elliptic fibration that is connected to the genus one fibration via an extremal transition.

It turns out that there is a significant overlap between the geometries in Table~\ref{tab:codim5higgsList} and Table~\ref{tab:codim3higgsList}.
In some cases, for example geometry $5_a$, we obtain a new fundamental period for the mirror, although the genus one fibrations themselves are Wall equivalent.
We also observe that some geometries can only be obtained via transition from the codimension five complete intersections while other geometries are connected only to a codimension three complete intersection.

\subsection{Genus one fibrations with 5-sections over $B=\mathbb{P}^2$}
We will now combine the lists of topological invariants for genus one fibrations, that we obtained in the previous two sections, and construct the data of associated GLSMs.
Localizing the sphere partition functions, we can then compare the fundamental periods to the ones obtained via extremal transitions.
This verifies that the GLSMs indeed correspond to the desired geometries and we can then deduce the geometric data of the genus one fibrations.
Note that the data for all of the $23$ geometries is listed in Appendix~\ref{sec:fibrationsOverP2}.

Except for $11_{ab}$, $7_b$ and $12_b$ we find that all of the geometries appear in pairs that, by transforming the associated Picard-Fuchs systems using $z_1\rightarrow z_1^{-1}$, can be checked to share the same moduli space~\footnote{In some cases the variable transformation involves an additional prefactor or power of $z_2$.}.
For $11_{ab}$ the same geometry appears twice in the same moduli space, which is compatible with the multiplicities of $I_2$ singular fibers $n_{\pm1}=n_{\pm2}=110$, that should be exchanged for the members of a pair.
Moreover, in the case of $7_b$ and $12_b$ one can transform the Picard-Fuchs system and extract the topological invariants of a genus one fibration that we did not obtain via an extremal transition from one of the elliptic fibrations.
This leads to a total number of 23 geometries.

For 21 of those geometries we can, using the base independent expressions~\eqref{eqn:genericMultis} for $n_{\pm1}, n_{\pm2}$ and $\chi$, find a GLSM with the field content~\eqref{eqn:p2fieldcontent} that reproduces the topological invariants.
By localizing the sphere partition functions, we check that those also reproduce the correct fundamental periods.
Note that, as discussed in Section~\ref{sec:glsm}, the weakly coupled phase of the GLSM always gives a fibration of complete intersection in Grassmannians $G(2,5)$ while the strongly coupled phase corresponds to a fibration of Pfaffian curves in $\mathbb{P}^4$.
Our naming convention is such that a geometry with label $n_a$ corresponds to the weakly coupled phase while $n_b$ is obtained from the strongly coupled one.
The field contents for the GLSMs are listed in Table~\ref{tab:23fibrations}.

This leaves us with two geometries, namely $3_a$ and $3_b$, that can seemingly not be realized using a GLSM with gauge group $U(2)\times U(1)$.
No integral values for the charges $q(p_i),q(q_i)$ lead to the invariants $n_{\pm1}=110$, $n_{\pm2}=115$ or $n_{\pm1}=115$, $n_{\pm2}=110$ via the formula~\eqref{eqn:genericMultis}.
It turns out that this is for geometric reasons, namely that the construction of the fibration involves the tangent bundle on the base $\mathbb{P}^2$.
This bundle is not toric and the sections can not be constructed using fields in the usual GLSM realization of $\mathbb{P}^2$ with an Abelian gauge symmetry.
The solution is to identify $\mathbb{P}^2\equiv G(2,3)$ and realize the latter with a non-Abelian GLSM.
This leads to the GLSM with gauge group $U(2)\times U(2)$ and field content~\eqref{eqn:glsmU2U2matter} that upon localization indeed reproduces the correct fundamental period.


%% file: section_6.tex
\section{Outlook}
While we were aiming for a rather exhaustive discussion of $5$-section geometries with certain properties there are many directions for further research.

The story of genus one fibrations with $N$-sections certainly does not stop at $N=5$.
For instance, in~\cite{Hori:2013gga} an elliptic curve has been constructed using non-Abelian GLSMs whose properties indicate that it can be used to compute genus one fibrations with $6$-sections.
Moreover, there are lists of second-order Picard-Fuchs-type differential operators~\cite{zagier,az} for most of which the associated geometries are unknown.
It is worth noting that the fundamental periods for the mirrors of families of genus one curves with $N$-sections for $N\le 4$ are generalized hypergeometric functions while the coefficients for $N=5$ are Ap\'{e}ry numbers.
In fact, the fundamental period of the mirror of the Pfaffian and Grassmannian genus one curves encodes exactly the sequence of numbers that has been used by Ap\'{e}ry to prove the irrationality of $\zeta(2)$~\cite{apery}.
This hints at a connection between genus one fibered Calabi-Yau threefolds with $N$-sections and number theory that is mostly unexplored.
However, it is closely connected to another observation.

It has been conjectured that families of genus one curves with $N$-sections are mirror dual to families of curves with torsional sections~\cite{Klevers:2014bqa,Braun:2014qka,Oehlmann:2016wsb,Cvetic:2016ner}.
At least for $N=1,\ldots,4$~\cite{Cota:2019cjx} and, with the results from our analysis, $N=5$, this follows from the monodromies in the stringy K\"ahler moduli space of the fiber, which generate $\Gamma_1(N)$.
Homological mirror symmetry then implies that the mirror family also exhibits $\Gamma_1(N)$ monodromy in the complex structure moduli space and as a consequence factors through the modular curve $X_1(N)$.
However, it has been shown in~\cite{Hajouji:2019vxs} that Calabi-Yau threefolds can only have $N$-torsional sections for $N\le 6$.
This suggests that also genus one fibered Calabi-Yau threefolds with $N$-sections for $N>6$ might not exist.
Moreover, the modular curves for $\Gamma_1(N)$ with $N>12$ have genus greater than zero while the moduli spaces of consistent theories of quantum gravity are conjectured to be simply connected~\cite{Ooguri:2006in}. 
This was used in~\cite{Dierigl:2020lai} to derive independent bounds on the possible torsional sections in genus one fibered Calabi-Yau manifolds.
An independent argument that the stringy K\"ahler monodromies for genus one fibrations with $N$-sections reduces to $\Gamma_1(N)$ in the limit of large fiber volume would therefore have striking implications.

Another interesting question is whether one can achieve a complete classification of Calabi-Yaus threefolds that are complete intersections in Grassmannian bundels and Pfaffian varieties. As we have seen, these need not necessarily be genus one fibrations, and there are non-trivial connections between them via various kinds of topological transitions.  

A further -- related -- direction concerns extensions of homological projective duality.
We have seen that examples we have constructed appear as phases of the same GLSM which would imply that the geometries are related by relative homological projective duality and that the associated derived categories are equivalent.
Related to that, the derived categories of coherent sheaves associated to these geometries deserve to be better understood both from the mathematical perspective and the GLSM where an extension of~\cite{Herbst:2008jq} to GLSMs with non-Abelian gauge groups is necessary. 

One of the novel phenomena that we have observed is the vector valued transformation of the topological string partition function on dual genus one fibrations with $5$-sections under the $\Gamma_0(5)$ action of the transfer matrix.
It is natural to ask, if this is part of a more general set of modular relations that connects all of the geometries in the Tate-Shafarevich group.
This question will be addressed in a follow up paper and it turns out that such a structure indeed exist~\cite{Schimannek:2021ab}.


%% file: appendix_0.tex
\section{On the computation of the Hodge numbers}
\label{sec:comp-hodge-numb-1}

\subsection{Calculus with Schur functors}
\label{sec:calculus-with-schur}


In this appendix we collect a number of formulas for working with
Schur functors. For details
see~\cite{Fulton:1991ab,Macdonald:2015ab,Weyman:2003ab}. They are used
in the explicit calculations of the examples in
Sections~\ref{sec:examples-Gr} and~\ref{sec:examples-Pf}.

The following formulas are used in the decomposition into irreducible
components of Schur functors applied to vector bundles obtained from
linear algebra operations. Let $E,F$ be vector bundles. 
The decomposition of $\mS_\nu$ on a direct sum is 
\begin{equation}
  \label{eq:S_direct_sum}
  \mS_\nu (E \oplus F) = \bigoplus_{\lambda,\mu} N^\nu_{\lambda\mu} \mS_\lambda E \otimes \mS_\mu F,
\end{equation}
where the sum is over all partitions $\lambda,\mu$ contained in $\nu$
such that the sum of the numbers partitioned by $\lambda$ and $\mu$ is
the number partitioned by $\nu$, and the multiplicities
$N_{\lambda\mu}^\nu$ can be determined using the
Littlewood--Richardson rule. These
multiplicities also appear in the decomposition of the tensor product
of two Schur functors
\begin{equation}
  \label{eq:68}
  \mS_\lambda E\otimes \mS_\mu E = \bigoplus_{\nu} N^\nu_{\lambda\mu}
  \mS_\nu E
\end{equation}
As a special case, one obtains
\begin{equation}
  \label{eq:wedge_direct_sum}
  \sideset{}{^p}\bigwedge \bigoplus_{i=1}^n E_i= \bigoplus_{\sum_{i=1}^n p_i = p} \bigotimes_{i=1}^n \wedge^{p_i}E_i.
\end{equation}
Another important special case is Pieri's rule
\begin{equation}
  \label{eq:Pieri}
  \mS_\lambda E \otimes \wedge^m E  = \bigoplus_\mu \mS_\mu E,
\end{equation}
where the Young diagram of the partitions $\mu$ is obtained by adding $m$ boxes to the
Young diagram of the partition $\lambda$ such that at most one box is
added per row. 

The action of $\mS_\nu$ on an external tensor product of two vector bundles $E$, $F$ of ranks $e$ and $f$, respectively, is
\begin{equation}
  \label{eq:S_tensor}
  \mS_\nu( E \otimes F) = \bigoplus_{\lambda,\mu} C_{\lambda\mu\nu}
  \mS_\lambda E \otimes \mS_\mu F,
\end{equation}
where $|\lambda|=|\mu|=|\nu|=d$ and the $C_{\lambda\mu\nu}$ are
\begin{equation}
  \label{eq:69}
  C_{\lambda\mu\nu} = \sum_{\rho \vdash n} z_\rho^{-1}
      \chi^\lambda_\rho \chi^\mu_{\rho} \chi^\nu_\rho
\end{equation}
where $\chi^\lambda_\mu$ is the value of the character of the
irreducible $S_n$ representation $V^\lambda$ on the conjuagy class
$K_\mu$ of $S_n$ and $z_\rho = \prod_{j=1}^r j^{m_j} m_j!$ for $\rho=
(1^{m_1},2^{m_2},\dots,r^{m_r})$. As an important special case, we obtain Cauchy's formula:
\begin{equation}
  \label{eq:wedge_tensor}
    \wedge^k(E\otimes F) = \bigoplus_{\lambda \in \Sigma^k_{v,w}} \mS_\lambda E \otimes \mS_{\lambda'} F  
\end{equation}
where $\Sigma^k_{e,f}$ is the set of partitions $\lambda$ of $k$ of
length $e$ such that $\lambda_j \leq f$, i.e.  with at most $\dim E$ rows
and at most $\dim F$ columns, and $\lambda'$ denotes the conjugate
partition of $\lambda$.

Finally, we need the decomposition of the
composition of two Schur functors $\mS_\lambda\mS_\mu E$ into irreducible components. This is known as plethysm. For
general partitions $\lambda$ and $\mu$  one has
\begin{equation}
  \mS_\lambda(\mS_\mu E) = \bigoplus_{\nu} M_{\lambda\mu}^\nu \mS_\nu E,
\end{equation}
where the multiplicities $M_{\lambda\mu}^\nu$ are generally difficult to determine. One efficient way to do it is as follows. Irreducible representations of $\tG\tL(E)$ are determined by their characters and the character of
$\mS_\lambda E$ is the Schur function $s_\lambda$:
\begin{equation}
  \label{eq:10}
  \chi_{\mS_\lambda E} (g) = s_{\lambda}(x_1,\dots,x_v), \qquad
  \forall\; g \in \tG\tL(E),
\end{equation}
where $x_1,\dots,x_v$ are the eigenvalues of $g$. This entails
\begin{equation}
  \label{eq:11}
  s_\lambda\circ s_\mu = \sum_\nu M^\nu_{\lambda\mu} s_\nu,
\end{equation}
where the plethysm of the Schur functions is performed by expressing
$s_\lambda,s_\mu$ in terms of the basis of the power sums $p_i$, 
using $p_i \circ p_j = p_{ij}$, and reexpressing the result in terms
of the basis $s_\nu$. The coefficients $M_{\lambda\mu}^\nu$ can be
computed using for instance the package SF in Maple~\cite{Stembridge:2005sf}.

Finally, one can compute the rank of
$\mS_\lambda E$ using the Weyl character formula. 
\begin{equation}
  \label{eq:WeylCharacter}
  \rank \mS_\lambda E = s_\lambda(1,\dots,1) = \prod_{1 \leq i < j \leq
    e} \frac{\lambda_i -\lambda_j + j-i}{j-i},
\end{equation}
where, as above, $e=\rank E$. More generally, there is a formula for
the Chern character of a Schur functor $\mS_\lambda E$. This is needed
in e.g. in the steps from~\eqref{eq:resIYV} to~\eqref{eq:RRembedding}. Using the isomorphism
  $\chern^{-1}: \tH^*(X,\mQ) \cong \tK^0(X) \otimes \mQ$, this formula can be used to determine $\mS_\lambda E$ given $E$.  Alternatively, one can use the decomposition formulas for the direct sum and the tensor product to decompose $\mS_\lambda E$ into simpler pieces.

The Chern character of $\mS_\lambda E$ can be computed in terms of the Chern character of the symmetric powers $\Sym^k E$ as follows
\begin{equation}
  \label{eq:4}
  \chern \mS_\lambda E = \det (\chern( \Sym^{\lambda_i+j-i}E))_{i,j}.
\end{equation}
We provide here a short proof since we haven't found one in the standard literature.
In order to see this, we use the splitting principle and assume that
$E$ is of the form $\bigoplus_{k=1}^r L_k$ for some line bundles
$L_i$. Let $p \in \mZ[x_1,\dots,x_r]$ be a polynomial with nonnegative
integral coefficients of the form $p(x_1,\dots,x_r)=\sum_{I=(i_1,\dots,i_r)} a_I x_1^{i_1}\dots
x_r^{i_r}$. Then we define
\begin{equation}
p(L_1,\dots,L_r) = \bigoplus_{I=(i_1,\dots,i_r)} \left(L_1^{\otimes i_1} \otimes \dots
  \otimes  L_r^{\otimes i_r}\right)^{\oplus a_I}.
  \label{eq:2}
\end{equation}
Fe recall two bases of the ring of symmetric polynomials~\cite[\S I.3]{Macdonald:2015ab}:
The completely symmetric polynomials $h_k\in \mZ[x_1,\dots,x_r]$,
\begin{equation}
  \label{eq:6}
  h_k(x_1,\dots,x_r) = \sum_{1\leq i_1 \leq \dots \leq i_k\leq r}
  x_{i_1}\dots x_{i_k},
\end{equation}
and the Schur polynomials $s_\lambda\in \mZ[x_1,\dots,x_r]$ associated with a
partition $\lambda$ which we define in terms of the Giambelli formula
\begin{equation}
  \label{eq:7}
  s_\lambda(x_1,\dots,x_r) = \det (h_{\lambda_i+j-i}(x_1,\dots,x_r))_{i,j}.
\end{equation}
Then we can write 
\begin{equation}
  \label{eq:5}
  \mS_\lambda (\bigoplus_{k=1}^r L_k ) = (s_\lambda \circ
  h_1)(L_1,\dots,:L_r) = s_\lambda(L_1.\dots,L_r)
\end{equation}
where $\circ$ denotes plethysm~\cite[\S I.8]{Macdonald:2015ab}. The second equation
follows because $h_1$ is the identity with respect to $\circ$.
Taking the Chern character on both sides of~\eqref{eq:5} then yields
\begin{equation}
  \label{eq:8}
   \chern \mS_\lambda (\bigoplus_{i=1}^r L_i ) = \chern  \det
   (h_{\lambda_i+j-i}(L_1,\dots,L_r))_{i,j} = \chern  \det
   (\Sym^{\lambda_i+j-i}(\bigoplus_{k=1}^r L_k))_{i,j}
\end{equation}
Since $\chern$ is a ring homomorphism $\chern\det =\det \chern$ which yields
the formula~\eqref{eq:4}.

\subsection{Further examples}
\label{sec:further-examples}

\subsubsection*{Example $X_3$}
\label{sec:genus-1-example-3}

For this example, we take $E = \rO_{G/\mP^2}(1) \otimes \pi^*E'$ with
\begin{equation}
  E' = Q \oplus \rO_{\mP^2}(1)^{\oplus 2} \oplus \rO_{\mP^2},
\end{equation}
where we view $\mP^2=\Gr_1(\mC^3)$ as the dual Grassmannian and $T_{\Gr_1(\mC^3)} = Q\otimes
S\spcheck = Q(1)$. Although this example involves a rank $2$ bundle on
the base $\mP^2$, it turns out that he computation of the Hodge
numbers is no different from, say $X_1$. Indeed, we find
\begin{equation}
  \begin{aligned}
    \wedge^{4}E'{}\spcheck\otimes E'{}\spcheck &= \Sym^2Q\spcheck(-2) \oplus
    Q\spcheck(-3)^{\oplus 3} \oplus Q\spcheck(-2)^{\oplus 3} \\
    &\phantom{=} \oplus
    \rO_{\mP^2}(-4)^{\oplus 2} \oplus \rO_{\mP^2}(-3)^{\oplus 6}
    \oplus \rO_{\mP^2}(-2)^{\oplus 2}\\
    \wedge^{5}E'{}\spcheck\otimes E'{}\spcheck &= Q\spcheck(-3) \oplus
    \rO_{\mP^2}(-4)^{\oplus 2}  \oplus \rO_{\mP^2}(-3).\\
  \end{aligned}
\end{equation}
which leads to different nonvanishing contributions to the Leray spectral sequence
\begin{equation}
  \begin{aligned}
    E_2^{6,2} &= \tH^2(\mP^2, 
    Q\spcheck(-3)^{\oplus 3} \oplus
    \rO_{\mP^2}(-4)^{\oplus 2} \oplus \rO_{\mP^2}(-3)^{\oplus 6}) = \mC^{21}, & i &=-4\\
  E_2^{6,2} &= \tH^2(\mP^2, Q\spcheck(-3)^{\oplus 10} \oplus
    \rO_{\mP^2}(-4)^{\oplus 20}  \oplus \rO_{\mP^2}(-3)^{\oplus 10}) = \mC^{100}, & i &=-5.\\
\end{aligned}
\end{equation}
This yields a nontrivial differential in the Koszul spectral sequence
\begin{equation}
  \diff{}{}_1 : E_1^{-5,8} \cong \mC^{100} \to E_1^{-4,8} \cong \mC^{21}.
\end{equation}
One can show that $\coker \diff{}{}_1 = 0$, hence
\begin{equation}
  \tH^q(X,N_{X/P}\spcheck) =
  \begin{cases}
    \mC^{79} & q=3,\\
     0 & \text{otherwise.}\\
  \end{cases}
\end{equation}
Again, this agrees with the computation of $\chi(X,N_{X/P}\spcheck) =
-79$ by the Hirzebruch--Riemann--Roch theorem. Proceeding as before,
this yields 
\begin{equation}
  \begin{aligned}
    h^{1,1}(X) = h^{1}(X,\Omega^1_X) &= 2, &h^{2,1}(X) = h^{1}(X,\Omega^2_X) &= 47 
  \end{aligned}
\end{equation}

\subsubsection*{Example $X_2'$}
\label{sec:exampl-flop-trans}

We start with $P=\mP^4$, $F=\rO_{\mP^4}^{\oplus 4}$ as
in~\eqref{eq:20} and set $G = \Tot(
\Gr_2(F) )$. We take $E$ from~\eqref{eq:29}
\begin{equation}
  E = S\spcheck \otimes \pi^*\rO_{\mP^4}(1) \oplus \rO_G(1) \otimes \pi^*\rO_{\mP^4}(1)^{\oplus 3}, 
\end{equation}
This example is more involved than the previous ones as it involves
the nontrivial rank 2 bundle $S\spcheck$. Again, we need to
determine
\begin{equation}
  E_1^{i,j} = \tH^j(G,\wedge^{-i} E \otimes B), \qquad B =
E\spcheck, \Omega^1_{G/P}, \pi^*\Omega^1_P.
\label{eq:47}
\end{equation}
These cohomology groups are in turn
determined by the Leray spectral sequence for $\pi: G\to P$ with 
\begin{equation}
  E_2^{s,t} = \tH^t(P, \tR^s\pi_*(\wedge^{-i}E\spcheck\otimes B) ), \qquad i=-5,\dots,0.
\end{equation}

By the identity~\eqref{eq:wedge_direct_sum} we have
\begin{equation}
  \label{eq:48}
  \begin{aligned}
    \wedge^0 E &= \rO_{G} \\
    \wedge^1E &= S\spcheck \otimes \pi^*\rO_{\mP^4}(1) \oplus \rO_G(1)
    \otimes \pi^*\rO_{\mP^4}(1)^{\oplus 3} \\
    \wedge^2E &= \rO_G(1) \otimes \pi^*\rO_{\mP^4}(2) \oplus
    S\spcheck(1) \otimes \pi^*\rO_{\mP^4}(2)^{\oplus 3} \oplus \rO_G(2) \otimes \pi^*\rO_{\mP^4}(2)^{\oplus 3} \\
    \wedge^3E &= \rO_G(2) \otimes \pi^*\rO_{\mP^4}(3)^{\oplus 3} \oplus
    S\spcheck(2) \otimes \pi^*\rO_{\mP^4}(3)^{\oplus 3} \oplus
    \rO_G(3) \otimes \pi^*\rO_{\mP^4}(3) \\
    \wedge^4E &=  \rO_G(3) \otimes \pi^*\rO_{\mP^4}(4)^{\oplus 3} \oplus
    S\spcheck(3) \otimes \pi^*\rO_{\mP^4}(4)\\
    \wedge^5 E &= \rO_G(4) \otimes \pi^*\rO_{\mP^4}(5).
  \end{aligned}
\end{equation}
If we dualize these expressions and tensor them by $E\spcheck$ and by $\Omega^1_{G/\mP^4} =
S\otimes Q\spcheck$, respectively, we see that besides
$\rO_{G/\mP^4}(\ell)$ and $S(\ell)$ also $\Sym^2S(\ell)$
appears. Therefore, we need the following results from the
Borel--Weil--Bott theorem in order to apply the projection formula to
$\wedge^kE\spcheck \otimes B$.
\begin{equation}
  \begin{aligned}
    \tR^s\pi_*\rO_G(\ell) &=
    \begin{cases}
      \mS_{(0,0,0,0)} F\spcheck = \rO_{\mP^4}, & (s,\ell)=(0,0)\\
      \mS_{(2,2,2,2)} F\spcheck = \rO_{\mP^4}, & (s,\ell)=(4,-4)\\
      \mS_{(3,3,2,2)} F\spcheck = \wedge^2F\spcheck =
      \rO_{\mP^4}^{\oplus 6}, & (s,\ell)=(4,-5)\\
      0 & \text{otherwise}
    \end{cases}\\
    \tR^s\pi_*S(\ell) &=
    \begin{cases}
       \mS_{(3,3,2,2)} F\spcheck = F\spcheck =
      \rO_{\mP^4}^{\oplus 4}, & (s,\ell)=(4,-4)\\
      0 & \text{otherwise}
    \end{cases}\\
    \tR^s\pi_*\Sym^2S(\ell) &=
    \begin{cases}
      \mS_{(1,1,1,1)} F\spcheck = \rO_{\mP^4}, & (s,\ell)=(2,-1)\\
      0 & \text{otherwise}
    \end{cases}
  \end{aligned}
\end{equation}
Therefore, the only nonvanishing contributions to the Leray spectral
sequence with $B=\pi^*\Omega^1_{\mP^4}$ are (using~\eqref{eq:Bottformula})
\begin{equation}
  \begin{aligned}
    E_2^{0,1} &= \tH^1(\mP^4, \Omega^1_{\mP^4} ) = \mC & i&=0,\\
    E_2^{4,4} &= \tH^4(\mP^4, \Omega^1_{\mP^4}(-5) ) = \mC^{24} & i &= -5,\\
  \end{aligned}
\end{equation}
This yields
\begin{equation}
  \tH^j(G, \wedge^{-i}E\otimes \pi^*\Omega^1_{P})  =
  \begin{cases}
    \mC & (i,j) = (0,1)\\
    \mC^{24} & (i,j) = (-5,8)\\
    0 & \text{ otherwise.}
  \end{cases}
\end{equation}
Similarly, the possibly only nonvanishing contributions to the Leray spectral
sequence for $B=E\spcheck$ are
\begin{equation}
  \begin{aligned}
    E_2^{2,t} &= \tH^t(\mP^4, \rO_{\mP^4}(-3)^{\oplus 3} ) = 0 & i&=-2,\\
    E_2^{4,4} &= \tH^4(\mP^4, \rO_{\mP^4}(-5)^{\oplus 12} \oplus
    \rO_{\mP^4}(-5)^{\oplus 10}) = \mC^{22} & i &= -4,\\
    E_2^{4,4} &= \tH^4(\mP^4, \rO_{\mP^4}(-6)^{\oplus 4} \oplus \rO_{\mP^4}(-6)^{\oplus 18}) = \mC^{110} & i &= -5.\\
  \end{aligned}
\end{equation}
This yields a nontrivial differential in the Koszul spectral sequence
\begin{equation}
  \diff{}{}_1 : E_1^{-5,8} \cong \mC^{110} \to E_1^{-4,8} \cong \mC^{22}
\end{equation}
One can show that $\coker \diff{}{}_1 = 0$, hence
\begin{equation}
  \tH^q(X,N_{X/P}\spcheck) =
  \begin{cases}
    \mC^{88} & q=3.\\
     0 & \text{otherwise.}\\
  \end{cases}
\end{equation}
This agrees with the computation of $\chi(X,N_{X/P}\spcheck) = -88$ by
Hirzebruch--Riemann--Roch.

Next, we need the following results from the
Borel--Weil--Bott theorem in order to apply the projection formula to
$\wedge^kE\spcheck \otimes \Omega^1_{G/\mP^4}$.
\begin{equation}
  \begin{aligned}
    \tR^s\pi_*Q\spcheck(\ell) &=
    \begin{cases}
      \mS_{(2,2,2,1)} F\spcheck = \rO_{\mP^4}^{\oplus 4}, & (s,\ell)=(4,-4)\\
      0 & \text{otherwise}
    \end{cases}\\
    \tR^s\pi_*S(\ell) \otimes Q\spcheck&=
    \begin{cases}
      \mS_{(0,0,0,0)} F\spcheck = \rO_{\mP^4}, & (s,\ell)=(1,0)\\
      \mS_{(1,1,1,1)} F\spcheck = \rO_{\mP^4}, & (s,\ell)=(3,-2)\\
      \mS_{(3,2,2,1)} F\spcheck = \rO_{\mP^4}^{\oplus 15}, & (s,\ell)=(4,-4)\\
      0 & \text{otherwise}
    \end{cases}\\
    \tR^s\pi_*\Sym^2S(\ell) \otimes Q\spcheck&=
    \begin{cases}
      \mS_{(2,1,1,1)} F\spcheck = \rO_{\mP^4}^{\oplus 4}, & (s,\ell)=(3,-2)\\
      0 & \text{otherwise}
    \end{cases}    
  \end{aligned}
\end{equation}
Therefore, the only nonvanishing contributions to the Leray spectral
sequence are
\begin{equation}
  \begin{aligned}
    E_2^{1,0} &= \tH^0(\mP^4, \rO_{\mP^4} ) = \mC & i&=0,\\
    E_2^{4,4} &= \tH^4(\mP^4, \rO_{\mP^4}(-5)^{\oplus 15}) = \mC^{15} & i &= -5.\\
  \end{aligned}
\end{equation}
Hence, we find
\begin{equation}
  \tH^j(G, \wedge^{-i}E\spcheck\otimes \Omega^1_{G/P}) =
  \begin{cases}
    \mC & (i,j) = (0,1)\\
    \mC^{15} & (i.j) = (-5,8)\\
    0, & (i,j) \text{ otherwise.}
  \end{cases}
\end{equation}
As before, from the long exact cohomology sequences (for each $i$)
associated to~\eqref{eq:50} we find
\begin{equation}
  \tH^j(G, \wedge^{-i}E\spcheck\otimes \Omega^1_{G}) =
  \begin{cases}
    \mC^2 & (i,j) = (0,1)\\
    \mC^{39} & (i,j) = (-5,8) \\
    0, & \text{ otherwise.}
  \end{cases}
\end{equation}
Therefore we get from the Koszul spectral sequence for $\Omega^1_G$
\begin{equation}
  \tH^q(X,\Omega^1_G|_X) =
  \begin{cases}
    \mC^{2} & q=1.\\
    \mC^{39} & q=3\\
     0 & \text{otherwise}\\
  \end{cases}
\end{equation}
Finally, the Hodge numbers are
\begin{equation}
  \begin{aligned}
    h^{1,1}(X) = h^{1}(X,\Omega^1_X) &= 2, &h^{2,1}(X) = h^{1}(X,\Omega^2_X) &= 49.
  \end{aligned}
\end{equation}

\subsubsection*{Example $Y_1$}
\label{sec:genus-1-example-5}

Here we consider the Pfaffian Calabi--Yau variety $Y_1$ constructed in
Section~\ref{sec:examples-Pf}. Recall that there the ambient variety
was the projective bundle  $V=\Tot(\mP(E') \to \mP^2)$ with $E'=\rO_{\mP^2}(-1)^{\oplus 3}\oplus
\rO_{\mP^2}^{\oplus 2}$. $Y_1$ is the Pfaffian
$Y=\Pf(\varphi)$ in $V$ of a general skew--symmetric morphism $\varphi: F \to
F\spcheck \otimes L$ with $F=\pi^*(\rO_{\mP^2}^{\oplus 5})$. We only
exhibit the differences to the example $Y_2$ given in Section~\ref{sec:comp-hodge-numb}.

By the theorem of Borel--Weil--Bott we find
\begin{equation}
  \label{eq:64}
  \begin{aligned}
    \tR^s\pi_*L^{\otimes a} &=
    \begin{cases}
      \mS_{(0,0,0,0,0)} E\spcheck = \rO_{\mP^2}, & (s,a)=(0,0)\\
      \mS_{(1,1,1,1,1)} E\spcheck = \rO_{\mP^2}(-3), & (s,a)=(4,-5)\\
      \mS_{(2,1,1,1,1)} E\spcheck = \rO_{\mP^2}(-4)^{\oplus 3} \oplus \rO_{\mP^2}(-3)^{\oplus 2}, & (s,a)=(4,-6)\\
      0 & \text{otherwise}
    \end{cases}\\
      \tR^s\pi_*(L^{\otimes a}\otimes S \otimes Q\spcheck) &= 
  \begin{cases}
    \mS_{(0,0,0,0,0)} E\spcheck = \rO_{\mP^2}, & (s,a)=(1,0)\\
    \mS_{(2,1,1,1,0)} E\spcheck = \rO_{\mP^2}(-2)^{\oplus 6} \oplus
    \rO_{\mP^2}(-3)^{\oplus 12} \\
    \phantom{\mS_{(2,1,1,1,0)} E\spcheck = }\oplus \rO_{\mP^2}(-4)^{\oplus 6}, & (s,a)=(4,-5)\\
    0 & \text{otherwise}.
  \end{cases}
  \end{aligned}
\end{equation}
For $B=\Omega^1_{V/\mP^2}$ and $B=\pi^*\Omega^1_{\mP^2}$ we find the same result as in the previous
example. Therefore,
\begin{equation}
  \tH^q(Y,\Omega^1_V|_{Y}) =
  \begin{cases}
    \mC^{2} & q=1.\\
    \mC^{32} & q=3\\
     0 & \text{otherwise}.\\
  \end{cases}
\end{equation}
For the contributions from the normal bundle, we find
from~\eqref{eq:56} for $m=1$ 
\begin{equation}
  \begin{aligned}
  E_2^{4,2} &= \tH^2(\mP^2, \rO_{\mP^2}(-3) ) = \mC, & i &=-3\,,
  \end{aligned}
\end{equation}
and from~\eqref{eq:58} for $m=2$
\begin{equation}
  \begin{aligned}
  E_2^{4,2} &= \tH^2(\mP^2,
  \rO_{\mP^2}(-3)^{\oplus 24} ) = \mC^{24}, & i &=-2\\
  E_2^{4,2} &= \tH^2(\mP^2, \rO_{\mP^2}(-3)^{\oplus 20} \oplus \rO_{\mP^2}(-4)^{\oplus 30}  ) = \mC^{110}, & i &=-3.
  \end{aligned}
\end{equation}
Hence, we find
\begin{align}
  \tH^q(V,\rI_Y ) &=
  \begin{cases}
    \mC & q=4\\
    0 & \text{ otherwise.}
  \end{cases}\\
  \tH^q(V,\rI_Y^2 ) &=
  \begin{cases}
    \mC^{86} & q=4\\
    0 & \text{otherwise,}
  \end{cases}
\end{align}
from which we conclude that
\begin{equation}
  \tH^q(X,N_{Y/V}\spcheck) =
  \begin{cases}
    \mC^{85} & q=3.\\
     0 & \text{otherwise.}\\
  \end{cases}
\end{equation}
This agrees with the computation of $\chi(X,N_{X/P}\spcheck) = -85$ by
Hirzebruch--Riemann--Roch. Finally
\begin{equation}
  \begin{aligned}
    h^{1,1}(Y) = h^{1}(Y,\Omega^1_Y) &= 2, &h^{2,1}(Y) = h^{1}(Y,\Omega^2_Y) &= 47 \,.
  \end{aligned}
\end{equation}


%% file: appendix_1.tex
\section{Modular forms and Jacobi forms}
\label{sec:appModularForms}
In this Appendix we will summarize the properties of modular forms for $\Gamma_1(5)$, as well as the ring of weak Jacobi forms.
Both are used to perform the modular bootstrap in Section~\ref{sec:gvandmod}.
But first let us recall the definition of a modular form and the basic examples for $SL(2,\mathbb{Z})$.

A modular form of weight $k$ on a subgroup $\Gamma\subseteq \Gamma_1=SL(2,\mathbb{Z})$ is a holomorphic function on the upper half-plane $f:\,\mathbb{H}\rightarrow\mathbb{C}$, that satisfies the transformation law
\begin{align}
	f\left(\frac{a\tau+b}{c\tau+d}\right)=(c\tau+d)^kf(\tau)\,,\quad\text{for }\left(\begin{array}{cc}c&b\\c&d\end{array}\right)\in\Gamma\,,
\end{align}
and admits a Fourier expansion around $\tau\rightarrow i\infty$ of the form
\begin{align}
	f(\tau)=\sum\limits_{n=0}^\infty c(n)q^n\,,
\end{align}
with $q=\exp(2\pi i \tau)$.
The Eisenstein series $E_k(\tau)$ with $k\ge 4$ are modular forms of weight $k$ and can be defined as
\begin{align}
	E_{k}(\tau)=1+\frac{2}{\zeta(1-k)}\sum\limits_{n=1}^\infty\frac{n^{k-1}q^n}{1-q^n}\,,
	\label{eqn:eisenstein}
\end{align}
with $\zeta(s)$ being the Riemann zeta function.
The ring $M_k(\Gamma_1)$ of modular forms of weight $k$ on $SL(2,\mathbb{Z})$ is generated by $E_4(\tau)$ and $E_6(\tau)$, i.e.
\begin{align}
	M_k(\Gamma_1)=\mathbb{C}[E_4,E_6]\,.
\end{align}
The weight $2$ Eisenstein series $E_2$ is not a modular form but a quasi modular form.
It satisfies the transformation law
\begin{align}
	E_2\left(\frac{a\tau+b}{c\tau+d}\right)=(c\tau+d)^2E_2(\tau)-\frac{6i}{\pi}c(c\tau+d)\,.
\end{align}
We also need the Dedekind eta function
\begin{align}
	\eta(\tau)=q^{\frac{1}{24}}\prod\limits_{n=1}^\infty(1-q^n)\,.
\end{align}
This is again not a modular form but satisfies the relations
\begin{align}
	\eta(\tau+1)=e^{\frac{\pi i}{12}}\eta(\tau)\,,\quad \eta\left(-\frac{1}{\tau}\right)=\sqrt{-i\tau}\eta(\tau)\,.
\end{align}
It is easy to see, that $\Delta(\tau)=\eta^{24}(\tau)$ is a modular form of weight $12$.
We note that if $f(\tau)$ is a modular form on $\Gamma_1$, then $f(5\tau)$ is a modular form of the same weight on $\Gamma_1(5)$.

\subsection{Modular forms for $\Gamma_1(5)$ and $\Gamma_0(5)$}
\label{sec:basis-modular-forms}

In this appendix we summarize the main properties of modular forms for
$\Gamma_1(5)$ and $\Gamma_0(5)$ that we use in the main text. For
further properties, see e.g.~\cite{Beukers:1983ab}

We set $\Gamma_1 = \SL(2,\mZ)$. The group
\begin{equation}
  \label{eq:102}
  \Gamma_0(5) =\{ \abcd \in \Gamma_1 \mid c \equiv 0 \mod 5\}
\end{equation}
is a subgroup of $\Gamma_1$ of index $6$, has exactly two cusps, $0$ and $\infty$, 2 elliptic
points of order $2$ at $\tau=\frac{2}{5}+\frac{i}{5}, \tau=\frac{3}{5}+\frac{i}{5}$, and no elliptic point of order $3$. Hence, its
genus is $0$, and $\dim M_k(\Gamma_0(5)) = 1 + 2\lfloor\frac{k}{4}\rfloor$, if $k$ is even and nonnegative, otherwise $\dim M_k(\Gamma_0(5))=0$.
The group
\begin{equation}
  \label{eq:103}
  \Gamma_1(5) =\{ \gamma \in \Gamma_1 \mid \gamma \equiv \bb1 \mod 5\}
\end{equation}
is a subgroup of $\Gamma_0(5)$ of index $4$. It has no
elliptic points, and $4$ cusps at $0$,
$\frac{1}{5}$, $\frac{1}{2}$, and $\infty$. Hence, its genus is $0$,
and $\dim M_k(\Gamma_1(5)) = k+1$, $k\geq0$.

Let $T=\left(\substack{1\;1\\0\;1}\right)$, $A_1 = \left(\substack{2\;-1\\5\;-2}\right)$, $A_2=\left(\substack{3\;-2\\5\,-3}\right)$. Then
\[
  \begin{aligned}
    \Gamma_0(5) &= \left\langle T, A_1, A_2\right\rangle, &
    \Gamma_1(5) &= \left\langle T, A_1A_2, A_1TA_1^{-1}\right\rangle.
  \end{aligned}
\]

There are two odd Dirichlet characters mod 5 which we denote by $\chi_1,\chi_2$ whose values are
\begin{align}
  \begin{array}[h!]{c|cccc}
    n \mod 5 & 1 & 2 & 3 & 4\\
    \hline
    \chi_1(n) & 1 & i & -i & -1\\
    \chi_2(n) & 1 & -i & i & -1\\
  \end{array}\,.
\end{align}
The space of modular forms of weight 1 for $\Gamma_1(5)$ is of the form
\begin{align}
  M_1(\Gamma_1(5)) = M_1(\Gamma_0(5),\chi_1) \oplus M_1(\Gamma_0(5),\chi_2)\,.
\end{align}
The spaces $M_1(\Gamma_0(5),\chi)$ are of dimension 1 and are generated by the Eisenstein series
\begin{align}
  E_1^{\chi}(\tau) = 1 + \frac{2}{L(0,k)}\sum_{n=1}^\infty \chi(n) \frac{q^n}{1-q^n}\,,
\end{align}
where
\begin{align}
  L(0,\chi) = -\frac{1}{5} \sum_{k=1}^4 \chi(k) k\,,
\end{align}
is the value at $s=0$ of the Dirichlet L--series with character $\chi$. The first few terms of the Fourier expansions of two real linear conbinations are
\begin{align}
	\begin{split}
	  E_{1,1}=&\frac{1}{2}\left(E^{\chi_1}_1+E^{\chi_2}_1\right) =1+3\,q+4\,{q}^{2}+2\,{q}^{3}+{q}^{4}+3\,{q}^{5}+6\,{q}^{6}+4\,{q}^{7} -{q}^{9}+O ({q}^{10} )\,, \\
	  E_{1,2}=&\frac{1}{2i}\left(E^{\chi_2}_1-E^{\chi_1}_1\right) = q-2\,{q}^{2}+4\,{q}^{3}-3\,{q}^{4}+{q}^{5}+2\,{q}^{6}-2\,{q}^{7}+3\,{ q}^{9}+O ( {q}^{10} )\,.
	\end{split}
\end{align}
Combined with the dimension formula above, we have
\[
  M_*(\Gamma_1(5)) \cong \mC[ E_1^{\chi_1},  E_1^{\chi_2}]
\]
Let $E^1_2 = E_1^{\chi_1}E_1^{\chi_2}, E^1_{4,1} = (E_1^{\chi_1})^4,
E^1_{4,2} = (E_1^{\chi_2})^4$. Then
\[
  M_*(\Gamma_0(5)) \cong \mC[E^1_2,E^1_{4,1},E^1_{4,2}] / (E^1_{4,1} E^1_{4,2} - (E_2^1)^4)
\]
Let $(a;q)_n = \prod_{k=0}^{n-1} (1-aq^k)$ denote the $q$--Pochhammer symbol. Consider the following theta constants
\begin{align}
  \begin{split}
    A(\tau) &= \frac{q^{\frac{1}{5}}}{(q;q)_\infty^{\frac{3}{5}}} \sum_{n=-\infty}^{\infty} (-1)^n q^{\frac{5n^2-3n}{2}}\,,\\
    B(\tau) &= \frac{1}{(q;q)_\infty^{\frac{3}{5}}} \sum_{n=-\infty}^{\infty} (-1)^n q^{\frac{5n^2-n}{2}}\,.
  \end{split}
\end{align}
There are the following relations to the Eisenstein series $E_1^\chi$:
\begin{align}
  \begin{split}
  E_1^{\chi_1} &= B^5 - iA^5\,,\\
  E_2^{\chi_2} &= B^5 + iA^5\,.\\  
  \end{split}
\end{align}
A Hauptmodul for $\Gamma_1(5)$ is
\begin{align}
  t = \frac{A^5}{B^5} =q-5\,{q}^{2}+15\,{q}^{3}-30\,{q}^{4}+40\,{q}^{5}-26\,{q}^{6}-30\,{q}^ {7}+125\,{q}^{8}-220\,{q}^{9}+O ( {q} ^{10} )\,.
\end{align}
This Hauptmodul also has the product expansion
\begin{align}
  t = q\prod_{n=1}^\infty (1-q^n)^{5 \legendre{n}{5}}\,,
\end{align}
where $\legendre{n}{5}$ denotes the Legendre symbol. It
maps the cusps $0, \frac{1}{5}, \frac{1}{2}, \infty$ to the points
$\frac{1}{2}(-11+5\sqrt{5})$, $\frac{1}{2}(-11-5\sqrt{5})$, $\infty$,
$0$, respectively. In fact, the map $\fH \to \mC\setminus \{0,
\frac{1}{2}(-11\pm5\sqrt{5})\}, \tau \to t(\tau)$ is a universal
covering with covering group $\Gamma_1(5)$.

The following modular forms play an important role in the modular
bootstrap and the elliptic genus
\begin{align}
  \label{eq:99}
  \Delta_{10} &=E_{1,1}^{\,6}E_{1,2}^{\,4} \in M_{10}(\Gamma_1(5))\,,\\
  \label{eq:100}
  \Delta'_{10} &=E_{1,1}^{\,4}E_{1,2}^{\,6}\in M_{10}(\Gamma_1(5))\,,\\
  \label{eq:101}           
  E_2^{(5)}&=E_{1,1}^{\,2}+6E_{1,1}E_{1,2}+E_{1,2}^{\,2}\in M_{2}(\Gamma_1(5))\,.
\end{align}



\subsection{The ring of weak Jacobi forms}
\label{sec:appendixJacobiForms}
The theory of Jacobi forms has been developed in~\cite{eichler1985theory}, which is also the standard reference on the topic.
A \textit{weak Jacobi form} of \textit{weight} $k$ and \textit{index} $m$ on a subgroup $\Gamma\subseteq \Gamma_1$ is a holomorphic function $\phi:\,\mathbb{H}\times\mathbb{C}$ that satisfies the modular transformation law
\begin{align}
	\phi\left(\frac{a\tau+b}{c\tau+d},\,\frac{z}{c\tau+d}\right)=(c\tau+d)^k\exp\left(2\pi i\frac{mcz^2}{c\tau+d}\right)\phi(\tau,z)\,,
\end{align}
as well as the elliptic transformation law
\begin{align}
	\phi\left(\tau,z+\lambda\tau+\mu\right)=\exp\left(-2\pi im\left[\lambda^2\tau+2\lambda z\right]\right)\phi(\tau,z)\,,
	\label{eqn:elliptictrans}
\end{align}
for every element
\begin{align}
	\left(\begin{array}{cc}c&b\\c&d\end{array}\right)\in\Gamma\,.
\end{align}
It also needs to admit a Fourier expansion of the form
\begin{align}
	\phi(\tau,z)=\sum\limits_{n=0}^\infty\sum\limits_{r=-\infty}^\infty c(n,r)q^n\zeta^r\,,
\end{align}
with $q=\exp(2\pi i\tau)$, $\zeta=\exp(2\pi i z)$.
If the coefficients satisfy the condition that $c(n,r)=0$ unless $n\ge r^2/4m$, then $\phi$ is called a \textit{Jacobi form}.
Of particular importance are the $SL(2,\mathbb{Z})$ Jacobi forms
\begin{align}
	\begin{split}
		\phi_{-2,1}(\tau,z)=&-\frac{\theta_1(\tau,z)^2}{\eta(\tau)^6}=(2\pi i z)^2+\frac{1}{12}E_2(\tau)(2\pi i z)^4+\mathcal{O}(z^6)\,,\\
		\phi_{0,1}(\tau,z)=&4\left[\frac{\theta_2(\tau,z)^2}{\theta_2(\tau,0)^2}+\frac{\theta_3(\tau,z)^2}{\theta_3(\tau,0)^2}+\frac{\theta_4(\tau,z)^2}{\theta_4(\tau,0)^2}\right]=12+E_2(\tau)(2\pi i z)^2+\mathcal{O}(z^4)\,,
	\end{split}
\end{align}
where $\phi_{k,m}$ has weight $k$ and index $m$.
The Jacobi theta functions are defined as
\begin{align}
	\begin{split}
		\theta_1(\tau,z)=&\vartheta_{\frac{1}{2}\frac{1}{2}}(\tau,z)\,,\quad \theta_2(\tau,z)=\vartheta_{\frac{1}{2}0}(\tau,z)\,,\\
		\theta_3(\tau,z)=&\vartheta_{00}(\tau,z)\,,\quad \theta_4(\tau,z)=\vartheta_{0\frac{1}{2}}(\tau,z)\,,
	\end{split}
	\label{eqn:jacobitheta}
\end{align}
where we have used
\begin{align}
	\vartheta_{ab}(\tau,z)=&\sum\limits_{n=-\infty}^\infty e^{\pi i(n+a)^2\tau+2\pi i z(n+a)+2\pi i b(n+a)}\,.
\end{align}
We denote the space of ordinary modular forms of weight $k$ on $\Gamma$ by $M_{k}\left(\Gamma\right)$.
Then $\phi_{-2,1}$ and $\phi_{0,1}$ generate the ring of weak Jacobi forms of weight $k$ and index $m$ on $\Gamma_1$
\begin{align}
	J^{\text{\textit{weak}}}_{k,m}\left(\Gamma_1\right)=\oplus_{j=0}^m M_{k+2j}\left(\Gamma_1\right)\left[\phi_{-2,1}(\tau,z)^j,\phi_{0,1}(\tau,z)^{m-j}\right]\,.
\end{align}
As discussed in~\cite{Cota:2019cjx}, we can obtain $\Gamma_1(N)$ modular forms by restricting the parameters of $\phi_{-2,1}$ and $\phi_{0,1}$.
Of particular importance for us are the relations
\begin{align}
\begin{split}
	\phi_{0,1}(5\tau,\tau)=&\left[q\Delta_{10}\right]^{-\frac15} E_2^{(5)}\,,\quad\phi_{-2,1}(5\tau,\tau)=\left[q\Delta_{10}\right]^{-\frac15}\,.
\end{split}
\end{align}
where $\Delta_{10}$ and $E_2^{(5)}$ are given in~\eqref{eq:99}
and~\eqref{eq:101}, respective;ly.


%% file: appendix_2.tex
\section{Fibrations over $\mathbb{P}^2$}
\label{sec:fibrationsOverP2}
In this section we summarize the data of 23 genus one fibrations with $5$-sections over $\mathbb{P}^2$.
This set of geometries has been obtained in Section~\ref{sec:classification}, by systematically constructing elliptic fibration that exhibit Higgs transitions in F-theory which break the gauge group to $\mathbb{Z}_5$.
Using the base independent multiplicities of reducible fibers~\eqref{eqn:fullyGenericMultis} we determined the associated bundles that define the fibrations geometrically, based on the general construction developed in Section~\ref{sec:geometry}.
From this we have also deduced a set of GLSM charges and calculated the sphere partition function, as discussed in Section~\ref{sec:glsm}, to verify that it reproduces the correct fundamental period of the mirror.
The topological invariants as well as the parameters of the fundamental periods of the mirrors and the GLSM charges are listed in Table~\ref{tab:23fibrations}.
Picard-Fuchs operators are provided in Section~\ref{sec:pfops} and genus zero Gopakumar-Vafa for low degress can be found in Section~\ref{sec:gvinvs}.

The data in Table~\ref{tab:23fibrations} is interpreted as follows.
All of the geometries have $h^{1,1}=2$ and together with the Euler characteristic $\chi$ this also fixes $h^{1,2}$.
The invariants $n_{\pm q},\,q=1,2$ are the multiplicities of $I_2$ fibers with a component that intersects the $5$-section $q$ times.
In an F-theory compactification, those correspond to the number of hypermultiplets that contain half-hypermultiplets with $\mathbb{Z}_5$ charges $\pm q$.
For every geometry we let $J_1,J_2$ correspond to a K\"ahler cone basis with $J_2$ being the vertical divisor, that arises from the hyperplane class in $\mathbb{P}^2$, and $J_1$ is, up to multiples of $J_2$, the $5$-section.
The triple intersection numbers are then defined as $c_{ijk}=J_i\cdot J_j\cdot J_k$ and the intersections with the second Chern class are denoted by $b_i=c_2\cdot J_i,\,i=1,2$.
Every fibration over $\mathbb{P}^2$ then has
\begin{align}
	c_{122}=5\,,\quad c_{222}=0\,,\quad b_2=36\,,
\end{align}
and to avoid redundancy those numbers are omitted from the list.

The fundamental period of the mirror Calabi-Yau takes the generic form
\begin{align}
	\begin{split}
	w_0=&\sum\limits_{\lambda_1,\lambda_2=0}^\infty\frac{{_3}F_2(d_1\lambda_2-\lambda_1,-\lambda_1,1-d_2\lambda_2+\lambda_1;1-e_1\lambda_2,1-e_2\lambda_2;1)}{\Gamma(1-e_1\lambda_2)\Gamma(1-e_2\lambda_2)\Gamma(1+\lambda_2)^3}\\
		&\cdot\prod\limits_{i=1}^3\frac{\Gamma(1+a_i\lambda_2+\lambda_1)}{\Gamma(1+b_i\lambda_2+\lambda_1)}z_1^{\lambda_1}(z_2 z_1^{-\epsilon})^{\lambda_2}\,.
	\end{split}
\end{align}
Here we have introduced $\epsilon$ such that $z_1,z_2$ are related to the complexified K\"ahler parameters in the expansion $\omega=t^iJ_i$ via mirror maps
\begin{align}
	t^i=\frac{1}{2\pi i}\log(z_i)+\mathcal{O}(z)\,.
\end{align}
Up to additional transformations of the form $z_2\rightarrow z_2z_1^{-\epsilon'}$, these periods match the results obtained from calculating the sphere partition functions of the GLSMs with gauge group $G=U(2)\times U(1)$ and field content
\begin{align}
	\begin{array}{c|ccccccccc|c}
		&p_{i,\,i=1,\dots,5}&x_{i,\,i=1,\dots,5}&b_1&b_2&b_3&FI\\\hline
		U(2)&\det^{-1}&\square&0&0&0&\zeta\\
		U(1)&q(p_i)&q(x_i)&1&1&1&\zeta_1
	\end{array}\,.
\end{align}
There is one exception to this and the GLSM associated to $3_a$ and $3_b$ is discussed in Section~\ref{sec:glsmModel3}.
The corresponding GLSM has gauge group $U(2)\times U(2)$, which is necessary to realize the tangent bundle of $\mathbb{P}^2$ that is involved in the geometric construction of the fibrations.

\begin{landscape}
{
\begin{table}[h!]
\begin{align*}
\begin{array}{|c|cccccc|rrrrrrrrrr|c|rrrrr|rrrrr|}\hline
	\#&\chi&n_{\pm1}&n_{\pm2}&c_{111}&c_{112}&b_1&a_1&a_2&a_3&b_1&b_2&b_3&d_1&d_2&e_1&e_2&\epsilon&\multicolumn{5}{c}{q(p_1)\dots q(p_5)}&\multicolumn{5}{|c|}{q(x_1)\dots q(x_5)}\\\hline
	1_a & -90 & 100 & 125 & 10 & 15 & 64 & 0 & 1 & 2 & 0 & 0 & 0 & 0 & 0 & 0 & 0 & 0 &\multirow{2}{*}{$-2$}&\multirow{2}{*}{$-1$}&\multirow{2}{*}{$ 0$}&\multirow{2}{*}{$ 0$}&\multirow{2}{*}{$ 0$}&\multirow{2}{*}{$ 0$}&\multirow{2}{*}{$ 0$}&\multirow{2}{*}{$ 0$}&\multirow{2}{*}{$ 0$}&\multirow{2}{*}{$ 0$}\\
	1_b & -90 & 125 & 100 & 5 & 5 & 38 & 0 & 0 & 0 & -2 & -1 & 0 & 0 & 0 & 0 & 0 & 1 &  &  &  &  &  &  &  &  &  &  \\\hline
	2_a & -90 & 105 & 120 & 15 & 15 & 66 & 1 & 1 & 1 & 0 & 0 & 0 & 0 & 0 & 0 & 0 & 0 &\multirow{2}{*}{$-1$}&\multirow{2}{*}{$-1$}&\multirow{2}{*}{$-1$}&\multirow{2}{*}{$ 0$}&\multirow{2}{*}{$ 0$}&\multirow{2}{*}{$ 0$}&\multirow{2}{*}{$ 0$}&\multirow{2}{*}{$ 0$}&\multirow{2}{*}{$ 0$}&\multirow{2}{*}{$ 0$}\\
	2_b & -90 & 120 & 105 & 0 & 5 & 36 & 0 & 0 & 0 & -1 & -1 & -1 & 0 & 0 & 0 & 0 & 1 &  &  &  &  &  &  &  &  &  &  \\\hline
	3_a & -90 & 110 & 115 & 20 & 15 & 68 & 0 & 1 & 1 & -1 & 0 & 0 & 0 & 0 & 0 & 0 & 0 &\multicolumn{10}{|c|}{\multirow{2}{*}{GLSM differs; see Section~\ref{sec:glsmModel3}.}}\\
	3_b & -90 & 115 & 110 & 25 & 15 & 70 & 0 & 0 & 1 & -1 & -1 & 0 & 0 & 0 & 0 & 0 & 0 &\multicolumn{10}{|c|}{}\\\hline
	4_a & -94 & 105 & 118 & 51 & 21 & 90 & 0 & 0 & 1 & 0 & 0 & 0 & 0 & 1 & 1 & 1 & 0 &\multirow{2}{*}{$-2$}&\multirow{2}{*}{$-1$}&\multirow{2}{*}{$-1$}&\multirow{2}{*}{$-1$}&\multirow{2}{*}{$ 0$}&\multirow{2}{*}{$ 1$}&\multirow{2}{*}{$ 0$}&\multirow{2}{*}{$ 0$}&\multirow{2}{*}{$ 0$}&\multirow{2}{*}{$ 0$}\\
	4_b & -94 & 118 & 105 & 10 & 9 & 52 & 0 & 0 & 0 & -2 & -1 & 0 & 1 & 0 & 0 & 0 & 1 &  &  &  &  &  &  &  &  &  &  \\\hline
	5_a & -94 & 108 & 115 & 42 & 19 & 84 & 0 & 0 & 1 & -1 & 0 & 0 & 0 & 1 & 0 & 1 & 0 &\multirow{2}{*}{$-1$}&\multirow{2}{*}{$ 0$}&\multirow{2}{*}{$ 0$}&\multirow{2}{*}{$ 0$}&\multirow{2}{*}{$ 0$}&\multirow{2}{*}{$-1$}&\multirow{2}{*}{$ 0$}&\multirow{2}{*}{$ 0$}&\multirow{2}{*}{$ 0$}&\multirow{2}{*}{$ 0$}\\
	5_b & -94 & 115 & 108 & 13 & 11 & 58 & -1 & 0 & 0 & -2 & 0 & 0 & 0 & 1 & 1 & 1 & 1 &  &  &  &  &  &  &  &  &  &  \\\hline
	6_a & -94 & 110 & 113 & 8 & 11 & 56 & 0 & 0 & 0 & -1 & 0 & 0 & 0 & 1 & 1 & 1 & 1 &\multirow{2}{*}{$-1$}&\multirow{2}{*}{$-1$}&\multirow{2}{*}{$-1$}&\multirow{2}{*}{$-1$}&\multirow{2}{*}{$-1$}&\multirow{2}{*}{$ 1$}&\multirow{2}{*}{$ 0$}&\multirow{2}{*}{$ 0$}&\multirow{2}{*}{$ 0$}&\multirow{2}{*}{$ 0$}\\
	6_b & -94 & 113 & 110 & 5 & 9 & 50 & 0 & 0 & 0 & -1 & -1 & 0 & 0 & 1 & 0 & 1 & 1 &  &  &  &  &  &  &  &  &  &  \\\hline
	7_a & -96 & 103 & 119 & 94 & 27 & 112 & 1 & 0 & 0 & 0 & 0 & 0 & 1 & 2 & 1 & 1 & 0 &\multirow{2}{*}{$-1$}&\multirow{2}{*}{$ 1$}&\multirow{2}{*}{$ 1$}&\multirow{2}{*}{$ 1$}&\multirow{2}{*}{$ 1$}&\multirow{2}{*}{$-1$}&\multirow{2}{*}{$-1$}&\multirow{2}{*}{$-1$}&\multirow{2}{*}{$ 0$}&\multirow{2}{*}{$ 0$}\\
	7_b & -96 & 119 & 103 & 26 & 13 & 68 & 0 & 0 & 1 & -2 & 0 & 0 & 0 & -1 & 0 & 0 & 0 &  &  &  &  &  &  &  &  &  &  \\\hline
	8_a & -96 & 104 & 118 & 65 & 23 & 98 & 0 & 0 & 1 & -1 & 0 & 0 & 1 & 1 & 0 & 1 & 0 &\multirow{2}{*}{$-1$}&\multirow{2}{*}{$ 0$}&\multirow{2}{*}{$ 0$}&\multirow{2}{*}{$ 1$}&\multirow{2}{*}{$ 1$}&\multirow{2}{*}{$-1$}&\multirow{2}{*}{$-1$}&\multirow{2}{*}{$ 0$}&\multirow{2}{*}{$ 0$}&\multirow{2}{*}{$ 0$}\\
	8_b & -96 & 118 & 104 & 7 & 7 & 46 & 0 & 0 & 0 & -2 & -1 & 0 & 0 & 1 & 0 & 0 & 1 &  &  &  &  &  &  &  &  &  &  \\\hline
	9_a & -96 & 108 & 114 & 33 & 17 & 78 & 0 & 0 & 1 & -1 & 0 & 0 & 0 & 0 & 0 & 1 & 0 &\multirow{2}{*}{$-2$}&\multirow{2}{*}{$-2$}&\multirow{2}{*}{$-1$}&\multirow{2}{*}{$-1$}&\multirow{2}{*}{$-1$}&\multirow{2}{*}{$ 1$}&\multirow{2}{*}{$ 1$}&\multirow{2}{*}{$ 0$}&\multirow{2}{*}{$ 0$}&\multirow{2}{*}{$ 0$}\\
	9_b & -96 & 114 & 108 & 21 & 13 & 66 & 0 & 0 & 0 & -2 & 0 & 0 & 1 & 1 & 0 & 1 & 1 &  &  &  &  &  &  &  &  &  &  \\\hline
	10_a & -96 & 109 & 113 & 16 & 13 & 64 & 0 & 0 & 0 & -1 & -1 & 0 & 1 & 1 & 0 & 1 & 1 &\multirow{2}{*}{$ 0$}&\multirow{2}{*}{$ 0$}&\multirow{2}{*}{$ 0$}&\multirow{2}{*}{$ 0$}&\multirow{2}{*}{$ 1$}&\multirow{2}{*}{$-1$}&\multirow{2}{*}{$-1$}&\multirow{2}{*}{$ 0$}&\multirow{2}{*}{$ 0$}&\multirow{2}{*}{$ 0$}\\
	10_b & -96 & 113 & 109 & 2 & 7 & 44 & 0 & 0 & 0 & -1 & -1 & -1 & 0 & 1 & 0 & 0 & 1 &  &  &  &  &  &  &  &  &  &  \\\hline
 11_{ab} & -100 & 110 & 110 & 29 & 15 & 74 & 0 & 0 & 0 & 0 & 0 & 0 & 0 & 2 & 2 & 1 & 1 &\multirow{1}{*}{$-1$}&\multirow{1}{*}{$-1$}&\multirow{1}{*}{$-1$}&\multirow{1}{*}{$ 0$}&\multirow{1}{*}{$ 0$}&\multirow{1}{*}{$ 1$}&\multirow{1}{*}{$-1$}&\multirow{1}{*}{$ 0$}&\multirow{1}{*}{$ 0$}&\multirow{1}{*}{$ 0$}\\\hline
	12_a & -104 & 108 & 110 & 9 & 9 & 54 & 0 & 0 & -1 & 0 & 0 & -2 & 0 & 0 & 1 & 1 & 1 &\multirow{2}{*}{$-1$}&\multirow{2}{*}{$-1$}&\multirow{2}{*}{$ 0$}&\multirow{2}{*}{$ 0$}&\multirow{2}{*}{$ 1$}&\multirow{2}{*}{$-1$}&\multirow{2}{*}{$-1$}&\multirow{2}{*}{$ 0$}&\multirow{2}{*}{$ 0$}&\multirow{2}{*}{$ 1$}\\
	12_b & -104 & 110 & 108 & 17 & 11 & 62 & 0 & 0 & 0 & -1 & 0 & 0 & 0 & 1 & 0 & 2 & 1 &  &  &  &  &  &  &  &  &  &  \\\hline
\end{array}
\end{align*}
	\caption{The topological invariants, parameters of the mirror fundamental period and GLSM charges for 23 genus one fibrations with $5$-section over $\mathbb{P}^2$.}
	\label{tab:23fibrations}
\end{table}
}
\end{landscape}

\subsection{Picard-Fuchs operators}
\label{sec:pfops}
\paragraph{$1_{a}$, $(\chi,n_{\pm1},n_{\pm2})=(-90,100,125)$}
{\small
\begin{align*}
\begin{autobreak}
\MoveEqLeft
		\mathcal{D}_1=\theta_1^2
-3\theta_1\theta_2
+7\theta_2^2
+z_1\bigl(-3
-11\theta_1
-11\theta_1^2\bigl)
-z_2\bigl(1
+\theta_1
+2\theta_2\bigl)
\bigl(14
+15\theta_1
+14\theta_2\bigl)
-z_1^2\bigl(1
+\theta_1
+\theta_2\bigl)
\bigl(1
+\theta_1
+2\theta_2\bigl)
\end{autobreak}\\
\begin{autobreak}
\MoveEqLeft
		\mathcal{D}_2=\theta_2^3
-z_2\bigl(1
+\theta_1
+\theta_2\bigl)
\bigl(1
+\theta_1
+2\theta_2\bigl)
\bigl(2
+\theta_1
+2\theta_2\bigl)
\end{autobreak}
\end{align*}
}

\paragraph{$1_{b}$, $(\chi,n_{\pm1},n_{\pm2})=(-90,125,100)$}
{\small
\begin{align*}
\begin{autobreak}
\MoveEqLeft
		\mathcal{D}_1=\theta_1\bigl(\theta_1
-\theta_2\bigl)
+z_1\bigl(-3
-11\theta_1
-11\theta_1^2
-11\theta_2
-22\theta_1\theta_2
-11\theta_2^2\bigl)
+z_1^2\bigl(-1
-2\theta_1
-\theta_1^2
-5\theta_2
-5\theta_1\theta_2
-11\theta_2^2\bigl)
-z_1z_2\bigl(\theta_1
-\theta_2\bigl)
\bigl(1
+15\theta_1
+\theta_2\bigl)
\end{autobreak}\\
\begin{autobreak}
\MoveEqLeft
		\mathcal{D}_2=z_1\theta_2^3
-z_2\theta_1\bigl(-1
+\theta_1
-\theta_2\bigl)
\bigl(\theta_1
-\theta_2\bigl)
\end{autobreak}
\end{align*}
}

\paragraph{$2_{a}$, $(\chi,n_{\pm1},n_{\pm2})=(-90,105,120)$}
{\small
\begin{align*}
\begin{autobreak}
\MoveEqLeft
		\mathcal{D}_1=\theta_1^2
-3\theta_1\theta_2
+6\theta_2^2
+z_1\bigl(-3
-11\theta_1
-11\theta_1^2\bigl)
+z_2\bigl(-6
-15\theta_1
-11\theta_1^2
-12\theta_2
-12\theta_1\theta_2
-12\theta_2^2\bigl)
+z_1^2\bigl(-1
-2\theta_1
-\theta_1^2
-3\theta_2
-3\theta_1\theta_2
-3\theta_2^2\bigl)
+z_1z_2\bigl(3
+11\theta_1
+11\theta_1^2\bigl)
+z_2^2\bigl(6
+15\theta_1
+10\theta_1^2
+12\theta_2
+15\theta_1\theta_2
+6\theta_2^2\bigl)
\end{autobreak}\\
\begin{autobreak}
\MoveEqLeft
		\mathcal{D}_2=\theta_2^3
-z_2\bigl(1
+\theta_1
+\theta_2\bigl)^3
\end{autobreak}
\end{align*}
}

\paragraph{$2_{b}$, $(\chi,n_{\pm1},n_{\pm2})=(-90,120,105)$}
{\small
\begin{align*}
\begin{autobreak}
\MoveEqLeft
		\mathcal{D}_1=\theta_1^2
-\theta_1\theta_2
+\theta_2^2
+z_1\bigl(-3
-11\theta_1
-11\theta_1^2
-11\theta_2
-22\theta_1\theta_2
-11\theta_2^2\bigl)
+z_2\bigl(-3
-11\theta_1
-11\theta_1^2
-11\theta_2
-22\theta_1\theta_2
-11\theta_2^2\bigl)
+z_1^2\bigl(-1
-2\theta_1
-\theta_1^2
-5\theta_2
-5\theta_1\theta_2
-10\theta_2^2\bigl)
+z_1z_2\bigl(-2
-7\theta_1
-11\theta_1^2
-7\theta_2
-10\theta_1\theta_2
-11\theta_2^2\bigl)
+z_2^2\bigl(-1
-5\theta_1
-10\theta_1^2
-2\theta_2
-5\theta_1\theta_2
-\theta_2^2\bigl)
\end{autobreak}\\
\begin{autobreak}
\MoveEqLeft
		\mathcal{D}_2=\theta_2^3
-z_2\bigl(1
+\theta_1
+\theta_2\bigl)
\bigl(3
+11\theta_1
+11\theta_1^2
+11\theta_2
+22\theta_1\theta_2
+11\theta_2^2\bigl)
+z_1z_2\bigl(-2
-5\theta_1
-4\theta_1^2
-\theta_1^3
-11\theta_2
-17\theta_1\theta_2
-6\theta_1^2\theta_2
-25\theta_2^2
-15\theta_1\theta_2^2
-20\theta_2^3\bigl)
+z_2^2\bigl(-2
-11\theta_1
-25\theta_1^2
-5\theta_2
-17\theta_1\theta_2
-15\theta_1^2\theta_2
-4\theta_2^2
-6\theta_1\theta_2^2
-\theta_2^3\bigl)
\end{autobreak}
\end{align*}
}

\paragraph{$3_{a}$, $(\chi,n_{\pm1},n_{\pm2})=(-90,110,115)$}
{\small
\begin{align*}
\begin{autobreak}
\MoveEqLeft
		\mathcal{D}_1=\theta_1^2
-3\theta_1\theta_2
+5\theta_2^2
+z_1\bigl(-3
-11\theta_1
-11\theta_1^2\bigl)
-z_2\bigl(\theta_1
-\theta_2\bigl)
\bigl(5
+7\theta_1
+5\theta_2\bigl)
+z_1^2\bigl(-1
-2\theta_1
-\theta_1^2
-3\theta_2
-3\theta_1\theta_2
-4\theta_2^2\bigl)
-4z_1^2z_2\bigl(1
+\theta_1
+\theta_2\bigl)^2
\end{autobreak}\\
\begin{autobreak}
\MoveEqLeft
		\mathcal{D}_2=\theta_2^3
-z_2\bigl(\theta_1
-\theta_2\bigl)
\bigl(1
+\theta_1
+\theta_2\bigl)^2
\end{autobreak}
\end{align*}
}

\paragraph{$3_{b}$, $(\chi,n_{\pm1},n_{\pm2})=(-90,115,110)$}
{\small
\begin{align*}
\begin{autobreak}
\MoveEqLeft
		\mathcal{D}_1=\theta_1^2
-3\theta_1\theta_2
+4\theta_2^2
+z_1\bigl(-3
-11\theta_1
-11\theta_1^2\bigl)
-4z_2\bigl(\theta_1
-\theta_2\bigl)^2
+z_1^2\bigl(-1
-2\theta_1
-\theta_1^2
-3\theta_2
-3\theta_1\theta_2
-5\theta_2^2\bigl)
-z_1^2z_2\bigl(2
+7\theta_1
-5\theta_2\bigl)
\bigl(1
+\theta_1
+\theta_2\bigl)
\end{autobreak}\\
\begin{autobreak}
\MoveEqLeft
		\mathcal{D}_2=\theta_2^3
-z_2\bigl(\theta_1
-\theta_2\bigl)^2
\bigl(1
+\theta_1
+\theta_2\bigl)
\end{autobreak}
\end{align*}
}

\paragraph{$4_{a}$, $(\chi,n_{\pm1},n_{\pm2})=(-94,105,118)$}
{\small
\begin{align*}
\begin{autobreak}
\MoveEqLeft
		\mathcal{D}_1=25\theta_1^2
-105\theta_1\theta_2
+186\theta_2^2
+z_1\bigl(-75
-275\theta_1
-275\theta_1^2
+165\theta_2
+330\theta_1\theta_2
-81\theta_2^2\bigl)
-2z_2\bigl(23\theta_1^2
-78\theta_1\theta_2
+93\theta_2^2\bigl)
+z_1^2\bigl(-25
-50\theta_1
-25\theta_1^2
-45\theta_2
-45\theta_1\theta_2
-31\theta_2^2\bigl)
-2z_1z_2\bigl(37
+63\theta_1
+15\theta_1^2
+98\theta_2
+131\theta_1\theta_2
+28\theta_2^2\bigl)
+z_1^2z_2\bigl(48
+79\theta_1
-31\theta_2\bigl)
\bigl(1
+\theta_1
+\theta_2\bigl)
+4z_1z_2^2\bigl(38
+63\theta_1
-25\theta_2\bigl)
\bigl(1
+\theta_1
+\theta_2\bigl)
\end{autobreak}\\
\begin{autobreak}
\MoveEqLeft
		\mathcal{D}_2=125z_1\theta_2^3
-2z_2\theta_1\bigl(25\theta_1^2
-80\theta_1\theta_2
+81\theta_2^2\bigl)
+z_1z_2\bigl(1
+\theta_1
+\theta_2\bigl)
\bigl(150
+550\theta_1
+525\theta_1^2
-330\theta_2
-580\theta_1\theta_2
+81\theta_2^2\bigl)
+4z_1z_2^2\bigl(13
+24\theta_1
-11\theta_2\bigl)
\bigl(1
+\theta_1
+\theta_2\bigl)
\bigl(2
+\theta_1
+\theta_2\bigl)
\end{autobreak}
\end{align*}
}

\paragraph{$4_{b}$, $(\chi,n_{\pm1},n_{\pm2})=(-94,118,105)$}
{\small
\begin{align*}
\begin{autobreak}
\MoveEqLeft
		\mathcal{D}_1=25\theta_1^2
-45\theta_1\theta_2
+31\theta_2^2
+z_1\bigl(-75
-275\theta_1
-275\theta_1^2
-165\theta_2
-330\theta_1\theta_2
-81\theta_2^2\bigl)
-z_2\bigl(\theta_1
-\theta_2\bigl)
\bigl(31
+79\theta_1
+31\theta_2\bigl)
+z_1^2\bigl(-25
-50\theta_1
-25\theta_1^2
-105\theta_2
-105\theta_1\theta_2
-186\theta_2^2\bigl)
-2z_1z_2\bigl(-11
-33\theta_1
+15\theta_1^2
-33\theta_2
-131\theta_1\theta_2
+28\theta_2^2\bigl)
+2z_1^2z_2\bigl(23
+46\theta_1
+23\theta_1^2
+78\theta_2
+78\theta_1\theta_2
+93\theta_2^2\bigl)
+4z_1z_2^2\bigl(\theta_1
-\theta_2\bigl)
\bigl(25
+63\theta_1
+25\theta_2\bigl)
\end{autobreak}\\
\begin{autobreak}
\MoveEqLeft
		\mathcal{D}_2=125\theta_2^3
-z_2\bigl(\theta_1
-\theta_2\bigl)
\bigl(125
+500\theta_1
+525\theta_1^2
+250\theta_2
+580\theta_1\theta_2
+81\theta_2^2\bigl)
-2z_1z_2\bigl(1
+\theta_1\bigl)
\bigl(25
+50\theta_1
+25\theta_1^2
+80\theta_2
+80\theta_1\theta_2
+81\theta_2^2\bigl)
-4z_2^2\bigl(-1
+\theta_1
-\theta_2\bigl)
\bigl(\theta_1
-\theta_2\bigl)
\bigl(11
+24\theta_1
+11\theta_2\bigl)
\end{autobreak}
\end{align*}
}

\paragraph{$5_{a}$, $(\chi,n_{\pm1},n_{\pm2})=(-94,108,115)$}
{\small
\begin{align*}
\begin{autobreak}
\MoveEqLeft
		\mathcal{D}_1=25\theta_1^2
-95\theta_1\theta_2
+151\theta_2^2
+z_1\bigl(-75
-275\theta_1
-275\theta_1^2
+110\theta_2
+220\theta_1\theta_2
-26\theta_2^2\bigl)
+z_2\bigl(-63\theta_1^2
+179\theta_1\theta_2
-151\theta_2^2\bigl)
+z_1^2\bigl(-25
-50\theta_1
-25\theta_1^2
-55\theta_2
-55\theta_1\theta_2
-56\theta_2^2\bigl)
+z_1z_2\bigl(27
+54\theta_1
+40\theta_1^2
+57\theta_2
+18\theta_1\theta_2
+95\theta_2^2\bigl)
-z_1^2z_2\bigl(57
+113\theta_1
-56\theta_2\bigl)
\bigl(1
+\theta_1
+\theta_2\bigl)
+z_1z_2^2\bigl(70
+139\theta_1
-69\theta_2\bigl)
\bigl(1
+\theta_1
+\theta_2\bigl)
\end{autobreak}\\
\begin{autobreak}
\MoveEqLeft
		\mathcal{D}_2=125z_1\theta_2^3
+z_2\theta_1\bigl(25\theta_1^2
-70\theta_1\theta_2
+56\theta_2^2\bigl)
-z_1z_2\bigl(1
+\theta_1
+\theta_2\bigl)
\bigl(75
+275\theta_1
+325\theta_1^2
-110\theta_2
-360\theta_1\theta_2
+138\theta_2^2\bigl)
-z_1z_2^2\bigl(11
+24\theta_1
-13\theta_2\bigl)
\bigl(1
+\theta_1
+\theta_2\bigl)
\bigl(2
+\theta_1
+\theta_2\bigl)
\end{autobreak}
\end{align*}
}

\paragraph{$5_{b}$, $(\chi,n_{\pm1},n_{\pm2})=(-94,115,108)$}
{\small
\begin{align*}
\begin{autobreak}
\MoveEqLeft
		\mathcal{D}_1=25\theta_1^2
-55\theta_1\theta_2
+56\theta_2^2
+z_1\bigl(-75
-275\theta_1
-275\theta_1^2
-110\theta_2
-220\theta_1\theta_2
-26\theta_2^2\bigl)
-z_2\bigl(\theta_1
-\theta_2\bigl)
\bigl(56
+113\theta_1
+56\theta_2\bigl)
+z_1^2\bigl(-25
-50\theta_1
-25\theta_1^2
-95\theta_2
-95\theta_1\theta_2
-151\theta_2^2\bigl)
+z_1z_2\bigl(-13
-26\theta_1
-40\theta_1^2
-39\theta_2
+18\theta_1\theta_2
-95\theta_2^2\bigl)
+z_1^2z_2\bigl(-63
-126\theta_1
-63\theta_1^2
-179\theta_2
-179\theta_1\theta_2
-151\theta_2^2\bigl)
+z_1z_2^2\bigl(\theta_1
-\theta_2\bigl)
\bigl(69
+139\theta_1
+69\theta_2\bigl)
\end{autobreak}\\
\begin{autobreak}
\MoveEqLeft
		\mathcal{D}_2=-125\bigl(-275\theta_1^2
+275\theta_1^3
+605\theta_1\theta_2
-325\theta_1^2\theta_2
-616\theta_2^2\bigl)
+125z_1\bigl(825\theta_1
+3025\theta_1^2
+3025\theta_1^3
+840\theta_2
+4290\theta_1\theta_2
+5500\theta_1^2\theta_2
+1232\theta_2^2
+2750\theta_1\theta_2^2
+6993\theta_2^3\bigl)
-z_2\bigl(\theta_1
-\theta_2\bigl)
\bigl(77000
+155375\theta_1
+44825\theta_1^2
+77000\theta_2
-8615\theta_1\theta_2
+8658\theta_2^2\bigl)
+3125z_1^2\bigl(1
+\theta_1\bigl)
\bigl(11
+22\theta_1
+11\theta_1^2
+53\theta_2
+53\theta_1\theta_2
+109\theta_2^2\bigl)
+z_1z_2\bigl(-8400
-881450\theta_1
-2590825\theta_1^2
-2136525\theta_1^3
+839880\theta_2
+820385\theta_1\theta_2
-453870\theta_1^2\theta_2
+1713704\theta_2^2
+1710579\theta_1\theta_2^2
+874125\theta_2^3\bigl)
+z_2^2\bigl(-1
+\theta_1
-\theta_2\bigl)
\bigl(\theta_1
-\theta_2\bigl)
\bigl(8658
+17359\theta_1
+8658\theta_2\bigl)
\end{autobreak}
\end{align*}
}

\paragraph{$6_{a}$, $(\chi,n_{\pm1},n_{\pm2})=(-94,110,113)$}
{\small
\begin{align*}
\begin{autobreak}
\MoveEqLeft
		\mathcal{D}_1=25\theta_1^2
-55\theta_1\theta_2
+81\theta_2^2
+z_1\bigl(-75
-275\theta_1
-275\theta_1^2
-110\theta_2
-220\theta_1\theta_2
-26\theta_2^2\bigl)
+z_2\bigl(-81
-269\theta_1
-281\theta_1^2
-162\theta_2
-210\theta_1\theta_2
-162\theta_2^2\bigl)
+z_1^2\bigl(-25
-50\theta_1
-25\theta_1^2
-95\theta_2
-95\theta_1\theta_2
-126\theta_2^2\bigl)
+z_1z_2\bigl(64
+251\theta_1
+243\theta_1^2
+63\theta_2
+186\theta_1\theta_2
-57\theta_2^2\bigl)
+z_2^2\bigl(81
+265\theta_1
+240\theta_1^2
+162\theta_2
+265\theta_1\theta_2
+81\theta_2^2\bigl)
\end{autobreak}\\
\begin{autobreak}
\MoveEqLeft
		\mathcal{D}_2=10125\theta_2^3
+z_2\bigl(-10125
-40500\theta_1
-59500\theta_1^2
-23525\theta_1^3
-30375\theta_2
-73625\theta_1\theta_2
-61645\theta_1^2\theta_2
-40500\theta_2^2
-31266\theta_1\theta_2^2
-20250\theta_2^3\bigl)
-7000z_1^2\theta_2^3
+z_1z_2\bigl(-4050
-18525\theta_1
-38900\theta_1^2
-31425\theta_1^3
-17415\theta_2
-39455\theta_1\theta_2
-36040\theta_1^2\theta_2
-28107\theta_2^2
-19357\theta_1\theta_2^2
-14742\theta_2^3\bigl)
+z_2^2\bigl(20250
+64391\theta_1
+54709\theta_1^2
-3432\theta_1^3
+50625\theta_2
+95657\theta_1\theta_2
+24709\theta_1^2\theta_2
+40500\theta_2^2
+31266\theta_1\theta_2^2
+10125\theta_2^3\bigl)
\end{autobreak}
\end{align*}
}

\paragraph{$6_{b}$, $(\chi,n_{\pm1},n_{\pm2})=(-94,113,110)$}
{\small
\begin{align*}
\begin{autobreak}
\MoveEqLeft
		\mathcal{D}_1=25\theta_1^2
-45\theta_1\theta_2
+56\theta_2^2
+z_1\bigl(-75
-275\theta_1
-275\theta_1^2
-165\theta_2
-330\theta_1\theta_2
-81\theta_2^2\bigl)
+z_2\bigl(-56
-235\theta_1
-243\theta_1^2
-112\theta_2
-300\theta_1\theta_2\bigl)
+z_1^2\bigl(-25
-50\theta_1
-25\theta_1^2
-105\theta_2
-105\theta_1\theta_2
-161\theta_2^2\bigl)
+z_1z_2\bigl(-93
-293\theta_1
-281\theta_1^2
-245\theta_2
-352\theta_1\theta_2
-233\theta_2^2\bigl)
+z_2^2\bigl(-56
-215\theta_1
-240\theta_1^2
-112\theta_2
-215\theta_1\theta_2
-56\theta_2^2\bigl)
\end{autobreak}\\
\begin{autobreak}
\MoveEqLeft
		\mathcal{D}_2=7000\theta_2^3
+z_2\bigl(-7000
-35000\theta_1
-55375\theta_1^2
-31425\theta_1^3
-21000\theta_2
-78125\theta_1\theta_2
-58235\theta_1^2\theta_2
-14000\theta_2^2
-41552\theta_1\theta_2^2\bigl)
-10125z_1^2\theta_2^3
+z_1z_2\bigl(-5600
-7925\theta_1
+11075\theta_1^2
+23525\theta_1^3
-26320\theta_2
-27515\theta_1\theta_2
+8930\theta_1^2\theta_2
-47824\theta_2^2
-21449\theta_1\theta_2^2
-27104\theta_2^3\bigl)
+z_2^2\bigl(-14000
-55323\theta_1
-65005\theta_1^2
-3432\theta_1^3
-35000\theta_2
-83771\theta_1\theta_2
-35005\theta_1^2\theta_2
-28000\theta_2^2
-28448\theta_1\theta_2^2
-7000\theta_2^3\bigl)
\end{autobreak}
\end{align*}
}

\paragraph{$7_{a}$, $(\chi,n_{\pm1},n_{\pm2})=(-96,103,119)$}
{\small
\begin{align*}
\begin{autobreak}
\MoveEqLeft
		\mathcal{D}_1=25\theta_1^2
-135\theta_1\theta_2
+259\theta_2^2
+z_1\bigl(-75
-275\theta_1
-275\theta_1^2
+330\theta_2
+660\theta_1\theta_2
-369\theta_2^2\bigl)
+z_2\bigl(-52\theta_1^2
+229\theta_1\theta_2
-259\theta_2^2\bigl)
+z_1^2\bigl(-25
-50\theta_1
-25\theta_1^2
-15\theta_2
-15\theta_1\theta_2
+\theta_2^2\bigl)
+z_1z_2\bigl(156
+572\theta_1
+416\theta_1^2
-531\theta_2
-272\theta_1\theta_2
-690\theta_2^2\bigl)
-z_1^2z_2\bigl(1
+\theta_1
+\theta_2\bigl)
\bigl(14
+3\theta_1
+16\theta_2\bigl)
+39z_1^3z_2\bigl(1
+\theta_1
+\theta_2\bigl)
\bigl(2
+\theta_1
+\theta_2\bigl)
+27z_1^2z_2^2\bigl(1
+\theta_1
+\theta_2\bigl)
\bigl(2
+\theta_1
+\theta_2\bigl)
\end{autobreak}\\
\begin{autobreak}
\MoveEqLeft
		\mathcal{D}_2=-8\bigl(13\theta_1^3
-91\theta_1^2\theta_2
+247\theta_1\theta_2^2
-216\theta_2^3\bigl)
+13z_1\bigl(24
+112\theta_1
+137\theta_1^2
+127\theta_1^3
-144\theta_2
-247\theta_1\theta_2
-625\theta_1^2\theta_2
-117\theta_2^2
+1197\theta_1\theta_2^2
-840\theta_2^3\bigl)
+27z_2\bigl(8\theta_1^2
+21\theta_1^3
-56\theta_1\theta_2
-139\theta_1^2\theta_2
+152\theta_2^2
+343\theta_1\theta_2^2
-280\theta_2^3\bigl)
-13z_1^2\bigl(-16
+77\theta_1
+397\theta_1^2
+421\theta_1^3
-184\theta_2
-1190\theta_1\theta_2
-1708\theta_1^2\theta_2
+832\theta_2^2
+2231\theta_1\theta_2^2
-944\theta_2^3\bigl)
-27z_1z_2\bigl(87
+382\theta_1
+463\theta_1^2
+246\theta_1^3
-498\theta_2
-1044\theta_1\theta_2
-950\theta_1^2\theta_2
+465\theta_2^2
+1169\theta_1\theta_2^2
-456\theta_2^3\bigl)
-729z_2^2\bigl(\theta_1
-2\theta_2\bigl)^3
-507z_1^3\bigl(1
+\theta_1
-\theta_2\bigl)^2
\bigl(1
+\theta_1
+\theta_2\bigl)
+351z_1^2z_2\bigl(1
+\theta_1
+\theta_2\bigl)
\bigl(-18
-34\theta_1
+23\theta_1^2
+52\theta_2
-76\theta_1\theta_2
+62\theta_2^2\bigl)
+729z_1z_2^2\bigl(1
+\theta_1
+\theta_2\bigl)
\bigl(3
+11\theta_1
+8\theta_1^2
-18\theta_2
-26\theta_1\theta_2
+21\theta_2^2\bigl)
\end{autobreak}
\end{align*}
}

\paragraph{$7_{b}$, $(\chi,n_{\pm1},n_{\pm2})=(-96,119,103)$}
{\small
\begin{align*}
\begin{autobreak}
\MoveEqLeft
		\mathcal{D}_1=25\theta_1^2
-65\theta_1\theta_2
+39\theta_2^2
+z_1\bigl(-75
-275\theta_1
-275\theta_1^2
-55\theta_2
-110\theta_1\theta_2
+16\theta_2^2\bigl)
-39z_2\bigl(-1
+\theta_1
-2\theta_2\bigl)
\bigl(\theta_1
-2\theta_2\bigl)
+z_1^2\bigl(-25
-50\theta_1
-25\theta_1^2
-85\theta_2
-85\theta_1\theta_2
-149\theta_2^2\bigl)
-z_1z_2\bigl(-11
+3\theta_1
-19\theta_2\bigl)
\bigl(\theta_1
-2\theta_2\bigl)
+z_1^2z_2\bigl(-260\theta_1
-416\theta_1^2
+519\theta_2
+560\theta_1\theta_2
+546\theta_2^2\bigl)
+z_1^3z_2\bigl(-52
-104\theta_1
-52\theta_1^2
-125\theta_2
-125\theta_1\theta_2
-82\theta_2^2\bigl)
-27z_1^2z_2^2\bigl(-1
+\theta_1
-2\theta_2\bigl)
\bigl(\theta_1
-2\theta_2\bigl)
\end{autobreak}\\
\begin{autobreak}
\MoveEqLeft
		\mathcal{D}_2=\bigl(5\theta_1
-3\theta_2\bigl)
\bigl(\theta_1
-2\theta_2\bigl)
\theta_2
+z_1\theta_2\bigl(-15
-55\theta_1
-55\theta_1^2
-11\theta_2
-22\theta_1\theta_2
+158\theta_2^2\bigl)
+3z_2\bigl(-2
+\theta_1
-2\theta_2\bigl)
\bigl(-1
+\theta_1
-2\theta_2\bigl)
\bigl(\theta_1
-2\theta_2\bigl)
+z_1^2\theta_2\bigl(-1
-\theta_1
+\theta_2\bigl)
\bigl(5
+5\theta_1
+22\theta_2\bigl)
-z_1z_2\bigl(-1
+\theta_1
-2\theta_2\bigl)
\bigl(\theta_1
-2\theta_2\bigl)
\bigl(157
+259\theta_1
+157\theta_2\bigl)
\end{autobreak}
\end{align*}
}

\paragraph{$8_{a}$, $(\chi,n_{\pm1},n_{\pm2})=(-96,104,118)$}
{\small
\begin{align*}
\begin{autobreak}
\MoveEqLeft
		\mathcal{D}_1=25\theta_1^2
-115\theta_1\theta_2
+204\theta_2^2
+z_1\bigl(-75
-275\theta_1
-275\theta_1^2
+220\theta_2
+440\theta_1\theta_2
-149\theta_2^2\bigl)
-12z_2\bigl(4\theta_1^2
-15\theta_1\theta_2
+17\theta_2^2\bigl)
+z_1^2\bigl(-25
-50\theta_1
-25\theta_1^2
-35\theta_2
-35\theta_1\theta_2
-14\theta_2^2\bigl)
+12z_1z_2\bigl(12
+25\theta_1
+15\theta_1^2
+13\theta_2
+5\theta_1\theta_2
+14\theta_2^2\bigl)
-28z_1^2z_2\bigl(\theta_1
-2\theta_2\bigl)
\bigl(1
+\theta_1
+\theta_2\bigl)
+144z_1z_2^2\bigl(\theta_1
-2\theta_2\bigl)
\bigl(1
+\theta_1
+\theta_2\bigl)
\end{autobreak}\\
\begin{autobreak}
\MoveEqLeft
		\mathcal{D}_2=690z_1\theta_2^3
+18z_2\theta_1^2\bigl(23\theta_1
-65\theta_2\bigl)
-445z_1^2\theta_2^3
-3z_1z_2\bigl(414
+1932\theta_1
+3082\theta_1^2
+1653\theta_1^3
-480\theta_2
-1116\theta_1\theta_2
-176\theta_1^2\theta_2
-1168\theta_2^2
-3171\theta_1\theta_2^2
+1862\theta_2^3\bigl)
+89z_1^2z_2\bigl(1
+\theta_1
+\theta_2\bigl)
\bigl(9
+33\theta_1
+34\theta_1^2
-30\theta_2
-60\theta_1\theta_2
+20\theta_2^2\bigl)
+72z_1z_2^2\bigl(21
+55\theta_1
-34\theta_2\bigl)
\bigl(\theta_1
-2\theta_2\bigl)
\bigl(1
+\theta_1
+\theta_2\bigl)
\end{autobreak}
\end{align*}
}

\paragraph{$8_{b}$, $(\chi,n_{\pm1},n_{\pm2})=(-96,118,104)$}
{\small
\begin{align*}
\begin{autobreak}
\MoveEqLeft
		\mathcal{D}_1=25\theta_1^2
-35\theta_1\theta_2
+14\theta_2^2
+z_1\bigl(-75
-275\theta_1
-275\theta_1^2
-220\theta_2
-440\theta_1\theta_2
-149\theta_2^2\bigl)
-28z_2\bigl(\theta_1
-\theta_2\bigl)
\bigl(1
+\theta_1
+2\theta_2\bigl)
+z_1^2\bigl(-25
-50\theta_1
-25\theta_1^2
-115\theta_2
-115\theta_1\theta_2
-204\theta_2^2\bigl)
-12z_1z_2\bigl(2
+5\theta_1
+15\theta_1^2
+8\theta_2
-5\theta_1\theta_2
+14\theta_2^2\bigl)
-12z_1^2z_2\bigl(4
+8\theta_1
+4\theta_1^2
+15\theta_2
+15\theta_1\theta_2
+17\theta_2^2\bigl)
+144z_1z_2^2\bigl(\theta_1
-\theta_2\bigl)
\bigl(1
+\theta_1
+2\theta_2\bigl)
\end{autobreak}\\
\begin{autobreak}
\MoveEqLeft
		\mathcal{D}_2=125\theta_2^3
-z_2\bigl(\theta_1
-\theta_2\bigl)
\bigl(250
+875\theta_1
+850\theta_1^2
+750\theta_2
+1410\theta_1\theta_2
+536\theta_2^2\bigl)
-3z_1z_2\bigl(1
+\theta_1\bigl)
\bigl(25
+50\theta_1
+25\theta_1^2
+90\theta_2
+90\theta_1\theta_2
+89\theta_2^2\bigl)
+72z_2^2\bigl(-1
+\theta_1
-\theta_2\bigl)
\bigl(\theta_1
-\theta_2\bigl)
\bigl(1
+\theta_1
+2\theta_2\bigl)
\end{autobreak}
\end{align*}
}

\paragraph{$9_{a}$, $(\chi,n_{\pm1},n_{\pm2})=(-96,108,114)$}
{\small
\begin{align*}
\begin{autobreak}
\MoveEqLeft
		\mathcal{D}_1=25\theta_1^2
-85\theta_1\theta_2
+124\theta_2^2
+z_1\bigl(-75
-275\theta_1
-275\theta_1^2
+55\theta_2
+110\theta_1\theta_2
+16\theta_2^2\bigl)
-4z_2\bigl(19\theta_1^2
-46\theta_1\theta_2
+31\theta_2^2\bigl)
+z_1^2\bigl(-25
-50\theta_1
-25\theta_1^2
-65\theta_2
-65\theta_1\theta_2
-64\theta_2^2\bigl)
-4z_1z_2\bigl(7
+12\theta_1
+5\theta_1^2
+21\theta_2
+19\theta_1\theta_2
+20\theta_2^2\bigl)
+64z_1^2z_2\bigl(1
+\theta_1
+\theta_2\bigl)^2
+64z_1z_2^2\bigl(1
+\theta_1
+\theta_2\bigl)^2
\end{autobreak}\\
\begin{autobreak}
\MoveEqLeft
		\mathcal{D}_2=10540\theta_2^3
+z_2\bigl(-4335\theta_1^2
-8313\theta_1^3
+8415\theta_1\theta_2
+14277\theta_1^2\theta_2
-21080\theta_2^3\bigl)
+63315z_1^2\theta_2^3
+z_1z_2\bigl(-4216
-4573\theta_1
+13447\theta_1^2
+7108\theta_1^3
-9052\theta_2
-10303\theta_1\theta_2
+28782\theta_1^2\theta_2
-19220\theta_2^2
-37797\theta_1\theta_2^2
+88164\theta_2^3\bigl)
-10540z_2^2\bigl(\theta_1
-\theta_2\bigl)^3
+9z_1^2z_2\bigl(4442
+11477\theta_1
-7035\theta_2\bigl)
\bigl(1
+\theta_1
+\theta_2\bigl)^2
+4z_1z_2^2\bigl(18941
+39792\theta_1
-22041\theta_2\bigl)
\bigl(1
+\theta_1
+\theta_2\bigl)^2
\end{autobreak}
\end{align*}
}

\paragraph{$9_{b}$, $(\chi,n_{\pm1},n_{\pm2})=(-96,114,108)$}
{\small
\begin{align*}
\begin{autobreak}
\MoveEqLeft
		\mathcal{D}_1=25\theta_1^2
-65\theta_1\theta_2
+64\theta_2^2
+z_1\bigl(-75
-275\theta_1
-275\theta_1^2
-55\theta_2
-110\theta_1\theta_2
+16\theta_2^2\bigl)
-64z_2\bigl(\theta_1
-\theta_2\bigl)^2
+z_1^2\bigl(-25
-50\theta_1
-25\theta_1^2
-85\theta_2
-85\theta_1\theta_2
-124\theta_2^2\bigl)
-4z_1z_2\bigl(-2\theta_1
+5\theta_1^2
+2\theta_2
-19\theta_1\theta_2
+20\theta_2^2\bigl)
+4z_1^2z_2\bigl(19
+38\theta_1
+19\theta_1^2
+46\theta_2
+46\theta_1\theta_2
+31\theta_2^2\bigl)
+64z_1z_2^2\bigl(\theta_1
-\theta_2\bigl)^2
\end{autobreak}\\
\begin{autobreak}
\MoveEqLeft
		\mathcal{D}_2=125\theta_2^3
-z_2\bigl(\theta_1
-\theta_2\bigl)
\bigl(125\theta_1
+200\theta_1^2
-125\theta_2
-70\theta_1\theta_2
-133\theta_2^2\bigl)
-z_1z_2\bigl(1
+\theta_1\bigl)
\bigl(25
+50\theta_1
+25\theta_1^2
+60\theta_2
+60\theta_1\theta_2
+39\theta_2^2\bigl)
-8z_2^2\bigl(-1
+\theta_1
-\theta_2\bigl)
\bigl(\theta_1
-\theta_2\bigl)^2
\end{autobreak}
\end{align*}
}

\paragraph{$10_{a}$, $(\chi,n_{\pm1},n_{\pm2})=(-96,109,113)$}
{\small
\begin{align*}
\begin{autobreak}
\MoveEqLeft
		\mathcal{D}_1=25\theta_1^2
-65\theta_1\theta_2
+89\theta_2^2
+z_1\bigl(-75
-275\theta_1
-275\theta_1^2
-55\theta_2
-110\theta_1\theta_2
+16\theta_2^2\bigl)
+z_2\bigl(-89\theta_1
-143\theta_1^2
+89\theta_2
+13\theta_1\theta_2
+178\theta_2^2\bigl)
+z_1^2\bigl(-25
-50\theta_1
-25\theta_1^2
-85\theta_2
-85\theta_1\theta_2
-99\theta_2^2\bigl)
+z_1z_2\bigl(-88
-301\theta_1
-291\theta_1^2
-104\theta_2
-151\theta_1\theta_2
-55\theta_2^2\bigl)
-z_2^2\bigl(\theta_1
-\theta_2\bigl)
\bigl(89
+167\theta_1
+89\theta_2\bigl)
\end{autobreak}\\
\begin{autobreak}
\MoveEqLeft
		\mathcal{D}_2=11125\theta_2^3
+z_2\bigl(-11125\theta_1
-26000\theta_1^2
-20725\theta_1^3
+11125\theta_2
-1375\theta_1\theta_2
-3965\theta_1^2\theta_2
+33375\theta_2^2
+7654\theta_1\theta_2^2
+22250\theta_2^3\bigl)
-4875z_1^2\theta_2^3
+z_1z_2\bigl(-4450
-2350\theta_1
+23275\theta_1^2
+30925\theta_1^3
-17355\theta_2
-20710\theta_1\theta_2
+1520\theta_1^2\theta_2
-25187\theta_2^2
-17312\theta_1\theta_2^2
-12282\theta_2^3\bigl)
+z_2^2\bigl(\theta_1
-\theta_2\bigl)
\bigl(-22250
-39654\theta_1
+2096\theta_1^2
-33375\theta_2
-18779\theta_1\theta_2
-11125\theta_2^2\bigl)
\end{autobreak}
\end{align*}
}

\paragraph{$10_{b}$, $(\chi,n_{\pm1},n_{\pm2})=(-96,113,109)$}
{\small
\begin{align*}
\begin{autobreak}
\MoveEqLeft
		\mathcal{D}_1=25\theta_1^2
-35\theta_1\theta_2
+39\theta_2^2
+z_1\bigl(-75
-275\theta_1
-275\theta_1^2
-220\theta_2
-440\theta_1\theta_2
-149\theta_2^2\bigl)
+z_2\bigl(-78
-281\theta_1
-291\theta_1^2
-234\theta_2
-431\theta_1\theta_2
-195\theta_2^2\bigl)
+z_1^2\bigl(-25
-50\theta_1
-25\theta_1^2
-115\theta_2
-115\theta_1\theta_2
-179\theta_2^2\bigl)
+z_1z_2\bigl(54
+197\theta_1
+143\theta_1^2
+121\theta_2
+299\theta_1\theta_2
-22\theta_2^2\bigl)
+z_2^2\bigl(1
+\theta_1
+2\theta_2\bigl)
\bigl(78
+167\theta_1
+78\theta_2\bigl)
\end{autobreak}\\
\begin{autobreak}
\MoveEqLeft
		\mathcal{D}_2=4875\theta_2^3
+z_2\bigl(-9750
-43875\theta_1
-69500\theta_1^2
-30925\theta_1^3
-39000\theta_2
-115250\theta_1\theta_2
-91255\theta_1^2\theta_2
-53625\theta_2^2
-72423\theta_1\theta_2^2
-24375\theta_2^3\bigl)
-11125z_1^2\theta_2^3
+z_1z_2\bigl(-5850
-21300\theta_1
-36175\theta_1^2
-20725\theta_1^3
-29835\theta_2
-65795\theta_1\theta_2
-58210\theta_1^2\theta_2
-55341\theta_2^2
-46591\theta_1\theta_2^2
-31356\theta_2^3\bigl)
+z_2^2\bigl(1
+\theta_1
+2\theta_2\bigl)
\bigl(19500
+43846\theta_1
+2096\theta_1^2
+29250\theta_2
+22971\theta_1\theta_2
+9750\theta_2^2\bigl)
\end{autobreak}
\end{align*}
}

\paragraph{$11_{ab}$, $(\chi,n_{\pm1},n_{\pm2})=(-100,110,110)$}
{\small
\begin{align*}
\begin{autobreak}
\MoveEqLeft
		\mathcal{D}_1=5\bigl(5\theta_1^2
-15\theta_1\theta_2
+16\theta_2^2\bigl)
-5z_1\bigl(15
+55\theta_1
+55\theta_1^2
-9\theta_2^2\bigl)
-8z_2\bigl(5\theta_1^2
-13\theta_1\theta_2
+10\theta_2^2\bigl)
-5z_1^2\bigl(5
+10\theta_1
+5\theta_1^2
+15\theta_2
+15\theta_1\theta_2
+16\theta_2^2\bigl)
+8z_1z_2\bigl(1
+2\theta_1\bigl)
\theta_2
-8z_1^2z_2\bigl(5
+10\theta_1
+5\theta_1^2
+13\theta_2
+13\theta_1\theta_2
+10\theta_2^2\bigl)
+64z_1z_2^2\bigl(1
+3\theta_1
+3\theta_1^2
+\theta_2
+\theta_2^2\bigl)
\end{autobreak}\\
\begin{autobreak}
\MoveEqLeft
		\mathcal{D}_2=-5\bigl(-5\theta_1^2
+5\theta_1^3
+15\theta_1\theta_2
-25\theta_1^2\theta_2
-16\theta_2^2
+46\theta_1\theta_2^2
+4\theta_2^3\bigl)
+5z_1\bigl(15\theta_1
+55\theta_1^2
+55\theta_1^3
-30\theta_2
-110\theta_1\theta_2
-110\theta_1^2\theta_2
-9\theta_1\theta_2^2
-38\theta_2^3\bigl)
+4z_2\bigl(25\theta_1^3
-3\theta_1\theta_2
-75\theta_1^2\theta_2
+9\theta_2^2
+81\theta_1\theta_2^2
-19\theta_2^3\bigl)
+5z_1^2\bigl(1
+\theta_1\bigl)
\bigl(5
+10\theta_1
+5\theta_1^2
+5\theta_2
+5\theta_1\theta_2
-14\theta_2^2\bigl)
+4z_1z_2\bigl(45
+270\theta_1
+555\theta_1^2
+400\theta_1^3
+19\theta_2
+69\theta_1\theta_2
+60\theta_1^2\theta_2
-38\theta_2^2
-78\theta_1\theta_2^2
-30\theta_2^3\bigl)
-96z_2^2\bigl(\theta_1
-\theta_2\bigl)^3
\end{autobreak}
\end{align*}
}

\paragraph{$12_{a}$, $(\chi,n_{\pm1},n_{\pm2})=(-104,108,110)$}
{\small
\begin{align*}
\begin{autobreak}
\MoveEqLeft
		\mathcal{D}_1=25\theta_1^2
-45\theta_1\theta_2
+36\theta_2^2
+z_1\bigl(-75
-275\theta_1
-275\theta_1^2
-165\theta_2
-330\theta_1\theta_2
-36\theta_2^2\bigl)
-36z_2\bigl(\theta_1
-\theta_2\bigl)
\bigl(2
+3\theta_1
+4\theta_2\bigl)
+z_1^2\bigl(-25
-50\theta_1
-25\theta_1^2
-105\theta_2
-105\theta_1\theta_2
-116\theta_2^2\bigl)
-36z_1z_2\bigl(1
+\theta_1
+2\theta_2\bigl)^2
\end{autobreak}\\
\begin{autobreak}
\MoveEqLeft
		\mathcal{D}_2=125\theta_2^3
+z_2\bigl(-250\theta_1
-700\theta_1^2
-525\theta_1^3
+250\theta_2
-40\theta_1\theta_2
-455\theta_1^2\theta_2
+732\theta_2^2
+464\theta_1\theta_2^2
+536\theta_2^3\bigl)
-z_1z_2\bigl(3
+2\theta_1\bigl)
\bigl(25
+50\theta_1
+25\theta_1^2
+105\theta_2
+105\theta_1\theta_2
+116\theta_2^2\bigl)
-36z_2^2\bigl(\theta_1
-\theta_2\bigl)
\bigl(1
+\theta_1
+2\theta_2\bigl)^2
\end{autobreak}
\end{align*}
}

\paragraph{$12_{b}$, $(\chi,n_{\pm1},n_{\pm2})=(-104,110,108)$}
{\small
\begin{align*}
\begin{autobreak}
\MoveEqLeft
		\mathcal{D}_1=25\theta_1^2
-55\theta_1\theta_2
+36\theta_2^2
+z_1\bigl(-75
-275\theta_1
-275\theta_1^2
-110\theta_2
-220\theta_1\theta_2
+19\theta_2^2\bigl)
-36z_2\bigl(\theta_1
-\theta_2\bigl)^2
+z_1^2\bigl(-25
-50\theta_1
-25\theta_1^2
-95\theta_2
-95\theta_1\theta_2
-106\theta_2^2\bigl)
+36z_1z_2\bigl(1
+3\theta_1
-\theta_2\bigl)
\bigl(1
+\theta_1
+2\theta_2\bigl)
\end{autobreak}\\
\begin{autobreak}
\MoveEqLeft
		\mathcal{D}_2=36\theta_2^3
+54z_1\theta_2^3
+z_2\bigl(-11\theta_1^2
-140\theta_1^3
+17\theta_1\theta_2
+204\theta_1^2\theta_2
+20\theta_1\theta_2^2
-108\theta_2^3\bigl)
+89z_1^2\theta_2^3
+z_1z_2\bigl(21
+209\theta_1
+533\theta_1^2
+345\theta_1^3
+62\theta_2
+394\theta_1\theta_2
+688\theta_1^2\theta_2
+57\theta_2^2
+21\theta_1\theta_2^2
+70\theta_2^3\bigl)
+36z_2^2\bigl(\theta_1
-\theta_2\bigl)^2
\bigl(1
+\theta_1
+2\theta_2\bigl)
\end{autobreak}
\end{align*}
}
\subsection{Gopakumar-Vafa invariants}\label{sec:gvinvs}
\paragraph{$1_{a}$, $(\chi,n_{\pm1},n_{\pm2})=(-90,100,125)$}
\begin{align*}
\begin{array}{c|cccccc}
	n^{(0)}_{d_1,d_2} & d_2=0 & 1 & 2 & 3 & 4 & 5 \\\hline
 d_1=0 & 0 & 50 & 5 & 0 & 0 & 0 \\
 1 & 100 & 360 & 360 & 100 & 0 & 0 \\
 2 & 125 & 3145 & 12615 & 18750 & 12615 & 3145 \\
 3 & 125 & 18050 & 229175 & 925765 & 1753865 & 1753865 \\
 4 & 100 & 88350 & 2850095 & 25692000 & 103046015 & 224158400 \\
 5 & 90 & 370650 & 27250390 & 485734300 & 3630824110 & 14418923560 \\
\end{array}
\end{align*}

\paragraph{$1_{b}$, $(\chi,n_{\pm1},n_{\pm2})=(-90,125,100)$}
\begin{align*}
\begin{array}{c|cccccc}
	n^{(0)}_{d_1,d_2} & d_2=0 & 1 & 2 & 3 & 4 & 5 \\\hline
 d_1=0 & 0 & 10 & -10 & 15 & -40 & 135 \\
 1 & 125 & 1845 & 625 & -1250 & 3750 & -14000 \\
 2 & 100 & 15200 & 232270 & 112750 & -298500 & 1098225 \\
 3 & 100 & 82100 & 6276315 & 74339025 & 41926075 & -110129250 \\
 4 & 125 & 360950 & 83987355 & 3654463000 & 35289701120 & 22055480425 \\
 5 & 90 & 1360380 & 773433535 & 84715572805 & 2539307225900 & 20790618733325 \\
\end{array}
\end{align*}

\paragraph{$2_{a}$, $(\chi,n_{\pm1},n_{\pm2})=(-90,105,120)$}
\begin{align*}
\begin{array}{c|cccccc}
	n^{(0)}_{d_1,d_2} & d_2=0 & 1 & 2 & 3 & 4 & 5 \\\hline
 d_1=0 & 0 & 30 & 0 & 0 & 0 & 0 \\
 1 & 105 & 330 & 105 & 0 & 0 & 0 \\
 2 & 120 & 2865 & 6585 & 2865 & 120 & 0 \\
 3 & 120 & 17400 & 151260 & 283755 & 151260 & 17400 \\
 4 & 105 & 87150 & 2141265 & 11044335 & 18347055 & 11044335 \\
 5 & 90 & 368670 & 22279830 & 256967580 & 974066175 & 1488072900 \\
\end{array}
\end{align*}

\paragraph{$2_{b}$, $(\chi,n_{\pm1},n_{\pm2})=(-90,120,105)$}
\begin{align*}
\begin{array}{c|cccccc}
	n^{(0)}_{d_1,d_2} & d_2=0 & 1 & 2 & 3 & 4 & 5 \\\hline
 d_1=0 & 0 & 120 & 105 & 105 & 120 & 90 \\
 1 & 120 & 2085 & 15690 & 83400 & 362850 & 1365060 \\
 2 & 105 & 15690 & 569475 & 9690270 & 107459880 & 901887570 \\
 3 & 105 & 83400 & 9690270 & 418812780 & 10086474180 & 164859436335 \\
 4 & 120 & 362850 & 107459880 & 10086474180 & 472152998265 & 13800385325580 \\
 5 & 90 & 1365060 & 901887570 & 164859436335 & 13800385325580 & 675995017391805 \\
\end{array}
\end{align*}

\paragraph{$3_{a}$, $(\chi,n_{\pm1},n_{\pm2})=(-90,110,115)$}
\begin{align*}
\begin{array}{c|cccccc}
	n^{(0)}_{d_1,d_2} & d_2=0 & 1 & 2 & 3 & 4 & 5 \\\hline
 d_1=0 & 0 & 15 & 0 & 0 & 0 & 0 \\
 1 & 110 & 280 & 10 & 0 & 0 & 0 \\
 2 & 115 & 2595 & 2765 & 75 & -10 & 0 \\
 3 & 115 & 16790 & 90195 & 56360 & 1075 & -150 \\
 4 & 110 & 85925 & 1525335 & 3771560 & 1653965 & 25775 \\
 5 & 90 & 366710 & 17693765 & 117782550 & 176866590 & 59562930 \\
\end{array}
\end{align*}

\paragraph{$3_{b}$, $(\chi,n_{\pm1},n_{\pm2})=(-90,115,110)$}
\begin{align*}
\begin{array}{c|cccccc}
	n^{(0)}_{d_1,d_2} & d_2=0 & 1 & 2 & 3 & 4 & 5 \\\hline
 d_1=0 & 0 & 5 & 0 & 0 & 0 & 0 \\
 1 & 115 & 210 & 0 & 0 & 0 & 0 \\
 2 & 110 & 2335 & 755 & 0 & 0 & 0 \\
 3 & 110 & 16220 & 45330 & 4590 & 0 & 0 \\
 4 & 115 & 84675 & 1001255 & 862855 & 36100 & 0 \\
 5 & 90 & 364770 & 13496020 & 42732450 & 16117885 & 323660 \\
\end{array}
\end{align*}

\paragraph{$4_{a}$, $(\chi,n_{\pm1},n_{\pm2})=(-94,105,118)$}
\begin{align*}
\begin{array}{c|cccccc}
	n^{(0)}_{d_1,d_2} & d_2=0 & 1 & 2 & 3 & 4 & 5 \\\hline
 d_1=0 & 0 & 1 & 0 & 0 & 0 & 0 \\
 1 & 105 & 70 & 0 & 0 & 0 & 0 \\
 2 & 118 & 735 & 11 & 0 & 0 & 0 \\
 3 & 118 & 5869 & 1655 & 0 & 0 & 0 \\
 4 & 105 & 33259 & 59392 & 1655 & 0 & 0 \\
 5 & 94 & 156986 & 1027240 & 244980 & 735 & 0 \\
\end{array}
\end{align*}

\paragraph{$4_{b}$, $(\chi,n_{\pm1},n_{\pm2})=(-94,118,105)$}
\begin{align*}
\begin{array}{c|cccccc}
	n^{(0)}_{d_1,d_2} & d_2=0 & 1 & 2 & 3 & 4 & 5 \\\hline
 d_1=0 & 0 & 11 & 0 & 0 & 0 & 0 \\
 1 & 118 & 734 & 55 & -22 & 3 & 0 \\
 2 & 105 & 7296 & 24495 & 1760 & -896 & 495 \\
 3 & 105 & 43789 & 894576 & 1977743 & 129969 & -65676 \\
 4 & 118 & 205204 & 14783398 & 140831918 & 233739551 & 14842534 \\
 5 & 94 & 814351 & 159104996 & 4339120010 & 25580682528 & 34111077016 \\
\end{array}
\end{align*}

\paragraph{$5_{a}$, $(\chi,n_{\pm1},n_{\pm2})=(-94,108,115)$}
\begin{align*}
\begin{array}{c|cccccc}
	n^{(0)}_{d_1,d_2} & d_2=0 & 1 & 2 & 3 & 4 & 5 \\\hline
 d_1=0 & 0 & 2 & 0 & 0 & 0 & 0 \\
 1 & 108 & 88 & 0 & 0 & 0 & 0 \\
 2 & 115 & 1097 & 13 & 0 & 0 & 0 \\
 3 & 115 & 8068 & 4557 & 0 & 0 & 0 \\
 4 & 108 & 45698 & 143264 & 8988 & 0 & 0 \\
 5 & 94 & 208354 & 2335054 & 1164454 & 8988 & 0 \\
\end{array}
\end{align*}

\paragraph{$5_{b}$, $(\chi,n_{\pm1},n_{\pm2})=(-94,115,108)$}
\begin{align*}
\begin{array}{c|cccccc}
	n^{(0)}_{d_1,d_2} & d_2=0 & 1 & 2 & 3 & 4 & 5 \\\hline
 d_1=0 & 0 & 13 & 0 & 0 & 0 & 0 \\
 1 & 115 & 510 & 26 & -4 & 0 & 0 \\
 2 & 108 & 5055 & 10608 & 516 & -156 & 39 \\
 3 & 108 & 31792 & 383024 & 510330 & 21654 & -5850 \\
 4 & 115 & 153957 & 6574173 & 36521697 & 35771110 & 1438707 \\
 5 & 94 & 626170 & 74112668 & 1165284722 & 3980245998 & 3091857310 \\
\end{array}
\end{align*}

\paragraph{$6_{a}$, $(\chi,n_{\pm1},n_{\pm2})=(-94,110,113)$}
\begin{align*}
\begin{array}{c|cccccc}
	n^{(0)}_{d_1,d_2} & d_2=0 & 1 & 2 & 3 & 4 & 5 \\\hline
 d_1=0 & 0 & 41 & 0 & 0 & 0 & 0 \\
 1 & 110 & 632 & 486 & 52 & 0 & 0 \\
 2 & 113 & 5449 & 29680 & 40521 & 15206 & 1318 \\
 3 & 113 & 32522 & 672004 & 3389134 & 6089576 & 4251622 \\
 4 & 110 & 155463 & 9213931 & 122021518 & 584124117 & 1230515498 \\
 5 & 94 & 628866 & 91886539 & 2682580356 & 27553828341 & 127127937012 \\
\end{array}
\end{align*}

\paragraph{$6_{b}$, $(\chi,n_{\pm1},n_{\pm2})=(-94,113,110)$}
\begin{align*}
\begin{array}{c|cccccc}
	n^{(0)}_{d_1,d_2} & d_2=0 & 1 & 2 & 3 & 4 & 5 \\\hline
 d_1=0 & 0 & 52 & 1 & 0 & 0 & 0 \\
 1 & 113 & 917 & 1318 & 486 & 41 & 0 \\
 2 & 110 & 7686 & 72260 & 202196 & 216954 & 94206 \\
 3 & 110 & 44710 & 1552038 & 14504210 & 54913627 & 99039844 \\
 4 & 113 & 206778 & 20307028 & 478454108 & 4413203985 & 19867607910 \\
 5 & 94 & 817708 & 193904333 & 9844099712 & 186449743672 & 1699388167752 \\
\end{array}
\end{align*}

\paragraph{$7_{a}$, $(\chi,n_{\pm1},n_{\pm2})=(-96,103,119)$}
\begin{align*}
\begin{array}{c|cccccc}
	n^{(0)}_{d_1,d_2} & d_2=0 & 1 & 2 & 3 & 4 & 5 \\\hline
 d_1=0 & 0 & 1 & 0 & 0 & 0 & 0 \\
 1 & 103 & 0 & 0 & 0 & 0 & 0 \\
 2 & 119 & 222 & 0 & 0 & 0 & 0 \\
 3 & 119 & 1768 & 0 & 0 & 0 & 0 \\
 4 & 103 & 12053 & 222 & 0 & 0 & 0 \\
 5 & 96 & 63214 & 32111 & 0 & 0 & 0 \\
\end{array}
\end{align*}

\paragraph{$7_{b}$, $(\chi,n_{\pm1},n_{\pm2})=(-96,119,103)$}
\begin{align*}
\begin{array}{c|cccccc}
	n^{(0)}_{d_1,d_2} & d_2=0 & 1 & 2 & 3 & 4 & 5 \\\hline
 d_1=0 & 0 & -2 & 0 & 0 & 0 & 0 \\
 1 & 119 & 119 & 0 & 0 & 0 & 0 \\
 2 & 103 & 3080 & 103 & 0 & 0 & 0 \\
 3 & 103 & 21866 & 21867 & 106 & 5 & 7 \\
 4 & 119 & 112336 & 1166863 & 112098 & -357 & -714 \\
 5 & 96 & 476228 & 20432340 & 20438949 & 496055 & 33141 \\
\end{array}
\end{align*}

\paragraph{$8_{a}$, $(\chi,n_{\pm1},n_{\pm2})=(-96,104,118)$}
\begin{align*}
\begin{array}{c|cccccc}
	n^{(0)}_{d_1,d_2} & d_2=0 & 1 & 2 & 3 & 4 & 5 \\\hline
 d_1=0 & 0 & 1 & 0 & 0 & 0 & 0 \\
 1 & 104 & 26 & 0 & 0 & 0 & 0 \\
 2 & 118 & 510 & -2 & 0 & 0 & 0 \\
 3 & 118 & 3848 & 104 & 0 & 0 & 0 \\
 4 & 104 & 23943 & 14496 & 1 & 0 & 0 \\
 5 & 96 & 116406 & 327394 & 3848 & 0 & 0 \\
\end{array}
\end{align*}

\paragraph{$8_{b}$, $(\chi,n_{\pm1},n_{\pm2})=(-96,118,104)$}
\begin{align*}
\begin{array}{c|cccccc}
	n^{(0)}_{d_1,d_2} & d_2=0 & 1 & 2 & 3 & 4 & 5 \\\hline
 d_1=0 & 0 & 14 & -2 & 0 & 0 & 0 \\
 1 & 118 & 1182 & 182 & -128 & 192 & -256 \\
 2 & 104 & 10567 & 79794 & 18215 & -15840 & 26784 \\
 3 & 104 & 60308 & 2460408 & 13359236 & 3604792 & -2472172 \\
 4 & 118 & 272895 & 36146298 & 771480045 & 3312059844 & 1023073275 \\
 5 & 96 & 1054802 & 357309870 & 20270306492 & 284607552762 & 1019641421694 \\
\end{array}
\end{align*}

\paragraph{$9_{a}$, $(\chi,n_{\pm1},n_{\pm2})=(-96,108,114)$}
\begin{align*}
\begin{array}{c|cccccc}
	n^{(0)}_{d_1,d_2} & d_2=0 & 1 & 2 & 3 & 4 & 5 \\\hline
 d_1=0 & 0 & 3 & 0 & 0 & 0 & 0 \\
 1 & 108 & 144 & 0 & 0 & 0 & 0 \\
 2 & 114 & 1584 & 156 & 0 & 0 & 0 \\
 3 & 114 & 11586 & 15312 & 144 & 0 & 0 \\
 4 & 108 & 62205 & 392922 & 111732 & 96 & 0 \\
 5 & 96 & 276867 & 5758020 & 7825032 & 653880 & 144 \\
\end{array}
\end{align*}

\paragraph{$9_{b}$, $(\chi,n_{\pm1},n_{\pm2})=(-96,114,108)$}
\begin{align*}
\begin{array}{c|cccccc}
	n^{(0)}_{d_1,d_2} & d_2=0 & 1 & 2 & 3 & 4 & 5 \\\hline
 d_1=0 & 0 & 3 & 0 & 0 & 0 & 0 \\
 1 & 114 & 273 & 0 & 0 & 0 & 0 \\
 2 & 108 & 3315 & 1248 & 0 & 0 & 0 \\
 3 & 108 & 22509 & 91326 & 10380 & 0 & 0 \\
 4 & 114 & 113814 & 2111280 & 2445468 & 112320 & 0 \\
 5 & 96 & 478305 & 28225656 & 132168096 & 64146600 & 1393050 \\
\end{array}
\end{align*}

\paragraph{$10_{a}$, $(\chi,n_{\pm1},n_{\pm2})=(-96,109,113)$}
\begin{align*}
\begin{array}{c|cccccc}
	n^{(0)}_{d_1,d_2} & d_2=0 & 1 & 2 & 3 & 4 & 5 \\\hline
 d_1=0 & 0 & 16 & 0 & 0 & 0 & 0 \\
 1 & 109 & 388 & 17 & -2 & 0 & 0 \\
 2 & 113 & 3596 & 5819 & 229 & -32 & 3 \\
 3 & 113 & 23215 & 192708 & 194097 & 7160 & -1338 \\
 4 & 109 & 115058 & 3230069 & 12737397 & 9328023 & 329094 \\
 5 & 96 & 480831 & 36674447 & 390988416 & 958309372 & 553565970 \\
\end{array}
\end{align*}

\paragraph{$10_{b}$, $(\chi,n_{\pm1},n_{\pm2})=(-96,113,109)$}
\begin{align*}
\begin{array}{c|cccccc}
	n^{(0)}_{d_1,d_2} & d_2=0 & 1 & 2 & 3 & 4 & 5 \\\hline
 d_1=0 & 0 & 83 & 16 & 1 & 0 & 0 \\
 1 & 113 & 1390 & 4852 & 8397 & 8397 & 4852 \\
 2 & 109 & 11049 & 212231 & 1564932 & 6194245 & 15358903 \\
 3 & 109 & 61274 & 3996235 & 84212217 & 863723586 & 5259927028 \\
 4 & 113 & 274853 & 47729220 & 2329302822 & 51019886835 & 629677247414 \\
 5 & 96 & 1058440 & 425021583 & 42180123053 & 1753282515287 & 39198944468076 \\
\end{array}
\end{align*}

\paragraph{$11_{ab}$, $(\chi,n_{\pm1},n_{\pm2})=(-100,110,110)$}
\begin{align*}
\begin{array}{c|cccccc}
	n^{(0)}_{d_1,d_2} & d_2=0 & 1 & 2 & 3 & 4 & 5 \\\hline
 d_1=0 & 0 & 1 & 0 & 0 & 0 & 0 \\
 1 & 110 & 156 & 0 & 0 & 0 & 0 \\
 2 & 110 & 2197 & 128 & 0 & 0 & 0 \\
 3 & 110 & 15808 & 23430 & 156 & 0 & 0 \\
 4 & 110 & 84095 & 713466 & 171742 & 100 & 0 \\
 5 & 100 & 363204 & 11054706 & 16620984 & 977416 & 156 \\
\end{array}
\end{align*}

\paragraph{$12_{a}$, $(\chi,n_{\pm1},n_{\pm2})=(-104,108,110)$}
\begin{align*}
\begin{array}{c|cccccc}
	n^{(0)}_{d_1,d_2} & d_2=0 & 1 & 2 & 3 & 4 & 5 \\\hline
 d_1=0 & 0 & 22 & -2 & 0 & 0 & 0 \\
 1 & 108 & 837 & 108 & 0 & 0 & 0 \\
 2 & 110 & 7346 & 37644 & 8370 & -1938 & 3072 \\
 3 & 110 & 44310 & 1115982 & 4075874 & 1226574 & -176874 \\
 4 & 108 & 205608 & 16782374 & 224504662 & 653204544 & 237607766 \\
 5 & 104 & 815788 & 172006188 & 5961656460 & 53124334044 & 130781850135 \\
\end{array}
\end{align*}

\paragraph{$12_{b}$, $(\chi,n_{\pm1},n_{\pm2})=(-104,110,108)$}
\begin{align*}
\begin{array}{c|cccccc}
	n^{(0)}_{d_1,d_2} & d_2=0 & 1 & 2 & 3 & 4 & 5 \\\hline
 d_1=0 & 0 & 1 & 0 & 0 & 0 & 0 \\
 1 & 110 & 402 & 0 & 0 & 0 & 0 \\
 2 & 108 & 4889 & 2598 & 0 & 0 & 0 \\
 3 & 108 & 31242 & 229396 & 35668 & 0 & 0 \\
 4 & 110 & 153143 & 5063042 & 10096232 & 605760 & 0 \\
 5 & 104 & 624326 & 63553122 & 548118454 & 435327210 & 12212466 \\
\end{array}
\end{align*}

%% file: appendix_3.tex
\section{Elliptic genera of E-strings}
\label{sec:estrings}
In this Appendix we construct genus one fibrations with $5$-sections over $\mathbb{F}_1$ and provide expressions for the elliptic genera of the corresponding E-strings.

The topological string partition function on elliptically fibered Calabi-Yau threefolds at fixed base degree can be interpreted as the elliptic genus of a six-dimensional string that arises from D3 branes that wrap the corresponding curve in the base~\cite{Klemm:1996hh}.
In particular, the elliptic genus of E-strings was found to take the form
\begin{align}
Z=-\frac{1}{\eta(\tau)^{12}\phi_{2,1}(\tau,\lambda)}\cdot\frac12\sum\limits_{i=2}^4\prod\limits_{j=1}^8 \theta_i\left(\tau,m_j\right)\,,
\end{align}
where the second factor is the $E_8$ theta function.
The dependence on the topological string coupling $\lambda$ expresses a spin refinement of the ordinary elliptic genus and the expression can be further refined by introducing a second string coupling~\cite{Gu:2017ccq}.
However, with the results from~\cite{Gu:2017ccq} the latter refinement is straightforward and we will not perform it explicitly in our discussion.

For generic elliptic fibrations over the Hirzebruch surface $\mathbb{F}_1$ the partition function $Z_B(\tau,\lambda)$, with $B$ being the base of the Hirzebruch surface, corresponds to the elliptic genus of the E-string with all mass parameters $m_i$ set to zero.
Subsequently, in~\cite{Cota:2019cjx} it was found that when considering genus one fibrations with $N\le4$-sections over $\mathbb{F}_1$ the topological string reproduces the elliptic genus of the E-string with mass parameters fixed at special values.
Some of those values had been considered earlier in~\cite{Eguchi:2002nx} as degenerations points of the E-string Seiberg-Witten curve~\cite{Eguchi:2002fc} at which particular subgroups of $E_8$ remain unbroken.

To obtain elliptic genera that arise from genus one fibrations with 5-sections we follow again the procedure outlined in Section~\ref{sec:classification} but with the Hirzebruch surface $\mathbb{F}_1$ as the base.
Since all of the steps can easily be generalized to fibrations over $\mathbb{F}_1$ we will only provide the results.
If the reader is interested in a list of the intermediate elliptic fibrations or the associated Gopakumar-Vafa invariants we are happy to provide them on request.
We start by constructing all elliptic fibrations of the codimension 3 fiber from Section~\ref{sec:codim3fiber} over Hirzebruch $\mathbb{F}_1$.
This provides us with a list of 405 fibrations of which 44 have $h^{1,1}=4$ and a sufficient amount of Higgs multiplets to perform the transition $U(1)\rightarrow\mathbb{Z}_5$.
We then calculate the genus zero free energy of those fibrations and restrict the parameters in accordance with the Higgs transition~\ref{sec:higgsU1Z5}.
The resulting invariants allow us to employ the modular bootstrap discussed in Section~\ref{sec:bootstrapreview}.

In this way we obtain five inequivalent elliptic genera
\begin{align}
\begin{split}
Z_B^{(1)}(\tau,\lambda)=&-\frac{1}{\eta(5\tau)^{12}\phi_{-2,1}(5\tau,\lambda)}\Delta_{10}^{-\frac{3}{10}}E_{1,1}^3E_{1,2}^3\left(3E_{1,1}+E_{1,2}\right)\,,\\
Z_B^{(2)}(\tau,\lambda)=&-\frac{1}{\eta(5\tau)^{12}\phi_{-2,1}(5\tau,\lambda)}\Delta_{10}^{-\frac{2}{10}}E_{1,1}^2E_{1,2}^3\left(8E_{1,1}+E_{1,2}\right)\,,\\
Z_B^{(3)}(\tau,\lambda)=&-\frac{1}{\eta(5\tau)^{12}\phi_{-2,1}(5\tau,\lambda)}\Delta_{10}^{-\frac{1}{10}}E_{1,1}^2E_{1,2}^2\left(2E_{1,1}-E_{1,2}\right)\,,\\
Z_B^{(4)}(\tau,\lambda)=&-\frac{5}{\eta(5\tau)^{12}\phi_{-2,1}(5\tau,\lambda)}E_{1,1}^2E_{1,2}^2\,.
\end{split}
\end{align}
By determining the Picard-Fuchs systems of the associated geometries and analytically continuing to the homologically projective dual phase we also obtain the corresponding elliptic genera
\begin{align}
\begin{split}
Z_B^{(1')}(\tau,\lambda)=&-\frac{1}{\eta(5\tau)^{12}\phi_{-2,1}(5\tau,\lambda)}{\Delta_{10}'}^{-\frac{3}{10}}E_{1,1}^3E_{1,2}^3\left(E_{1,1}-3E_{1,2}\right)\,,\\
Z_B^{(2')}(\tau,\lambda)=&-\frac{1}{\eta(5\tau)^{12}\phi_{-2,1}(5\tau,\lambda)}{\Delta_{10}'}^{-\frac{2}{10}}E_{1,1}^3E_{1,2}^2\left(E_{1,1}-8E_{1,2}\right)\,,\\
Z_B^{(3')}(\tau,\lambda)=&-\frac{1}{\eta(5\tau)^{12}\phi_{-2,1}(5\tau,\lambda)}{\Delta_{10}'}^{-\frac{1}{10}}E_{1,1}^2E_{1,2}^2\left(E_{1,1}+2E_{1,2}\right)\,,\\
Z_B^{(4')}(\tau,\lambda)=&-\frac{5}{\eta(5\tau)^{12}\phi_{-2,1}(5\tau,\lambda)}E_{1,1}^2E_{1,2}^2\,.
\end{split}
\end{align}
As discussed in Section~\ref{sec:transferrelation}, the dual expressions $Z_B^{(k)}$ and $Z_B^{(k')}$ are related via
\begin{align}
E_{1,1}\rightarrow -E_{1,2}\,,\quad E_{1,2}\rightarrow E_{1,1}\,.
\end{align}
Note that the numerator of $Z_B^{(4)}=Z_B^{(4')}$ is actually a $\Gamma_0(5)$ modular form and the elliptic genera are self dual.
Up to an overall power of $q$, all of the elliptic genera can be written in the general form
\begin{align}
	\phi^{(k)}(\tau,\lambda)=-\frac{1}{\eta(5\tau)^{12}\phi_{-2,1}(5\tau,\lambda)}\cdot\frac12\sum\limits_{i=2}^4\prod\limits_{j=1}^8 \theta_i\left(5\tau,v^{(k)}_j\cdot\tau\right)\,,
\end{align}
with Wilson loop parameters $\vec{v}^{\,(k)}$ given by
\begin{align}
	\begin{split}
		\vec{v}^{\,(1)}=&(0,0,0,0,0,4,4,4)\,,\quad\vec{v}^{\,(1')}=(0,0,0,0,0,2,2,2)\,,\\
		\vec{v}^{\,(2)}=&(0,0,0,2,2,2,2,4)\,,\quad\vec{v}^{\,(2')}=(0,0,0,2,4,4,4,4)\,,\\
		\vec{v}^{\,(3)}=&(0,0,0,2,2,4,4,4)\,,\quad\vec{v}^{\,(3')}=(0,0,0,2,2,2,4,4)\,,\\
		\vec{v}^{\,(4)}=&(0,0,0,2,2,2,2,2)\simeq \vec{v}^{\,(4')}=(0,0,0,4,4,4,4,4)\,.
	\end{split}
\end{align}
Thus at the level of the E-string mass parameters the duality manifests as an exchange of values fixed to $2\tau$ and $4\tau$.

The values are not unique but the elliptic genus is invariant under actions of the $E_8$ Weyl group.
In particular, $\vec{v}^{(4)}$ is equivalent to the choice  $\vec{v}=(0,0,0,1,1,1,1,4)$ that was related in~\cite{Eguchi:2002nx} to a degenerate point with unbroken $A_4\oplus A_4$ symmetry.
However, the other values have, to our knowledge, not been considered before.
It would be very interesting to further investigate the physics associated to the special values of the mass parameters and the duality that relates them.
We leave such a study for future work.


%% file: u2fibrations.bbl
\providecommand{\href}[2]{#2}\begingroup\raggedright\begin{thebibliography}{100}

\bibitem{Braun:2014oya}
V.~Braun and D.~R. Morrison, ``{F-theory on Genus-One Fibrations},''
  \href{http://dx.doi.org/10.1007/JHEP08(2014)132}{{\em JHEP} {\bfseries 08}
  (2014) 132},
\href{http://arxiv.org/abs/1401.7844}{{\ttfamily arXiv:1401.7844 [hep-th]}}.

\bibitem{Klevers:2014bqa}
D.~Klevers, D.~K. Mayorga~Pena, P.-K. Oehlmann, H.~Piragua, and J.~Reuter,
  ``{F-Theory on all Toric Hypersurface Fibrations and its Higgs Branches},''
  \href{http://dx.doi.org/10.1007/JHEP01(2015)142}{{\em JHEP} {\bfseries 01}
  (2015) 142},
\href{http://arxiv.org/abs/1408.4808}{{\ttfamily arXiv:1408.4808 [hep-th]}}.

\bibitem{Braun:2014qka}
V.~Braun, T.~W. Grimm, and J.~Keitel, ``{Complete Intersection Fibers in
  F-Theory},'' \href{http://dx.doi.org/10.1007/JHEP03(2015)125}{{\em JHEP}
  {\bfseries 03} (2015) 125}, \href{http://arxiv.org/abs/1411.2615}{{\ttfamily
  arXiv:1411.2615 [hep-th]}}.

\bibitem{Witten:1993yc}
E.~Witten, ``{Phases of N=2 theories in two-dimensions},''
  \href{http://dx.doi.org/10.1016/0550-3213(93)90033-L}{{\em Nucl. Phys. B}
  {\bfseries 403} (1993) 159--222},
  \href{http://arxiv.org/abs/hep-th/9301042}{{\ttfamily arXiv:hep-th/9301042}}.

\bibitem{Hori:2011pd}
K.~Hori, ``{Duality In Two-Dimensional (2,2) Supersymmetric Non-Abelian Gauge
  Theories},'' \href{http://dx.doi.org/10.1007/JHEP10(2013)121}{{\em JHEP}
  {\bfseries 10} (2013) 121}, \href{http://arxiv.org/abs/1104.2853}{{\ttfamily
  arXiv:1104.2853 [hep-th]}}.

\bibitem{Benini:2012ui}
F.~Benini and S.~Cremonesi, ``{Partition Functions of ${\mathcal{N}=(2,2)}$
  Gauge Theories on S$^{2}$ and Vortices},''
  \href{http://dx.doi.org/10.1007/s00220-014-2112-z}{{\em Commun. Math. Phys.}
  {\bfseries 334} no.~3, (2015) 1483--1527},
  \href{http://arxiv.org/abs/1206.2356}{{\ttfamily arXiv:1206.2356 [hep-th]}}.

\bibitem{Doroud:2012xw}
N.~Doroud, J.~Gomis, B.~Le~Floch, and S.~Lee, ``{Exact Results in D=2
  Supersymmetric Gauge Theories},''
  \href{http://dx.doi.org/10.1007/JHEP05(2013)093}{{\em JHEP} {\bfseries 05}
  (2013) 093}, \href{http://arxiv.org/abs/1206.2606}{{\ttfamily arXiv:1206.2606
  [hep-th]}}.

\bibitem{Jockers:2012dk}
H.~Jockers, V.~Kumar, J.~M. Lapan, D.~R. Morrison, and M.~Romo, ``{Two-Sphere
  Partition Functions and Gromov-Witten Invariants},''
  \href{http://dx.doi.org/10.1007/s00220-013-1874-z}{{\em Commun. Math. Phys.}
  {\bfseries 325} (2014) 1139--1170},
\href{http://arxiv.org/abs/1208.6244}{{\ttfamily arXiv:1208.6244 [hep-th]}}.

\bibitem{Huang:2015sta}
M.-x. Huang, S.~Katz, and A.~Klemm, ``{Topological String on elliptic CY
  3-folds and the ring of Jacobi forms},''
  \href{http://dx.doi.org/10.1007/JHEP10(2015)125}{{\em JHEP} {\bfseries 10}
  (2015) 125}, \href{http://arxiv.org/abs/1501.04891}{{\ttfamily
  arXiv:1501.04891 [hep-th]}}.

\bibitem{Cota:2019cjx}
C.~F. Cota, A.~Klemm, and T.~Schimannek, ``{Topological strings on genus one
  fibered Calabi-Yau 3-folds and string dualities},''
  \href{http://dx.doi.org/10.1007/JHEP11(2019)170}{{\em JHEP} {\bfseries 11}
  (2019) 170}, \href{http://arxiv.org/abs/1910.01988}{{\ttfamily
  arXiv:1910.01988 [hep-th]}}.

\bibitem{sturmfels1996grobner}
B.~Sturmfels, {\em Grobner Bases and Convex Polytopes}.
\newblock Memoirs of the American Mathematical Society. American Mathematical
  Society, 1996.
\newblock \url{https://books.google.de/books?id=K-bxBwAAQBAJ}.

\bibitem{Batyrev:1998kx}
V.~V. Batyrev, I.~Ciocan-Fontanine, B.~Kim, and D.~van Straten, ``Conifold
  transitions and mirror symmetry for {C}alabi-{Y}au complete intersections in
  {G}rassmannians,''
  \href{http://dx.doi.org/10.1016/S0550-3213(98)00020-0}{{\em Nuclear Phys. B}
  {\bfseries 514} no.~3, (1998) 640--666},
  \href{http://arxiv.org/abs/alg-geom/9710022}{{\ttfamily
  arXiv:alg-geom/9710022}}.

\bibitem{Paul-KonstantinOehlmann:2019jgr}
P.-K. Oehlmann and T.~Schimannek, ``{GV-Spectroscopy for F-theory on genus-one
  fibrations},'' \href{http://arxiv.org/abs/1912.09493}{{\ttfamily
  arXiv:1912.09493 [hep-th]}}.

\bibitem{Cianci:2018vwv}
F.~M. Cianci, D.~K. Mayorga Pe\~na, and R.~Valandro, ``{High U(1) charges in
  type IIB models and their F-theory lift},''
  \href{http://dx.doi.org/10.1007/JHEP04(2019)012}{{\em JHEP} {\bfseries 04}
  (2019) 012}, \href{http://arxiv.org/abs/1811.11777}{{\ttfamily
  arXiv:1811.11777 [hep-th]}}.

\bibitem{Klaewer:2021ab}
D.~Kl\"a{}wer, ``{Modular Curves and the Refined Distance Conjecture},''. To
  appear.

\bibitem{arithmeticMIT}
A.~Sutherland, ``{Introduction to Arithmetic geometry},''.
  \url{{https://ocw.mit.edu/courses/mathematics/18-782-introduction-to-arithmetic-geometry-fall-2013/lecture-notes/}}.
  lecture notes.

\bibitem{fisher1}
T.~Fisher, ``The invariants of a genus one curve,''
  \href{http://dx.doi.org/10.1112/plms/pdn021}{{\em Proc. Lond. Math. Soc. (3)}
  {\bfseries 97} no.~3, (2008) 753--782},
  \href{http://arxiv.org/abs/math/0610318}{{\ttfamily arXiv:math/0610318
  [math.NT]}}.

\bibitem{Kimura:2019bzv}
Y.~Kimura, ``{Discrete gauge groups in certain F-theory models in six
  dimensions},'' \href{http://dx.doi.org/10.1007/JHEP07(2019)027}{{\em JHEP}
  {\bfseries 07} (2019) 027}, \href{http://arxiv.org/abs/1905.03775}{{\ttfamily
  arXiv:1905.03775 [hep-th]}}.

\bibitem{Fulton:1998ab}
W.~Fulton, \href{http://dx.doi.org/10.1007/978-1-4612-1700-8}{{\em Intersection
  theory}}, vol.~2 of {\em Ergebnisse der Mathematik und ihrer Grenzgebiete. 3.
  Folge. A Series of Modern Surveys in Mathematics}.
\newblock Springer-Verlag, Berlin, second~ed., 1998.

\bibitem{Trautmann:2007ab}
G.~Trautmann, ``Introduction to intersection theory.'' Lecture Notes at the
  University of Kaiserslautern, 2007.

\bibitem{Fulton:1991ab}
W.~Fulton and J.~Harris,
  \href{http://dx.doi.org/10.1007/978-1-4612-0979-9}{{\em Representation
  theory}}, vol.~129 of {\em Graduate Texts in Mathematics}.
\newblock Springer-Verlag, New York, 1991.
\newblock \url{https://doi.org/10.1007/978-1-4612-0979-9}.

\bibitem{Kuechle:1995ab}
O.~K\"{u}chle, ``On {F}ano {$4$}-fold of index {$1$} and homogeneous vector
  bundles over {G}rassmannians,''
  \href{http://dx.doi.org/10.1007/BF02571923}{{\em Math. Z.} {\bfseries 218}
  no.~4, (1995) 563--575}.

\bibitem{Weyman:2003ab}
J.~Weyman, \href{http://dx.doi.org/10.1017/CBO9780511546556}{{\em Cohomology of
  vector bundles and syzygies}}, vol.~149 of {\em Cambridge Tracts in
  Mathematics}.
\newblock Cambridge University Press, Cambridge, 2003.

\bibitem{Okonek:2011ab}
C.~Okonek, M.~Schneider, and H.~Spindler, {\em Vector bundles on complex
  projective spaces}.
\newblock Modern Birkh\"auser Classics. Birkh\"auser/Springer Basel AG, Basel,
  2011.
\newblock Corrected reprint of the 1988 edition, With an appendix by S. I.
  Gelfand.

\bibitem{Mukai:1992ab}
S.~Mukai, \href{http://dx.doi.org/10.1017/CBO9780511662652.019}{``Polarized
  {$K3$} surfaces of genus {$18$} and {$20$},''} in {\em Complex projective
  geometry ({T}rieste, 1989/{B}ergen, 1989)}, vol.~179 of {\em London Math.
  Soc. Lecture Note Ser.}, pp.~264--276.
\newblock Cambridge Univ. Press, Cambridge, 1992.

\bibitem{Oguiso:1993ab}
K.~Oguiso, ``On algebraic fiber space structures on a {C}alabi-{Y}au
  {$3$}-fold,'' \href{http://dx.doi.org/10.1142/S0129167X93000248}{{\em
  Internat. J. Math.} {\bfseries 4} no.~3, (1993) 439--465}. With an appendix
  by Noboru Nakayama.

\bibitem{Katz:1992ab}
S.~Katz and S.~A. Str{\o}mme, {\em Schubert, a package for MAPLE}, 1992.
\newblock \url{https://github.com/jmokland/schubert/}.
\newblock
  \href{https://faculty.math.illinois.edu/~katz/schubert/}{\tt{https://faculty.math.illinois.edu/\textasciitilde
  katz/schubert/}}.

\bibitem{Klemm:2004km}
A.~Klemm, M.~Kreuzer, E.~Riegler, and E.~Scheidegger, ``Topological string
  amplitudes, complete intersection {C}alabi-{Y}au spaces and threshold
  corrections,'' \href{http://dx.doi.org/10.1088/1126-6708/2005/05/023}{{\em J.
  High Energy Phys.} no.~5, (2005) 023, 116},
  \href{http://arxiv.org/abs/hep-th/0410018}{{\ttfamily arXiv:hep-th/0410018}}.

\bibitem{Donagi:2007hi}
R.~Donagi and E.~Sharpe, ``G{LSM}s for partial flag manifolds,''
  \href{http://dx.doi.org/10.1016/j.geomphys.2008.07.010}{{\em J. Geom. Phys.}
  {\bfseries 58} no.~12, (2008) 1662--1692},
  \href{http://arxiv.org/abs/0704.1761}{{\ttfamily arXiv:0704.1761 [hep-th]}}.

\bibitem{Inoue:2019ab}
D.~Inoue, A.~Ito, and M.~Miura, ``Complete intersection {C}alabi-{Y}au
  manifolds with respect to homogeneous vector bundles on {G}rassmannians,''
  \href{http://dx.doi.org/10.1007/s00209-018-2163-5}{{\em Math. Z.} {\bfseries
  292} no.~1-2, (2019) 677--703}.

\bibitem{Inoue:2019jle}
D.~Inoue, ``{Calabi--Yau 3-folds from projective joins of del Pezzo
  manifolds},'' \href{http://arxiv.org/abs/1902.10040}{{\ttfamily
  arXiv:1902.10040 [math.AG]}}.

\bibitem{Prince:2019vsu}
T.~Prince, ``{Smoothing Calabi-Yau toric hypersurfaces using the Gross-Siebert
  algorithm},'' \href{http://arxiv.org/abs/1909.02140}{{\ttfamily
  arXiv:1909.02140 [math.AG]}}.

\bibitem{Pragacz:1988ab}
P.~Pragacz, ``Enumerative geometry of degeneracy loci,'' {\em Ann. Sci.
  \'{E}cole Norm. Sup. (4)} {\bfseries 21} no.~3, (1988) 413--454.
  \url{http://www.numdam.org/item?id=ASENS_1988_4_21_3_413_0}.

\bibitem{Fulton:1998bc}
W.~Fulton and P.~Pragacz, \href{http://dx.doi.org/10.1007/BFb0096380}{{\em
  Schubert varieties and degeneracy loci}}, vol.~1689 of {\em Lecture Notes in
  Mathematics}.
\newblock Springer-Verlag, Berlin, 1998.

\bibitem{Okonek:1994ab}
C.~Okonek, ``Notes on varieties of codimension {$3$} in {$\mathbb{P}^N$},''
  \href{http://dx.doi.org/10.1007/BF02567467}{{\em Manuscripta Math.}
  {\bfseries 84} no.~3-4, (1994) 421--442}.

\bibitem{Buchsbaum:1977ab}
D.~A. Buchsbaum and D.~Eisenbud, ``Algebra structures for finite free
  resolutions, and some structure theorems for ideals of codimension {$3$},''
  \href{http://dx.doi.org/10.2307/2373926}{{\em Amer. J. Math.} {\bfseries 99}
  no.~3, (1977) 447--485}.

\bibitem{Boffi:2006ab}
G.~Boffi and D.~A. Buchsbaum, {\em Threading homology through algebra: selected
  patterns}.
\newblock Oxford Mathematical Monographs. The Clarendon Press, Oxford
  University Press, Oxford, 2006.

\bibitem{Kanazawa:2012ab}
A.~Kanazawa, ``Pfaffian {C}alabi-{Y}au threefolds and mirror symmetry,''
  \href{http://dx.doi.org/10.4310/CNTP.2012.v6.n3.a3}{{\em Commun. Number
  Theory Phys.} {\bfseries 6} no.~3, (2012) 661--696},
  \href{http://arxiv.org/abs/1006.0223}{{\ttfamily arXiv:1006.0223 [math.AG]}}.

\bibitem{Fisher:2010ab}
T.~Fisher, ``Pfaffian presentations of elliptic normal curves,''
  \href{http://dx.doi.org/10.1090/S0002-9947-09-04876-4}{{\em Trans. Amer.
  Math. Soc.} {\bfseries 362} no.~5, (2010) 2525--2540}.

\bibitem{Caldararu:2002ab}
A.~C\u{a}ld\u{a}raru, ``Derived categories of twisted sheaves on elliptic
  threefolds,'' \href{http://dx.doi.org/10.1515/crll.2002.022}{{\em J. Reine
  Angew. Math.} {\bfseries 544} (2002) 161--179},
  \href{http://arxiv.org/abs/math/0012083}{{\ttfamily arXiv:math/0012083
  [math.AG]}}.

\bibitem{Benedetti:2020ab}
V.~Benedetti, S.~A. Filipini, L.~Manivel, and F.~Tanturri, ``Orbital degeneracy
  loci and applications,''
  \href{http://dx.doi.org/10.2422/2036-2145.201804_016}{{\em Ann. Sc. Norm.
  Super. Pisa Cl. Sci. (5)} {\bfseries XXI} (2020) 169--206},
  \href{http://arxiv.org/abs/1704.01436}{{\ttfamily arXiv:1704.01436
  [math.AG]}}.

\bibitem{Jockers:2012zr}
H.~Jockers, V.~Kumar, J.~M. Lapan, D.~R. Morrison, and M.~Romo, ``{Nonabelian
  2D Gauge Theories for Determinantal Calabi-Yau Varieties},''
  \href{http://dx.doi.org/10.1007/JHEP11(2012)166}{{\em JHEP} {\bfseries 11}
  (2012) 166}, \href{http://arxiv.org/abs/1205.3192}{{\ttfamily arXiv:1205.3192
  [hep-th]}}.

\bibitem{Tjotta:2001ab}
E.~N. Tj\o{}tta, ``Quantum cohomology of a {P}faffian {C}alabi-{Y}au variety:
  verifying mirror symmetry predictions,''
  \href{http://dx.doi.org/10.1023/A:1017585802461}{{\em Compositio Math.}
  {\bfseries 126} no.~1, (2001) 79--89},
  \href{http://arxiv.org/abs/math.AG/9906119}{{\ttfamily arXiv:math.AG/9906119
  [math.AG]}}.

\bibitem{Kuznetsov:2007ab}
A.~Kuznetsov, ``Homological projective duality,''
  \href{http://dx.doi.org/10.1007/s10240-007-0006-8}{{\em Publ. Math. Inst.
  Hautes \'{E}tudes Sci.} no.~105, (2007) 157--220},
  \href{http://arxiv.org/abs/math/0507292}{{\ttfamily arXiv:math/0507292}}.

\bibitem{Kuznetsov:2014ab}
A.~Kuznetsov, ``Semiorthogonal decompositions in algebraic geometry,'' in {\em
  Proceedings of the {I}nternational {C}ongress of {M}athematicians---{S}eoul
  2014. {V}ol. {II}}, pp.~635--660.
\newblock Kyung Moon Sa, Seoul, 2014.
\newblock \href{http://arxiv.org/abs/1404.3143}{{\ttfamily arXiv:1404.3143
  [math.AG]}}.

\bibitem{Gelfand:2000ab}
I.~M. Gelfand, M.~M. Kapranov, and A.~V. Zelevinsky, {\em Discriminants,
  resultants and multidimensional determinants}.
\newblock Modern Birkh\"{a}user Classics. Birkh\"{a}user Boston, Inc., Boston,
  MA, 2008.

\bibitem{Borisov:2009ab}
L.~Borisov and A.~C\u{a}ld\u{a}raru, ``The {P}faffian-{G}rassmannian derived
  equivalence,'' \href{http://dx.doi.org/10.1090/S1056-3911-08-00496-7}{{\em J.
  Algebraic Geom.} {\bfseries 18} no.~2, (2009) 201--222},
  \href{http://arxiv.org/abs/math/0608404}{{\ttfamily arXiv:math/0608404
  [math.AG]}}.

\bibitem{Kuznetsov:2007ac}
A.~G. Kuznetsov, ``Hyperplane sections and derived categories,''
  \href{http://dx.doi.org/10.1070/IM2006v070n03ABEH002318}{{\em Izv. Ross.
  Akad. Nauk Ser. Mat.} {\bfseries 70} no.~3, (2006) 23--128},
  \href{http://arxiv.org/abs/math/0507300}{{\ttfamily arXiv:math/0507300}}.

\bibitem{Hosono:2011np}
S.~Hosono and H.~Takagi, ``{Mirror symmetry and projective geometry of Reye
  congruences I},''
  \href{http://dx.doi.org/10.1090/S1056-3911-2013-00618-9}{{\em J. Alg. Geom.}
  {\bfseries 23} no.~2, (2014) 279--312},
\href{http://arxiv.org/abs/1101.2746}{{\ttfamily arXiv:1101.2746 [math.AG]}}.

\bibitem{Stembridge:2005sf}
J.~Stembridge, {\em SF package}, 2005.
\newblock \url{https://www.math.lsa.umich.edu/~jrs/maple.html#SF}.
\newblock A Maple package for symmetric functions.

\bibitem{Schubert2Source}
D.~R. Grayson, M.~E. Stillman, S.~A. Str{\o}mme, D.~Eisenbud, and C.~Crissman,
  ``{Schubert2: characteristic classes for varieties without equations.
  Version~0.7}.'' A \emph{Macaulay2} package available at
  \url{https://github.com/Macaulay2/M2/tree/master/M2/Macaulay2/packages}.
\newblock \href{
  https://www2.macaulay2.com/Macaulay2/doc/Macaulay2-1.15/share/doc/Macaulay2/Schubert2/html/}{Documentation}.

\bibitem{griffiths1978principles}
P.~Griffiths and J.~Harris, \href{http://dx.doi.org/10.1002/9781118032527}{{\em
  Principles of algebraic geometry}}.
\newblock Wiley Classics Library. John Wiley \& Sons, Inc., New York, 1994.

\bibitem{Borcea:1983ab}
C.~Borcea, ``Smooth global complete intersections in certain compact
  homogeneous complex manifolds,''
  \href{http://dx.doi.org/10.1515/crll.1983.344.65}{{\em J. Reine Angew. Math.}
  {\bfseries 344} (1983) 65--70}.

\bibitem{Hartshorne:1983ab}
R.~Hartshorne, {\em Algebraic geometry}.
\newblock Graduate Texts in Mathematics, No. 52. Springer-Verlag, New
  York-Heidelberg, 1977.

\bibitem{Blumenhagen:2010pv}
R.~Blumenhagen, B.~Jurke, T.~Rahn, and H.~Roschy, ``{Cohomology of Line
  Bundles: A Computational Algorithm},''
  \href{http://dx.doi.org/10.1063/1.3501132}{{\em J. Math. Phys.} {\bfseries
  51} (2010) 103525}, \href{http://arxiv.org/abs/1003.5217}{{\ttfamily
  arXiv:1003.5217 [hep-th]}}.

\bibitem{cohomCalg:Implementation}
``{cohomCalg package}.'' Download link, 2010.
\newblock \url{https://github.com/BenjaminJurke/cohomCalg}. High-performance
  line bundle cohomology computation based on \cite{Blumenhagen:2010pv}.

\bibitem{Gonciulea:1996ab}
N.~Gonciulea and V.~Lakshmibai, ``Degenerations of flag and {S}chubert
  varieties to toric varieties,''
  \href{http://dx.doi.org/10.1007/BF02549207}{{\em Transform. Groups}
  {\bfseries 1} no.~3, (1996) 215--248}.

\bibitem{Rossi:2006ab}
M.~Rossi, ``Geometric transitions,''
  \href{http://dx.doi.org/10.1016/j.geomphys.2005.09.005}{{\em J. Geom. Phys.}
  {\bfseries 56} no.~9, (2006) 1940--1983},
  \href{http://arxiv.org/abs/math/0412514}{{\ttfamily arXiv:math/0412514
  [math.AG]}}.

\bibitem{Gomis:2012wy}
J.~Gomis and S.~Lee, ``{Exact Kahler Potential from Gauge Theory and Mirror
  Symmetry},'' \href{http://dx.doi.org/10.1007/JHEP04(2013)019}{{\em JHEP}
  {\bfseries 04} (2013) 019}, \href{http://arxiv.org/abs/1210.6022}{{\ttfamily
  arXiv:1210.6022 [hep-th]}}.

\bibitem{Gerchkovitz:2014gta}
E.~Gerchkovitz, J.~Gomis, and Z.~Komargodski, ``{Sphere Partition Functions and
  the Zamolodchikov Metric},''
  \href{http://dx.doi.org/10.1007/JHEP11(2014)001}{{\em JHEP} {\bfseries 11}
  (2014) 001}, \href{http://arxiv.org/abs/1405.7271}{{\ttfamily arXiv:1405.7271
  [hep-th]}}.

\bibitem{Gomis:2015yaa}
J.~Gomis, P.-S. Hsin, Z.~Komargodski, A.~Schwimmer, N.~Seiberg, and S.~Theisen,
  ``{Anomalies, Conformal Manifolds, and Spheres},''
  \href{http://dx.doi.org/10.1007/JHEP03(2016)022}{{\em JHEP} {\bfseries 03}
  (2016) 022}, \href{http://arxiv.org/abs/1509.08511}{{\ttfamily
  arXiv:1509.08511 [hep-th]}}.

\bibitem{Gerhardus:2015sla}
A.~Gerhardus and H.~Jockers, ``{Dual pairs of gauged linear sigma models and
  derived equivalences of Calabi–Yau threefolds},''
  \href{http://dx.doi.org/10.1016/j.geomphys.2016.12.005}{{\em J. Geom. Phys.}
  {\bfseries 114} (2017) 223--259},
\href{http://arxiv.org/abs/1505.00099}{{\ttfamily arXiv:1505.00099 [hep-th]}}.

\bibitem{Hori:2013gga}
K.~Hori and J.~Knapp, ``{Linear sigma models with strongly coupled phases - one
  parameter models},'' \href{http://dx.doi.org/10.1007/JHEP11(2013)070}{{\em
  JHEP} {\bfseries 11} (2013) 070},
\href{http://arxiv.org/abs/1308.6265}{{\ttfamily arXiv:1308.6265 [hep-th]}}.

\bibitem{Caldararu:2017usq}
A.~Caldararu, J.~Knapp, and E.~Sharpe, ``{GLSM realizations of maps and
  intersections of Grassmannians and Pfaffians},''
  \href{http://dx.doi.org/10.1007/JHEP04(2018)119}{{\em JHEP} {\bfseries 04}
  (2018) 119}, \href{http://arxiv.org/abs/1711.00047}{{\ttfamily
  arXiv:1711.00047 [hep-th]}}.

\bibitem{Knapp:2019cih}
J.~Knapp and E.~Sharpe, ``{GLSMs, joins, and nonperturbatively-realized
  geometries},'' \href{http://dx.doi.org/10.1007/JHEP12(2019)096}{{\em JHEP}
  {\bfseries 12} (2019) 096}, \href{http://arxiv.org/abs/1907.04350}{{\ttfamily
  arXiv:1907.04350 [hep-th]}}.

\bibitem{Hori:2006dk}
K.~Hori and D.~Tong, ``{Aspects of Non-Abelian Gauge Dynamics in
  Two-Dimensional N=(2,2) Theories},''
  \href{http://dx.doi.org/10.1088/1126-6708/2007/05/079}{{\em JHEP} {\bfseries
  05} (2007) 079}, \href{http://arxiv.org/abs/hep-th/0609032}{{\ttfamily
  arXiv:hep-th/0609032}}.

\bibitem{Beukers:1983ab}
F.~Beukers, ``Irrationality of {$\pi ^{2}$}, periods of an elliptic curve and
  {$\Gamma _{1}(5)$},'' in {\em Diophantine approximations and transcendental
  numbers ({L}uminy, 1982)}, vol.~31 of {\em Progr. Math.}, pp.~47--66.
\newblock Birkh\"{a}user, Boston, Mass., 1983.

\bibitem{Chen:2020iyo}
Z.~Chen, J.~Guo, and M.~Romo, ``{A GLSM view on Homological Projective
  Duality},'' \href{http://arxiv.org/abs/2012.14109}{{\ttfamily
  arXiv:2012.14109 [hep-th]}}.

\bibitem{Hori:2016txh}
K.~Hori and J.~Knapp, ``{A pair of Calabi-Yau manifolds from a two parameter
  non-Abelian gauged linear sigma model},''
  \href{http://arxiv.org/abs/1612.06214}{{\ttfamily arXiv:1612.06214
  [hep-th]}}.

\bibitem{Zhdanov1998}
O.~N. Zhdanov and A.~K. Tsikh, ``Studying the multiple mellin-barnes integrals
  by means of multidimensional residues,''
  \href{http://dx.doi.org/10.1007/BF02677509}{{\em Siberian Mathematical
  Journal} {\bfseries 39} no.~2, (Apr, 1998) 245--260}.
  \url{https://doi.org/10.1007/BF02677509}.

\bibitem{Candelas:1994hw}
P.~Candelas, A.~Font, S.~H. Katz, and D.~R. Morrison, ``{Mirror symmetry for
  two parameter models. 2.},''
  \href{http://dx.doi.org/10.1016/0550-3213(94)90155-4}{{\em Nucl. Phys. B}
  {\bfseries 429} (1994) 626--674},
  \href{http://arxiv.org/abs/hep-th/9403187}{{\ttfamily arXiv:hep-th/9403187}}.

\bibitem{Aganagic:2006wq}
M.~Aganagic, V.~Bouchard, and A.~Klemm, ``{Topological Strings and (Almost)
  Modular Forms},'' \href{http://dx.doi.org/10.1007/s00220-007-0383-3}{{\em
  Commun. Math. Phys.} {\bfseries 277} (2008) 771--819},
  \href{http://arxiv.org/abs/hep-th/0607100}{{\ttfamily arXiv:hep-th/0607100}}.

\bibitem{Gunaydin:2006bz}
M.~Gunaydin, A.~Neitzke, and B.~Pioline, ``{Topological wave functions and heat
  equations},'' \href{http://dx.doi.org/10.1088/1126-6708/2006/12/070}{{\em
  JHEP} {\bfseries 12} (2006) 070},
  \href{http://arxiv.org/abs/hep-th/0607200}{{\ttfamily arXiv:hep-th/0607200}}.

\bibitem{Andreas:2000sj}
B.~Andreas, G.~Curio, D.~H. Ruiperez, and S.-T. Yau, ``{Fourier-Mukai transform
  and mirror symmetry for D-branes on elliptic Calabi-Yau},''
  \href{http://arxiv.org/abs/math/0012196}{{\ttfamily arXiv:math/0012196}}.

\bibitem{Andreas:2001ve}
B.~Andreas, G.~Curio, D.~Hernandez~Ruiperez, and S.-T. Yau, ``{Fiber wise T
  duality for D-branes on elliptic Calabi-Yau},''
  \href{http://dx.doi.org/10.1088/1126-6708/2001/03/020}{{\em JHEP} {\bfseries
  03} (2001) 020}, \href{http://arxiv.org/abs/hep-th/0101129}{{\ttfamily
  arXiv:hep-th/0101129}}.

\bibitem{Schimannek:2019ijf}
T.~Schimannek, ``{Modularity from Monodromy},''
  \href{http://dx.doi.org/10.1007/JHEP05(2019)024}{{\em JHEP} {\bfseries 05}
  (2019) 024}, \href{http://arxiv.org/abs/1902.08215}{{\ttfamily
  arXiv:1902.08215 [hep-th]}}.

\bibitem{Klemm:2012sx}
A.~Klemm, J.~Manschot, and T.~Wotschke, ``{Quantum geometry of elliptic
  Calabi-Yau manifolds},'' \href{http://arxiv.org/abs/1205.1795}{{\ttfamily
  arXiv:1205.1795 [hep-th]}}.

\bibitem{Alim:2012ss}
M.~Alim and E.~Scheidegger, ``{Topological Strings on Elliptic Fibrations},''
  \href{http://dx.doi.org/10.4310/CNTP.2014.v8.n4.a4}{{\em Commun. Num. Theor.
  Phys.} {\bfseries 08} (2014) 729--800},
  \href{http://arxiv.org/abs/1205.1784}{{\ttfamily arXiv:1205.1784 [hep-th]}}.

\bibitem{DelZotto:2017mee}
M.~Del~Zotto, J.~Gu, M.-X. Huang, A.-K. Kashani-Poor, A.~Klemm, and
  G.~Lockhart, ``{Topological Strings on Singular Elliptic Calabi-Yau 3-folds
  and Minimal 6d SCFTs},''
  \href{http://dx.doi.org/10.1007/JHEP03(2018)156}{{\em JHEP} {\bfseries 03}
  (2018) 156}, \href{http://arxiv.org/abs/1712.07017}{{\ttfamily
  arXiv:1712.07017 [hep-th]}}.

\bibitem{Lee:2018urn}
S.-J. Lee, W.~Lerche, and T.~Weigand, ``{Tensionless Strings and the Weak
  Gravity Conjecture},'' \href{http://dx.doi.org/10.1007/JHEP10(2018)164}{{\em
  JHEP} {\bfseries 10} (2018) 164},
  \href{http://arxiv.org/abs/1808.05958}{{\ttfamily arXiv:1808.05958
  [hep-th]}}.

\bibitem{Lee:2018spm}
S.-J. Lee, W.~Lerche, and T.~Weigand, ``{A Stringy Test of the Scalar Weak
  Gravity Conjecture},''
  \href{http://dx.doi.org/10.1016/j.nuclphysb.2018.11.001}{{\em Nucl. Phys. B}
  {\bfseries 938} (2019) 321--350},
  \href{http://arxiv.org/abs/1810.05169}{{\ttfamily arXiv:1810.05169
  [hep-th]}}.

\bibitem{Oberdieck:2016nvt}
G.~Oberdieck and J.~Shen, ``{Curve counting on elliptic Calabi-Yau threefolds
  via derived categories},'' \href{http://arxiv.org/abs/1608.07073}{{\ttfamily
  arXiv:1608.07073 [math.AG]}}.

\bibitem{Haghighat:2013gba}
B.~Haghighat, A.~Iqbal, C.~Koz\c{c}az, G.~Lockhart, and C.~Vafa,
  ``{M-Strings},'' \href{http://dx.doi.org/10.1007/s00220-014-2139-1}{{\em
  Commun. Math. Phys.} {\bfseries 334} no.~2, (2015) 779--842},
  \href{http://arxiv.org/abs/1305.6322}{{\ttfamily arXiv:1305.6322 [hep-th]}}.

\bibitem{Haghighat:2014vxa}
B.~Haghighat, A.~Klemm, G.~Lockhart, and C.~Vafa, ``{Strings of Minimal 6d
  SCFTs},'' \href{http://dx.doi.org/10.1002/prop.201500014}{{\em Fortsch.
  Phys.} {\bfseries 63} (2015) 294--322},
  \href{http://arxiv.org/abs/1412.3152}{{\ttfamily arXiv:1412.3152 [hep-th]}}.

\bibitem{Candelas:1990rm}
P.~Candelas, X.~C. de~la Ossa, P.~S. Green, and L.~Parkes, ``A pair of
  {C}alabi-{Y}au manifolds as an exactly soluble superconformal theory,''
  \href{http://dx.doi.org/10.1016/0550-3213(91)90292-6}{{\em Nuclear Phys. B}
  {\bfseries 359} no.~1, (1991) 21--74}.

\bibitem{Hosono:1993qy}
S.~Hosono, A.~Klemm, S.~Theisen, and S.-T. Yau, ``{Mirror symmetry, mirror map
  and applications to Calabi-Yau hypersurfaces},''
  \href{http://dx.doi.org/10.1007/BF02100589}{{\em Commun. Math. Phys.}
  {\bfseries 167} (1995) 301--350},
  \href{http://arxiv.org/abs/hep-th/9308122}{{\ttfamily arXiv:hep-th/9308122}}.

\bibitem{Hosono:1994ax}
S.~Hosono, A.~Klemm, S.~Theisen, and S.-T. Yau, ``{Mirror symmetry, mirror map
  and applications to complete intersection Calabi-Yau spaces},''
  \href{http://dx.doi.org/10.1016/0550-3213(94)00440-P}{{\em AMS/IP Stud. Adv.
  Math.} {\bfseries 1} (1996) 545--606},
  \href{http://arxiv.org/abs/hep-th/9406055}{{\ttfamily arXiv:hep-th/9406055}}.

\bibitem{Gopakumar:1998ii}
R.~Gopakumar and C.~Vafa, ``{M theory and topological strings. 1.},''
  \href{http://arxiv.org/abs/hep-th/9809187}{{\ttfamily arXiv:hep-th/9809187}}.

\bibitem{Gopakumar:1998jq}
R.~Gopakumar and C.~Vafa, ``{M theory and topological strings. 2.},''
  \href{http://arxiv.org/abs/hep-th/9812127}{{\ttfamily arXiv:hep-th/9812127}}.

\bibitem{Kashani-Poor:2019jyo}
A.-K. Kashani-Poor, ``{Determining F-theory matter via Gromov-Witten
  invariants},'' \href{http://arxiv.org/abs/1912.10009}{{\ttfamily
  arXiv:1912.10009 [hep-th]}}.

\bibitem{Bershadsky:1993ta}
M.~Bershadsky, S.~Cecotti, H.~Ooguri, and C.~Vafa, ``{Holomorphic anomalies in
  topological field theories},''
  \href{http://dx.doi.org/10.1016/0550-3213(93)90548-4}{{\em AMS/IP Stud. Adv.
  Math.} {\bfseries 1} (1996) 655--682},
  \href{http://arxiv.org/abs/hep-th/9302103}{{\ttfamily arXiv:hep-th/9302103}}.

\bibitem{Bershadsky:1993cx}
M.~Bershadsky, S.~Cecotti, H.~Ooguri, and C.~Vafa, ``{Kodaira-Spencer theory of
  gravity and exact results for quantum string amplitudes},''
  \href{http://dx.doi.org/10.1007/BF02099774}{{\em Commun. Math. Phys.}
  {\bfseries 165} (1994) 311--428},
  \href{http://arxiv.org/abs/hep-th/9309140}{{\ttfamily arXiv:hep-th/9309140}}.

\bibitem{Vafa:1995ta}
C.~Vafa, ``{A Stringy test of the fate of the conifold},''
  \href{http://dx.doi.org/10.1016/0550-3213(95)00279-2}{{\em Nucl. Phys. B}
  {\bfseries 447} (1995) 252--260},
  \href{http://arxiv.org/abs/hep-th/9505023}{{\ttfamily arXiv:hep-th/9505023}}.

\bibitem{10.2140/gt.2001.5.287}
J.~Bryan and R.~Pandharipande, ``B{PS} states of curves in {C}alabi-{Y}au
  3-folds,'' \href{http://dx.doi.org/10.2140/gt.2001.5.287}{{\em Geom. Topol.}
  {\bfseries 5} (2001) 287--318},
  \href{http://arxiv.org/abs/math/0009025}{{\ttfamily arXiv:math/0009025}}.

\bibitem{doan2021gopakumarvafa}
A.~Doan, E.-N. Ionel, and T.~Walpuski, ``{The Gopakumar-Vafa finiteness
  conjecture},'' \href{http://arxiv.org/abs/2103.08221}{{\ttfamily
  arXiv:2103.08221 [math.SG]}}.

\bibitem{Aspinwall:2004jr}
P.~S. Aspinwall,
  \href{http://dx.doi.org/10.1142/9789812775108_0001}{``{D-branes on Calabi-Yau
  manifolds},''} in {\em {Theoretical Advanced Study Institute in Elementary
  Particle Physics (TASI 2003): Recent Trends in String Theory}}.
\newblock 3, 2004.
\newblock \href{http://arxiv.org/abs/hep-th/0403166}{{\ttfamily
  arXiv:hep-th/0403166}}.

\bibitem{Andreas:2004uf}
B.~Andreas and D.~Hern\'{a}ndez~Ruip\'{e}rez, ``Fourier {M}ukai transforms and
  applications to string theory,'' {\em RACSAM. Rev. R. Acad. Cienc. Exactas
  F\'{\i}s. Nat. Ser. A Mat.} {\bfseries 99} no.~1, (2005) 29--77,
  \href{http://arxiv.org/abs/math/0412328}{{\ttfamily arXiv:math/0412328}}.

\bibitem{Herbst:2008jq}
M.~Herbst, K.~Hori, and D.~Page, ``{Phases Of N=2 Theories In 1+1 Dimensions
  With Boundary},'' \href{http://arxiv.org/abs/0803.2045}{{\ttfamily
  arXiv:0803.2045 [hep-th]}}.

\bibitem{Hori:2013ika}
K.~Hori and M.~Romo, ``{Exact Results In Two-Dimensional (2,2) Supersymmetric
  Gauge Theories With Boundary},''
  \href{http://arxiv.org/abs/1308.2438}{{\ttfamily arXiv:1308.2438 [hep-th]}}.

\bibitem{never}
R.~Eager, K.~Hori, J.~Knapp, and M.~Romo unpublished.

\bibitem{beijing}
R.~Eager, K.~Hori, J.~Knapp, and M.~Romo, ``{Beijing Lectures on the Grade
  Restriction Rule},'' {\em Chin. Ann. Math} {\bfseries 38B(4)} (2017) 1--12.

\bibitem{donovan2020stringy}
W.~Donovan, ``{Stringy K\"ahler moduli for the Pfaffian-Grassmannian
  correspondence},'' \href{http://arxiv.org/abs/2009.12630}{{\ttfamily
  arXiv:2009.12630 [math.AG]}}.

\bibitem{sagemath}
{The Sage Developers}, {\em {S}ageMath, the {S}age {M}athematics {S}oftware
  {S}ystem ({V}ersion 9.0)}, 2021.
\newblock \url{https://www.sagemath.org}.

\bibitem{Vafa:1996xn}
C.~Vafa, ``{Evidence for F theory},''
  \href{http://dx.doi.org/10.1016/0550-3213(96)00172-1}{{\em Nucl. Phys. B}
  {\bfseries 469} (1996) 403--418},
  \href{http://arxiv.org/abs/hep-th/9602022}{{\ttfamily arXiv:hep-th/9602022}}.

\bibitem{Morrison:1996na}
D.~R. Morrison and C.~Vafa, ``{Compactifications of F theory on Calabi-Yau
  threefolds. 1},'' \href{http://dx.doi.org/10.1016/0550-3213(96)00242-8}{{\em
  Nucl. Phys. B} {\bfseries 473} (1996) 74--92},
  \href{http://arxiv.org/abs/hep-th/9602114}{{\ttfamily arXiv:hep-th/9602114}}.

\bibitem{Morrison:1996pp}
D.~R. Morrison and C.~Vafa, ``{Compactifications of F theory on Calabi-Yau
  threefolds. 2.},'' \href{http://dx.doi.org/10.1016/0550-3213(96)00369-0}{{\em
  Nucl. Phys. B} {\bfseries 476} (1996) 437--469},
  \href{http://arxiv.org/abs/hep-th/9603161}{{\ttfamily arXiv:hep-th/9603161}}.

\bibitem{Weigand:2018rez}
T.~Weigand, ``{F-theory},'' {\em PoS} {\bfseries TASI2017} (2018) 016,
  \href{http://arxiv.org/abs/1806.01854}{{\ttfamily arXiv:1806.01854
  [hep-th]}}.

\bibitem{Cvetic:2018bni}
M.~Cveti\v{c} and L.~Lin, ``{TASI Lectures on Abelian and Discrete Symmetries
  in F-theory},'' \href{http://dx.doi.org/10.22323/1.305.0020}{{\em PoS}
  {\bfseries TASI2017} (2018) 020},
  \href{http://arxiv.org/abs/1809.00012}{{\ttfamily arXiv:1809.00012
  [hep-th]}}.

\bibitem{Bonetti:2011mw}
F.~Bonetti and T.~W. Grimm, ``{Six-dimensional (1,0) effective action of
  F-theory via M-theory on Calabi-Yau threefolds},''
  \href{http://dx.doi.org/10.1007/JHEP05(2012)019}{{\em JHEP} {\bfseries 05}
  (2012) 019}, \href{http://arxiv.org/abs/1112.1082}{{\ttfamily arXiv:1112.1082
  [hep-th]}}.

\bibitem{Grassi:2000we}
A.~Grassi and D.~R. Morrison, ``{Group representations and the Euler
  characteristic of elliptically fibered Calabi-Yau threefolds},''
  \href{http://arxiv.org/abs/math/0005196}{{\ttfamily arXiv:math/0005196}}.

\bibitem{Grassi:2011hq}
A.~Grassi and D.~R. Morrison, ``{Anomalies and the Euler characteristic of
  elliptic Calabi-Yau threefolds},''
  \href{http://dx.doi.org/10.4310/CNTP.2012.v6.n1.a2}{{\em Commun. Num. Theor.
  Phys.} {\bfseries 6} (2012) 51--127},
  \href{http://arxiv.org/abs/1109.0042}{{\ttfamily arXiv:1109.0042 [hep-th]}}.

\bibitem{Sadov:1996zm}
V.~Sadov, ``{Generalized Green-Schwarz mechanism in F theory},''
  \href{http://dx.doi.org/10.1016/0370-2693(96)01134-3}{{\em Phys. Lett. B}
  {\bfseries 388} (1996) 45--50},
  \href{http://arxiv.org/abs/hep-th/9606008}{{\ttfamily arXiv:hep-th/9606008}}.

\bibitem{Katz:1996ht}
S.~H. Katz, D.~R. Morrison, and M.~Plesser, ``{Enhanced gauge symmetry in type
  II string theory},''
  \href{http://dx.doi.org/10.1016/0550-3213(96)00331-8}{{\em Nucl. Phys. B}
  {\bfseries 477} (1996) 105--140},
  \href{http://arxiv.org/abs/hep-th/9601108}{{\ttfamily arXiv:hep-th/9601108}}.

\bibitem{Witten:1996qb}
E.~Witten, ``{Phase transitions in M theory and F theory},''
  \href{http://dx.doi.org/10.1016/0550-3213(96)00212-X}{{\em Nucl. Phys. B}
  {\bfseries 471} (1996) 195--216},
  \href{http://arxiv.org/abs/hep-th/9603150}{{\ttfamily arXiv:hep-th/9603150}}.

\bibitem{Bershadsky:1996nh}
M.~Bershadsky, K.~A. Intriligator, S.~Kachru, D.~R. Morrison, V.~Sadov, and
  C.~Vafa, ``{Geometric singularities and enhanced gauge symmetries},''
  \href{http://dx.doi.org/10.1016/S0550-3213(96)90131-5}{{\em Nucl. Phys. B}
  {\bfseries 481} (1996) 215--252},
  \href{http://arxiv.org/abs/hep-th/9605200}{{\ttfamily arXiv:hep-th/9605200}}.

\bibitem{Park:2011ji}
D.~S. Park, ``{Anomaly Equations and Intersection Theory},''
  \href{http://dx.doi.org/10.1007/JHEP01(2012)093}{{\em JHEP} {\bfseries 01}
  (2012) 093}, \href{http://arxiv.org/abs/1111.2351}{{\ttfamily arXiv:1111.2351
  [hep-th]}}.

\bibitem{Grimm:2015wda}
T.~W. Grimm, A.~Kapfer, and D.~Klevers, ``{The Arithmetic of Elliptic
  Fibrations in Gauge Theories on a Circle},''
  \href{http://dx.doi.org/10.1007/JHEP06(2016)112}{{\em JHEP} {\bfseries 06}
  (2016) 112}, \href{http://arxiv.org/abs/1510.04281}{{\ttfamily
  arXiv:1510.04281 [hep-th]}}.

\bibitem{Morrison:2012ei}
D.~R. Morrison and D.~S. Park, ``{F-Theory and the Mordell-Weil Group of
  Elliptically-Fibered Calabi-Yau Threefolds},''
  \href{http://dx.doi.org/10.1007/JHEP10(2012)128}{{\em JHEP} {\bfseries 10}
  (2012) 128}, \href{http://arxiv.org/abs/1208.2695}{{\ttfamily arXiv:1208.2695
  [hep-th]}}.

\bibitem{morrisonLecture}
D.~R. Morrison, ``{Wilson Lines in F-Theory},'' January, 1999.
\newblock Lecture at Harvard University (unpublished).

\bibitem{deBoer:2001wca}
J.~de~Boer, R.~Dijkgraaf, K.~Hori, A.~Keurentjes, J.~Morgan, D.~R. Morrison,
  and S.~Sethi, ``{Triples, fluxes, and strings},''
  \href{http://dx.doi.org/10.4310/ATMP.2000.v4.n5.a1}{{\em Adv. Theor. Math.
  Phys.} {\bfseries 4} (2002) 995--1186},
  \href{http://arxiv.org/abs/hep-th/0103170}{{\ttfamily arXiv:hep-th/0103170}}.

\bibitem{Morrison:2014era}
D.~R. Morrison and W.~Taylor, ``{Sections, multisections, and U(1) fields in
  F-theory},'' \href{http://arxiv.org/abs/1404.1527}{{\ttfamily arXiv:1404.1527
  [hep-th]}}.

\bibitem{Anderson:2014yva}
L.~B. Anderson, I.~n. Garc\'\i{}a-Etxebarria, T.~W. Grimm, and J.~Keitel,
  ``{Physics of F-theory compactifications without section},''
  \href{http://dx.doi.org/10.1007/JHEP12(2014)156}{{\em JHEP} {\bfseries 12}
  (2014) 156}, \href{http://arxiv.org/abs/1406.5180}{{\ttfamily arXiv:1406.5180
  [hep-th]}}.

\bibitem{Mayrhofer:2014laa}
C.~Mayrhofer, E.~Palti, O.~Till, and T.~Weigand, ``{On Discrete Symmetries and
  Torsion Homology in F-Theory},''
  \href{http://dx.doi.org/10.1007/JHEP06(2015)029}{{\em JHEP} {\bfseries 06}
  (2015) 029}, \href{http://arxiv.org/abs/1410.7814}{{\ttfamily arXiv:1410.7814
  [hep-th]}}.

\bibitem{dolgachev1992elliptic}
I.~Dolgachev and M.~Gross, ``{Elliptic three-folds I: Ogg-Shafarevich
  theory},'' {\em {Algebraic Geom.}} {\bfseries 06} (1994) 39--80,
  \href{http://arxiv.org/abs/9210009}{{\ttfamily arXiv:9210009 [alg-geom]}}.

\bibitem{Banks:2010zn}
T.~Banks and N.~Seiberg, ``{Symmetries and Strings in Field Theory and
  Gravity},'' \href{http://dx.doi.org/10.1103/PhysRevD.83.084019}{{\em Phys.
  Rev. D} {\bfseries 83} (2011) 084019},
  \href{http://arxiv.org/abs/1011.5120}{{\ttfamily arXiv:1011.5120 [hep-th]}}.

\bibitem{Hellerman:2010fv}
S.~Hellerman and E.~Sharpe, ``{Sums over topological sectors and quantization
  of Fayet-Iliopoulos parameters},''
  \href{http://dx.doi.org/10.4310/ATMP.2011.v15.n4.a7}{{\em Adv. Theor. Math.
  Phys.} {\bfseries 15} (2011) 1141--1199},
  \href{http://arxiv.org/abs/1012.5999}{{\ttfamily arXiv:1012.5999 [hep-th]}}.

\bibitem{Cvetic:2015moa}
M.~Cveti\v{c}, R.~Donagi, D.~Klevers, H.~Piragua, and M.~Poretschkin,
  ``{F-theory vacua with $\mathbb Z_3$ gauge symmetry},''
  \href{http://dx.doi.org/10.1016/j.nuclphysb.2015.07.011}{{\em Nucl. Phys. B}
  {\bfseries 898} (2015) 736--750},
  \href{http://arxiv.org/abs/1502.06953}{{\ttfamily arXiv:1502.06953
  [hep-th]}}.

\bibitem{Cvetic:2013qsa}
M.~Cvetic, D.~Klevers, H.~Piragua, and P.~Song, ``{Elliptic fibrations with
  rank three Mordell-Weil group: F-theory with U(1) x U(1) x U(1) gauge
  symmetry},'' \href{http://dx.doi.org/10.1007/JHEP03(2014)021}{{\em JHEP}
  {\bfseries 03} (2014) 021}, \href{http://arxiv.org/abs/1310.0463}{{\ttfamily
  arXiv:1310.0463 [hep-th]}}.

\bibitem{Oehlmann:2016wsb}
P.-K. Oehlmann, J.~Reuter, and T.~Schimannek, ``{Mordell-Weil Torsion in the
  Mirror of Multi-Sections},''
  \href{http://dx.doi.org/10.1007/JHEP12(2016)031}{{\em JHEP} {\bfseries 12}
  (2016) 031}, \href{http://arxiv.org/abs/1604.00011}{{\ttfamily
  arXiv:1604.00011 [hep-th]}}.

\bibitem{Batyrev:1994pg}
V.~V. Batyrev and L.~A. Borisov, ``{On Calabi-Yau complete intersections in
  toric varieties},'' \href{http://arxiv.org/abs/alg-geom/9412017}{{\ttfamily
  arXiv:alg-geom/9412017}}.

\bibitem{Cox:2000vi}
D.~A. Cox and S.~Katz, \href{http://dx.doi.org/10.1090/surv/068}{{\em Mirror
  symmetry and algebraic geometry}}, vol.~68 of {\em Mathematical Surveys and
  Monographs}.
\newblock American Mathematical Society, Providence, RI,
1999.
\newblock

\bibitem{Grassi:2021wii}
A.~Grassi and T.~Weigand, ``{Elliptic threefolds with high Mordell-Weil
  rank},'' \href{http://arxiv.org/abs/2105.02863}{{\ttfamily arXiv:2105.02863
  [math.AG]}}.

\bibitem{Kreuzer:2002uu}
M.~Kreuzer and H.~Skarke, ``{PALP: A Package for analyzing lattice polytopes
  with applications to toric geometry},''
  \href{http://dx.doi.org/10.1016/S0010-4655(03)00491-0}{{\em Comput. Phys.
  Commun.} {\bfseries 157} (2004) 87--106},
\href{http://arxiv.org/abs/math/0204356}{{\ttfamily math/0204356 [math.NA]}}.

\bibitem{Braun:2012vh}
A.~P. Braun, J.~Knapp, E.~Scheidegger, H.~Skarke, and N.-O. Walliser,
  \href{http://dx.doi.org/10.1142/9789814412551_0024}{``P{ALP}---a user
  manual,''} in {\em Strings, gauge fields, and the geometry behind},
  pp.~461--550.
\newblock World Sci. Publ., Hackensack, NJ, 2013.
\newblock \href{http://arxiv.org/abs/1205.4147}{{\ttfamily arXiv:1205.4147
  [math.AG]}}.

\bibitem{Park:2011wv}
D.~S. Park and W.~Taylor, ``{Constraints on 6D Supergravity Theories with
  Abelian Gauge Symmetry},''
  \href{http://dx.doi.org/10.1007/JHEP01(2012)141}{{\em JHEP} {\bfseries 01}
  (2012) 141}, \href{http://arxiv.org/abs/1110.5916}{{\ttfamily arXiv:1110.5916
  [hep-th]}}.

\bibitem{Batyrev:2005jc}
V.~Batyrev and M.~Kreuzer, ``Integral cohomology and mirror symmetry for
  {C}alabi-{Y}au 3-folds,'' in {\em Mirror symmetry. {V}}, vol.~38 of {\em
  AMS/IP Stud. Adv. Math.}, pp.~255--270.
\newblock Amer. Math. Soc., Providence, RI, 2006.
\newblock \href{http://arxiv.org/abs/math/0505432}{{\ttfamily
  arXiv:math/0505432}}.

\bibitem{caldararuThesis}
A.~H. Caldararu, {\em Derived categories of twisted sheaves on {C}alabi-{Y}au
  manifolds}.
\newblock ProQuest LLC, Ann Arbor, MI, 2000.
\newblock
  \url{http://www.math.wisc.edu/~andreic/publications/ThesisSingleSpaced.pdf}.
\newblock Thesis (Ph.D.)--Cornell University.

\bibitem{Caldararu:2010ljp}
A.~Caldararu, J.~Distler, S.~Hellerman, T.~Pantev, and E.~Sharpe,
  ``{Non-birational twisted derived equivalences in abelian GLSMs},''
  \href{http://dx.doi.org/10.1007/s00220-009-0974-2}{{\em Commun. Math. Phys.}
  {\bfseries 294} (2010) 605--645},
  \href{http://arxiv.org/abs/0709.3855}{{\ttfamily arXiv:0709.3855 [hep-th]}}.

\bibitem{Arras:2016evy}
P.~Arras, A.~Grassi, and T.~Weigand, ``{Terminal Singularities, Milnor Numbers,
  and Matter in F-theory},''
  \href{http://dx.doi.org/10.1016/j.geomphys.2017.09.001}{{\em J. Geom. Phys.}
  {\bfseries 123} (2018) 71--97},
  \href{http://arxiv.org/abs/1612.05646}{{\ttfamily arXiv:1612.05646
  [hep-th]}}.

\bibitem{Cvetic:2013uta}
M.~Cveti\v{c}, A.~Grassi, D.~Klevers, and H.~Piragua, ``{Chiral
  Four-Dimensional F-Theory Compactifications With SU(5) and Multiple
  U(1)-Factors},'' \href{http://dx.doi.org/10.1007/JHEP04(2014)010}{{\em JHEP}
  {\bfseries 04} (2014) 010}, \href{http://arxiv.org/abs/1306.3987}{{\ttfamily
  arXiv:1306.3987 [hep-th]}}.

\bibitem{Honecker:2006qz}
G.~Honecker and M.~Trapletti, ``{Merging Heterotic Orbifolds and K3
  Compactifications with Line Bundles},''
  \href{http://dx.doi.org/10.1088/1126-6708/2007/01/051}{{\em JHEP} {\bfseries
  01} (2007) 051}, \href{http://arxiv.org/abs/hep-th/0612030}{{\ttfamily
  arXiv:hep-th/0612030}}.

\bibitem{Batyrev:1997tv}
V.~V. Batyrev and L.~A. Borisov,
  \href{http://dx.doi.org/10.1007/s002220050093}{``Dual cones and mirror
  symmetry for generalized {C}alabi-{Y}au manifolds,''} in {\em Mirror
  symmetry, {II}}, vol.~1 of {\em AMS/IP Stud. Adv. Math.}, pp.~71--86.
\newblock Amer. Math. Soc., Providence, RI, 1997.
\newblock \href{http://arxiv.org/abs/alg-geom/9402002}{{\ttfamily
  arXiv:alg-geom/9402002}}.

\bibitem{zagier}
D.~Zagier, \href{http://dx.doi.org/10.1090/crmp/047/22}{``Integral solutions of
  {A}p\'{e}ry-like recurrence equations,''} in {\em Groups and symmetries},
  vol.~47 of {\em CRM Proc. Lecture Notes}, pp.~349--366.
\newblock Amer. Math. Soc., Providence, RI, 2009.
\newblock
  \url{{http://people.mpim-bonn.mpg.de/zagier/files/tex/AperylikeRecEqs/fulltext.pdf}}.

\bibitem{az}
G.~Almkvist and W.~Zudilin, ``Differential equations, mirror maps and zeta
  values,'' in {\em Mirror Symmetry V}, vol.~38 of {\em AMS/IP Studies in Adv.
  Math.}, pp.~481--516.
\newblock Intern. Press \& Amer. Math. Soc., 2007.
\newblock \href{http://arxiv.org/abs/math/0402386}{{\ttfamily
  arXiv:math/0402386 [math.NT]}}.

\bibitem{apery}
R.~Ap\'ery, ``Irrationalit\'e de $\zeta 2$ et $\zeta 3$,'' in {\em Journ\'ees
  Arithm\'etiques de Luminy}, no.~61 in Ast\'erisque.
\newblock Soci\'et\'e math\'ematique de France, 1979.

\bibitem{Cvetic:2016ner}
M.~Cvetic, A.~Grassi, and M.~Poretschkin, ``{Discrete Symmetries in
  Heterotic/F-theory Duality and Mirror Symmetry},''
  \href{http://dx.doi.org/10.1007/JHEP06(2017)156}{{\em JHEP} {\bfseries 06}
  (2017) 156}, \href{http://arxiv.org/abs/1607.03176}{{\ttfamily
  arXiv:1607.03176 [hep-th]}}.

\bibitem{Hajouji:2019vxs}
N.~Hajouji and P.-K. Oehlmann, ``{Modular Curves and Mordell-Weil Torsion in
  F-theory},'' \href{http://dx.doi.org/10.1007/JHEP04(2020)103}{{\em JHEP}
  {\bfseries 04} (2020) 103}, \href{http://arxiv.org/abs/1910.04095}{{\ttfamily
  arXiv:1910.04095 [hep-th]}}.

\bibitem{Ooguri:2006in}
H.~Ooguri and C.~Vafa, ``{On the Geometry of the String Landscape and the
  Swampland},'' \href{http://dx.doi.org/10.1016/j.nuclphysb.2006.10.033}{{\em
  Nucl. Phys. B} {\bfseries 766} (2007) 21--33},
  \href{http://arxiv.org/abs/hep-th/0605264}{{\ttfamily arXiv:hep-th/0605264}}.

\bibitem{Dierigl:2020lai}
M.~Dierigl and J.~J. Heckman, ``{Swampland cobordism conjecture and non-Abelian
  duality groups},'' \href{http://dx.doi.org/10.1103/PhysRevD.103.066006}{{\em
  Phys. Rev. D} {\bfseries 103} no.~6, (2021) 066006},
  \href{http://arxiv.org/abs/2012.00013}{{\ttfamily arXiv:2012.00013
  [hep-th]}}.

\bibitem{Schimannek:2021ab}
T.~Schimannek, ``{Modularity, the Tate-Shafarevich group and Gopakumar-Vafa
  invariants with discrete charges},''. To appear.

\bibitem{Macdonald:2015ab}
I.~G. Macdonald, {\em Symmetric functions and {H}all polynomials}.
\newblock Oxford Classic Texts in the Physical Sciences. The Clarendon Press,
  Oxford University Press, New York, second~ed., 2015.

\bibitem{eichler1985theory}
M.~Eichler and D.~Zagier,
  \href{http://dx.doi.org/10.1007/978-1-4684-9162-3}{{\em The theory of
  {J}acobi forms}}, vol.~55 of {\em Progress in Mathematics}.
\newblock Birkh\"{a}user Boston, Inc., Boston, MA, 1985.

\bibitem{Klemm:1996hh}
A.~Klemm, P.~Mayr, and C.~Vafa, ``{BPS states of exceptional noncritical
  strings},'' \href{http://dx.doi.org/10.1016/S0920-5632(97)00422-2}{{\em Nucl.
  Phys. B Proc. Suppl.} {\bfseries 58} (1997) 177},
  \href{http://arxiv.org/abs/hep-th/9607139}{{\ttfamily arXiv:hep-th/9607139}}.

\bibitem{Gu:2017ccq}
J.~Gu, M.-x. Huang, A.-K. Kashani-Poor, and A.~Klemm, ``{Refined BPS invariants
  of 6d SCFTs from anomalies and modularity},''
  \href{http://dx.doi.org/10.1007/JHEP05(2017)130}{{\em JHEP} {\bfseries 05}
  (2017) 130}, \href{http://arxiv.org/abs/1701.00764}{{\ttfamily
  arXiv:1701.00764 [hep-th]}}.

\bibitem{Eguchi:2002nx}
T.~Eguchi and K.~Sakai, ``{Seiberg-Witten curve for E string theory
  revisited},'' \href{http://dx.doi.org/10.4310/ATMP.2003.v7.n3.a3}{{\em Adv.
  Theor. Math. Phys.} {\bfseries 7} no.~3, (2003) 419--455},
  \href{http://arxiv.org/abs/hep-th/0211213}{{\ttfamily arXiv:hep-th/0211213}}.

\bibitem{Eguchi:2002fc}
T.~Eguchi and K.~Sakai, ``{Seiberg-Witten curve for the E string theory},''
  \href{http://dx.doi.org/10.1088/1126-6708/2002/05/058}{{\em JHEP} {\bfseries
  05} (2002) 058}, \href{http://arxiv.org/abs/hep-th/0203025}{{\ttfamily
  arXiv:hep-th/0203025}}.

\end{thebibliography}\endgroup
